\documentclass[twocolumn,aps,rmp,amsmath,amssymb,longbibliography,10pt]{revtex4-1}

\usepackage{graphicx}
\usepackage{xcolor}
\usepackage{longtable}
\usepackage{bm}
\usepackage{lineno}
\usepackage{hyperref}
\usepackage{enumitem}
\hypersetup{backref=true,pagebackref=true,hyperindex=true,colorlinks=true,breaklinks=true,urlcolor=
red,linkcolor=blue,bookmarks=true,bookmarksopen=true}

\usepackage{amsmath}
\usepackage{graphicx}
\usepackage{xcolor}
\usepackage{longtable}
\usepackage{bm}
\usepackage{lineno}
\usepackage{hyperref}
\usepackage{enumitem}
\usepackage{natbib}
\usepackage{import}
\usepackage{siunitx}
\usepackage[version=3]{mhchem}

\usepackage{dsfont}
\usepackage{color} 
\usepackage{dcolumn}  
\usepackage{array}

\usepackage{commath}
\usepackage{bbold}
\usepackage{textcomp}

\begin{document}

\title{Topological Aspects of Antiferromagnets}

\author{V. Bonbien$^{1}$}
\email{bonbien.varga@kaust.edu.sa}
\author{Fengjun Zhuo$^{1}$}
\author{A. Salimath$^{2}$}
\author{O. Ly$^{1}$}
\author{A. Abbout$^{3}$}
\author{A. Manchon$^{4}$}
\email{manchon@cinam.univ-mrs.fr}
\affiliation{$^1$King Abdullah University of Science and Technology (KAUST), Physical Science and Engineering Division (PSE), Thuwal, 23955-6900, Saudi Arabia.}
\affiliation{$^2$Department of Engineering Sciences, UiA, Grimstad, Norway.} 
\affiliation{$^3$ Department of Physics, King Fahd University of Petroleum and Minerals (KFUPM), Dhahran 31261, Saudi Arabia.}
\affiliation{$^4$ Aix Marseille Univ, CNRS, CINAM, Marseille, France.}
\date{\today}

\begin{abstract}

The long fascination antiferromagnetic materials have exerted on the scientific community over about a century has been entirely renewed recently with the discovery of several unexpected phenomena including various classes of anomalous spin and charge Hall effects and unconventional magnonic transport, but also homochiral magnetic entities such as skyrmions. With these breakthroughs, antiferromagnets standout as a rich playground for the investigation of novel topological behaviors, and as promising candidate materials for disruptive low-power microelectronic applications. Remarkably, the newly discovered phenomena are all related to the topology of the magnetic, electronic or magnonic ground state of the antiferromagnets. This review exposes how non-trivial topology emerges at different levels in antiferromagnets and explores the novel mechanisms that have been discovered recently. We also discuss how novel classes of quantum magnets could enrich the currently expanding field of antiferromagnetic spintronics and how spin transport can in turn favor a better understanding of exotic quantum excitations.

\end{abstract}

\pacs{}
\maketitle
\tableofcontents

\section{Introduction\label{s:0}} 



\subsection{The blossoming of antiferromagnetic spintronics}
Antiferromagnets are materials with a local magnetic order characterized by a vanishing net magnetization. Their history, initiated in the 30s with the pioneering works of N\'eel \cite{Neel1932,Neel1936}, Bitter \cite{Bitter1938} and van Vleck \cite{vanVleck1941}, is remarkably long and rich. The research in antiferromagnetic materials has left a deep footprint in several subfields of condensed matter physics \cite{Jungwirth2016,Baltz2018,Jungwirth2018}. They display rich dynamical behavior in the GHz and THz frequency range \cite{Fiebig2008}, host exotic, possibly fractional \cite{Mourigal2013}, spin excitations and are being widely used in magnetic devices to pin the magnetization direction of adjacent ferromagnets through exchange bias \cite{Nogues1999}. The interest in antiferromagnetic materials has been revitalized recently with the prediction \cite{Nunez2006,Zelezny2014} and observation \cite{Wadley2016} of current-driven switching of the antiferromagnetic order on the one hand, and of anomalous Hall effect \cite{Surgers2014,Nakatsuji2015,Suzuki2016,Nayak2016,Higo2018b} on the other hand, both effects thought to exist only in ferromagnets. These discoveries open thrilling perspectives for disruptive applications: since antiferromagnets exhibit ultrafast dynamics and are robust against reasonably large external magnetic field ($\sim$ a few Teslas), they represent a competitive alternative to ferromagnets to store and manipulate information. These properties are intimately related to the emergence of an internal effective magnetic field, the Berry curvature of the ground state wave functions \cite{Berry1984,Sundaram1999,Xiao2010b}, a fundamental property that quantifies the phase accumulated by the electron wave function when evolving in a given parameter space. The presence this effective field can lead, in certain cases, to entirely novel classes of materials such as antiferromagnetic topological insulators \cite{Mong2010} and semimetals \cite{Wan2011}, as well as their magnonic counterpart. \par

The extraordinary properties of antiferromagnets are not limited to electronic and magnonic transport. Under certain conditions, antiferromagnets' soft modes also exhibit numerous fascinating features including unconventional magnetic textures such as solitons \cite{Kosevic1986} or skyrmions \cite{Barker2016,Zhang2016l}. Because the antiferromagnetic spin configuration often involves competing interactions, antiferromagnets are sometimes classified as frustrated magnets \cite{Lacroix2011,Batista2016}, a class of materials displaying fractional excitations obeying fermionic or even anyonic statistics \cite{Balents2010,Zhou2017,Savary2017}. Recent research has been targeting the emergence of such topological fractional excitations, demonstrating the existence of topologically protected edge modes \cite{Kasahara2018}.\par

All the properties mentioned above involve, at a level or another, the concept of {\em topology}, i.e. the emergence of a global characteristic (usually, a topological index) that qualifies the system independently of its local details. The global topology of a given system is associated with the curvature of the local geometry of the system's ground state. The nature of this (Berry) curvature, be it in the momentum space, in real space or in both, is dictated by the symmetries of the antiferromagnets and can therefore be extremely rich, due to the vast diversity of the magnetic textures possible. These various aspects, electronic and magnonic transport properties, topological magnetic textures and fractional excitations are usually investigated by different communities and it is therefore difficult to gain an overview of the topological properties of antiferromagnets. In this review, we discuss the topological aspects of antiferromagnets, and explain how topology builds up and gives rise to all these unconventional properties. Our intention is to build bridges between the various research communities (spintronics, quantum magnetism, THz physics, materials science) in order to stimulate interdisciplinary collaboration aiming to explore novel fundamental and practical aspects of topological antiferromagnets.

\subsection{Topology and geometry}

Before entering in the core of our subject, it is important to clarify the meaning of "topology", a term that has gained impressive popularity in the past decade. 
Ordered phases of matter such as crystals, quasicrystals or nematic phases, but also superconductivity, magnetism, ferroelectricity etc., are described by a microscopic order parameter that characterizes their internal structure and is associated with specific symmetry breaking. This order parameter (and its associated symmetry breaking) changes abruptly across a phase transition, following Ginzburg-Landau theory \cite{Ginzburg1950}. As pointed out by X.-G. Wen \cite{Wen1989,Wen2002}, certain systems exhibit an order that is not easily characterized by symmetry breaking. This is the case of the fractional quantum Hall systems \cite{Tsui1982,Kalmeyer1987}, quantum spin liquids \cite{Wen1989,Balents2010}, and the so-called topological materials \cite{Hasan2010,Qi2010,Moore2010,Ortmann2015} but also classical magnetic skyrmion lattices \cite{Nagaosa2013}. In these systems, the (quantum) order is better characterized by a {\em global} "topological" index rather than by its local geometry. A given topological index defines a topological universality class and its specific definition depends on the class of systems under consideration. In the systems we are interested in, it is related to the {\em curvature} of the ground state in a specific parameter space. Concretely, the topological index of a given quantum state $|n\rangle$ in momentum space is given by the Berry phase \cite{Berry1984}, defined ${\cal C}_n=\int d{\bf k} \langle \partial_{\bf k}n|\times|\partial_{\bf k}n\rangle$, where ${\bf k}$ is the momentum. In a classical magnet, the topological index of the spatial magnetic texture is given by ${\cal C}=\int d\Omega (\partial_x{\bf m}\times \partial_y{\bf m})\cdot{\bf m}$, where ${\bf m}$ is the magnetic order parameter and $\Omega$ is the volume of the magnet. These are just two examples of the broad family of topological indices and we encourage the reader interested in deeper discussions of these aspects to refer to more specialized articles \cite{Wen2002,Qi2008,Yu2011c,Nagaosa2013}.\par

What we want to stress out is that the topological indices defined above are related to the integration of a {\em curvature} over the volume of the parameter space. Therefore, in the case of Bloch states, the topological index is interpreted as a (global) quantized gauge flux endowed by a quantum state and its associated curvature is equivalent to an emergent (local) vector potential (see Section \ref{s:sym}). This vector potential is the cornerstone of numerous important transport mechanisms such as the intrinsic anomalous and spin Hall effects \cite{Nagaosa2010,Sinova2015} (and their counterpart in photonic, magnonic and phononic systems). This is the reason why it has become quite common to refer to these effects as "topological", although they do not involve the states' topology but rather their geometry in momentum space. In this review, we will adopt this loose convention and discuss phenomena taking place in antiferromagnets and stemming from local Berry curvarture, as well as physical states exhibiting non-trivial global topology.



\subsection{Readers' guide}

This review focuses on equilibrium and nonequilibrium phenomena that relate to the non-trivial topology of antiferromagnets. It does not intend to review the vast field of antiferromagnetic spintronics and we encourage the readers to refer to the existing reviews in this field \cite{Jungwirth2016,Baltz2018,Jungwirth2018}. The audience we target includes materials scientists, condensed matter physicists, as well as applied scientists. The various sections have been designed in a self-contained manner so that, depending on one’s degree of preparation and motivation, the reader may directly jump to the section of his/her interest. The first section gives a detailed description of the crystal symmetries and resulting topology, applied to antiferromagnets. It is rather written as an introduction to these important but abstract notions. Hence, experts in the topological quantum materials or readers interested in more practical aspects might simply skip it. Section \ref{s:ahe} (unconventional electronic and magnonic transport), Section \ref{s:topaf} (topological antiferromagnetic materials), and Section \ref{s:textures} (topological antiferromagnetic textures) present concrete paradigms of the non-trivial topological phenomena encountered in antiferromagnets. These sections are written for a broad scientific audience and we paid particular attention to provide quantitative estimates of the physical effects as well as to provide an extensive (but never complete) description of candidate materials. As a matter of fact, the development of the research on antiferromagnets strongly benefits from the synthesis of novel materials and we hope that these discussions can act as a source of inspiration for the materials science and chemistry communities, encouraging enhanced collaborative efforts with condensed matter physicists. The last section, Section \ref{s:topexc}, discusses the emergence of topologically non-trivial magnetic excitations in quantum antiferromagnets. This section is written for non-specialists interested in expanding their field of interest towards exotic quantum magnets. This discussion presents itself as a practical summary of existing comprehensive reviews \cite{Balents2010,Lacroix2011,Zhou2017,Savary2017,Batista2016,Knolle2019b}, emphasizing transport properties that are appealing to the spin transport community. By doing so, we wish the present review emboldens researchers familiar with frustrated magnetism on the one hand and those familiar with spintronics on the other hand to accelerate multifaceted collaborations across their field of expertise. The perspectives given in Section \ref{s:persp} intend to outline the promises of this expanding field of  research and nurture interdisciplinary endeavors on these fascinating materials.


\section{Symmetries and Topology in Antiferromagnets \label{s:sym}} 

\subsection{Fundamentals of the Magnetic Group}

\subsubsection{Classical Point Groups and Space Groups}
Crystal structures are most conveniently analyzed and classified by the symmetries they carry. The mathematical structure incorporating the symmetries of an object is known as a group and the discipline involving the systematic study of such objects is referred to as group theory. The mathematical theory of symmetries as the study of certain groups has been extensively developed and found to be of paramount importance for the description of material properties ranging from mechanical to electronic, magnetic and optical ones \cite{Dresselhaus2008,Cracknell2009}.

An ideal crystal is a periodically repeating pattern and can be broken down into two constituents. A lattice, containing a series of equivalent points, and a (usually multi-atomic) motif that is superimposed on the lattice, while maintaining periodicity. Since all lattice points of a periodic lattice are equivalent, every such point has the same neighborhood as other points and so the lattice itself can be characterized by listing the symmetry operations that keep each of them fixed. Such \textit{point} symmetries can be rotations $C_n$, reflections $\mathcal{M}$, inversion $\mathcal{P}$ and their various combinations. Rotations cannot be arbitrary and are restricted by periodicity to be rotations of $360^{\circ}/n$ where $n\in\{1,2,3,4,6\}$. These symmetry operations can be combined into point groups and classify the lattice into \textit{lattice systems}. In two dimensions, there are 4 lattice systems; the monoclinic, orthorhombic, tetragonal and hexagonal with point groups $C_2,D_2,D_4$ and $D_6$, respectively, whereas in three dimensions, there are 7 lattice systems; the triclinic, monoclinic, orthorhombic, tetragonal, rhombohedral, hexagonal and cubic with corresponding point groups $C_i,C_{2h},D_{2h},D_{4h},D_{3d},D_{6h}$ and $O_h$. A further refinement can be added by realizing that there might be several inequivalent lattices with the same point group and therefore in the same lattice system. Indeed, for example, a quick glance at the primitive cubic and body or face centered cubic lattices reveals that all of them have the same point symmetries and are thus in the same cubic lattice system, yet are inequivalent. Enumerating the different lattices in a lattice system, we find that, in total, there are 5 inequivalent lattices in two dimensions and 14 in three dimensions, with all of them being referred to as the Bravais lattices.

The second constituent of a crystal is the set of atoms that are added to the lattice, called the motif. The simplest case is when the motif comprises one single atom. Here the entire crystal is accurately described by the same point group as its lattice system. However, when the motif comprises several atoms, placing different atoms or groups of atoms at different lattice sites lowers the point symmetry of the crystal to a subgroup of its lattice point group. It is thus worthwhile to think of the lattice point group as the maximal point symmetry the lattice can support; then, adding atoms to form a crystal can lower this "maximal" point symmetry to a specific subgroup. In three dimensions there are 32 crystallographic point groups $\mathcal{G}_p$ and the crucial question that arises is: "how do we assign these groups to the different lattice systems?", or in other words, "how far can we lower the symmetry while remaining in the same lattice system?". For example, the cubic group $O_h$ describes the cubic lattice system, but all other crystallographic point groups are its subgroups. One resolution is a classification scheme of point groups based on the highest symmetry rotations they contain leading to the idea of a \textit{crystal system}, seven of which are defined: triclinic, monoclinic, orthorhombic, tetragonal, trigonal, hexagonal and cubic. Note the difference from the 7 lattice systems. For the cubic case, we can say that all subgroups of $O_h$ that have at least 4 different axes of $120^{\circ}$ rotations --- $O,T_d,T_h,T$ --- are crystals in the cubic \textit{crystal system} and can also be assigned to the cubic \textit{lattice system}. Similarly, the point groups having an axis with at least $60^{\circ}$ rotation --- $D_{6h},C_{6v}$ etc. --- are said to form the hexagonal crystal system and are assigned to the hexagonal lattice system, whereas the point groups with an axis of at least $120^{\circ}$ rotation --- $D_{3d},C_3$ etc. --- form the trigonal crystal system, but are assigned to both the rhombohedral and hexagonal lattice systems. Similar rules are used to assign all other point groups to the different crystal and lattice systems \cite{Hammond2015}.

There are certain apparent symmetries of a crystal that have been casually swept under the rug and thus omitted from our discussion above. These correspond to symmetry operations that do not leave a point fixed, and can be thought of, in most cases, as translations. Indeed, the underlying lattice of a crystal, in itself, carries a translational symmetry since it is a periodic arrangement of equivalent points. Defining a lattice translation as an operation carrying a lattice point to another one, we find that the set of lattice translations form a group, $\mathcal{L}$. Constructing the crystal by adding atoms to the lattice can yield further symmetries, known as compound symmetries, that are neither lattice translations, nor point symmetries but combinations of a sublattice translation and a point operation, where under sublattice translation, we refer to a translation that is not a lattice translation. There are two possibilities: glide reflections and screw rotations, in which the point operations are reflections and rotations, respectively. All the symmetries of a given crystal can thus be brought together to form an object known as a space group $\mathcal{G}$ and in three dimensions, there are 230 such groups \cite{Burns2013}. The space group of a crystal has a very large number of elements since it includes all of its lattice translations that are symmetries by definition. \\

It would be much more straightforward to handle a group with a significantly lower number of elements and to this end, we could think of making all space group elements that differ by any lattice translation, equivalent. This means, that if $g_1,g_2\in \mathcal{G}$ are elements of the space group and $l\in\mathcal{L}\subset\mathcal{G}$ is a lattice translation, then if $g_2=g_1+l$, we consider $g_2$ and $g_1$ to be equivalent, $g_2\sim g_1$, and consequently elements of the same equivalence class. The set of these equivalence classes forms a group, the quotient of $\mathcal{G}$ by $\mathcal{L}$, that we label as $\mathcal{G}/\mathcal{L}$. Indeed, we have "divided out" the lattice translations from the space group and are left with a group containing point and perhaps compound operations. This is an enormous simplification, however we ought to remain cautious. Consider the case of no compound operations. Here, all the elements of the quotient $\mathcal{G}/\mathcal{L}$ are purely point operations. Such space groups are known as symmorphic space groups and there are 73 of them in three dimensions. On the other hand, should $\mathcal{G}/\mathcal{L}$ contain compound operations, it is referred to as a nonsymmorphic space group of which there are 157 in three dimensions. In fact, for the nonsymmorphic case, the point group $\mathcal{G}_p$ obtained from $\mathcal{G}/\mathcal{L}$ by setting all sublattice translations to zero is definitely a subgroup of the \textit{lattice} point group, but is \textit{not} a subgroup of the space group $\mathcal{G}$ of the crystal and consequently, does \textit{not} describe the symmetry of the crystal. This is a very important point that leads to significant confusion, and will be crucial for the understanding of magnetic groups to be introduced below. In order to grasp this subtle point we highlight it further from another perspective. 

Space group elements are commonly denoted as $\{R|\boldsymbol{\tau}\}$, referred to as Seitz symbols, where $R$ is a point operation and $\boldsymbol{\tau}$ is a translation vector. The action of the space group element $\{R|\boldsymbol{\tau}\}\in\mathcal{G}$ on a point $\textbf{r}$ is then $\{R|\boldsymbol{\tau}\}\textbf{r} = R\textbf{r}+\boldsymbol{\tau}$, meaning that composing two space group elements yields $\{R_2|\boldsymbol{\tau}_2\}\{R_1|\boldsymbol{\tau}_1\}=\{R_2R_1|R_2\boldsymbol{\tau}_1+\boldsymbol{\tau}_2\}$. A straightforward conclusion from the notation is that a pure point operation can be written as $\{R|0\}$ whereas a pure translation is denoted by $\{E|\boldsymbol{\tau}\}$, where $E$ is the identity point operation leaving everything unchanged. Every translation $\boldsymbol{\tau}$ can be written as a combination of a lattice $\textbf{L}$ and a sublattice $\boldsymbol{\tau}_s$ translation: $\boldsymbol{\tau}=\textbf{L}+\boldsymbol{\tau}_s$. We want to "divide out" the lattice translations, and introduce equivalence classes. Indeed, we have $[\{R|\boldsymbol{\tau}_s+\textbf{L}\}]=[\{R|\boldsymbol{\tau}_s\}]$, where $[\dots]$ denotes an equivalence class and the set of these classes form $\mathcal{G}/\mathcal{L}$. If $\boldsymbol{\tau}_s=0$ for all $R$ the elements of the quotient group are $[\{R|0\}]$, that are purely point operations and so the point group $\mathcal{G}_p$ formed by the elements $R$ is a subgroup of $\mathcal{G}$ and fully describes the crystal. However, for nonsymmorphic space groups we can have $\boldsymbol{\tau}_s\neq 0$ for certain point operations $R$, and together they form compound operations $[\{R|{\boldsymbol\tau}_s\}]$. This means that the point group $\mathcal{G}_p$ arrived at by taking $\boldsymbol{\tau}_s=0$ will \textit{not} be a subgroup of $\mathcal{G}$. Furthermore, while being compatible with the point symmetry of the lattice, it does not describe the symmetry of the crystal, rather that is accomplished by the quotient $\mathcal{G}/\mathcal{L}$.

\begin{figure}
\includegraphics[height=4cm]{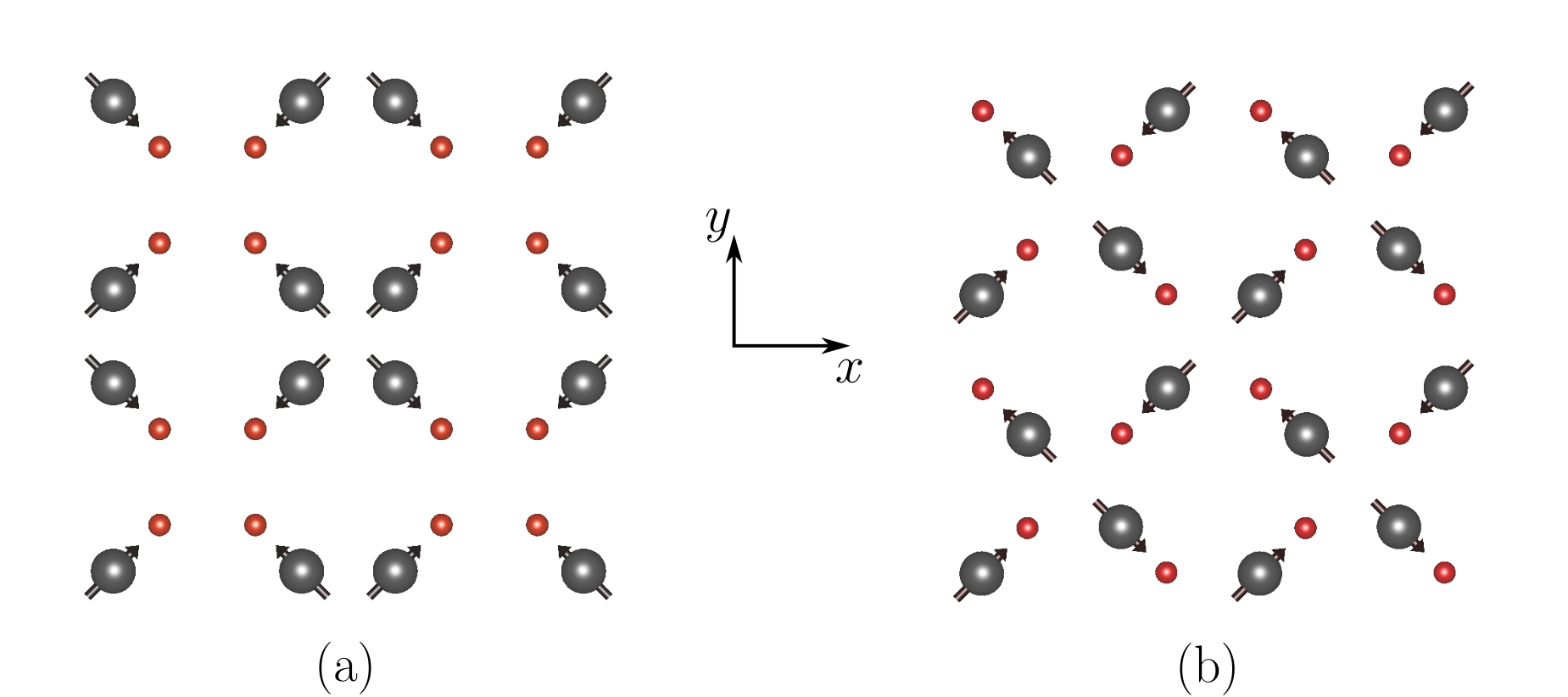}
\caption{Examples of antiferromagnetic crystal structures on a tetragonal lattice. Both (a) and (b) have $D_4\,(422)$ as point groups, but $P422$ and $P42_12$ as space groups. [Crystal structures made with the VESTA package \cite{Momma2011}]}
\label{fig:p4p4g}
\end{figure}

We show a concrete example of this on Fig. \ref{fig:p4p4g}. Two antiferromagnetic structures are presented on a tetragonal lattice with the same point group $D_4 (422)$. However, their space groups are different, being $P422$ and $P42_12$ for structures (a) and (b) respectively. The elements of $D_4$ are $\{E,C_4,C_4^2,C_4^3,C_2,C_4C_2,C_4^2C_2,C_4^3C_2\}$, where $C_4$ is a $90^{\circ}$ rotation about a vertical axis, and $C_4^n$ refers to performing this operation $n$ times in succession, whereas $C_2$ is a $180^{\circ}$ rotation about the $x$ axis. The elements of the quotient groups $P422/\mathcal{L}$ and $P42_12/\mathcal{L}$ are respectively $\{[R_a,0]\}$, where $R_a\in D_4$ and $\{[R_b^{(1)},0]\}$ if $R_b^{(1)}\in \{E,C_4,C_4^2,C_4^3\}$ and $\{[R_b^{(2)},\boldsymbol{\tau}_s]\}$ if $R_b^{(2)}\in \{C_2,C_4C_2,C_4^2C_2,C_4^3C_2\}$, with $\boldsymbol\tau_s=(1/2,1/2,0)$. The space group of structure (a), $P422/\mathcal{L}$, does not contain any compound operations, only point operations, and so the point group is a subgroup of the full space group. On the other hand, it is apparent that not all $D_4$ elements are symmetries of structure (b), thereby the point group is not a subgroup of the space group, but compounding certain $D_4$ operations with a sublattice translation yields the proper symmetries of the crystal.

As we saw above, taking all translations to zero in a space group yields a point group, in fact all space groups can be derived from a point group obtained in this way. This procedure can lead to some deep mathematics with the different space groups classified by the use of the theory of group cohomology \cite{Schwarzenberger1980}.

\subsubsection{Magnetic Point Groups and Space Groups}

The classical point and space groups introduced above can be applied with a high degree of efficiency to the understanding of the properties of crystals, but as is usually the case, nature has further suprises up her sleeve. A large number of crystalline materials exhibit magnetic ordering and the classical crystallographic symmetry groups can be extended in a straightforward manner to describe some of these magnetic crystals \cite{Cracknell1975,Burns2013,Birss1964}.

The elements of the classical groups act on points $\textbf{r}=(x,y,z)$ in space. However, the presence of magnetic ordering supposes that certain points in the crystal, at which an atom with a nonzero magnetic moment is present, can have a further property: the direction of this moment. In the simplest case this property can take on two values depending on its alignment with a particular direction: parallel or antiparallel. Thus, a fourth coordinate $s$ that takes on only two values can be introduced, and the new 'space' on which the group operations will act is $(\textbf{r},s)=(x,y,z,s)$. Abstractly, we can think of the two values of $s$ as two colours: black and white. Furthermore, a new point operation $\mathcal{T}$ can be defined, that transforms $s$ to its other value or black to white and vice-versa $\mathcal{T}:s\to -s$, analogously to how inversion reverses spatial coordinates $\mathcal{P}:\textbf{r}\to -\textbf{r}$. Since $s$ refers to the alignment of the magnetic moment at a given point, the operation that reverses its direction is time-reversal, thereby yielding a clear interpretation of $\mathcal{T}$.

Extending the action space and adding the time-reversal operation $\mathcal{T}$ to the set of point symmetries greatly increases the number of possible point and space groups, many of which can describe the symmetries of crystals with magnetic ordering. These extended groups are known as magnetic point groups and magnetic space groups. These extended, magnetic groups have an interesting and varied history, and we refer to \cite{Wills2017,Cracknell1975,Burns2013} for a detailed overview.

We start with the magnetic point groups. Just as the classical point groups, these contain operations that leave a point fixed, and, depending on the role of the time-reversal operation $\mathcal{T}$, are classified into three different types.

 Type I magnetic point groups, $\mathcal{G}_p^{I}$, are those which do not contain the time-reversal operation $\mathcal{T}$ in any capacity, and are thus equivalent to the classical point groups: $\mathcal{G}_p^{I}\sim\mathcal{G}_p$, therefore we have 32 of them. In this case, the fourth coordinate $s$ can change its value at a point if a reflection over a plane containing the magnetic moment --- an axial vector --- is done, since $s$ refers to the direction of the moment. This restricts the choice of point groups that can support magnetic ordering considerably. Type I magnetic point groups can describe some ferromagnetic and antiferromagnetic crystals. The structures on Fig. \ref{fig:p4p4g} are two examples of the latter.
 
 Type II magnetic point groups, $\mathcal{G}_p^{II}$, also known as grey groups, contain $\mathcal{T}$ as an operation in itself. This requires that $s$ takes on both of its values simultaneously at every point (hence the name grey groups, since $s$ is neither black nor white, but both). The reason for this, is that $\mathcal{T}$ being a point symmetry operation, should not change the value of $s$ at any point, otherwise it will not be a \textit{point} symmetry. However, if $s$ takes on both of its values at every point, it cannot describe magnetic ordering, since the direction of the ordered moments is encoded in one of the values of $s$. Indeed, grey groups can only describe diamagnetic and paramagnetic crystals \cite{Cracknell1975}. Indeed, since the atoms in a diamagnetic crystal do not carry a magnetic moment, they do not carry a specific value of $s$. In a paramagnetic crystal, because the magnetic moments are randomly oriented, they can be considered as not taking on a definite value of $s$. Consequently, neither diamagnetic nor paramagnetic crystals are affected by $\mathcal{T}$. As for the group structure of the grey groups, since $s$ does not take on a definite value, it is not subject to change by either $\mathcal{T}$ nor any of the classical point symmetry operations in $\mathcal{G}_p$, therefore a combination of the two $\mathcal{T}\mathcal{G}_p$ should leave it invariant also. Thus, both $\mathcal{G}_p$ and $\mathcal{T}\mathcal{G}_p$ are contained in $\mathcal{G}_p^{II}$, meaning that we can formally write $\mathcal{G}_p^{II}=\mathcal{G}_p+\mathcal{T}\mathcal{G}_p$, and there are just as many of them as classical point groups: 32.
 
Finally, Type III magnetic point groups, $\mathcal{G}_p^{III}$, sometimes referred to as black and white groups, contain $\mathcal{T}$ only in combination with other non-trivial classical point group operations. To construct this, consider a classical point group $\mathcal{G}_p$. Now, take a proper subgroup $\mathcal{H}_p$ of $\mathcal{G}_p$ that contains half the number of elements of $\mathcal{G}_p$, a so-called halving subgroup, and combine the remaining elements with $\mathcal{T}$. We can formally write the resulting black and white point group as $\mathcal{G}_p^{III}=\mathcal{H}_p+\mathcal{T}(\mathcal{G}_p-\mathcal{H}_p)$. The reason for demanding that half the elements be classical point operations and the other half combined ones, or in other words, that there be an \textit{equal} number of classical and combined operations, is for $\mathcal{G}_p^{III}$ to have the structure of a group. In this case, the number of options are not as restricted as they happened to be for the Type I groups, since even though a reflection could change the value of $s$ at a point the existence of combined operations could flip it back. Indeed, there are 58 black and white point groups and apart from certain ferro and ferrimagnets they can also describe several different kinds of antiferromagnetic crystals, as we will illustrate with some examples in the next section.
 
Having discussed magnetic point symmetries, we now extend our bag of symmetry operations with translations and see how they all combine together to yield the magnetic space groups.

Extending the point operations of the magnetic point groups with translations yields three types of magnetic space groups. Type I magnetic point groups $\mathcal{G}_p^I$ are equivalent to the classical point groups, therefore the Type I magnetic space groups $\mathcal{G}^I$ are equivalent to the classical space groups and there are 230 of them. Similarly, Type II magnetic point groups, the grey groups, $\mathcal{G}_p^{II}$ are an extension of the classical point groups $\mathcal{G}_p$ as $\mathcal{G}_p^{II}=\mathcal{G}_p+\mathcal{T}\mathcal{G}_p$, so the Type II magnetic space groups will be $\mathcal{G}^{II}=\mathcal{G}+\mathcal{T}\mathcal{G}$, where $\mathcal{G}$ is a classical space group derived from $\mathcal{G}_p$, and there are likewise 230 of them. Finally, Type III or black and white magnetic space groups can be constructed in the same way as the corresponding Type III magnetic point groups: $\mathcal{G}^{III}=\mathcal{H}+\mathcal{T}(\mathcal{G}-\mathcal{H})$, where $\mathcal{H}$ is a space group containing half the elements of the space group $\mathcal{G}$. There are 674 Type III magnetic space groups.

Till now, we have assumed that the underlying lattice of the magnetic crystal is always one of the Bravais lattices describing a repeating set of completely equivalent points with the difference in the magnetic moment directions at the points resulting from the atoms or groups of atoms added to the lattice when forming the crystal. The magnetic point and space groups discussed so far were formed to describe crystals built on such lattices. However, it is possible to introduce a magnetic lattice in which the lattice points themselves can be black or white \cite{Cracknell1975}. These magnetic lattices are important for the description of certain antiferromagnets, since the sublattices of black and white points can be considered as the two opposing sublattices of the antiferromagnet. In three dimensions, 22 such magnetic Bravais lattices can be formed and the space groups describing crystals built on them are known as Type IV magnetic space groups, with there being 517 of them. The group structure of a Type IV magnetic space group $\mathcal{G}^{IV}$ can be seen by relating it to one of the classical space groups $\mathcal{G}$ as follows. The magnetic Bravais lattice can be divided into two sublattices: black and white. These two sublattices are connected by a sublattice translation $\boldsymbol{\tau}_0$, meaning that every lattice point on the black sublattice can be arrived at with a translation by $\boldsymbol{\tau}_0$ from a lattice point on the white sublattice. The two sublattices themselves are one of the classical Bravais lattices and the crystals built on them separately can be described by one of the classical space groups. Thus the Type IV magnetic space group can be formally written as $\mathcal{G}^{IV}=\mathcal{G}+\mathcal{T}\{E|\boldsymbol{\tau}_0\}\mathcal{G}$.

Taking a tally of all the different magnetic point and space groups we have the following numbers. The possible Bravais lattices add up to $2\times 14+22=50$, whereas the number of magnetic point and space groups are $2\times 32 +58 =122$ and $2\times 230 + 674 +517 = 1651$, respectively.

\subsubsection{Application to antiferromagnetic order}

We now delve into more detail regarding the relevance of the discussed magnetic symmetry groups to the description of antiferromagnetic order. Only Type I,Type III magnetic point groups and Type I, Type III, Type IV magnetic space groups can describe antiferromagnetic crystals. There seems to be considerable confusion regarding this in the literature \cite{Dresselhaus2008,Smejkal2018}, especially with respect to Type IV space groups, mainly stemming from the role of compound and combined symmetry operations, where we refer to the composition of a classical point operation and a sublattice translation as a compound operation and the composition of $\mathcal{T}$ with any other as a combined operation. Recall the discussion on classical space groups above, during which we narrowed down the number of elements by dividing out the lattice translations and obtained the quotient group $\mathcal{G}/\mathcal{L}$ that describes the given crystal. The point group $\mathcal{G}_p$ was a subgroup of the space group $\mathcal{G}$, if, and only if $\mathcal{G}/\mathcal{L}$ did not contain compound operations. A similar quotient can be formed for magnetic groups. The cases of the first three types, $\mathcal{G}^{I},\mathcal{G}^{II}$ and $\mathcal{G}^{III}$ are simple, since they are all built on one of the classical Bravais lattices containing equivalent points, and we can simply divide out by the group of lattice translations $\mathcal{L}$ of the given lattice to obtain $\mathcal{G}^{I}/\mathcal{L},\mathcal{G}^{II}/\mathcal{L}$ and $\mathcal{G}^{III}/\mathcal{L}$. If these groups do not contain compound operations, then, and only then are the magnetic point groups $\mathcal{G}_p^{I},\mathcal{G}_p^{II}$ and $\mathcal{G}_p^{III}$, subgroups of their respective magnetic space groups and we can describe the crystal symmetry with the latter. Otherwise we have to stick with the quotient groups. 

On the other hand, Type IV magnetic space groups are a different breed. First of all, they were built on magnetic Bravais lattices that are made up of non-equivalent point. However, we would still want to divide out the large number of lattice translations. Indeed, since we have $\mathcal{G}^{IV}=\mathcal{G}+\mathcal{T}\{E|\boldsymbol{\tau}_0\}\mathcal{G}$, with $\mathcal{G}$ being a classical space group built on a classical Bravais lattice consisting of equivalent points, it is possible to consider the black and white sublattices in their own right, and divide out by the lattice translations $\mathcal{L}_s$ of these sublattices. Formally, both the black and white sublattices are the same and the classical space group built on them is $\mathcal{G}$, with the only difference between the sublattices being in the fourth coordinate $s$ of their respective lattice points. The resulting quotient is then found to be $\mathcal{G}^{IV}/\mathcal{L}_s=\mathcal{G}/\mathcal{L}_s+\mathcal{T}\{E|\boldsymbol{\tau}_0\}\mathcal{G}/\mathcal{L}_s$. This group does not contain any translations between the equivalent points of a sublattice, and so translational degrees of freedom have been divided out. A crucial question is whether this quotient group, without the translational degrees of freedom, can be made identical to a magnetic point group containing only point operations, as was the case with the Type I, Type II and Type III magnetic space groups. The answer is that it \textit{cannot}. Suppose that $\mathcal{G}$ does not have compound operations, or in other words, is symmorphic. Dividing $\mathcal{G}$ out by the translations $\mathcal{L}_s$ yields a point group $\mathcal{G}/\mathcal{L}_s \sim \mathcal{G}_p$ that is a subgroup of $\mathcal{G}$. In this case, we have $\mathcal{G}^{IV}/\mathcal{L}_s\sim\mathcal{G}_p+\mathcal{T}\{E|\boldsymbol{\tau}_0\}\mathcal{G}_p$. This is not a group containing solely point operations, since it clearly contains a translation $\{E|\boldsymbol{\tau}_0\}$; the one that moves points between the black and white sublattices. A common argument then \cite{Dresselhaus2008,Smejkal2018} is to take $\boldsymbol{\tau}_0=0$ yielding $\mathcal{G}_p+\mathcal{T}\mathcal{G}_p$, which is a Type II magnetic point group, or grey point group, and then claim that grey point groups can describe antiferromagnetic order. There are several problems with this argument. The first, most obvious one that was already highlighted when we introduced grey point groups, is that they contain the time-reversal operation, a \textit{point} operation, by itself and not only in combination with others. Hence, if bare time-reversal is considered as a symmetry operation applied to a crystal containing magnetic order, it flips every magnetic moment, clearly a contradiction, since a symmetry should leave everything unchanged. Thus, any magnetic group containing $\mathcal{T}$ as a lone operation cannot describe crystals with magnetic order. Indeed, it was discussed at length earlier that grey point groups cannot describe magnetic order. However, the far more crucial problem with the argument is that by taking $\boldsymbol{\tau}_0=0$ we are effectively changing the underlying lattice. In fact, $\boldsymbol{\tau}_0$ is the translation taking a white lattice point to a black one and vice versa, thus making it vanish simply means that we are merging the black and white sublattices, thereby forming a classical Bravais lattice. From the point of view of the fourth coordinate $s$, we are merging a black point with one value of the coordinate with a white point taking the other value, thus the resulting merged point will take on both values of $s$ simultaneously. Changing the underlying lattice yields a completely different structure, so we simply cannot assign a magnetic point group such as a grey group to a Type IV magnetic space group. 

Even though it is possible to assign a magnetic space group to a large number of magnetic crystals there are certain complicated antiferromagnetic arrangements, such as antiferromagnets with conical or spiral patterns, that cannot be described by the magnetic groups discussed till now \cite{Cracknell1975}. There have been several proposals for extending these magnetic groups to include further symmetry operations \cite{Cracknell2009}, with a relatively recent proposal introducing the operation of rotation-reversal receiving considerable attention \cite{Gopalan2011}. The rationale behind this new operation is the following. Vectors can be categorized in two different ways depending on their spatio-temporal properties. The first is their behavior under spatial inversion, yielding polar and axial vectors, the second is their behavior under time-reversal. A time-dependent polar vector, such as velocity, is reversed by both time-reversal and spatial inversion whereas a time-dependent axial vector, such as the magnetic moment resulting from a circular current, is only reversed by time-reversal. On the other hand, a time-independent polar vector, such as position, is only reversed by spatial inversion. The question is, what is the operation that reverses a time-independent axial vector, such as a rotation axis? The answer is: a new operation dubbed rotation-reversal. Adding rotation-reversal to the two existing reversal operations, namely spatial inversion and time-reversal, results in a myriad of new symmetry groups numbering upwards of 15,000. These Gopalan-Litvin groups could have great potential in being used to describe certain antiferromagnetic patterns. 

Summarizing this subsection, we discussed the different types of classical and magnetic point and space groups, with the magnetic ones being able to describe certain magnetic crystals. The ones relevant to antiferromagnetic order are the Type I and Type III magnetic point groups and Type I, Type III, Type IV magnetic space groups. The detailed theory is presented in several excellent textbooks \cite{Cracknell1975,Burns2013,Cracknell2009}. In the next subsection we go on to apply these groups to some concrete examples of antiferromagnets and see the conclusions that can be drawn with respect to the electronic properties from such an analysis. 

\subsection{Band degeneracy in antiferromagnets\label{ssec:degaf}}


Because of the absence of a net magnetization at the level of the magnetic unit cell, one might na\"ively imagine that the band structure of antiferromagnets is systematically spin-degenerate. This is in fact true for some simple examples, such as the collinear bipartite antiferromagnet on a square lattice. However, this is in general incorrect and recent theories \cite{Zelezny2017b,Hayami2019,Ahn2019,Smejkal2020,Yuan2020,Yuan2020b} have pointed out explicitly that the band structure of several classes of antiferromagnets exhibits spin-splitting and spin-momentum locking, even in the absence of spin-orbit coupling. These features arise upon breaking certain symmetries that would otherwise ensure spin degeneracy.

\subsubsection{Conditions for spin degeneracy}

\begin{figure}
	\includegraphics[width=8cm]{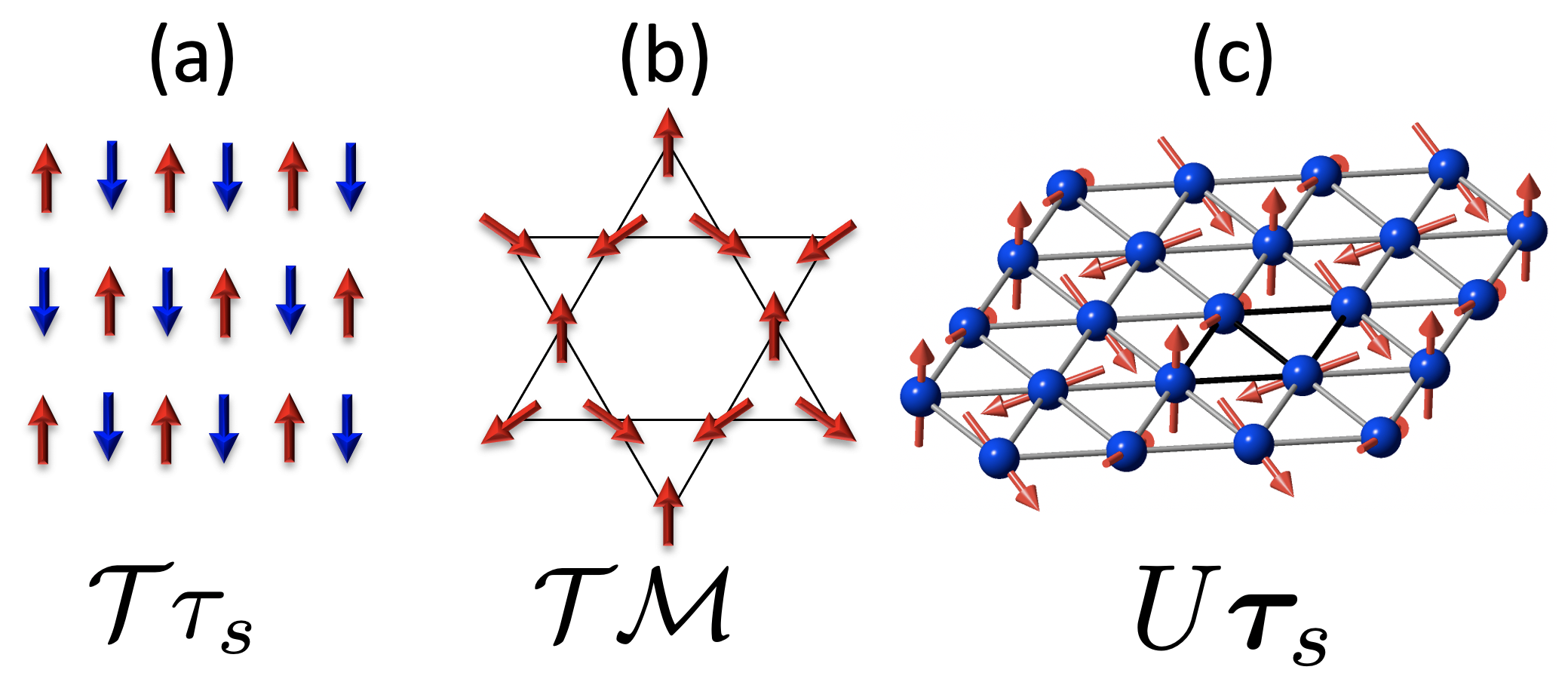}
	\caption{\label{AF-degeneracy} 
	Examples of antiferromagnets exhibiting doubly spin degenerate band structure due to combined symmetry operations.}
\end{figure}

In nonmagnetic crystals, time-reversal symmetry $\mathcal{T}$ ensures that states with opposite spin and opposite momenta possess the same energy. If, in addition, the crystal possesses inversion symmetry $\mathcal{P}$, the combined $\mathcal{PT}$ ensures the complete spin degeneracy of the band structure across the momentum space. This is Kramers' theorem \cite{Kramers1930,Wigner1932}. Antiferromagnets, like any other magnetic system, break $\mathcal{T}$ as well as spin rotation symmetry $U$ [for spin-1/2, this is called a SU(2) rotation] and therefore Kramers' theorem needs to be revisited. In fact, time-reversal symmetry can be effectively restored if a crystal symmetry operation ${\cal O}$, combined with $\mathcal{T}$, is a good symmetry of the magnetic system. For instance, as long as one can find a lattice translation ${\bm\tau}_s$ such that $\mathcal{T}{\bm\tau}_s$ is a good symmetry of the crystal, then the spin degeneracy is preserved. $\mathcal{T}{\bm\tau}_s$ is sometimes called nonsymmorphic time-reversal symmetry. This is the case of the paradigmatic collinear bipartite cubic antiferromagnet depicted in Fig. \ref{AF-degeneracy}(a): the combination of time-reversal symmetry $\mathcal{T}$ with the translation connecting the two sublattices ${\bm\tau}_s$ is a good symmetry of the system so that $\mathcal{PT}{\bm\tau}_s$ acts as an analog to Kramer's theorem and ensures the double degeneracy of the band structure. The same happens in the case of the noncollinear, coplanar antiferromagnet depicted in Fig. \ref{AF-degeneracy}(b). The mirror symmetry ${\cal M}$ of the kagom\'e plane restores time reversal as discussed in \onlinecite{Chen2014}.\par

In addition, if one can find a spin rotation symmetry $U$ such that, combined by a translation vector of the lattice ${\bm\tau}_s$, the combination $U{\bm\tau}_s$ is a symmetry of the system, then the spin degeneracy is also preserved for all wave vectors. Of course, this condition is only valid in the absence of spin-orbit coupling, otherwise $U$ cannot be defined (see below). This case is illustrated by the noncollinear non-coplanar antiferromagnet with triple-Q spin texture depicted in Fig. \ref{AF-degeneracy}(c). In this case the four spins point towards the vertices of a tetrahedron so that one can define a common spin rotation operation $U$ that ensures double degeneracy as discussed in \onlinecite{Martin2008,Ndiaye2019}. In summary, an antiferromagnet that preserves both $\mathcal{PT'}$ (where $\mathcal{T'}$ is the {\em effective} time reversal) and $U{\bm\tau}_s$ exhibits spin-degenerate band structure in the absence of spin-orbit coupling. Breaking either of these two combined transformations can lead to spin-dependent band structure. 

\subsubsection{Momentum-dependent spin-splitting}

\begin{figure}
\includegraphics[width=8cm]{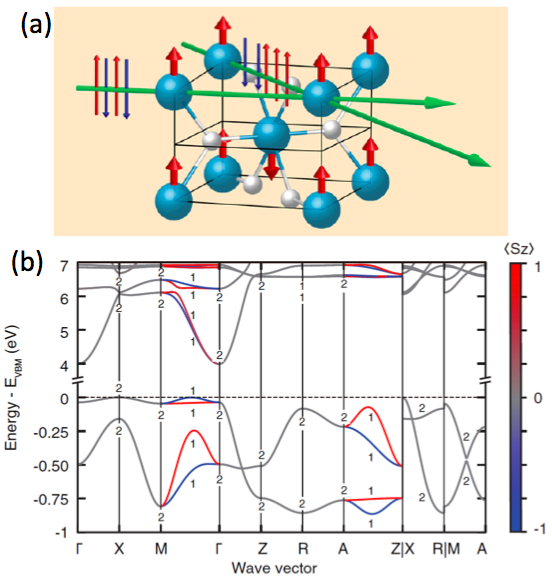}
\caption{(a) In certain antiferromagnets (here, MnF$_2$), a flow of electrons can become spin-polarized depending of the momentum direction. From \onlinecite{Yuan2020}, adapted by APS/Alan Stonebraker. (b) Spin-projected band structure of MnF$_2$ calculated from first principles. From \onlinecite{Yuan2020}.}
\label{fig:FigPolarization}
\end{figure}

We define spin-splitting as the energy separation between bands of opposite spin at a given point in momentum space. In the absence of spin-orbit coupling, spin-splitting occurs as long as both $\mathcal{PT}$ and $U{\bm\tau}_s$ are broken, i.e., in antiferromagnets that either belong to type I and type III magnetic space groups \cite{Hayami2019,Yuan2020}. An example is given by MnF$_2$, a well-known collinear antiferromagnet that exhibits spin-splitting only along certain directions of the Brillouin zone, as illustrated on Fig. \ref{fig:FigPolarization}. The spin remains aligned along the direction of the antiferromagnetic order, so the spin-splitting is similar to that obtained in ferromagnets. Such a spin-splitting has also been predicted in RuO$_2$ \cite{Ahn2019,Smejkal2020}. Notice that spin-splitting is most likely to occur in collinear antiferromagnets rather than in noncollinear ones since spin is only a good quantum number in the former.

In the presence of spin-orbit coupling, breaking $\mathcal{PT}$ becomes a sufficient condition to obtain spin-splitting as it well known in the case of the nonmagnetic Rashba gas \cite{Manchon2015}. In addition, since spin-orbit coupling breaks the spin rotation symmetry $U$, $U{\bm\tau}_s$ is automatically broken in the presence of spin-orbit coupling. Therefore, antiferromagnets from Type TypeIV magnetic space groups can display spin-splitting, as obtained in BiCoO$_3$ \cite{Yamauchi2019}.

\subsubsection{Spin texture in momentum space}

\begin{figure}
\includegraphics[width=8cm]{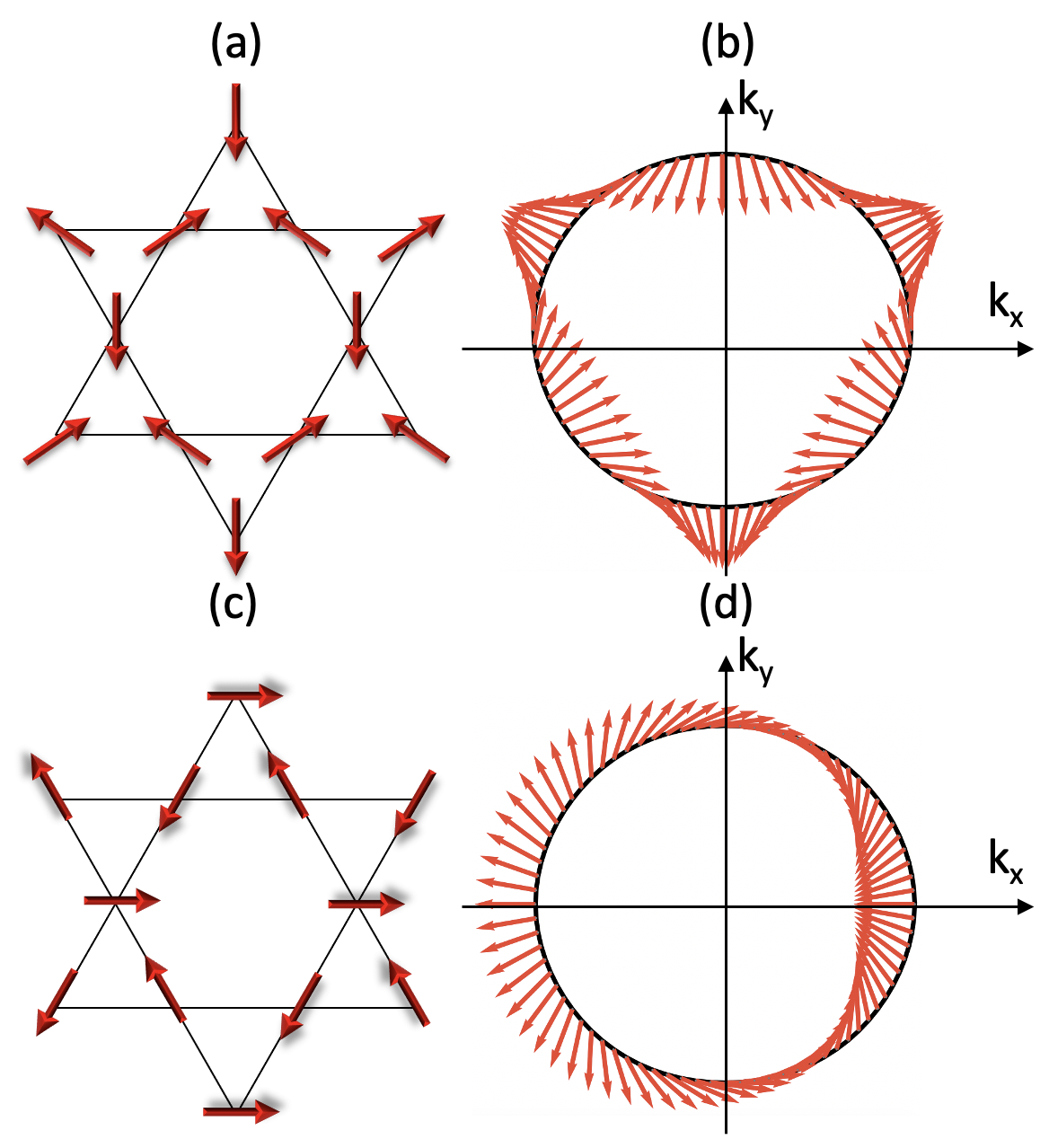}
\caption{(Color online) Spin texture in momentum space for two noncollinear coplanar antiferromagnets on a kagom\'e lattice. (a) $\Gamma_{4g}$ and (b) $\Gamma_{5g}$ magnetic configuration and (c,d) corresponding spin texture in momentum space. Inspired from \onlinecite{Zelezny2017b}.}
\label{fig:FigTexture}
\end{figure}

Materials where the spin and lattice degrees of freedom are coupled to each other tend to display a spin texture in momentum space. In other words, the orientation of the electron's spin varies across the Brillouin zone. This spin-momentum connection is responsible for the emergence of Berry curvature and anomalous spin transport, as further discussed below. This spin texture can occur either under the presence of spin-orbit coupling (for a review, see \cite{Manchon2015}), or - more interestingly - in the presence of {\em noncollinear} magnetic ordering. This leads to new classes of anomalous transport, tagged the magnetic spin Hall effect \cite{Zelezny2017b}, that will be discussed in Section \ref{s:ahe}. Figure \ref{fig:FigTexture} shows an example of such a spin-momentum locking in two noncollinear kagom\'e antiferromagnets. 


%

\subsection{Properties of the Conductivity Tensor} 

In this section we focus on how the magnetic and crystalline symmetries of a material restrict the electric charge and spin conductivities, quantities that describe the related transport effects. This introduction will serve as a basis to the discussion of the various anomalous (electronic and magnonic) transport effects discovered recently in antiferromagnets that is developed in Section \ref{s:ahe}.

\subsubsection{Charge Conductivity}
\label{chargeCond}

The charge conductivity $\sigma$ is defined as the second-rank tensor coefficient relating the electric field $\textbf{E}$ to the charge current density $\textbf{J}$,

\begin{equation}
\label{eq:conDef}
J_i=\sum_{j}\sigma_{ij}E_j.
\end{equation}

Several restrictions can be deduced from this definition itself. First of all, looking at the action of spatial inversion $\mathcal{P}$ on both sides we see that $\sigma$ must be even $\mathcal{P}:\sigma\to\sigma$, and thus a polar tensor, since both $\textbf{E}$ and $\textbf{J}$ are odd. Similarly, acting with time-reversal on both sides we find that $\sigma$ must be odd $\mathcal{T}:\sigma\to-\sigma$, since the current $\textbf{J}$ is odd, whereas the electric field $\textbf{E}$ is even. However, this latter argument is naive, because we are applying the time-reversal operation to the entire system including the current and the crystal, so this is not equivalent to the time-reversal operation that is an element of the crystal's magnetic symmetry group. Indeed, suppose $\textbf{m}$ labels the magnetic structure of the crystal. Then, the conductivity can be taken as a function $\sigma_{ij}(\textbf{m})$ and, considered  as an element of the magnetic symmetry group that reverses $\textbf{m}$, $\mathcal{T}$ acts as $\mathcal{T}^{\text{sym}}\sigma_{ij}(\textbf{m})=\sigma_{ij}(-\textbf{m})$. The charge current is the response of the system to an electric field, and from the microscopic point of view it is the expectation value of the current operator. This expectation value is dependent on the magnetic structure through the Hamiltonian so we can write the current as a function $J_i(\textbf{m})$, transforming as $\mathcal{T}^{\text{sym}}J_{i}(\textbf{m})=J_{i}(-\textbf{m})$. The constitutive relation \eqref{eq:conDef} can be written as

\begin{equation}
\label{eq:conMdef}
J_i(\textbf{m})=\sum_{j}\sigma_{ij}(\textbf{m})E_j \xrightarrow{\mathcal{T}^{\text{sym}}} J_i(-\textbf{m})=\sum_{j}\sigma_{ij}(-\textbf{m})E_j ,
\end{equation}

and transforms consistently under $\mathcal{T}^{\text{sym}}$. The charge current quantifies the flow of charges and is sensitive to the direction of time. When considering the full time dependence, both as a flow and as a function of $\textbf{m}$, the current changes its sign too under time-reversal: $J_i(\textbf{m})\xrightarrow{\mathcal{T}^{\text{full}}}-J_i(-\textbf{m})$. Note that $\mathcal{T}^{\text{full}}$ and $\mathcal{T}^{\text{sym}}$ are both time-reversal operations but act differently. They can be thought of as different representations of the same abstract time-reversal operator, however such a discussion is beyond the scope of this review.\\

The operation $\mathcal{T}^{\text{full}}$ allows us to decompose the current into two terms $J_{i}(\textbf{m})=J^{\text{nd}}_{i}(\textbf{m})+J^{\text{d}}_{i}(\textbf{m})$, where $J^{\text{nd}}_{i}(\textbf{m})$ is even whereas $J^{\text{d}}_{i}(\textbf{m})$ is odd under $\mathcal{T}^{\text{full}}$:

\begin{subequations}
\begin{align}
\nonumber
J^{\text{nd}}_{i}(\textbf{m})&=\frac{J_i(\textbf{m})+\mathcal{T}^{\text{full}}J_i(\textbf{m})}{2}=\frac{J_i(\textbf{m})-J_i(-\textbf{m})}{2}
\\
\label{eq:Jndiss}
&=\sum_{j}\frac{\sigma_{ij}(\textbf{m})-\sigma_{ij}(-\textbf{m})}{2}E_j=\sum_j \sigma^{\text{nd}}_{ij}(\textbf{m})E_j,
\\
\nonumber
J^{\text{d}}_{i}(\textbf{m})&=\frac{J_i(\textbf{m})-\mathcal{T}^{\text{full}}J_i(\textbf{m})}{2}=\frac{J_i(\textbf{m})+J_i(-\textbf{m})}{2}
\\
\label{eq:Jdiss}
&=\sum_{j}\frac{\sigma_{ij}(\textbf{m})+\sigma_{ij}(-\textbf{m})}{2}E_j=\sum_j\sigma^{\text{d}}_{ij}(\textbf{m})E_j.
\end{align}
\end{subequations}

The even term $J^{\text{nd}}_{i}(\textbf{m})$ behaves as $\mathcal{T}^{\text{full}}J^{\text{nd}}(\textbf{m})=J^{\text{nd}}(\textbf{m})$ and is thus not sensitive to the direction of time. This means that it cannot dissipate energy so the corresponding conductivity $\sigma^{\text{nd}}_{ij}(\textbf{m})$ describes a non-dissipative response. On the other hand, the odd term  $J^{\text{d}}_{i}(\textbf{m})$ behaving as $\mathcal{T}^{\text{full}}J^{\text{d}}(\textbf{m})=-J^{\text{d}}(\textbf{m})$, changes sign and, as a consequence, is purely dissipative. The conductivity $\sigma^{\text{d}}_{ij}(\textbf{m})$ thus describes dissipative response.\\

Suppose we take $\textbf{m}=0$ corresponding to the case of nonmagnetic materials. This situation is usually referred to as time-reversal symmetric, where the time-reversal is understood to be $\mathcal{T}^{\text{sym}}$, an element of the magnetic symmetry group. From \eqref{eq:Jndiss} and \eqref{eq:Jdiss} we see that $\sigma^{\text{nd}}_{ij}(\textbf{m}=0)=0$, so the response is purely dissipative and time-reversal symmetry must be broken to obtain a non-dissipative response.\\

The time-reversal properties of a response are closely related to its symmetries as a tensor. Such relations were first given by Onsager and for the charge conductivity they read \cite{Cracknell1975,Nagaosa2010}:

\begin{equation}
\label{eq:Onsager}
\sigma_{ij}(\textbf{m})=\sigma_{ji}(-\textbf{m}).
\end{equation}

Let us proceed to decompose $\sigma$ into symmetric and antisymmetric parts $\sigma_{ij}(\textbf{m})=\sigma^s_{ij}(\textbf{m})+\sigma^{a}_{ij}(\textbf{m})$, with

\begin{subequations}
\begin{align}
\label{eq:SigmaSymm}
\sigma^s_{ij}(\textbf{m})=&\frac{\sigma_{ij}(\textbf{m})+\sigma_{ji}(\textbf{m})}{2}=\frac{\sigma_{ij}(\textbf{m})+\sigma_{ij}(-\textbf{m})}{2}
\\
\nonumber
=&\sigma^{\text{d}}_{ij}(\textbf{m}),
\\
\label{eq:SigmaAsymm}
\sigma^a_{ij}(\textbf{m})=&\frac{\sigma_{ij}(\textbf{m})-\sigma_{ji}(\textbf{m})}{2}=\frac{\sigma_{ij}(\textbf{m})-\sigma_{ij}(-\textbf{m})}{2}
\\
\nonumber
=&\sigma^{\text{nd}}_{ij}(\textbf{m}),
\end{align}
\end{subequations}

where, in each expression, Onsager's relations (\ref{eq:Onsager}) were used for the second equality and the definitions of $\sigma^{\text{d}}_{ij}(\textbf{m}),\,\sigma^{\text{nd}}_{ij}(\textbf{m})$ from \eqref{eq:Jdiss}, \eqref{eq:Jndiss} for the third equality.\\
We see very clearly, that the symmetric part of the conductivity provides the dissipative response, whereas the antisymmetric part provides the non-dissipative response. This correspondence is usually explained via energy arguments \cite{Landau1960,Nagaosa2010}, however it follows in a straightforward manner from Onsager's relations. Furthermore, we saw above that the breaking of time-reversal symmetry as $\mathcal{T}^{\text{sym}}$ is required to obtain a non-dissipative response, and the same can be said about the antisymmetric part of the conductivity. On the other hand, only the symmetric, purely dissipative part contributes in the presence of time-reversal symmetry.\\

The existence of two time-reversal operations $\mathcal{T}^{\text{full}}$ and $\mathcal{T}^{\text{sym}}$ caused significant confusion in the literature, since they gave rise to the important dilemma: how should $\sigma$ transform under time-reversal? Acting on both sides of the constitutive relations \eqref{eq:conMdef} we have $\mathcal{T}^{\text{full}}\sigma(\textbf{m})=-\sigma(-\textbf{m})$ and $\mathcal{T}^{\text{sym}}\sigma(\textbf{m})=\sigma(-\textbf{m})$. Thus, if we do not clearly separate the two operations, a contradiction might arise since $\sigma$ does not transform in the same way under them. As described in \cite{Grimmer1993} the discussion of which transformation property to choose as the "right" one remained riddled with controversy for decades. The resolution is, of course, that both are "right" and simply correspond to different ways of representing the time-reversal operation on the conductivity. Following our earlier usage, henceforth under time-reversal we mean $\mathcal{T}^{\text{sym}}$ and continue to denote it as $\mathcal{T}$.\\

\subsubsection{Neumann's Principle}

The symmetries derived from the magnetic and crystalline structure of the material give further relations between different components of the conductivity tensor, thereby restricting its form considerably. The observation that the mathematical object describing macroscopically measurable quantities should satisfy the symmetries of the crystal is commonly referred to as Neumann's principle \cite{Birss1964,Cracknell1975,Dresselhaus2008,Burns2013}. In its classical form, this principle is usually implemented using the classical point group of the crystal, by requiring that the action of the point group should not change the given quantity: $P\sigma=\sigma$, for some element $P$ of the point group. The conductivity is a second rank polar tensor, meaning that this implementation of Neumann's principle can be expressed in component form as

\begin{equation}
\label{eq:NeumannClassical}
\sum_{k,l}P_{ik}P_{jl}\sigma_{kl}=\sigma_{ij}.
\end{equation}

Being a polar tensor, $\sigma$ does not change sign under an improper rotation and consequently has the same transformation properties under proper and improper rotations. There are two ways to proceed. First, we could introduce the proper part $R$ of an improper rotation as $P=\mathcal{P}R$, where $\mathcal{P}$ is spatial inversion. Then, we simply replace $P$ with a pure rotation $R$ in (\ref{eq:NeumannClassical}) since inversion has no effect on the components of a second rank polar tensor; both indices refer to polar vectors and since inversion flips both, there is no overall effect. We thus need only consider point groups stripped of all their improper rotations. The second way would be to use the fact that $\sigma$ is insensitive to inversion and only consider point groups containing inversion as an element. The older literature \cite{Kleiner1966,Grimmer1991,Grimmer1993} follows the first way and refers to point groups stripped of their improper rotations as Laue groups, whereas more recent works \cite{Seeman2015,Wimmer2016} follow the second path and call inversion-containing point groups as Laue groups. Nevertheless, in both cases there are 11 classical Laue groups and $2\times 11+10=32$ magnetic Laue groups. 

\begin{figure}
\includegraphics[height=4cm]{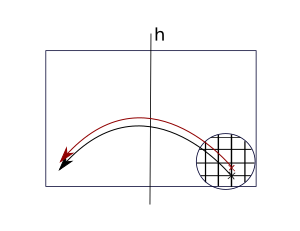}
\caption{The reflection of points separated by sublattice translations about an axis $h$ a macroscopic distance away is in general considered as reflecting the same point.}
\label{fig:Neumann}
\end{figure}

By considering only the point group in the implementation of Neumann's principle above, we are, in fact, performing an approximation. In general, the symmetry of the crystal is described by the space group and only for symmorphic space groups can we assume that point operations alone provide a complete description. Yet, the principle is implemented in the same way for nonsymmorphic space groups containing compound operations. The reason for this, is that the quantity under consideration is \textit{macroscopic}: while a point operation, such as a reflection, connects points that are separated by a macroscopic number of unit cells, a sublattice translation does not and its effect on macroscopic properties, as depicted on Fig. \ref{fig:Neumann} should be negligible, i.e. we should not be able to distinguish between $\{R|0\}$ and $\{R|\boldsymbol{\tau}_s\}$ on this scale. In fact, this is not always the case, since upon performing a sublattice translation over a macroscopic number of unit cells certain quantities could pick up a phase from each displaced unit cell, which could be measurable when added up. It follows that even if a point group analysis would result in certain terms being cancelled, they could very well be nonvanishing.

In the presence of magnetic symmetries Neumann's principle for the conductivity has to be extended in order to accomodate the symmetries containing $\mathcal{T}$. An early proposal based on a microscopic formulation was given in \cite{Kleiner1966} with a more recent extension in \cite{Seeman2015}. A concise argument for Neumann's principle in this case without resorting to microscopics was presented in \cite{Grimmer1993} as follows. We consider two equivalent tranformations done separately. For the first one, we rotate the underlying crystal $\textbf{m}$ by $R_1$, while leaving the coordinate system, representing our lab frame in which the conductivity tensor is defined, intact whereas for the second one we rotate the coordinate system by $R_1$, while leaving the crystal intact. The result of performing these two transformations should be the same, with the components of the tensor becoming

\begin{equation}
\label{eq:NeumannExtendedR}
\sigma_{ij}(R_1\textbf{m})=\sum_{k,l}\left(R_1^{-1}\right)_{ik}\left(R_1^{-1}\right)_{jl}\sigma_{kl}(\textbf{m}).
\end{equation}

Using the same argument, but this time for a rotation $R_2$ combined with time-reversal $\mathcal{T}R_2$, and that the effect of $\mathcal{T}$ on $\textbf{m}$ is to simply reverse it; $\mathcal{T}R_2\textbf{m}=-R_2\textbf{m}$, we have

\begin{equation}
\label{eq:NeumannExtendedTR}
\sigma_{ij}(\mathcal{T}R_2\textbf{m})=\sigma_{ij}(-R_2\textbf{m})=\sum_{k,l}\left(R_2^{-1}\right)_{ik}\left(R_2^{-1}\right)_{jl}\sigma_{kl}(-\textbf{m}).
\end{equation}

Now suppose, that in (\ref{eq:NeumannExtendedR}), $R_1$ is a \textit{classical} point symmetry of the magnetic crystal or equivalently a classical element of the magnetic point group. Then, this operation should leave the magnetic order intact $R_1\textbf{m}=\textbf{m}$, and using the orthonality $R_1^{-1}=R_1^T$ of $R_1$, (\ref{eq:NeumannExtendedR}) becomes

\begin{equation}
\label{eq:Neumann1}
\begin{split}
\sigma_{ij}(\textbf{m})=&\sum_{k,l}(R_1^T)_{ik}(R_1^T)_{jl}\sigma_{lk}(\textbf{m})
\\
&=\sum_{k,l}(R_1)_{ki}(R_1)_{lj}\sigma_{lk}(\textbf{m}).
\end{split}
\end{equation}
Similarly, if $\mathcal{T}R_2$ is a point symmetry and therefore an element of the magnetic point group, then (\ref{eq:NeumannExtendedTR}) yields

\begin{equation}
\label{eq:Neumann2}
\sigma_{ij}(\textbf{m})=\sum_{k,l}(R_2)_{ki}(R_2)_{lj}\sigma_{lk}(\textbf{m}),
\end{equation} 

where, apart from the orthogonality of $R_2$, Onsager's relations (\ref{eq:Onsager}) were used. Note the subtle difference between (\ref{eq:Neumann1}) and (\ref{eq:Neumann2}) in the ordering of the indices of $\sigma$. The former one is equivalent to the classical statement of Neumann's principle applicable to classical point symmetries, whereas the latter is a modified version for point symmetries involving $\mathcal{T}$. These two identities present significant constraints on the conductivity tensor by providing relations between its components. With the trick of reducing the point groups to Laue groups explained above, the scope of the symmetry analysis based on this extended Neumann's principle can be narrowed down from 122 to 32 magnetic point groups, a considerable margin, with the possible forms of the conductivity tensor for all magnetic Laue groups listed in \cite{Kleiner1966,Grimmer1993,Seeman2015,Wimmer2016}. 

\subsubsection{Anomalous Hall Effect}

Nondissipative transport is the dream of any scientist involved in improving the efficiency of electronic devices by reducing energy loss. As we saw above, in the static limit, only the symmetric part of the conductivity $\sigma^s_{ij}$ plays a part in dissipation while the antisymmetric part $\sigma^a_{ij}$ gives rise to nondissipative effects. The anomalous Hall effect \cite{Nagaosa2010} arises as a transverse charge current induced by an electric field perpendicular to it and the corresponding conductivity is defined as

\begin{equation}
\label{eq:AHEcond}
\sigma^{AHE}_{ij}=\frac{\sigma_{ij}-\sigma_{ji}}{2}=\sigma^a_{ij},
\end{equation}

meaning that there is a possibility for nondissipative transport arising through this effect. A crucial question is then to find the crystals that could support such an effect. We already discussed that the presence of time-reversal symmetry restricts $\sigma_{ij}$ to its symmetric part $\sigma_{ij}=\sigma^s_{ij}$. This is consistent with the extended Neumann's principle (\ref{eq:Neumann2}) applicable to symmetries containing $\mathcal{T}$. Indeed, assuming $\mathcal{T}$ is an element of the magnetic point group describing the crystal, i.e. the group is a grey group, and taking $R_2=E$, the identity operation, in (\ref{eq:Neumann2}) we obtain $\sigma_{ij}(\textbf{m})=\sigma_{ji}(\textbf{m})$. The same argument can be made for crystals with the combined symmetry $\mathcal{T}\mathcal{P}$, where $\mathcal{P}$ is spatial inversion, since recall that we can strip off the improper part of all point operations when using (\ref{eq:Neumann2}) and $\mathcal{P}$ is the improper part of $\mathcal{P}E$. Thus, time-reversal symmetry must be broken for $\sigma^a_{ij}$ to be nonzero, and consequently, nonmagnetic crystals described with grey groups cannot support such an effect, unless, of course, an external magnetic field is applied. Similarly, no magnetic crystal with a magnetic point group containing $\mathcal{T}\mathcal{P}$ as an element can have a non-vanishing $\sigma^a_{ij}$. 

Further restrictions can be made by exploiting the fact that $\sigma^a_{ij}$ is antisymmetric. Indeed, any antisymmetric second rank polar tensor in 3 dimensions is dual to an axial vector. This is an example of Hodge duality and can be expressed with the totally antisymmetric Levi-Civita symbol $\varepsilon_{ijk}$ as $A_i=\varepsilon_{ijk}\sigma^a_{jk}$ or in matrix form

\begin{equation}
\nonumber
\sigma^a=\begin{pmatrix}
0&\sigma^a_{xy}&-\sigma^a_{zx}\\\\
-\sigma^a_{xy}&0&\sigma^a_{yz}\\\\
\sigma^a_{zx}&-\sigma^a_{yz}&0
\end{pmatrix}
\to
\begin{pmatrix}
\sigma^a_{yz}\\\\
\sigma^a_{zx}\\\\
\sigma^a_{xy}
\end{pmatrix}=\textbf{A}.
\end{equation}

Keeping Neumann's principle in mind, we need $g\sigma^a=\sigma^a$ for $g$ an element of the magnetic group describing the crystal symmetry. However, since $\sigma^a$ is dual to an axial vector $\textbf{A}$, we can rephrase this as $g\textbf{A}=\textbf{A}$, meaning that only those groups are allowed whose elements leave an axial vector invariant. The magnetization of a ferromagnet is described by an axial vector $\textbf{M}$ and only those groups can describe ferromagnetism whose elements don't change $\textbf{M}$, otherwise they would not be magnetic symmetries since they change the magnetic structure described by $\textbf{M}$. Such point groups are called ferromagnetic and we can conclude that $\sigma^a$ can only be present in crystals described by one of these ferromagnetic groups. We ought to be cautious with the naming. Just because such groups are referred to as ferromagnetic does not mean that they \textit{only} describe ferromagnets, rather that they \textit{can}. In fact, certain antiferromagnets are also described by these ferromagnetic groups and it is only such antiferromagnets that can give rise to nondissipative transport described by the antisymmetric $\sigma^a(\textbf{m})$. There are 5 ferromagnetic Laue groups of Type I and 5 of Type III. These correspond to 13 point groups of Type I and 18 of Type III yielding 31 in total, while giving rise to 275 magnetic space groups of Type I and Type III \cite{Cracknell1975}. 

Antiferromagnetic crystals can also be described by Type IV magnetic space groups written as $\mathcal{G}^{IV}=\mathcal{G}+\mathcal{T}\{E|\boldsymbol{\tau}_0\}\mathcal{G}$, however these are unique in the sense that we cannot really associate a magnetic point group to them in the usual way by taking all translations to zero, since, as pointed out earlier, assuming $\boldsymbol{\tau}_0=0$ changes the underlying magnetic Bravais lattice, on which the space group itself is built, to a classical one thereby changing the entire structure of the crystal. These groups do \textit{not} contain the \textit{point} operation $\mathcal{T}$ as a lone element, rather only in combination with the translation between the black and white sublattices $\mathcal{T}\{E|\boldsymbol{\tau}_0\}$, meaning that, contrary to the usual arguments, in principle they could support a nonzero $\sigma^a$. To see this, we simply think of implementing Neumann's principle with the full set of space operations and not only point operations as we did till now. Indeed, the action of the combined operation $\mathcal{T}\{E|\boldsymbol{\tau}_0\}$ on the magnetic structure $\textbf{m}$ can be written as $\mathcal{T}\{E|\boldsymbol{\tau}_0\}\textbf{m}=-\textbf{m}+\boldsymbol{\tau}_0$, with Neumann's principle yielding $\mathcal{T}\{E|\boldsymbol{\tau}_0\}\sigma(\textbf{m})=\sigma(\textbf{m})\Longrightarrow \sigma(\textbf{m})=\sigma(-\textbf{m}+\boldsymbol{\tau}_0)$, so using (\ref{eq:SigmaAsymm})

\begin{equation}
\sigma^a(\textbf{m})=\frac{\sigma(\textbf{m})-\sigma(-\textbf{m})}{2}=\frac{\sigma(\textbf{m})-\sigma(\textbf{m}+\boldsymbol{\tau}_0)}{2}.
\end{equation}

This is not vanishing in general, but the sublattice translation is usually considered negligible for macroscopic properties since it does not connect points separated by macroscopic distances (Fig. \ref{fig:Neumann}). On the other hand, it could be a possibility that a phase accumulates through sublattice translations performed over a macroscopic range leading to a nonzero $\sigma^a$, this is however, an avenue for future research.

\subsubsection{Spin Conductivity}

Moving on to spin transport we introduce the spin conductivity in a manner analogous to the charge conductivity \cite{Seeman2015,Maekawa2017}. Denote the spin current as $J_{s,i}$, where $s\in\{x,y,z\}$ refers to the direction of spin polarization and $i\in\{x,y,z\}$ to the direction of 'flow'. We define the spin conductivity as the tensor coefficient relating the spin current to the applied electric field
\begin{equation}
J_{s,i}=\sum_j\sigma_{(s,i)j}E_j.
\end{equation}

It is important to emphasize that the spin current $J_{s,i}$ does \textit{not} refer to a physical 'flow' of charges that are spin polarized, rather it refers to the flow of spin itself. The behavior of $\sigma_{(s,i)j}$ under spatial inversion is straightforward to see. $i$ and $j$ are polar indices, whereas $s$ is an axial index (it labels spin, an axial vector), meaning that $\sigma_{(s,i)j}$ transforms as a third rank axial tensor and does not change sign under $\mathcal{P}$. Following the discussion for the charge conductivity in section \ref{chargeCond}, we take $\mathcal{T}$ to be the operation that only reverses the magnetic structure $\textbf{m}$ and $\mathcal{T}^{\text{full}}$ to be the one that in addition reverses all time flows. The latter acts on the current as $\mathcal{T}^{\text{full}}J_{s,i}(\textbf{m})=J_{s,i}(-\textbf{m})$, since both spin and velocity are reversed. We see that the action of both $\mathcal{T}^{\text{full}}$ and $\mathcal{T}$ is the same on the spin Current and we can once more decompose the spin current into non-dissipative and dissipative components

\begin{subequations}
\begin{align}
\nonumber
J^{\text{nd}}_{s,i}(\textbf{m})&=\frac{J_{s,i}(\textbf{m})+\mathcal{T}^{\text{full}}J_{s,i}(\textbf{m})}{2}=\frac{J_{s,i}(\textbf{m})+J_{s,i}(-\textbf{m})}{2}
\\
\label{eq:Jsndiss}
&=\sum_{j}\frac{\sigma_{(s,i)j}(\textbf{m})+\sigma_{(s,i)j}(-\textbf{m})}{2}E_j
\\\nonumber
&=\sum_j \sigma^{\text{nd}}_{(s,i)j}(\textbf{m})E_j,
\\
\nonumber
J^{\text{d}}_{s,i}(\textbf{m})&=\frac{J_{s,i}(\textbf{m})-\mathcal{T}^{\text{full}}J_{s,i}(\textbf{m})}{2}=\frac{J_{s,i}(\textbf{m})-J_{s,i}(-\textbf{m})}{2}
\\
\label{eq:Jsdiss}
&=\sum_{j}\frac{\sigma_{(s,i)j}(\textbf{m})-\sigma_{(s,i)j}(-\textbf{m})}{2}E_j
\\\nonumber
&=\sum_j\sigma^{\text{d}}_{(s,i)j}(\textbf{m})E_j.
\end{align}
\end{subequations}

In the presence of time-reversal symmetry ($\textbf{m}=0$), we have $\sigma^{\text{d}}_{(s,i)j}(\textbf{m}=0)=0$ and the spin current is purely non-dissipative. This is in contrast to the charge current, in which case time-reversal breaking was necessary for a non-dissipative response.\\
 
Being a linear response coefficient, $\sigma_{(s,i)j}$ should satisfy an Onsager relation 
\begin{equation}
\label{eq:SpinOnsager}
\sigma_{(s,i)j}(\textbf{m})=-\sigma_{j(s,i)}(-\textbf{m}).
\end{equation}

We can define symmetry and antisymmetry as behaviour under the exchange of $(s,i)$ and $j$ and proceed to separate the spin conductivity into symmetric and antisymmetric parts

\begin{subequations}
\begin{align}
\label{eq:SpinSigmaSymm}
\nonumber
\sigma^s_{(s,i)j}(\textbf{m})&=\frac{\sigma_{(s,i)j}(\textbf{m})+\sigma_{j(s,i)}(\textbf{m})}{2}
\\
&=\frac{\sigma_{(s,i)j}(\textbf{m})-\sigma_{(s,i)j}(-\textbf{m})}{2}
\\\nonumber
&=\sigma^{\text{d}}_{(s,i)j}(\textbf{m}),
\\\nonumber
\\
\label{eq:SpinSigmaAsymm}
\nonumber
\sigma^a_{(s,i)j}(\textbf{m})&=\frac{\sigma_{(s,i)j}(\textbf{m})-\sigma_{j(s,i)}(\textbf{m})}{2}
\\
&=\frac{\sigma_{(s,i)j}(\textbf{m})+\sigma_{(s,i)j}(-\textbf{m})}{2}
\\\nonumber
&=\sigma^{\text{nd}}_{(s,i)j}(\textbf{m}),
\end{align}
\end{subequations}

where (\ref{eq:SpinOnsager}) was used. The action of $\mathcal{T}$ on the spin conductivity is defined as $\mathcal{T}\sigma_{(s,i)j}(\textbf{m})=\sigma_{(s,i)j}(-\textbf{m})$, from which we can infer the behavior of the symmetric and antisymmetric parts under $\mathcal{T}$ to be

\begin{subequations}
\begin{align}
\label{eq:SpinSigmaSymmTR}
&\mathcal{T}\sigma_{(s,i)j}^s(\textbf{m})=\sigma_{(s,i)j}^s(-\textbf{m})=-\sigma_{(s,i)j}^s(\textbf{m}),
\\
\label{eq:SpinSigmaAsymmTR}
&\mathcal{T}\sigma_{(s,i)j}^a(\textbf{m})=\sigma_{(s,i)j}^a(-\textbf{m})=\sigma_{(s,i)j}^a(\textbf{m}).
\end{align}
\end{subequations}

Thus we see that in the presence of time-reversal symmetry ($\textbf{m}=0$), only $\sigma_{(s,i)j}^a$ can yield a spin current and from \eqref{eq:SpinSigmaAsymm} this will be non-dissipative.\\

The restrictions on the spin conductivity tensor imposed by crystal symmetry can be taken into account by implementing Neumann's principle for $\sigma_{(s,i)j}$ \cite{Seeman2015}. We can follow the same line of reasoning as we did for the charge conductivity when we got Eqs. (\ref{eq:Neumann1}) and (\ref{eq:Neumann2}) from Eqs. (\ref{eq:NeumannExtendedR}) and (\ref{eq:NeumannExtendedTR}) with one crucial difference: $\sigma_{(s,i)j}$ is an \textit{axial} tensor. Neumann's principle can then be implemented for some point operation $R$ that is a symmetry as

\begin{subequations}
\begin{align}
\label{eq:SpinNeumann1}
&\sigma_{(s,i)j}(\textbf{m})=\sum_{s',k,l}\text{det(R)}R_{ss'}R_{ki}R_{lj}\sigma_{(s',k)l}(\textbf{m}),
\\\nonumber
\\\nonumber
&\sigma_{(s,i)j}(\textbf{m})=\sum_{s',k,l}\text{det(R)}R_{ss'}R_{ki}R_{lj}\sigma_{(s',k)l}(-\textbf{m})
\\
\label{eq:SpinNeumann2}
&\qquad\qquad\,=-\sum_{s',k,l}\text{det(R)}R_{ss'}R_{ki}R_{lj}\sigma_{l(s',k)}(\textbf{m}),
\end{align}
\end{subequations}

where the first equation is valid for classical point operations, whereas the second one is for operations combined with $\mathcal{T}$. This is consistent with our analysis that lead to (\ref{eq:SpinSigmaAsymmTR}), since assuming $\mathcal{T}$ is a symmetry we take $R=E$ in (\ref{eq:SpinNeumann2}) corresponding to $\mathcal{T}R=\mathcal{T}E=\mathcal{T}$ and get $\sigma_{(s,i)j}=-\sigma_{j(s,i)}$, just what we expect in the presence of time-reversal symmetry. We can get a further simple constraint. Suppose that the combined operation $\mathcal{T}\mathcal{P}$ is a symmetry. From Eq. (\ref{eq:SpinNeumann2}) we again see that $\sigma_{(s,i)j}(\textbf{m})=-\sigma_{j(s,i)}(\textbf{m})$ and thus only the antisymmetric part $\sigma_{(s,i)j}^a(\textbf{m})$ can contribute.\\

The spin conductivity $\sigma_{(s,i)j}$ is a third rank axial tensor, so it does not change sign under inversion. Thus again it is sufficient to examine the magnetic Laue groups and the possible forms of the tensor are listed in \cite{Seeman2015,Wimmer2016}.

\subsubsection{Spin Hall Effect\label{ss:spincond}}

An important example of a spin transport phenomenon where $\sigma_{(s,i)j}^a$ is relevant arises in the context of the spin Hall effect \cite{Maekawa2017}. This is an analogue of the charge Hall effect, where instead of a transverse charge current, the electric field induces a transverse spin current. In the spin conductivity $\sigma_{(s,i)j}$, $i$ and $j$ refer to the current and applied field directions, respectively, to get the transverse response we can thus consider the spin Hall conductivity by antisymmetrizing these

\begin{equation}
\begin{split}
\sigma^{SHE}_{(s,i)j}&=\frac{\sigma_{(s,i)j}-\sigma_{(s,j)i}}{2}
\\
&=\frac{\sigma_{(s,i)j}^s+\sigma_{(s,i)j}^a}{2}-\frac{\sigma_{(s,j)i}^s+\sigma_{(s,j)i}^a}{2}
\\
&=\frac{\sigma_{(s,i)j}^s-\sigma_{(s,j)i}^s}{2}+\frac{\sigma_{(s,i)j}^a-\sigma_{(s,j)i}^a}{2}.
\end{split}
\end{equation}

Both of the Onsager based symmetric and antisymmetric parts (\ref{eq:SpinSigmaSymm}), (\ref{eq:SpinSigmaAsymm}) contribute to the spin Hall effect and we can infer, based on their behavior under time-reversal in (\ref{eq:SpinSigmaSymmTR}), (\ref{eq:SpinSigmaAsymmTR}), that the effect can be non-vanishing regardless of time-reversal symmetry being present or broken. This is in stark contrast to the anomalous Hall effect, where time-reversal breaking is necessary. Note that $\sigma^{SHE}_{j(s,i)}$ is the inverse spin Hall conductivity.\\

The form of the spin Hall conductivity tensor is also subject to the constraints imposed by crystal symmetry. The restrictions on the general spin conductivity (\ref{eq:SpinNeumann1}), (\ref{eq:SpinNeumann2}) can be expanded considerably for the spin Hall conductivity as follows. Indeed, $\sigma^{SHE}_{(s,i)j}$ is antisymmetric in $i,j$, we can thus define a dual second rank tensor as $T_{(s,k)}=\varepsilon_{ijk}\sigma^{SHE}_{(s,i)j}$. Both $\sigma^{SHE}_{(s,i)j}$ and the Levi-Civita symbol $\varepsilon_{ijk}$ are third rank axial tensors, thus $T_{(s,k)}$ is in fact a second rank \textit{polar} tensor. This means, that for the spin Hall conductivity we can use the implementation of Neumann's principle corresponding to second order polar tensors, the same as the full charge conductivity tensor. As an example, consider a 2-dimensional crystal and apply an electric field in the $y$ direction. The spin Hall conductivity describing the $z$ polarized spin Hall current in the $x$ direction is $\sigma^{SHE}_{(z,x)y}$. This corresponds to the tensor component $T_{(z,z)}$, thus only those crystals can support such a spin Hall effect that can keep the $zz$ component of a second rank polar tensor invariant. It can be noted that it is not possible to narrow down the possible point groups as much as we did for the charge Hall effect that lead to the ferromagnetic groups. Indeed, an antiferromagnet that is not described by a ferromagnetic group can, in principle, still support a spin Hall effect, while not having an anomalous Hall effect. This gives rise to much broader prospects for finding and utilizing the spin Hall effect, rather than the charge Hall effect in antiferromagnets.\\

Along similar lines, a detailed symmetry analysis can be performed for all the different kinds of linear response tensors. Two final examples of recent interest in spintronics are the tensor giving the spin polarization response to an electric field \cite{Zelezny2017} and the torkance tensor quantifying the torque response to an electric field \cite{Wimmer2016}.

\subsection{Abelian and Non-Abelian Berry Curvatures}

\begin{figure}
\includegraphics[height=6cm]{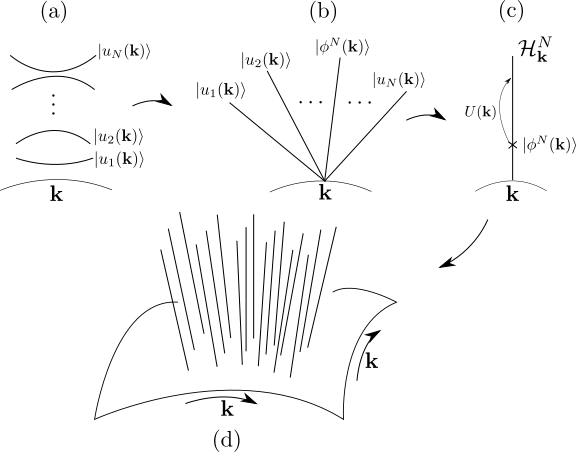}
\caption{The fibre bundle structure of the Bloch functions over the Brillouin zone. (a) depicts $N$ Bloch bands $|u_{\alpha}(\textbf{k})\rangle$ at a point $\textbf{k}$. (b) shows how we can consider these $N$ bands as a basis of an $N$-dimensional complex vector space which a generic element $|\phi^N(\textbf{k})\rangle$, while (c) presents the $N$-dimensional vector space $\mathcal{H}^N_{\textbf{k}}$ as a line with each point of the line representing an $N$-dimensional normalized, lattice-periodic state $|\phi^N(\textbf{k})\rangle$, whereas $U(\textbf{k})$ is a unitary transformation between such vectors. (d) shows how we have such an $N$-dimensional complex vector space at each $\textbf{k}$-point, and can be considered as a typical fibre of the fibre bundle with the base space being the Brillouin zone.}
\label{fig:fibre1}
\end{figure}

Shortly after the discovery of the quantized Hall effect \cite{Klitzing1980}, the chiral edges states were connected to the \textit{topology} of the space of Bloch states defined on the Brillouin zone \cite{Thouless1982}. Describing the topology of a space, a global property, using its \textit{geometry}, a local characterization, goes back to the 19th century and was first expressed in the form of the Gauss-Bonnet theorem for a two-dimensional surface, where the integral of the curvature describing the local geometry is related to the Euler characteristic, a topological property \cite{Nakahara2003}. This theorem was extended by Chern to certain general spaces known as fibre bundles \cite{Nakahara2003}. As realized in \cite{Simon1983}, the space of Bloch states over the Brillouin zone of a crystal can be considered as a fibre bundle (Fig. \ref{fig:fibre1}), and it is possible to define a curvature of this space that is a specific realization of a Berry curvature. Different variations of this curvature have been shown to play an important role in a large variety of physical effects, from the quantized Hall and anomalous Hall effects to the spin Hall effect and intrinsic spin-orbit torque \cite{Nagaosa2010,Maekawa2017,Xiao2010b,Manchon2019}.

\subsubsection{Berry Connection}

The Berry curvature is a geometric quantity and is the curvature of the Berry connection, thus to discuss the former we need an understanding of the latter. We now present the preliminaries required to define the Berry connection. Let $\textbf{k}$ be a point of the Brillouin zone. Bloch's theorem ensures that an electronic state in a crystal can be written as $\langle\textbf{r}|\psi_n(\textbf{k})\rangle = e^{i\textbf{k}\cdot\textbf{r}}\langle\textbf{r}|u_n(\textbf{k})\rangle$, where $\langle\textbf{r}|u_n(\textbf{k})\rangle=\langle\textbf{r}+\textbf{R}|u_n(\textbf{k})\rangle$ with $\textbf{R}$ being a lattice vector and $n$ the band index. We can see that this puts a boundary condition on the states $|\psi_n(\textbf{k})\rangle$ as $\langle\textbf{r}+\textbf{R}|\psi_n(\textbf{k})\rangle=e^{i\textbf{k}\cdot\textbf{R}}\langle\textbf{r}|\psi_n(\textbf{k})\rangle$. This condition depends on $\textbf{k}$ so the states $|\psi_n(\textbf{k})\rangle$ are in different Hilbert spaces for different $\textbf{k}$. However, since the condition on $|u_n(\textbf{k})\rangle$ is $\textbf{k}$ independent --- they are lattice periodic for all $\textbf{k}$ ---, we can consider them to live in copies $\mathcal{H}_{\textbf{k}}$ of the same Hilbert space $\mathcal{H}$ consisting of normalized lattice periodic states and can define the Bloch Hamiltonian $H(\textbf{k})= e^{-i\textbf{k}\cdot\textbf{r}}He^{i\textbf{k}\cdot\textbf{r}}$ whose eigenstates they will be, $H(\textbf{k})|u_n(\textbf{k})\rangle=\varepsilon_n(\textbf{k})|u_n(\textbf{k})\rangle$.  Now consider an $N$-dimensional subspace  $\mathcal{H}^N_{\textbf{k}}\subset\mathcal{H}_{\textbf{k}}$ that is well-separated from the rest of the bands and is spanned by Bloch states $\{|u_1(\textbf{k})\rangle,\dots |u_{\alpha}(\textbf{k})\rangle,\dots,|u_N(\textbf{k})\rangle\}$, where $\alpha\in\{1,\dots,N\}$ is the band index within the subspace. An arbitrary, normalized lattice-periodic state $|\phi^N(\textbf{k})\rangle\in\mathcal{H}^N_{\textbf{k}}$ can be written as a linear combination of the Bloch states

\begin{equation}
\label{eq:phiBasis}
|\phi^N(\textbf{k})\rangle=\sum_{n=1}^N\phi^N_{\alpha}(\textbf{k})|u_{\alpha}(\textbf{k})\rangle,
\end{equation}

 with $\langle \phi^N(\textbf{k})|\phi^N(\textbf{k})\rangle=1$.  Acting with a $\textbf{k}$-dependent unitary transformation $U^N(\textbf{k}): |\phi^N(\textbf{k})\rangle\to U^N(\textbf{k})|\phi^N(\textbf{k})\rangle$, in this case an $N\times N$ matrix, provides another normalized state $U^N(\textbf{k})|\phi^N(\textbf{k})\rangle\in\mathcal{H}^N_{\textbf{k}}$ since $\langle\phi^N(\textbf{k})|(U^N(\textbf{k}))^{\dagger}U^N(\textbf{k})|\phi^N(\textbf{k})\rangle=1$. The unitary transformation $U^N(\textbf{k})$ acts on the Hamiltonian restricted to $\mathcal{H}^N_{\textbf{k}}$ as a similarity transformation $H^N(\textbf{k}) \to U^N(\textbf{k})H^N(\textbf{k})(U^N(\textbf{k}))^{\dagger}$ and consequently does not alter its spectrum of energy eigenvalues within $\mathcal{H}^N_{\textbf{k}}$. Acting with $U^N(\textbf{k})$ on the Bloch states $|u_{\alpha}(\textbf{k})\rangle \to U^N(\textbf{k})|u_{\alpha}(\textbf{k})\rangle$ amounts to a change of basis in each $\mathcal{H}^N_{\textbf{k}}$ and owing to $U^N(\textbf{k})$ being unitary the resulting basis maintains its orthonormal nature. We can now construct a larger space by attaching $\mathcal{H}^N_{\textbf{k}}$ to each $\textbf{k}$ (Fig. \ref{fig:fibre1}) and with the action of $U(\textbf{k})$ on each copy, this becomes a fibre bundle over the Brillouin zone. Suppose we take $N=1$ corresponding to $\mathcal{H}^1_{\textbf{k}}$ spanned by a single band, say $|u_{\alpha}(\textbf{k})\rangle$. The unitary transformation $U^1(\textbf{k})=e^{i\varphi_{\alpha}(\textbf{k})}$ is a 1x1 unitary matrix which is just a phase factor. This means, that the order of applying multiple such transformations does not matter, the operations commute and form what is known as a commutative, or abelian group usually labelled as $\mathcal{U}(1)$. On the other hand, if we consider multiple bands $N>1$, the unitary transformations will be matrices, and these do not necessarily commute which is why they form a non-abelian group, the unitary group $\mathcal{U}(N)$. Physically, the abelian case is important if we do not worry about degeneracies and only look at each band in isolation, whereas the non-abelian case is crucial when band degeneracies and multi-band effects are non-negligible. 

\begin{figure}
\includegraphics[height=3cm, width=8cm]{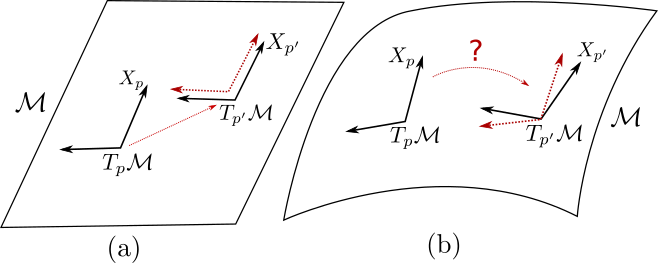}
\caption{(color online) The difficulty of identifying the tangent spaces at different points of a curved surface. (a) depicts two tangent planes $\text{T}_{p}\mathcal{M}$ and $\text{T}_{p'}\mathcal{M}$ at points $p$ and $p'$ of a flat space. We can trivially identify the two tangent planes  by translating the tangent vectors (shown as red). (b) presents the failure of this trivial identification if the embedded surface is curved. The vector resulting from "translating" a tangent vector from $p$ to $p'$ (shown in red) will not be a tangent vector at $p'$.}
\label{fig:curvedspace}
\end{figure}

Now that we have constructed the bundle of Bloch states over the Brillouin zone, what if we wanted to compare a state $|\phi^N(\textbf{k})\rangle$ at $\textbf{k}$ with a state $|\phi^N(\textbf{k}+d\textbf{k})\rangle$ at $\textbf{k}+d\textbf{k}$? This is a non-trivial question, since the states at $\textbf{k}$ are in $\mathcal{H}^N_{\textbf{k}}$ whereas those at $\textbf{k}+d\textbf{k}$ are in $\mathcal{H}^N_{\textbf{k}+d\textbf{k}}$ and these spaces are different. To be precise, they are in different copies of the same space, but these copies at different $\textbf{k}$-points cannot be naturally identified. An intuitive analogy is the case of tangent vectors $X_p$ at different points $p$ of a surface $\mathcal{M}$, as depicted on Fig. \ref{fig:curvedspace}. The vectors $X_p$ form a tangent plane $T_p\mathcal{M}$ at each point $p$ of $\mathcal{M}$. If $\mathcal{M}$ is a flat plane, case (a) on Fig. \ref{fig:curvedspace}, we can simply translate $X_p$ to another point $p'$ and obtain a tangent vector $X_{p'}$ in the tangent plane $T_{p'}\mathcal{M}$ at $p'$. On the other hand, if $\mathcal{M}$ is curved, case (b) on Fig. \ref{fig:curvedspace}, there is no natural way to move $X_p$ from $T_p\mathcal{M}$ to $X_{p'}$ at $T_{p'}\mathcal{M}$. Thus even though the tangent planes $T_p\mathcal{M}$ at all points $p$ are copies of the same plane $\mathbb{R}^2$, they cannot, in general, be naturally identified. The tangent plane $T_p\mathcal{M}$ at $p$ is the analogue of our $N$-dimensional Hilbert space $\mathcal{H}^N_{\textbf{k}}$ at $\textbf{k}$, and just as the notion of the "geometric" curvature of the curved surface can be heuristically looked at as the failure of identifying the tangent planes at different points, it is possible to define a curvature of the space of Bloch states over the Brillouin zone that quantifies the failure of identifying the Hilbert spaces at different points. It is important to stress that the latter curvature is not a result of the underlying Brillouin zone being "curved". In fact, the Brillouin zone is intrinsically flat, since it is just a cube with periodic boundary conditions. The visualization as a curved torus is a result of its embedding in three dimensional space, and the resulting curvature is extrinsic and dependent on the embedding. The key point is that "curvature" can be looked at as the failure of naturally identifying the fibre spaces at different points of the base space, just as how this failure for the tangent planes describes the "geometric" curvature. Therefore, to compare elements of the Hilbert spaces at two different points, we need a way to move the elements from the space at one point to the one at another, thereby connecting them. This is facilitated by a connection, and we now construct it.

Define a transporter $T^N(\textbf{k}+d\textbf{k},\textbf{k})$ as an $N\times N$ matrix that takes a state $|\phi^N(\textbf{k})\rangle\in\mathcal{H}^N_{\textbf{k}}$ to $\mathcal{H}^N_{\textbf{k}+d\textbf{k}}$ infinitesimally close:

\begin{equation}
\label{eq:Transporter}
|\phi^N(\textbf{k}+\textbf{dk})\rangle^T=T^N(\textbf{k}+d\textbf{k},\textbf{k})|\phi^N(\textbf{k})\rangle.
\end{equation}

We drop the $N$ superscripts until further notice. The transporter has to satisfy certain properties. First of all, if $d\textbf{k}=0$ we want the transporter to be the identity matrix $T(\textbf{k},\textbf{k})=1$. Next, we know that the states at all $\textbf{k}$ can be transformed by a unitary transformation $U(\textbf{k})$ such that $|\phi(\textbf{k})\rangle^U=U(\textbf{k})|\phi(\textbf{k})\rangle \in \mathcal{H}^N_{\textbf{k}}$. How would the transporter behave under a unitary transformation? To arrive at this we require the transformed transporter to relate the transformed states as follows

\begin{equation}
U(\textbf{k}+d\textbf{k})|\phi(\textbf{k}+d\textbf{k})\rangle^T=T^U(\textbf{k}+d\textbf{k},\textbf{k})U(\textbf{k})|\phi(\textbf{k})\rangle.
\end{equation}

Comparing with \eqref{eq:Transporter}, we find 

\begin{equation}
\label{eq:TTransform}
T^U(\textbf{k}+d\textbf{k},\textbf{k})=U(\textbf{k}+d\textbf{k})T(\textbf{k}+d\textbf{k},\textbf{k})U^{\dagger}(\textbf{k}).
\end{equation}

For infinitesimal $d\textbf{k}$ we write the transporter as

\begin{equation}
\label{eq:ConnectionDefinition}
T(\textbf{k}+d\textbf{k},\textbf{k})=1-i\textbf{A}(\textbf{k})\cdot d\textbf{k}+O(d\textbf{k}^2),
\end{equation}

where we defined the infinitesimal transporter $\textbf{A}(\textbf{k})$. Plugging this definition into (\ref{eq:TTransform}) we find the transformation rule for the infinitesimal transporter

\begin{equation}
\label{eq:ConnectionTransform}
\textbf{A}^U(\textbf{k})=U(\textbf{k})\textbf{A}(\textbf{k})U^{\dagger}(\textbf{k})+i(\partial_{\textbf{k}}U(\textbf{k}))U^{\dagger}(\textbf{k}),
\end{equation}

where $\partial_\textbf{k}\equiv (\partial/\partial k^1,\dots,\partial/\partial k^N)$ is the gradient. It is important to note that the infinitesimal transporter $\textbf{A}=(A_1,\dots,A_d)$ is a \textit{vector}  with as many components as the dimension $d$ of the Brillouin zone and the components $A_i$ are $N\times N$ \textit{matrices}. Conversely, we could think of $\textbf{A}$ as a matrix whose elements are $d$-component vectors. We are now ready to compare states at $\textbf{k}$ points that are infinitesimally close. We define the covariant derivative of states as follows

\begin{equation}
\label{eq:CovDerivativeDefinition}
|\phi(\textbf{k})\rangle-|\phi(\textbf{k})\rangle^T = \mathcal{D}_{\textbf{k}}|\phi(\textbf{k})\rangle\cdot d\textbf{k}.
\end{equation} 

This makes sense, since we first transported the state from $\textbf{k}-d\textbf{k}$ to $\textbf{k}$ and only then compared it to the one at $\textbf{k}$. Using the definition of the transporter and infinitesimal transporter from Eqs. (\ref{eq:Transporter}) and (\ref{eq:ConnectionDefinition}) in Eq. (\ref{eq:CovDerivativeDefinition}), we find 

\begin{equation}
\label{eq:CovDerivative}
\mathcal{D}_{\textbf{k}}|\phi(\textbf{k})\rangle=(\partial_{\textbf{k}}+i\textbf{A}(\textbf{k}))|\phi(\textbf{k})\rangle.
\end{equation}

Arriving at the transformation property of the covariant derivative using (\ref{eq:ConnectionTransform}) is a simple exercise leading to $\mathcal{D}_{\textbf{k}}^U=U(\textbf{k})\mathcal{D}_{\textbf{k}}U^{\dagger}(\textbf{k})$. The infinitesimal transporter $\textbf{A}$ is what facilitates the comparison of states at nearby $\textbf{k}$ points, thereby providing a connection between the spaces at these points and is, in fact, known as a connection form. The covariant derivative maps a state $|\phi(\textbf{k})\rangle$ in $\mathcal{H}^N_{\textbf{k}}$ to another state in the same space and it would be conducive to find the components of the resulting state in the Bloch basis. (In what follows, we reinstate the $N$ superscripts.) In order to do this, expand $|\phi^N(\textbf{k})\rangle$ in a basis of Bloch states according to \eqref{eq:phiBasis} and act with the covariant derivative:

\begin{equation}
\label{eq:covBasis1}
\mathcal{D}_{j}|\phi^N(\textbf{k})\rangle=
\sum_{\alpha=1}^N(\mathcal{D}_j\phi_{\alpha}^N(\textbf{k}))|u_{\alpha}(\textbf{k})\rangle+\phi_{\alpha}^N(\textbf{k})\mathcal{D}_j|u_{\alpha}(\textbf{k})\rangle,
\end{equation}

where $\mathcal{D}_j$ refers to the covariant derivative in direction $j$. Its action on a scalar function is $\mathcal{D}_j\phi_{\alpha}^N(\textbf{k})=\partial_j\phi_{\alpha}^N(\textbf{k})$, the standard derivative, whereas its action on the basis elements, should provide another state that is in turn a linear combination of basis elements:

\begin{equation}
\label{eq:connectionBasis}
\mathcal{D}_j|u_{\alpha}(\textbf{k})\rangle=i\sum_{\beta=1}^N A^N_{j\beta\alpha}(\textbf{k})|u_{\beta}(\textbf{k})\rangle.
\end{equation}

With these, \eqref{eq:covBasis1} becomes

\begin{equation}
\mathcal{D}_{j}|\phi^N(\textbf{k})\rangle=\sum_{\alpha=1}^N\bigg(\partial_j\phi^N_{\alpha}(\textbf{k})+i\sum_{\beta=1}^NA^N_{j\alpha\beta}(\textbf{k})\phi^N_{\beta}(\textbf{k})\bigg)|u_{\alpha}(\textbf{k})\rangle.
\end{equation}

Comparing with \eqref{eq:CovDerivative}, it is clear that $A^N_{j\alpha\beta}$ are the components of the connection form in the Bloch basis and from \eqref{eq:connectionBasis} we see that these components are related to how the basis, or frame changes covariantly as we move along the $\textbf{k}$-points of the Brillouin zone. Indeed, from \eqref{eq:connectionBasis} we have

\begin{equation}
\label{eq:connectionBasis2}
A^N_{j\alpha\beta}(\textbf{k})=-i\langle u_{\alpha}(\textbf{k})|\mathcal{D}_j|u_{\beta}(\textbf{k})\rangle.
\end{equation} 

However, that is all we can say; the specifics of the connection form are not thrown at us and choices have to be made. First, we can realize that an inner product between the states is available at each $\textbf{k}$, so a natural requirement for the transporter would be to maintain the modulus of the state it transports

\begin{equation}
^T\langle \phi^N(\textbf{k}+d\textbf{k})|\phi^N(\textbf{k}+d\textbf{k})\rangle^T = \langle \phi^N(\textbf{k})|\phi^N(\textbf{k})\rangle.
\end{equation}

As can be seen by using the definition (\ref{eq:Transporter}), this requirement is satisfied if $T^N$ is unitary $T^N(\textbf{k}+d\textbf{k},\textbf{k})(T^N(\textbf{k}+d\textbf{k},\textbf{k}))^{\dagger}=1$. The infinitesimal version of this condition gives for the connection form $(\textbf{A}^N)^{\dagger}=\textbf{A}^N$, or in other words, that the connection form is Hermitian. If we are only looking at a single band, then the transporter is just a complex number and not an $N\times N$ matrix, meaning that the connection form $\textbf{A}$ is just a vector of numbers and not a vector of $N\times N$ matrices, so Hermiticity implies that the connection form is a vector of \textit{real} numbers. However, while being Hermitian is a considerable restriction on what the connection can be, it leaves room for further constraints.

In order to arrive at a concrete connection we use the fact that we have been working in $N$-dimensional subspaces $\mathcal{H}^N_{\textbf{k}}$ of the total, $N_{\text{tot}}$-dimensional Hilbert space $\mathcal{H}$. Note that $N_{\text{tot}}$ could be infinite, however, in practical calculations, only a finite number of bands are taken to span the total Hilbert space. The states of the Bloch basis $|u_{\alpha}(\textbf{k})\rangle$ with $\alpha\in\{1,\dots,N\}$ spanning $\mathcal{H}^N_{\textbf{k}}$ are not arbitrary, but eigenstates of the Bloch Hamiltonian, and we already know them as $N_{\text{tot}}$-component vectors in the standard basis of the total space $\mathcal{H}$. Indeed, the standard basis consists of the $N_{\text{tot}}$-component column vectors

\begin{equation}
|e_1\rangle=
\begin{pmatrix}1\\0\\\vdots\\0\\\vdots\\0\end{pmatrix},\dots,
|e_n\rangle=
\begin{pmatrix}0\\0\\\vdots\\1\\\vdots\\0\end{pmatrix},\dots,
|e_{N_{\text{tot}}}\rangle=
\begin{pmatrix}0\\0\\\vdots\\0\\\vdots\\1\end{pmatrix},
\end{equation}

and we can write the Bloch basis states spanning $\mathcal{H}^N_{\textbf{k}}$ as linear combinations 

\begin{equation}
\label{eq:uDef}
|u_{\alpha}(\textbf{k})\rangle=\sum_{n=1}^{N_{\text{tot}}}u_{n\alpha}(\textbf{k})|e_n\rangle =\begin{pmatrix}u_{1\alpha}(\textbf{k})\\\vdots\\ u_{n\alpha}(\textbf{k})\\ \vdots\\ u_{N_{\text{tot}}\alpha}(\textbf{k})\end{pmatrix},
\end{equation}

where $u_{n\alpha}(\textbf{k})$ are the elements of an $N_{\text{tot}}\times N$ matrix. The components \eqref{eq:connectionBasis2} of the connection form  thus become

\begin{equation}
\label{eq:BerryConnComp}
\begin{split}
A^{N,B}_{j\alpha\beta}(\textbf{k})&=-i\sum_{n=1}^{N_{\text{tot}}}u^*_{n\alpha}(\textbf{k})\partial_{j}u_{n\beta}(\textbf{k})
\\
&\to-i\langle u_{\alpha}(\textbf{k})|\partial_ju_{\beta}(\textbf{k})\rangle,
\end{split}
\end{equation}

and we can recognize the components of Berry's connection labelled by the $B$ superscript. Crucially, note that $\alpha,\beta \in \{1,\dots,N\}$ are indices spanning an $N$-dimensional \textit{sub}-space of the total space $\mathcal{H}$, whereas the inner product is taken over the entire space. It is straightforward to see that $A^{N,B}_{j\alpha\beta}(\textbf{k})=(A^{N,B}_{j\beta\alpha}(\textbf{k}))^*$ rendering the Berry connection form Hermitian.\\

\begin{figure}
\includegraphics[height=5cm, width=8cm]{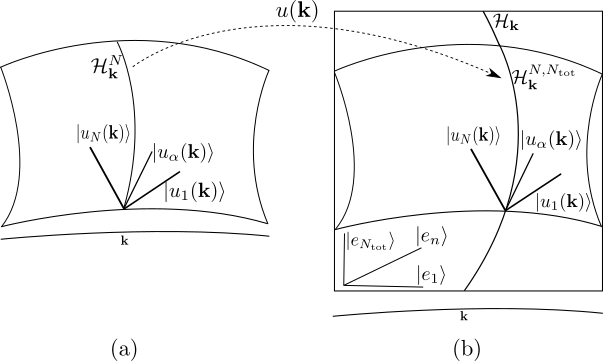}
\caption{A subspace as an embedding. (a) depicts a curved Hilbert bundle with $N$-dimensional fibres $\mathcal{H}^N_{\textbf{k}}$ and a local basis or frame of Bloch states. (b) shows the curved Hilbert bundle as embedded in a flat one with $N_{\text{tot}}$-dimensional fibres $\mathcal{H}_{\textbf{k}}$. The embedding map is $u(\textbf{k})$ and takes $\mathcal{H}^N_{\textbf{k}}$ to $\mathcal{H}^{N,N_{\text{tot}}}_{\textbf{k}}\subset\mathcal{H}_{\textbf{k}}$ (see text).}
\label{fig:bundleMap}
\end{figure}

Further insight can be gained by adapting the discussion of \cite{Atiyah1978} to our case and taking a closer look at the components $u_{n\alpha}(\textbf{k})$ of the Bloch states. These are the elements of an $N_{\text{tot}}\times N$ matrix $u(\textbf{k})$ and the latter can be interpreted as an embedding map $u(\textbf{k}):\mathcal{H}^N_{\textbf{k}}\to \mathcal{H}^{N,N_{\text{tot}}}_{\textbf{k}}\subset\mathcal{H}$, where $\mathcal{H}^{N,N_{\text{tot}}}_{\textbf{k}}$ is $\mathcal{H}^N_{\textbf{k}}$ thought of as embedded in $\mathcal{H}_{\textbf{k}}$ (see Fig. \ref{fig:bundleMap}). The adjoint $u^{\dagger}(\textbf{k}):\mathcal{H}\to \mathcal{H}^N_{\textbf{k}}$ is an $N\times N_{\text{tot}}$ matrix and maps a state of the total space to $\mathcal{H}^N_{\textbf{k}}$. We thus have the combined maps $u^{\dagger}(\textbf{k}) u(\textbf{k}):\mathcal{H}^N_{\textbf{k}}\to\mathcal{H}^N_{\textbf{k}}$ and $u(\textbf{k})u^{\dagger}(\textbf{k}):\mathcal{H}_{\textbf{k}}\to \mathcal{H}^{N,N_{\text{tot}}}_{\textbf{k}}$. Following from the orthonormality of Bloch states, the definition \eqref{eq:uDef} yields $u^{\dagger}(\textbf{k}) u(\textbf{k})=1$, the identity on $\mathcal{H}^N_{\textbf{k}}$, meaning that $(u(\textbf{k})u^{\dagger}(\textbf{k}))(u(\textbf{k})u^{\dagger}(\textbf{k}))=u(\textbf{k})u^{\dagger}(\textbf{k})$. In other words, the map $u(\textbf{k})u^{\dagger}(\textbf{k})\equiv P(\textbf{k})$ satisfies $P^2(\textbf{k})=P(\textbf{k})$ and is thus a projector.\\
Next, consider a state $|\phi^N(\textbf{k})\rangle \in \mathcal{H}^N_{\textbf{k}}$ and apply the map $u(\textbf{k})$ to obtain $u(\textbf{k})|\phi^N(\textbf{k})\rangle \in \mathcal{H}^{N,N_{\text{tot}}}_{\textbf{k}}$. We can now take the partial derivative of the latter to get $\partial_{j}(u|\phi^N\rangle)\in \mathcal{H}_{\textbf{k}}$, where we dropped the $\textbf{k}$ arguments for clarity. Note that the partial derivative is not necessarily in $\mathcal{H}^{N,N_{\text{tot}}}_{\textbf{k}}$ so we have to apply the projector $P\partial_{j}(u|\phi^N\rangle) \in\mathcal{H}^{N,N_{\text{tot}}}_{\textbf{k}}$. Expanding, we have

\begin{equation}
\begin{split}
P\partial_{j}(u|\phi^N\rangle)&=uu^{\dagger}((\partial_j u)|\phi^N\rangle+u\partial_j|\phi^N\rangle)
\\
&=u(\partial_j+u^{\dagger}\partial_ju)|\phi^N\rangle.
\end{split}
\end{equation} 

We can read off the covariant derivative
\begin{equation}
\mathcal{D}_i|\phi^N(\textbf{k})\rangle=(\partial_j+u^{\dagger}(\textbf{k})\partial_ju(\textbf{k}))|\phi^N(\textbf{k})\rangle \in \mathcal{H}^N_{\textbf{k}}.
\end{equation}

Comparing with \eqref{eq:CovDerivative}, it is apparent that the Berry connection form is given by $A^{N,B}_j(\textbf{k})=-iu^{\dagger}(\textbf{k})\partial_ju(\textbf{k})$ with the components presented in \eqref{eq:BerryConnComp}.\\

As a final remark, consider the case of $N=N_{\text{tot}}$. Now the map $u(\textbf{k})$ becomes an $N_{\text{tot}}\times N_{\text{tot}}$ unitary matrix $U(\textbf{k})$ and the projector degenerates to the identity. The Berry connection form can then be written as $A^{N_{\text{tot}},B}_j(\textbf{k})=-iU^{\dagger}(\textbf{k})\partial_jU(\textbf{k})$. Consider a unitary transformation by $U(\textbf{k})$ itself. The connection form transforms under the latter according to \eqref{eq:ConnectionTransform} and we obtain

\begin{equation}
A^{N_{\text{tot}},B}_j\to U(-iU^{\dagger}\partial_jU)U^{\dagger}+i(\partial_{j}U)U^{\dagger}=0,
\end{equation} 

meaning that the connection can be made to vanish. Echoing our earlier discussion, a vanishing connection results in the infinitesimal transporter \eqref{eq:ConnectionDefinition} being the identity, thereby allowing the standard translation of states between spaces above different points of the Brillouin zone, rendering the space of states curvature free. It is thus crucial to consider a proper subspace with $N<N_{\text{tot}}$ in order to obtain a non-trivial Berry connection.\\

The Berry connection is intimately related to a physical observable, the position operator. The latter is usually defined as the gradient operator $\textbf{X}\equiv i\partial_{\textbf{k}}$. Suppose we perform a $\textbf{k}$-dependent unitary transformation, then $\textbf{X}\to X+iU(\textbf{k})\partial_{\textbf{k}}U^{\dagger}(\textbf{k})$. A physical observable should transform covariantly and we can define the physical position operator for the $N$-dimensional subspace to be the covariant derivative $\textbf{r}=i(\partial_{\textbf{k}}+i\textbf{A}^{N,B})$. This manner of defining the position operator was first proposed by \citet{Blount1962} and is directly related to measurable physical effects \cite{Nagaosa2010,Xiao2010b}. From this interpretation, it is also clear how the connection form should transform under time-reversal. Position transforms as $\mathcal{T}\textbf{r}\mathcal{T}^{\dagger}=\textbf{r}$, thus 
\begin{equation}
\mathcal{T}\textbf{A}^{N,B}(\textbf{k})\mathcal{T}^{\dagger}=\textbf{A}^{N,B}(-\textbf{k}),
\end{equation}
where we used the fact that $\textbf{k}$ is the crystal momentum and changes sign under $\mathcal{T}$.

\subsubsection{Berry Curvature\label{ss:Berrycurv}}

Taking a look at the expression \eqref{eq:CovDerivative} of the covariant derivative, the connection form $\textbf{A}$ is what separates it from being a plain partial derivative. Should the connection form be made to vanish --- a possibility when $\textbf{A}=-iU^{\dagger}\partial_{\textbf{k}}U$ for a unitary matrix $U$ as discussed above --- the Hilbert spaces attached to different points can be trivially identified and the change of the states between different $\textbf{k}$ points can be quantified by the partial derivative. Since the failure of this trivial identification is what leads to the "curvature" of the space, the "further" the covariant derivative is from being a partial derivative, the more curved the space is. The way to quantify this, is to use the fact that partial derivatives in different directions commute: $[\partial_{i},\partial_j]=\partial_i\partial_j-\partial_j\partial_i=0$. We can thus measure how "far" the covariant derivative is from being a partial derivative by taking the commutator of its action in different directions $[\mathcal{D}_i,\mathcal{D}_j]$, and this quantity is known as the curvature of the connection, since it quantifies how "far" the connection is from the possibility of it vanishing. Using expression (\ref{eq:CovDerivative}) for the covariant derivative, we  have

\begin{equation}
F_{ij}(\textbf{k})\equiv -i[\mathcal{D}_i,\mathcal{D}_j]=\partial_iA_j-\partial_jA_i+i[A_i,A_j].
\end{equation}

It is straightforward to show that the curvature vanishes, as it should, if we take $A_j=-iU^{\dagger}\partial_{j}U$ for the connection form. Assuming that the connection is given by the Berry connection form $\textbf{A}^{N,B}(\textbf{k})$ in \eqref{eq:BerryConnComp} for some $N$-dimensional subspace, we obtain the Berry curvature $F^{N,B}_{ij}(\textbf{k})$. Recall that for the case of a single band, the connection form is a vector of real numbers and not of matrices, so the commutator of the connection components is zero, and we arrive at the abelian Berry curvature:

\begin{equation}
\label{eq:AbelianBerry}
F_{ij}^{1,B}=\partial_iA^{1,B}_j-\partial_jA^{1,B}_i
\end{equation}

In the general case of multiple bands, such as degeneracies, the connection components will be matrices and we cannot neglect the commutator, meaning that the curvature of this space spanned by the degenerate bands is described by a non-abelian Berry curvature.

The transformation rule for the curvature under unitary transformations can be found via (\ref{eq:ConnectionTransform}) for the connection form and is $(F_{ij}^{N,B})^U(\textbf{k})=U(\textbf{k})F^{N,B}_{ij}(\textbf{k})U^{\dagger}(\textbf{k})$. In the abelian case, $U(\textbf{k})$ is just a complex number and this becomes $(F_{ij}^{1,B})^U(\textbf{k})=F_{ij}^{1,B}(\textbf{k})$, meaning that it is invariant. Since these unitary transformations are redundancies in our descriptions - for instance in the abelian case it is simply changing the phase of the Bloch function -, only quantities that are not affected by them can contribute to physically measurable properties. Since $F_{ij}^{1,B}(\textbf{k})$ is invariant, it is one such quantity and contributes to conductivity, among other things. On the contrary the non-abelian Berry curvature is not invariant, but we can form its trace $\text{tr}(F^{N,B}_{ij}(\textbf{k}))$, where the trace is over the $N\times N$ matrices. This quantity is invariant due to the cyclicity of the trace. Further invariant quantities can be defined as integrals of the Berry connection over a closed path in the Brillouin zone, yielding the Berry phase, widely discussed in the literature \cite{Xiao2010b}.

According to Blount's prescription described above, the covariant derivative can be interpreted as the position operator $r_i=i\mathcal{D}_i$ and the Berry curvature becomes the commutator of the position operator in different directions, indeed $F^{N,B}_{ij}(\textbf{k})=-i[\mathcal{D}_i,\mathcal{D}_j]=i[r_i,r_j]$ \cite{Xiao2010b}. This result can provide a physical interpretation of the Berry curvature by thinking of localization. The eigenvalues of the coordinate operator give the locations of the electron in that particular coordinate direction. If this does not commute with the coordinate operator in another direction, it means that we cannot diagonalize both simultaneously, so only one of the coordinates can be given precisely, while the other cannot, resulting in not being able to localize the electron and tell its precise location \cite{Bacry1964,Berard2006}. The Berry curvature thus quantifies the resistance of the electron towards localizing, for example by the way it spreads transverse to the force applied on it resulting in the anomalous Hall effect.\\

The behaviour of the Berry curvature under time-reversal can be arrived at straightforwardly if we look at it as the commutator of the position operators. We have

\begin{equation}
\label{eq:BerryCurvTR}
\mathcal{T}F_{ij}^{N,B}(\textbf{k})\mathcal{T}^{\dagger}=-F_{ij}^{N,B}(-\textbf{k}),
\end{equation}

where the minus sign comes from the action of the anti-unitary $\mathcal{T}$ on $i$ in front of the commutator.\\

The Berry curvature is a \textit{local} quantity since it is explicitly dependent on $\textbf{k}$, and describes the local geometry of the space of Bloch states over the Brillouin zone. It is possible to get global information out of the curvature by, for example, integrating its trace over the Brillouin zone. However, since the curvature is a rank 2 tensor field with components $F^{N,B}_{ij}(\textbf{k})$, slightly deforming the space would change the directions of the vectors and tensors, and so this might be a global quantity, but it is too sensitive to the local geometry and consequently not topological. On the other hand, we can contract the tensor indices with the totally antisymmetric Levi-Civita symbol $\varepsilon_{ijk\dots}$, as follows.\\

In two dimensions we have $\varepsilon_{ij}\text{tr}(F^{N,B}_{ij}(\textbf{k}))$, where $i,j\in \{x,y\}$ and we sum over the repeated indices. Since all indices are summed over, this is a scalar, and integrating it over the two-dimensional Brillouin zone provides topological information about the space of Bloch functions. We have

\begin{equation}
c^N=\frac{1}{2\pi}\int d^2k\,\frac12\varepsilon_{ij}\text{tr}(F^{N,B}_{ij}(\textbf{k})).
\end{equation}

Indeed, this integral is an example of a topological invariant. It can be rewritten in a simpler form. As an $N\times N$ matrix, $F^{N,B}_{ij}$ is Hermitian so it is possible to diagonalize and split it into abelian curvatures on 1-dimensional complex bundles over the Brillouin zone. The trace is then the sum of the abelian curvatures. This is a special case of a general theorem referred to as the splitting principle for Chern classes \cite{Nakahara2003}, of which we looked at the case of the first Chern class. In general the Chern Classes cannot distinguish between an $N$-dimensional complex vector bundle and the sum of $N$ 1-dimensional complex vector bundles \cite{Nakahara2003}. We thus have

\begin{equation}
c^N=\frac{1}{2\pi}\sum_{\alpha=1}^N\int d^2k\,\frac12\varepsilon_{ij}F^{1,B,\alpha}_{ij}(\textbf{k})=\sum_{\alpha=1}^Nc^{\alpha}.
\end{equation}

The quantities $c^{\alpha}$ are integers known as first Chern numbers. Using \eqref{eq:AbelianBerry} we have $\varepsilon_{ij}F^{1,B}_{ij}=2(\partial_xA^{1,B}_{y}-\partial_yA^{1,B}_{x})=2({\bm\nabla}\times\textbf{A}^{1,B})_z$ so

\begin{equation}
c^N=\frac{1}{2\pi}\sum_{\alpha=1}^N\int d^2k\,({\bm\nabla}\times\textbf{A}^{1,B,\alpha})_z,
\end{equation}

and the integers $c^{\alpha}$ can be interpreted as a flux through the 2-dimensional Brillouin zone. Should the $N$-dimensional space be that of filled electronic valence bands, the integer $c^N$ is referred to as the TKNN-invariant \cite{Thouless1982,Xiao2010b,Hasan2010} and is directly related to the quantization of the Hall conductivity.\\

In three dimensions we have $\varepsilon_{ijk}\text{tr}(F_{jk}(\textbf{k}))$, where $i,j,k\in \{x,y,z\}$, and since this quantity has one free index $i$ it is a vector, so its integral cannot be topological, it is too sensitive to the local geometry. Indeed, in three-dimensional transport the intrinsic conductivity resulting from this integral is \textit{geometrical} and \textit{not} topological. Going further to four dimensions, it is again possible to define a scalar $\varepsilon_{ijkl}\text{tr}(F_{ij}(\textbf{k})F_{kl}(\textbf{k}))$, integrating this gives another topological invariant, the so-called second Chern number and it is important in the 4d quantized Hall effect \cite{Qi2008,Zilberberg2018,Lohse2018}.\\

From \eqref{eq:BerryCurvTR} we see that in the presence of time-reversal symmetry, the Berry curvature should satisfy $F_{ij}^{N,B}(\textbf{k})=-F_{ij}^{N,B}(-\textbf{k})$ and thereby the integral of its trace over the Brillouin zone should vanish. This means that the first Chern numbers are all zero and do not provide any topological information. Note however that the second Chern numbers related to the square of the Berry curvature do not necessarily vanish and yield a quantized Hall effect even in the presence of time-reversal, albeit in 4d. Nevertheless, It is possible to define a topological invariant even in the time-reversal symmetric case, the celebrated $\mathbb{Z}^2$ invariant \cite{Hasan2010}. The presence of $\mathcal{T}$-symmetry leads to Kramers degeneracy in electron systems (see section \ref{ssec:degaf}) and as a consequence every band becomes doubly degenerate with opposite spin-polarizations in the absence of spin-orbit coupling. Thus even though the total first Chern number $c_{\uparrow}+c_{\downarrow}=0$ vanishes, the difference $c_{\uparrow}-c_{\downarrow}$ does not and is known as the spin Chern number. A $\mathbb{Z}^2$ invariant $\nu$ can then be defined as the parity $\nu=\frac12(c_{\uparrow}-c_{\downarrow}) \text{mod} 2$ of the spin Chern number. Should we turn on spin-orbit coupling a more involved discussion is required since the double degeneracy of the bands is lifted and it only remains above certain special time-reversal invariant $\textbf{k}$-points. We refer to \cite{Hasan2010} for a review emphasizing the physical aspects and \cite{Kaufmann2016} for a mathematical perspective.\\

Our discussion of the Berry connection and curvature was built on the existence of Bloch states, a characteristic of periodic systems. While the physical applications briefly discussed here apply to electrons, any quasiparticle in a periodic system can experience such geometric effects. This is particularly the case of magnetic excitations such as magnons, addressed further below.\\

\subsection{Snapshot of selected antiferromagnetic crystals}
To bridge this first section with the physical properties discussed in the review, Table \ref{Table1} shows a list of selected antiferromagnetic compounds that are currently being intensively investigated for their topological properties. This list is far from comprehensive and only aims to illustrate the diversity of magnetic textures and associated topological properties one can encounter in antiferromagnets. For the sake of completeness, we also add NiO, a collinear layered antiferromagnet that does not display topological electronic or magnonic properties but has attracted substantial interest lately in the context of spintronics, including ultrafast dynamics \cite{Satoh2010}, magnonic spin transport \cite{Hahn2014,Wang2014d}, spin Seebeck effect \cite{Holanda2017,Prakash2016}, spin Hall magnetoresistance \cite{Hou2017,Baldrati2018a} and current-driven torque \cite{Moriyama2018b,Chen2018}.

\begin{table*}[ht!]
\begin{tabular}{c|ccccc}
Antiferromagnet & Chemical composition & Space group & Magnetic space group & Remarkable feature\\\\\hline\hline
\begin{minipage}{0.12\textwidth}
\includegraphics[width=\linewidth]{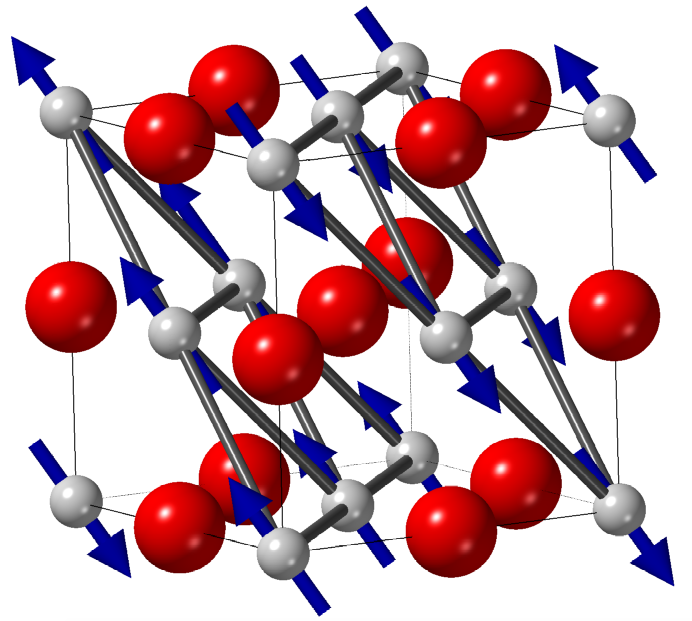}
\end{minipage}& NiO & F$m\bar{3}m$ & C$_{2c}2/m'(\text{C}_c2/c)$ &\begin{tabular}{c}
THz excitations\\Magnon transport\\ Spin Hall magnetoresistance
\end{tabular}\\\hline\\
\begin{minipage}{0.12\textwidth}
\includegraphics[width=\linewidth]{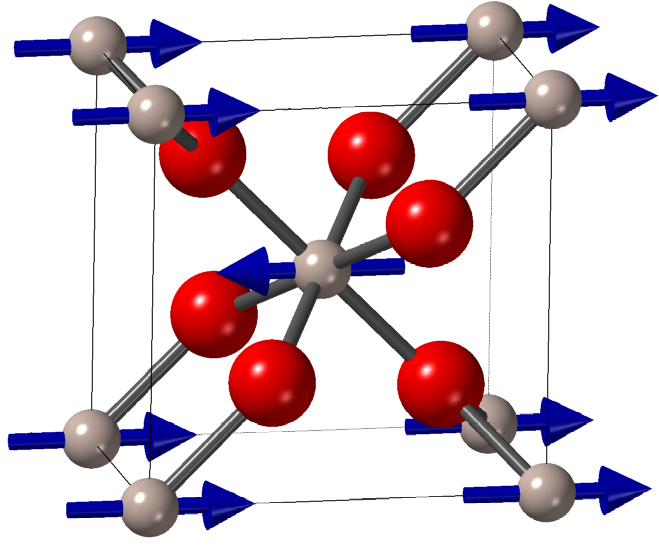}
\end{minipage}&RuO$_2$, MnF$_2$& P4$_2/mnm$& P4$_2'/mnm'$ &\begin{tabular}{c}
Crystal Hall effect \\Spin polarization \\Magnetic spin Hall effect
\end{tabular}\\\hline\\
\begin{minipage}{0.12\textwidth}
\includegraphics[width=\linewidth]{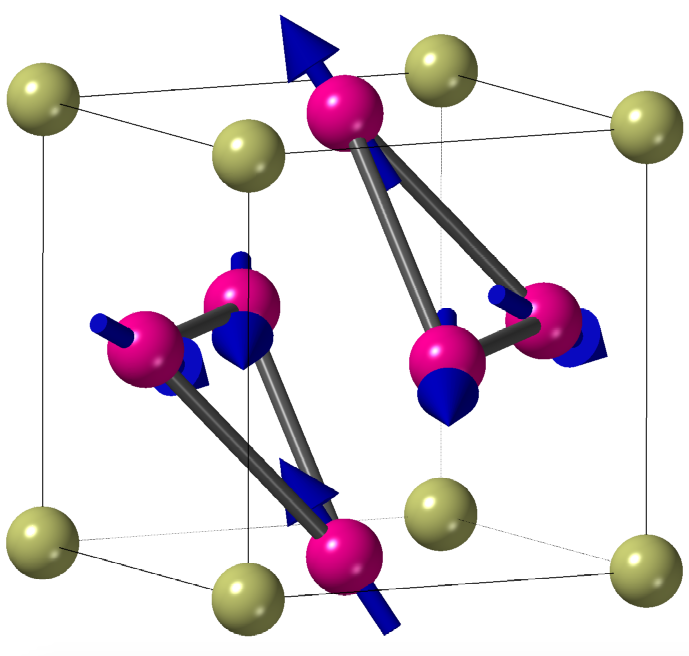}
\end{minipage}&IrMn$_3$& P$m\bar{3}m$& P$\bar{3}1m'$ & \begin{tabular}{c} Magnetooptical Kerr effect\\Anomalous Hall effect\\ Spin-momentum locking
\end{tabular}\\\hline\\
\begin{minipage}{0.15\textwidth}
\includegraphics[width=\linewidth]{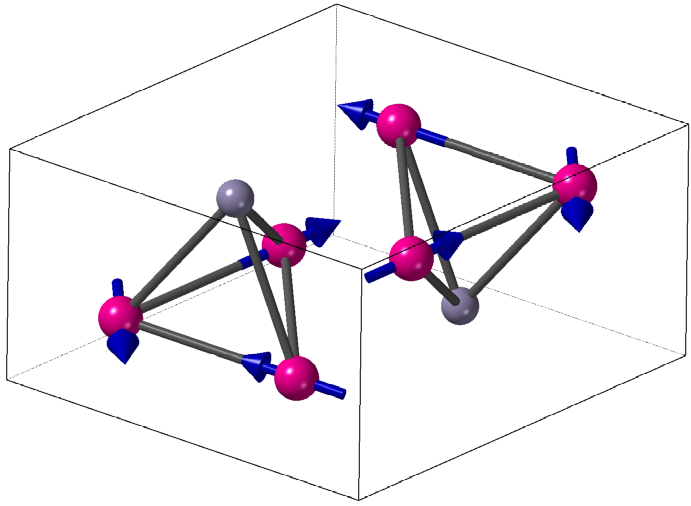}
\end{minipage}&Mn$_3$Sn&P6$_3mmc$&P$2'/c'$& \begin{tabular}{c} Magnetooptical Kerr effect\\Anomalous Hall effect\\ Magnetic spin Hall effect\\Spin-momentum locking
\end{tabular}\\\hline\\
\begin{minipage}{0.12\textwidth}
\includegraphics[width=\linewidth]{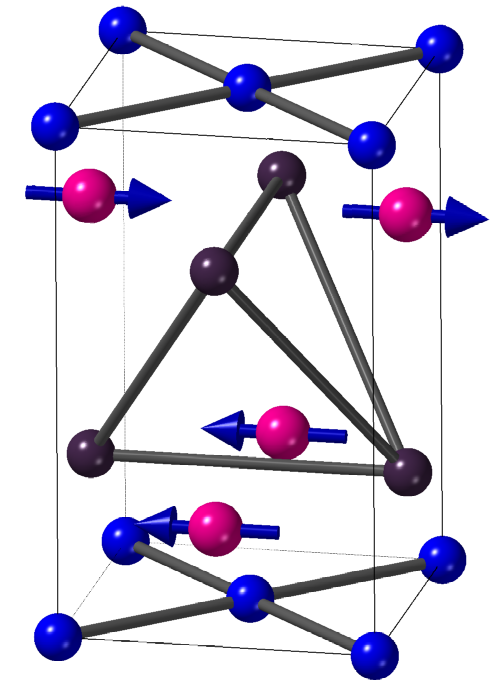}
\end{minipage}&CuMnAs& P$mmn$&P$m'mn$& Dirac semimetal\\\hline\\
\begin{minipage}{0.15\textwidth}
\includegraphics[width=\linewidth]{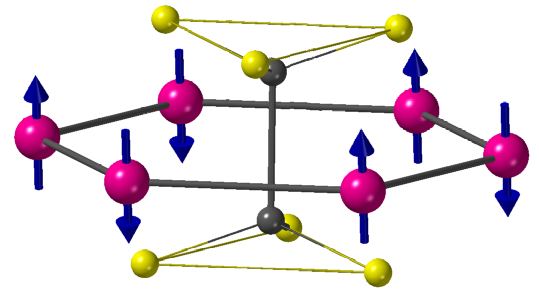}
\end{minipage}& MnPS$_3$& C$2/m$&C$2'/m$& Spin Nernst effect\\\hline\\
\begin{minipage}{0.15\textwidth}
\includegraphics[width=\linewidth]{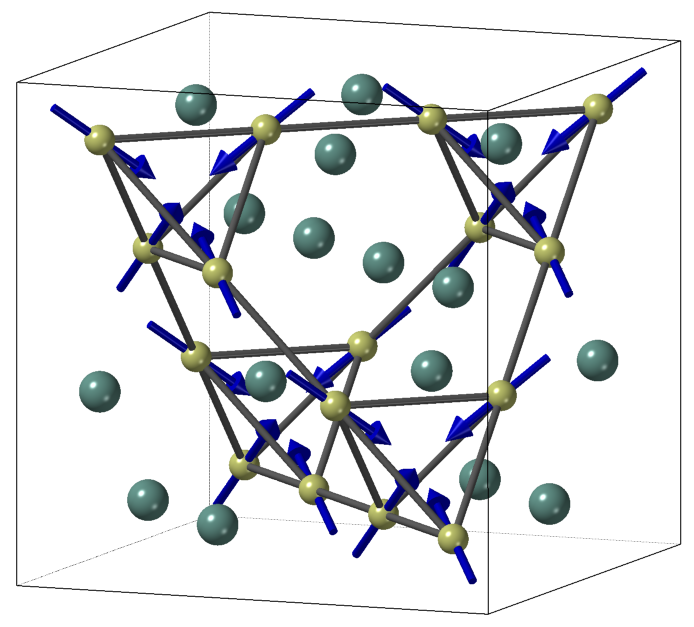}
\end{minipage}&Y$_2$Ir$_2$O$_7$&F$d\bar{3}m$&F$d\bar{3}m$& Weyl semimetal\\\hline\\
\begin{minipage}{0.15\textwidth}
\includegraphics[width=\linewidth]{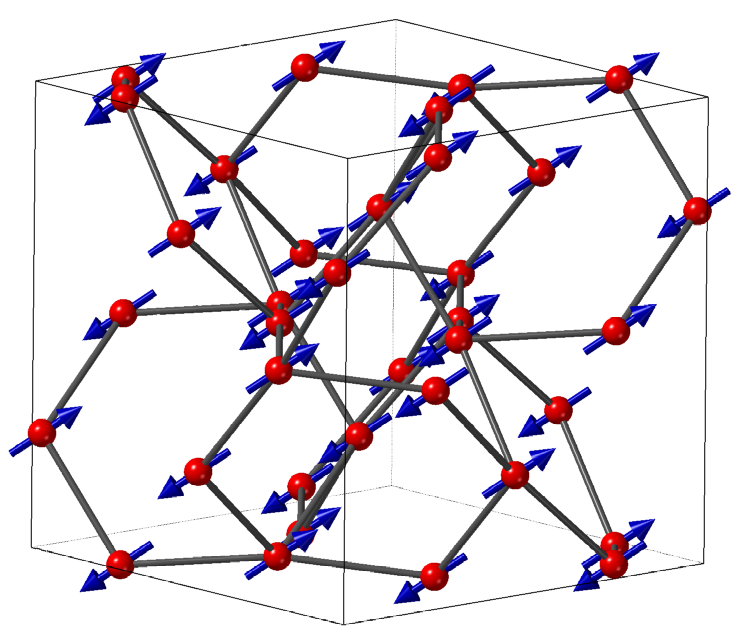}
\end{minipage}& Cu$_3$TeO$_6$ & I$a\bar{3}$& R$\bar3$'& Topological magnon bands
\end{tabular}\caption{Selected antiferromagnetic compounds that display remarkable topological properties. In the cases of Mn$_3$Sn, Y$_2$Ir$_2$O$_7$ and Cu$_3$TeO$_6$ only the magnetic ions and selected nonmagnetic ions are represented for readability. \label{Table1}}
\end{table*}

\section{Anomalous Transport: Electrons and Magnons \label{s:ahe}} 

In this section, we address the anomalous transport of electrons and magnons in antiferromagnets. In fact, because antiferromagnets can exist under a wide range of magnetic configurations (collinear, coplanar, non-coplanar etc.), the charge, spin and magnon transport properties display a remarkable variety that has only been uncovered very recently. As explained in details in the previous section, the key ingredient for these unconventional transport properties is the emergence of a Berry curvature in momentum space either due to spin-orbit coupling, to the magnetic texture or both. The present section aims to provide an up-to-date overview of the experimental and theoretical research along this direction.


\subsection{Anomalous Transport of Electrons}

Hall currents, defined as the flow of particle, energy or (pseudo)spin {\em transverse} to an injected current, have intrigued and inspired researchers for more than a century since the pioneering observations by E. H. Hall \cite{Hall1879,Hall1881}. Two main effects have attracted most attention in the past decades due to their connection with the topology of electronic band structures: the anomalous Hall effect \cite{Nagaosa2010} and the spin Hall effect \cite{Sinova2015}. These two effects share a number of common features. The charge (spin) current generated by the anomalous (spin) Hall effect reads
\begin{eqnarray}\label{eq:jch}
{\bf J}_{\rm H}^c&=&\theta_{\rm h}{\bf m}\times{\bf J}_{0}^c,\\\label{eq:jsh}
\mathcal{J}_{\rm H}^\gamma&=&(\hbar/2e)\theta_{\rm sh}{\bm\sigma}_\gamma\times{\bf J}_{0}^c,
\end{eqnarray}
where $\theta_{\rm h}$ ( $\theta_{\rm sh}$) is the ratio between the anomalous (spin) Hall conductivity and the longitudinal conductivity and referred to as the "Hall angle". The charge Hall current density is in unit of A$\cdot$m$^{-2}$, whereas the spin Hall current density is in units of $(\hbar/2)\;$s$^{-1}\cdot$m$^{-2}$. We also denote ${\bf m}$ as the magnetic order, ${\bm\sigma}_\gamma$ is the spin polarization direction of the spin current and ${\bf J}_{0}^c$ is the injected charge current. Notice that while ${\bf J}_{\rm H}^c$ is a vector, $\mathcal{J}_{\rm H}^s$ is a 3$\times$3 tensor in spin and real spaces. The physical mechanisms responsible for both effects are essentially the same and have been investigated thoroughly over the past decades (for extensive reviews, see \onlinecite{Nagaosa2010},  \onlinecite{Sinova2015}). Without entering into these details, we stress out that these mechanisms can be parsed into "intrinsic" and "extrinsic" mechanisms. The former is governed by the band structure only, and more specifically by the Berry curvature of the electron states, whereas the latter mechanisms (side-jump and skew scattering) stem from scattering against impurities and disorder. Only the "intrinsic" anomalous and spin Hall effects are of topological origin. Up till recently, the investigation of spin Hall effect has been limited to nonmagnetic metals and (more recently) ferromagnets, and anomalous Hall effect has been mostly limited to ferromagnets and uncompensated antiferromagnets (see, e.g., \cite{Asa2020}). The search for novel forms of anomalous transport in antiferromagnets has led to the identification of several new classes of Hall effects, beyond the conventional anomalous and spin Hall phenomena introduced above.

\subsubsection{Charge Hall effect}
\paragraph{Noncollinear antiferromagnets}
\label{subsec:AHE}

The fundamental requirement for anomalous Hall effect is time-reversal symmetry breaking. As discussed in Section \ref{ssec:degaf}, in magnetic crystals time-reversal symmetry is naturally broken by the presence of remanent finite magnetic moments. However, if there exists a crystal symmetry $\mathcal{O}$ or a lattice translation ${\bm\tau}_s$ such that, combined with time-reversal $\mathcal{T}$, the symmetry $\mathcal{TO}$ or $\mathcal{T}{\bm\tau}_s$ leaves the magnetic unit cell invariant, then time-reversal symmetry is {\em effectively} preserved. Examples of antiferromagnets exhibiting doubly degenerate band structure are given in see Fig. \ref{AF-degeneracy} and related discussion. Therefore, breaking this combined symmetry {\em may} result in effective time-reversal symmetry breaking and therefore, anomalous Hall effect shall be allowed. For instance, in the noncollinear, coplanar kagom\'e lattice [Fig. \ref{AF-degeneracy}(b)], the mirror symmetry with respect to the kagom\'e plane $\mathcal{M}$ plays the same role as $\mathcal{T}$, so that $\mathcal{TM}$ is an effective time-reversal symmetry. \citet{Chen2014} suggested that turning on the spin-orbit coupling breaks the mirror symmetry so that $\mathcal{TM}$ is effectively broken, resulting in a finite anomalous Hall effect. 

An alternative view on the anomalous Hall effect in antiferromagnet was put forward by \citet{Suzuki2017,Suzuki2019} in the context of cluster multipole expansion. Cluster multipole moments are defined as the multipole expansion of the total magnetic moment of a cluster of atoms in the magnetic unit cell. A cluster is a subset of atoms that transforms into another cluster in the magnetic unit cell upon crystal point symmetry. In antiferromagnets, the net magnetic moment of such a cluster is necessarily zero. In cluster multipole expansion theory, this net magnetic moment is nothing but the rank-$1$ multipole moment (equivalent to a magnetic dipole). However, the rank-$2$ multipole moment (the quadrupole) and rank-$3$ multipole moment (the octupole) might not vanish. \citet{Suzuki2017,Suzuki2019} noticed that in Mn$_3$Sn or Mn$_3$Ir, which both belong to $O_h$ representation, the octupole belongs to the same irreducible representation as the dipole moment. In other words, the octupole produces a finite magnetization that results in anomalous Hall effect in the presence of spin-orbit coupling. The analysis of multipole orders has been extended to other materials and their electromagnetic response, a topic that remains out of the scope of the present review \cite{Watanabe2017b,Watanabe2018b}. We also notice that finite anomalous Hall effect in a noncollinear, coplanar kagom\'e antiferromagnet was pointed out by \citet{Tomizawa2009} in the slightly different context of the "orbital Aharonov-Bohm effect" (see below).

\begin{figure}
	\includegraphics[width=9cm]{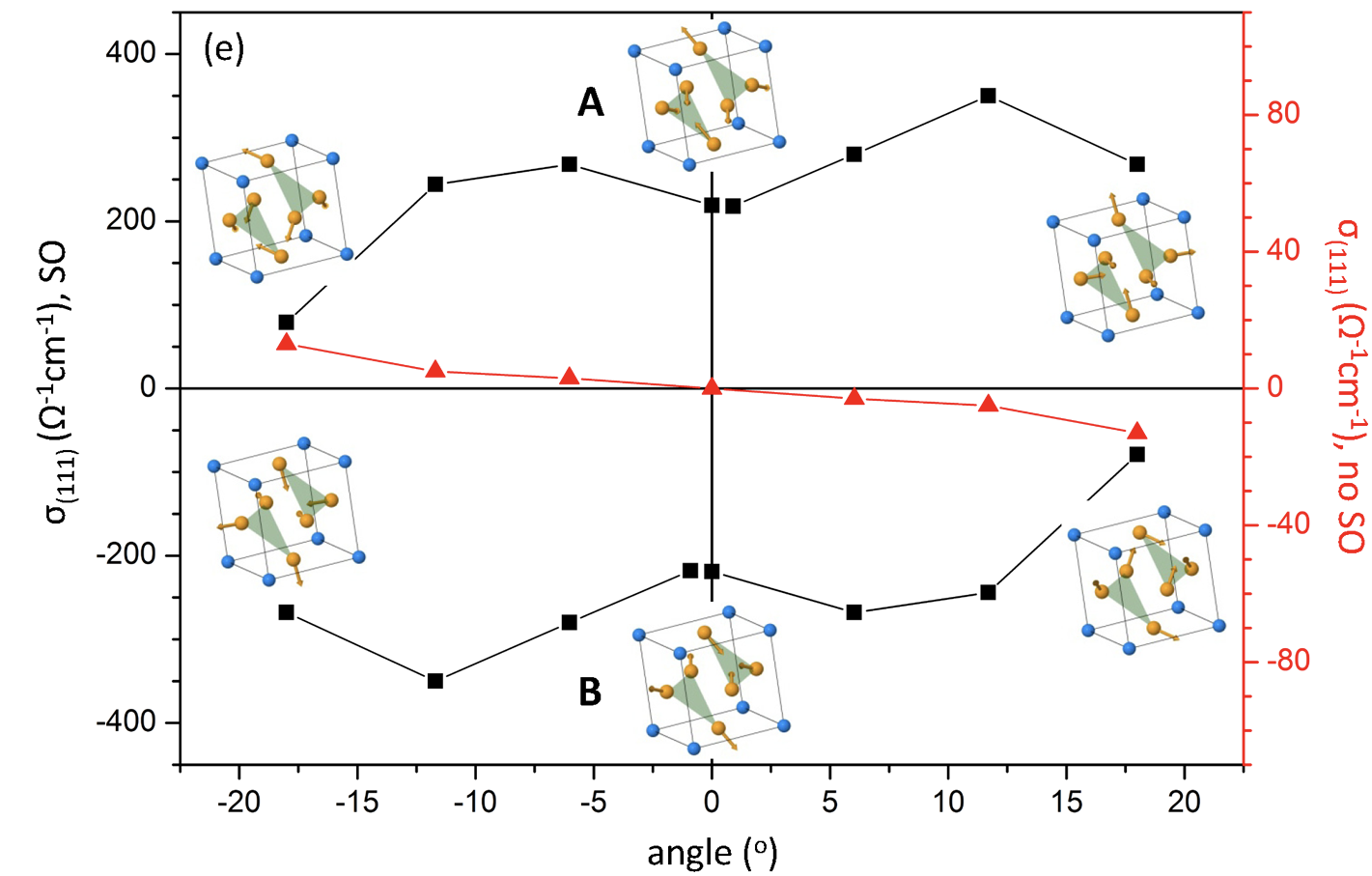}
	\caption{\label{fig:Chen2014} (Color online)
	Angular dependence of the anomalous Hall effect computed numerically for Mn$_3$Ir when tilting the magnetic moments out of the (111) plane. In the absence of spin-orbit coupling (red symbols), topological Hall effect arising from the noncoplanar magnetic ordering is observed. When turning on spin-orbit coupling (black symbols), the anomalous Hall effect remains finite even in the absence of tilting. We also notice that the Hall conductivity obtained for positive or negative tilting is asymmetric, an effect referred to as chiral Hall effect (see text). From \cite{Chen2014}. }
\end{figure}

\begin{figure}
	\includegraphics[width=9cm]{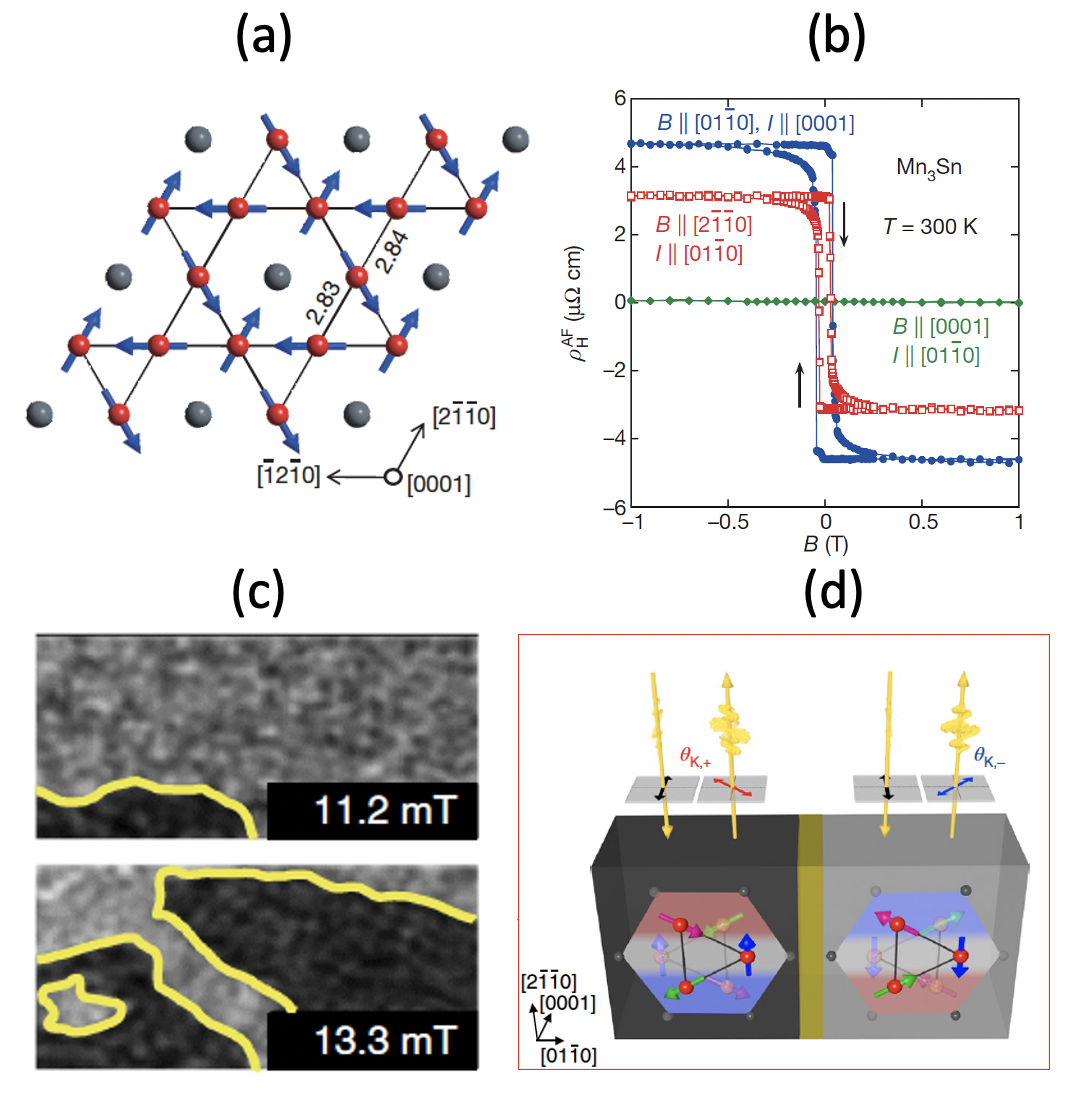}
	\caption{\label{fig:AHE-exp} (Color online)
	(a) Magnetic configuration and (b) anomalous Hall effect reported experimentally in Mn$_3$Sn for various field and current directions with respect to the crystallographic axes. From \cite{Nakatsuji2015}. (c) Imaging of the antiferromagnetic domains of Mn$_3$Sn using magnetooptical Kerr effect. The domain wall is represented by the yellow line and depicted schematically in (d). From \cite{Higo2018}.}
\end{figure}

The concrete realization of these ideas was first predicted in Mn$_3$Ir, a cubic antiferromagnet which consists in kagom\'e lattices with 120$^\circ$ magnetic moments, stacked along the (111) direction (see Table \ref{Table1}). In this compound, the stacking of the antiferromagnetic kagom\'e lattices (not the spin-orbit coupling) breaks the mirror symmetry, so that in the presence of spin-orbit coupling, anomalous Hall effect emerges \cite{Chen2014}, as reported on Fig. \ref{fig:Chen2014}. This prediction was soon extended to other members of the Mn$_3$X family \cite{Kubler2014,Zhang2017b} (X = Ge, Sn, Ga, Ir, Rh and Pt) with both cubic and hexagonal magnetic unit cells. The anomalous Hall effect was first reported in Mn$_3$Sn \cite{Nakatsuji2015,Higo2018b}, which exhibits an anomalous Hall conductivity of $ \sim 20\;(\Omega\cdot\text{cm})^{-1}$ at room temperature (see Fig. \ref{fig:AHE-exp}(a,b)) and $ \sim 100\;(\Omega\cdot\text{cm})^{-1}$ at low temperatures, and soon observed in Mn$_3$Ge with an anomalous Hall conductivity of $\sim 50\;(\Omega\cdot\text{cm})^{-1}$ at room temperature and $ \sim 500\;(\Omega\cdot\text{cm})^{-1}$ at 2 K was reported experimentally \cite{Nayak2016,Kiyohara2016}. Electrical switching of the anomalous Hall current has been achieved in Mn$_3$Pt deposited on the ferroelectric substrate BaTiO$_3$ \cite{Liu2018b}. The interfacial strain is tuned by an electric gate, which enables the small canting of the magnetic order to be achieved without external magnetic field. Finally, due to the noncollinear spin texture, an orbital magnetization emerges which substantially enhances the coupling to an external magnetic field \cite{Chen2020b}.

An interesting aspect of anomalous Hall effect is that it unlocks the magnetooptical Kerr effect: the reflection and transmission of an incident beam of light depends on the light polarization. Magnetooptical Kerr effect is a remarkably useful tool to image magnetic domains in ferromagnetic materials and should, henceforth, also exist in these antiferromagnets. As a matter of fact, the magnetooptical Kerr effect has been computed from first principles in Mn$_3$X compounds \cite{Feng2015} and reported experimentally in Mn$_3$Sn \cite{Higo2018}, see Fig. \ref{fig:AHE-exp}(c,d). The thermal counterpart of anomalous Hall effect, the anomalous Nernst effect, was also computed \cite{Guo2017} and reported in Mn$_3$Sn \cite{Ikhlas2017,Li2017d}. We emphasize that whereas the effective time-reversal symmetry breaking is due to the absence of a crystal symmetry operation that would restore time-reversal symmetry, the spin-orbit coupling remains a crucial ingredient for the anomalous Hall effect, Nernst effect and magnetooptical Kerr effect. Indeed, spin-orbit coupling provides the Berry curvature (in other words, the intrinsic Lorentz force) that drives the anomalous velocity. \par

A parent family of noncollinear, coplanar antiferromagnets that is currently attracting substantial interest covers the Mn$_3$AN antiperovskite compounds, where A=Ga, Sn, Ni. The magnetic texture can adopt $\Gamma_{5g}$ (similar as Mn$_3$Sn - see Fig. \ref{fig:FigTexture}(b)) or $\Gamma_{4g}$ configuration (similar as Mn$_3$Ir - see Fig. \ref{fig:FigTexture}(a)) depending on the temperature. These two magnetic phases are nearly degenerate and display different Hall signature. The former does not show anomalous Hall \cite{Zhou2019b,Gurung2019,Huyen2019} or Nernst effect \cite{Zhou2020}, whereas only the latter does. The anomalous Hall effect has been measured experimentally in Mn$_3$NiN \cite{Boldrin2019} as well as in Mn$_3$(Ni,Cu)N where $\Gamma_{4g}$ configuration could be stabilized \cite{Zhao2019}. This compound is particularly interesting because it exhibits large piezomagnetism, i.e., a net magnetization induced by the lattice distortion \cite{Boldrin2018}.

\begin{figure}
	\includegraphics[width=8cm]{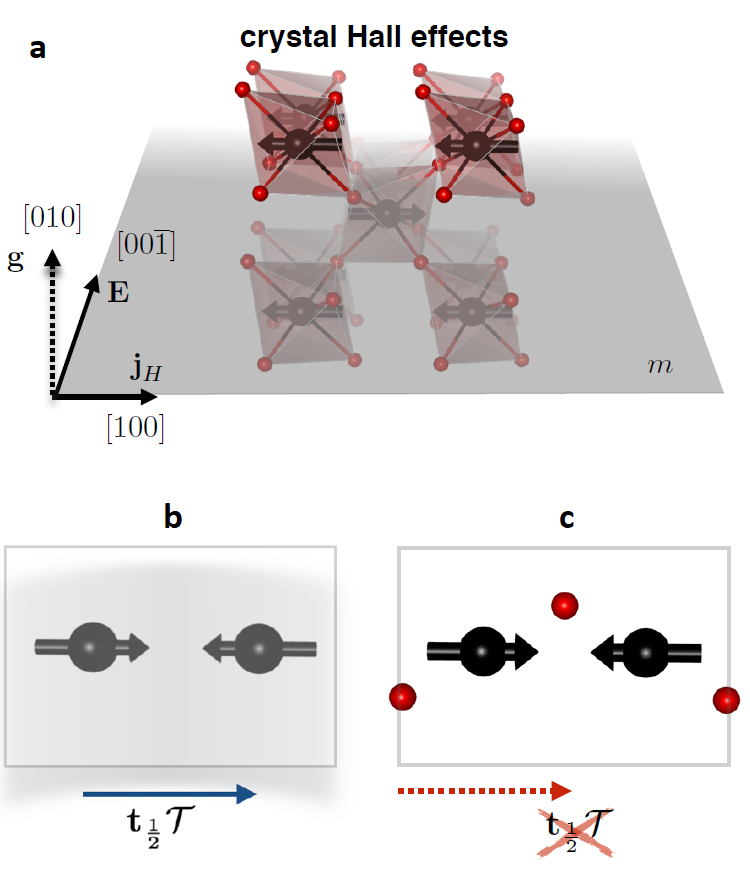}
	\caption{\label{Crystal}(Color online) Principle of crystal Hall effect in \ce{RuO2}. (a) \ce{RuO2} crystal structure (O: red spheres, Ru: black spheres). The black moments (on the \ce{Ru} atoms) define the collinear antiferromagnet structure. (b) One-dimensional model of a collinear antiferromagnet which preserves the $\bf{t_\frac{1}{2}\mathcal{T}}$ symmetry. (c) The presence of the non-magnetic atoms (red spheres) breaks the $\bf{t_\frac{1}{2}\mathcal{T}}$ symmetry. From \cite{Smejkal2020}.}
\end{figure}

\paragraph{Collinear antiferromagnets}
\label{subsec:CHE}
Along a similar spirit, \citet{Smejkal2020} suggested another mechanism for anomalous Hall effect in {\em collinear} antiferromagnets. The driving force is still the spin-orbit coupling, but the time-reversal symmetry breaking is ensured by the structural inequivalence of the magnetic sublattices set by their respective chemical environment. An example of such a configuration is given in Fig. \ref{Crystal}(b,c) for a two-sublattice system. In Fig. \ref{Crystal}(b), the two sublattices are connected by a translation $\bf{t_\frac{1}{2}}$ so that the magnetic crystal obeys $\bf{t_\frac{1}{2}\mathcal{T}}$ symmetry. Consequently, the antiferromagnet possesses an {\em effective} time-reversal symmetry and Hall effect is not allowed. However, if the two magnetic sublattices are structurally inequivalent, as in Fig \ref{Crystal}(c), $\bf{t_\frac{1}{2}\mathcal{T}}$ is broken and Hall effect is allowed. 

 \citet{Smejkal2020} proposed that such an anomalous Hall effect, tagged "crystal Hall effect", appears in the collinear rutile antiferromagnet \ce{RuO2}, where a very large value of the Hall conductivity of the order 1000 $(\Omega\cdot\text{cm})^{-1}$ was obtained from first principles. For the anomalous Hall effect to emerge, the N\'eel vector needs to lie in the (001) plane and along the [100] axis [see Fig. \ref{Crystal}(a)], otherwise, the endowed symmetries remain too high to permit a Hall effect (this is the case when the N\'eel vector is along the [001] axis). The remaining symmetries are usually not altered when the antiferromagnetic moments are canted in the [010] direction. The effect of the canting on the Hall conductivity for \ce{RuO2} is very small, which clearly contrasts with the anomalous Hall effect reported in, e.g., \ce{EuTiO3} \cite{Takahashi2018} and \ce{GdPtBi} \cite{Suzuki2016} compounds, which is mainly due to the field-induced canting. \par

\begin{figure}
	\includegraphics[width=7cm]{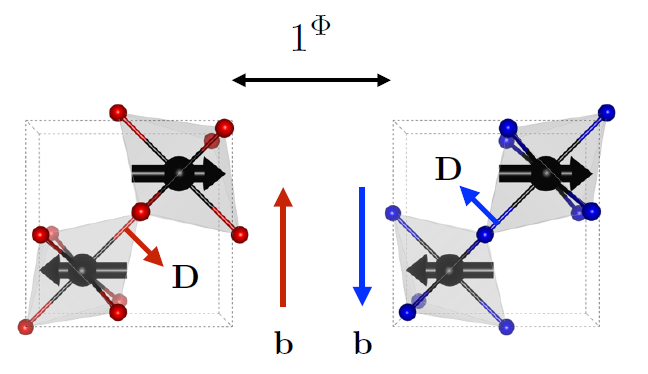}
	\caption{\label{Transformation} (Color online)
	Flipping the sign of the CHE by rearranging the positions of the oxygen atoms under a roto-reversal symmetry ($1^\Phi$) changing the oxygen atoms from the bleu positions to the red ones. $\bf{D}$, the Dzyaloshinskii-Moriya vector is changing sign. $\bf{b}$ is the Berry curvature which changes sign under the transformation. From \cite{Smejkal2020}.}
\end{figure}

The sign of the crystal Hall effect can be flipped by reverting the N\'eel vector of \ce{RuO2}. It is also possible to control the sign (flip the sign) of the anomalous Hall effect by rearranging the positions of the non-magnetic oxygen atoms in the \ce{RuO2} while keeping the moments \ce{Ru} fixed. This rearrangement fulfills a roto-reversal symmetry ($1^\Phi$) \cite{Gopalan2011} which transforms by a \ang{90} angle in the opposite sense the two oxygen octahedra such that the real space chirality is reversed as shown in Fig. \ref{Transformation}. The magnitude of the conductivity $\sigma_{xz}$ is preserved and only the sign changes.

\begin{figure}
	\includegraphics[width=6cm]{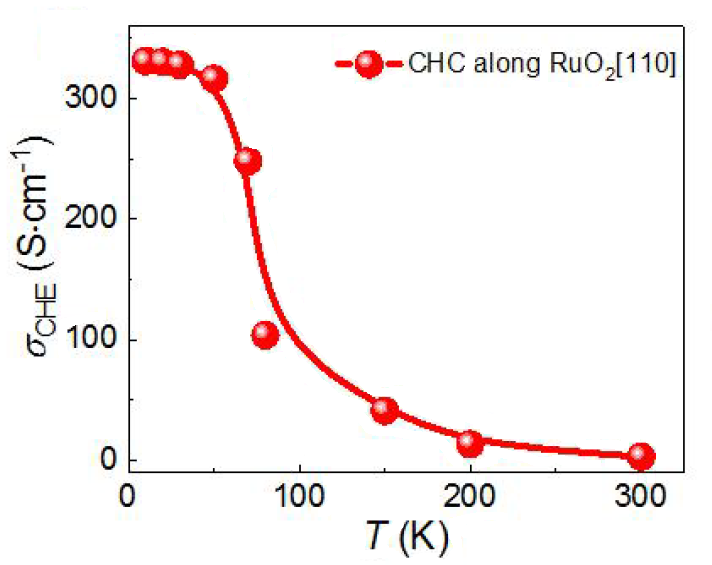}
	\caption{(Color online) The crystal Hall conductivity as a function of temperature for \ce{RuO2}/\ce{MgO}. From \cite{Feng2020}.}
	\label{Observation-CHE}
\end{figure}

Experimental measurements have been conducted on high quality epilayers of \ce{RuO2} grown on \ce{MgO} substrates, and a crystal Hall conductivity of 330 $(\Omega\cdot\text{cm})^{-1}$ has been reported at very low temperatures \cite{Feng2020}, displaying a strong temperature dependence (see Fig. \ref{Observation-CHE}). This anomalous Hall conductivity is three times larger than that reported in noncollinear antiferromagnets such as \ce{Mn3Sn} \cite{Nakatsuji2015}, and is comparable to that of \ce{Fe} thin films \cite{Sangiao2009}. The crystal Hall effect is expected to be present in many other classes of antiferromagnets, such as orthoferrites or perovskites. For instance, \citet{Smejkal2020} pointed out that the spontaneous Hall effect observed in \ce{CoNb3S6} \cite{Ghimire2018} is consistent with the crystal Hall effect mechanism, whereas the Hall effect reported in Ce-doped canted antiferromagnet \ce{CaMnO3} \cite{Vistoli2019} is at least partly due to crystal Hall effect.In the latter case, the spin canting is also likely to promote a topological Hall effect contribution \cite{Nakane2020}.

We emphasize that the search for anomalous Hall effect in both collinear and noncollinear antiferromagnets is only at its infancy and the exploration of the vast magnetic point group of antiferromagnets is wide open. Finally, the optical (magnetooptical Kerr and Faraday effects) and thermal counterparts (anomalous Nernst Hall effect) could also present interesting features that remain to be discovered and characterized.

\paragraph{Topological Hall effect}

\begin{figure}
	\includegraphics[width=8cm]{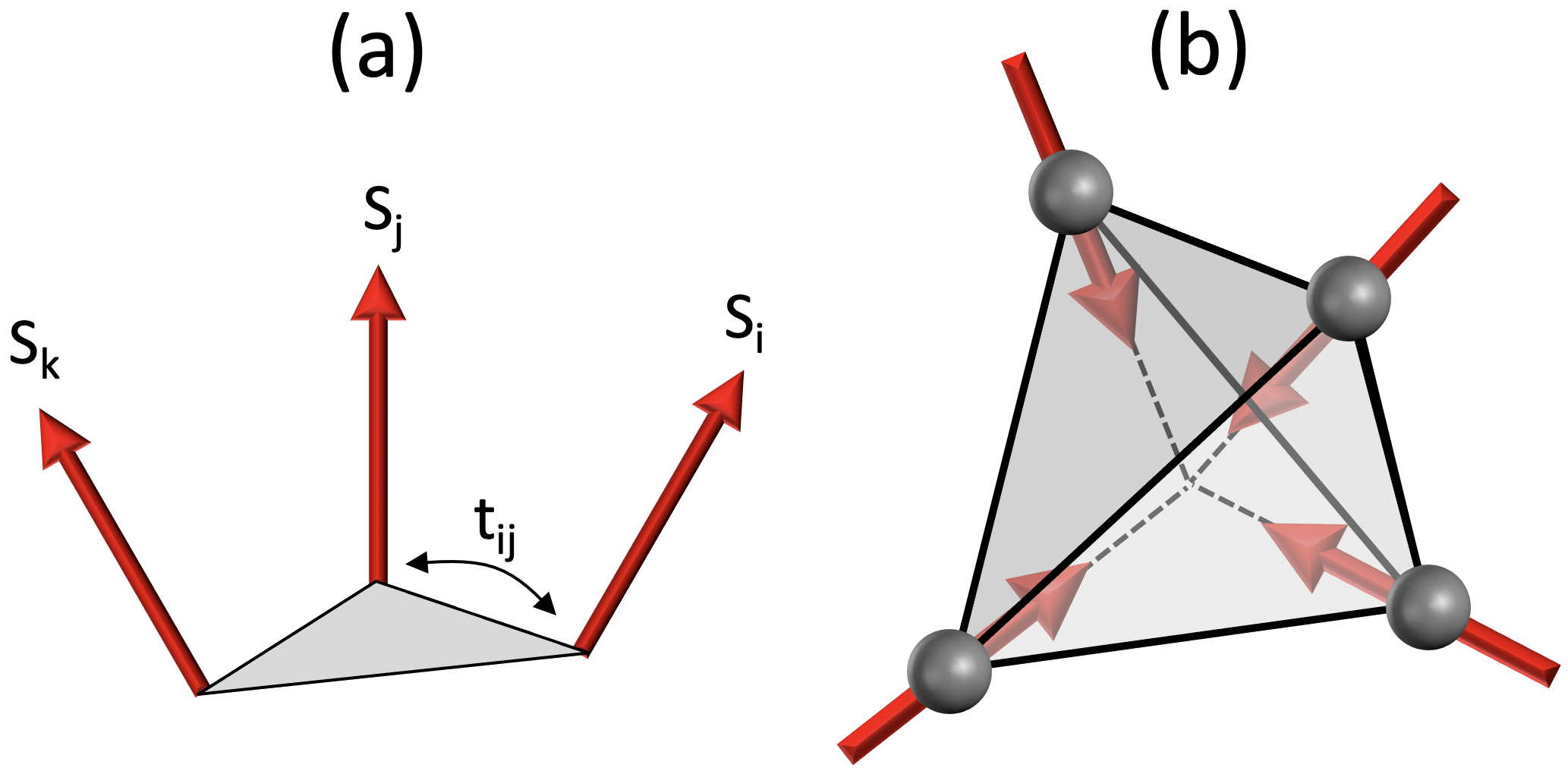}
	\caption{(Color online) (a) A triangular plaquette with non-coplanar magnetic moments at each vertex. Due to the non-coplanarity, an electron hopping around this plaquette acquires a finite Berry phase. (b) Magnetic tetrahedron with all-in configuration.}
	\label{fig:THE}
\end{figure}

In the previous two phenomena, anomalous and crystal Hall effects, the Berry curvature arises from spin-orbit coupling whereas specific crystal symmetries combined with the antiferromagnetic order effectively breaks the time-reversal symmetry at the level of the magnetic unit cell. Another means to generate Berry curvature is to take advantage of the non-coplanar alignment of neighboring magnetic moments. To see how the Berry curvature emerges, let us consider a triangular plaquette, depicted on Fig. \ref{fig:THE}(a), with noncollinear non-coplanar magnetic moments, ${\bf S}_i$ ($i=1,2,3$), at its vertices. A conduction electron is allowed to hop between neighboring sites $i$ and $j$ with a spin-independent hopping integral $t_{ij}$, and due to the strong exchange interaction, its spin angular momentum remains aligned on the local magnetic moment. As a consequence, the conduction electron wave function on site $i$ reads
\begin{eqnarray}
|\psi_i\rangle=\cos \frac{\theta_i}{2}|\uparrow\rangle+e^{i\phi_i}\sin\frac{\theta_i}{2}|\downarrow\rangle,
\end{eqnarray}
where $(\theta_i,\phi_i)$ are the polar coordinate of ${\bf S}_i$ (an overall phase associated with the gauge choice has been omitted for simplicity). Taking ${\bf S}_k$ as the common quantization axis of the plaquette, one can rewrite the spin-independent hopping parameters $t_{ij}$ by performing a gauge transformation in which spin $\bf{S}_i$ is rotated around a vector ${\bf n}_i=({\bf S}_i+{\bf S}_k)/|{\bf S}_i+{\bf S}_k|$. Therefore, the hopping $t_{ij}$ in the magnetic texture transforms as \cite{Ohgushi2000,Ndiaye2019}
\begin{eqnarray}
t_{ij}\rightarrow t_{ij}\left[{\bf n}_i\cdot{\bf n}_j+i\hat{\bm\sigma}\cdot({\bf n}_i\times{\bf n}_j)\right].
\end{eqnarray}
Upon this gauge transformation, the hopping term becomes spin-dependent. More precisely, the magnetic texture in real space locks the spin degree of freedom with the propagation direction of the conduction electron, mimicking the effect of spin-orbit coupling. Consequently, in the adiabatic limit an electronic spin hopping in a closed path between three magnetic moments picks up a Berry phase given by half the solid angle $\Omega$ covered by the three local moments. The solid angle is proportional to the spin chirality, defined as \cite{Lee1992,Miyashita1984,Shindou2001} 
\begin{eqnarray}
\kappa= \bf{S}_k\cdot\left(\bf{S}_i\times \bf{S}_j\right),
\end{eqnarray}
which is finite as long as the three magnetic moments are non-coplanar. The consequence of this real-space Berry curvature is that the charge carriers acquire an anomalous velocity without spin-orbit coupling \cite{Taguchi2001}. Whether this anomalous velocity induces anomalous Hall current or spin Hall current depends on effectively breaking the time-reversal symmetry, as discussed above. These effects have been tagged "topological Hall" and "topological spin Hall" effects, in reference to the non-trivial topology of the magnetic texture in real space. The topological Hall effect has been observed in a vast number of noncollinear ferromagnets such as pyrochlores \cite{Taguchi2001} (e.g., \ce{Nd2Mo2O7}), and chiral magnets supporting skyrmion crystals \cite{Neubauer2009,Lee2009b,Kanazawa2011}. In all these systems, a remanent net magnetization ensures the explicit time-reversal symmetry breaking.

The earliest proposal for topological Hall effect in a noncollinear non-coplanar antiferromagnet without net magnetization was suggested by \citet{Shindou2001}. The authors considered a triple-Q antiferromagnet, which features a tetrahedral magnetic unit cell with magnetic moments at its vertices that point either towards or away from the center of the tetrahedron, see Fig. \ref{fig:THE}(b). This type of configuration is stabilized by the so-called 4-spin interaction \cite{Takahashi1977,Akagi2012,Yasui2020} and is present in \ce{FeMn} alloys \cite{Endoh1971,Sakuma2000}, but also in NiS$_2$ \cite{Kikuchi1978,Matsuura2003} as well as at the Cu/Mn(111) surface \cite{Kurz2001}. Theoretical analysis of the Kondo lattice model shows that the 3Q configuration is stable for a broad range of parameters at 1/4 filling \cite{Akagi2010} and is stable against thermal and quantum fluctuations \cite{Kato2010,Akagi2013}. Because there exists a rotation operation that transforms the magnetic texture into itself, the band structure is doubly-degenerate and the spin-Berry phase is non-abelian \cite{Ndiaye2019} (see discussion in Section \ref{ssec:degaf}). The triple-Q magnetic texture supports orbital magnetization \cite{Wang2007,Hanke2016}, as well as topological Hall effect \cite{Shindou2001,Martin2008} that can become quantized when the chemical potential is tuned into the orbital gap, realizing an antiferromagnetic Chern insulator \cite{Ndiaye2019}. Quantum anomalous Hall effect has also been predicted in the double-exchange kagom\'e lattice \cite{Ishizuka2013,Chern2014}, as well as in checkerboard (G-type) lattice \cite{Venderbos2012}. Recently, it was proposed that ${\rm K_{0.5}RhO_2}$ could be such an antiferromagnetic kagom\'e Chern insulator \cite{Zhou2016}.

On the experimental side, the earliest signature of topological Hall effect in a compensated noncollinear antiferromagnet is probably the one reported by \citet{Machida2007,Machida2010} in the geometrically frustrated Kondo lattice of the pyrochlore iridate Pr$_2$Ir$_2$O$_7$. This material experiences a phase transition towards a chiral liquid state where the long-range ordering is prevented as discussed in further details in Section \ref{s:topaf}. The appearance of a spontaneous Hall effect in the absence of long-range magnetic ordering suggests that time-reversal symmetry is indeed broken on the macroscopic scale, which is considered as the signature of chiral spin liquids. In more conventional states of matter, topological Hall effect has been reported experimentally in Mn$_5$Si$_3$ by \citet{Surgers2014, Surgers2016}. We also mention that an "unconventional" Hall effect has been reported in the triangular lattice PdCrO$_2$ \cite{Takatsu2010}, which could be related to topological Hall effect.

\paragraph{Chiral Hall effect}

Whereas the Berry curvature at the origin of the anomalous and crystal Hall effects is due to the presence of spin texture in momentum space induced by spin-orbit coupling, the Berry curvature responsible for topological Hall effect is due to the magnetic texture in real space. A third scenario takes advantage of the interplay between spin texture in momentum space and spin texture in real space. Such a scenario was put forward by \citet{Tomizawa2009,Tomizawa2010} in the context the "unconventional" anomalous Hall effect reported in the metallic pyrochlores \ce{Nd2Mo2O7} \cite{Taguchi2001} and \ce{Pr2Ir2O7} \cite{Machida2007,Machida2010}. The authors point out that in a multi-orbital model, the noncollinearity between the magnetic moments induces a so-called "orbital Aharonov-Bohm" effect  - in other words, a Berry-curvature - that is proportional to both the spin-orbit coupling strength and the tilting angle. The emergence of a Berry curvature in the mixed real-momentum space was derived in the context of a $t_{2g}$ model, and was independently developed in a more general manner by \citet{Kipp2020} and applied to the antiferromagnet \ce{SrRuO3}. We stress out that a chiral Hall signal appears in \cite{Chen2014} when computing the anomalous Hall conductivity of Mn$_3$Ir as a function of the tilting angle of the magnetic moments away from (111) plane, see Fig. \ref{fig:Chen2014}. It reveals itself as a Hall signal that is {\em odd} in tilting angle when spin-orbit coupling is turned on. This effect, distinct from the topological Hall effect, complements the previous studies on momentum- and real-space Berry curvatures.
 
\paragraph{Nonlinear anomalous Hall effect}
In the paradigms discussed above, charge Hall effect emerges by breaking the combined symmetry operations that ensure effective time-reversal symmetry. In fact, another way to break time-reversal symmetry effectively is to consider the second-order response in electric field. The resulting {\em nonlinear} Hall effect emerges in systems with effective time-reversal symmetry but lacking spatial inversion symmetry. From a theoretical standpoint, this effect has been related to the Berry curvature dipole, i.e. the momentum derivative of the Berry curvature \cite{Sodemann2015}. The rectified (dc) current reads
\begin{eqnarray}\label{eq:nhe1}
{\bf J}&=&-\frac{e^3\tau}{2\hbar^2}\sum_n\int \frac{d^d{\bf k}}{(2\pi)^d}f^0_{{\bf k},n}({\bf E}\cdot\partial_{\bf k})({\bm \Omega}_{{\bf k},n}\times {\bf E}^*),\\
&=&\frac{e^3\tau}{2\hbar}\sum_n\int \frac{d^d{\bf k}}{(2\pi)^d}({\bf E}\cdot{\bf v}_{{\bf k},n})\partial_{\varepsilon_{{\bf k},n}}f^0_{{\bf k},n}({\bm \Omega}_{{\bf k},n}\times {\bf E}^*).\label{eq:nhe2}
\end{eqnarray}
Here, $d$ is the dimension of the system, $f^0$ is the equilibrium Fermi distribution, $\varepsilon_{{\bf k},n}$, ${\bf v}_{{\bf k},n}$ and ${\bm \Omega}_{{\bf k},n}$ are the dispersion relation, velocity and Berry curvature of state $n$ and the electric field is defined ${\rm Re}[{\bf E}e^{i\omega t}]$. Like Berry curvature, the Berry curvature dipole contribution is maximum close to avoided band crossing points in the band structure, suggesting that nonlinear Hall effect is indeed large in materials displaying Weyl points. Since this second-order theory is based on the first-order expansion of the wave packet velocity (${\bf v}_{{\bf k},n}=\partial_{\bf k}{\varepsilon_{{\bf k},n}}+{\bm \Omega}_{{\bf k},n}\times\dot{\bf k}$), further theoretical refining is necessary to capture the full physical effect. Equation \eqref{eq:nhe2} suggests the following phenomenological picture: in the presence of spin-orbit coupling and inversion symmetry breaking, an applied electric field generates a nonequilibrium effective magnetic field [${\bf B}\sim \tau({\bf E}\cdot{\bf v}_{{\bf k},n})\partial_{\varepsilon_{{\bf k},n}}f^0_{{\bf k},n}{\bm \Omega}_{{\bf k},n}$] at the first order (akin to the extrinsic Rashba-Edelstein effect). Once this effective field is present, a regular intrinsic anomalous Hall effect (${\bf J}\sim {\bf B}\times {\bf E}$) can be obtained. The nonlinear Hall effect has been reported experimentally in noncentrosymmetric, non-magnetic WTe$_2$ \cite{Ma2019, Kang2019}. 

Of most interest to the present review, \citet{Shao2020} recently investigated the nonlinear anomalous Hall effect in collinear antiferromagnets where inversion symmetry is broken. They specifically address the case of CuMnSb, where {\em linear} anomalous Hall effect is absent and {\em nonlinear} anomalous Hall effect is present due to glide symmetries with mirrors normal to [$\bar{1}$01], [0$\bar{1}$1] and [$\bar{1}$10] directions in the crystal lattice. This nonlinear Hall signal can be used to probe the N\'eel order direction. This proposal opens appealing perspectives for the experimental detection of antiferromagnetic order, but also for the exploration of nonlinear topological effects.
 
\subsubsection{Spin Hall effect}

Spin Hall effect \cite{Hoffman2013,Sinova2015}, the generation of a pure, charge neutral spin current transverse to the injected charge flow, was predicted in the early seventies \cite{Dyakonov1971}, revived in the early 2000's \cite{Hirsch1999,Zhang2000}, and reported experimentally in a wide range of nonmagnetic materials, from semiconductors \cite{Kato2004d,Wunderlich2005} to metals \cite{Valenzuela2006,Saitoh2006,Kimura2007} and superconductors \cite{Wakamura2015}, but also in ferromagnets \cite{Miao2013}. The investigation of spin Hall effect in antiferromagnets is only recent, as exposed below. In fact, the "conventional" spin Hall effect arising from the spin-Berry curvature induced by spin-orbit coupling is a very robust phenomenon and naturally emerges in antiferromagnets possessing heavy elements. Remarkably though, because antiferromagnets can display very diverse magnetic configurations that obey complex combined time-reversal symmetries, it turns out that the spin Hall currents may adopt much richer tensorial forms compared to the "conventional" one given by Eq. \eqref{eq:jsh}, as pointed out recently \cite{Seeman2015,Wimmer2016}.\par
\begin{figure}
	\includegraphics[width=\linewidth]{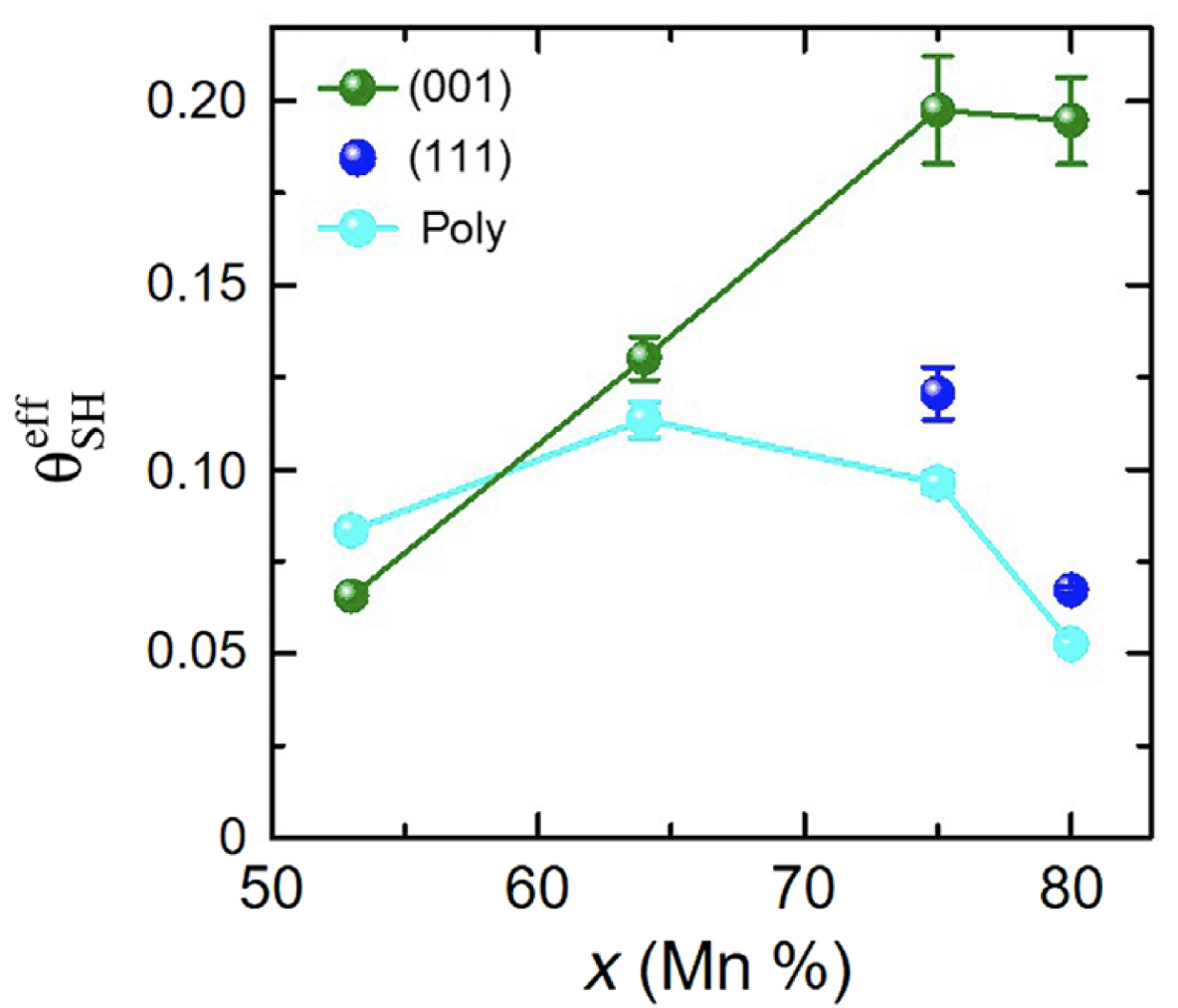}
	\caption{\label{fig:SHE-af} 
		Composition dependence of spin Hall conductivity of Mn$_x$Ir$_{1-x}$ measured along the (001) and (111) facets. The polycrystalline case is also given for comparison. From \cite{Zhang2016m}.}
\end{figure}
\paragraph{"Conventional" intrinsic spin Hall effect}
\label{subsec:SHE}

Initial theoretical and experimental investigations in antiferromagnets focused on the conventional spin Hall effect, arising from spin-orbit coupling and obeying Eq. \eqref{eq:jsh}. Experiments in metallic antiferromagnets such as sputtered \ce{Ir20Mn80} \cite{Mendes2014} and MnX compounds \cite{Zhang2014e} (X=Ir, Pt, Pd, Fe) reported decent spin Hall angles (i.e., the ratio between the Hall and longitudinal conductivities), between 0.8\% (MnFe) and 6\% (MnPt). These conductivities were in reasonable agreement with first principle simulations based on the intrinsic spin-Berry curvature. In the weak disorder limit, the conductivity of a spin Hall current polarized along $\gamma$, flowing along $\alpha$ and generated by an electric field applied along $\beta$, reads

\begin{equation}
\label{eqn:intrinsicSHE}
\sigma _{\alpha\beta}^{\gamma}
=-2e \hbar \sum_{{\bf k},m \neq n}^{\begin{tiny}\begin{tabular}{c}m\;{\rm unocc}\\
n\;{\rm occ}\end{tabular}\end{tiny}} \frac{\Im (\langle \psi_{n\boldsymbol{k}} |\mathcal{J}_{\alpha}^{\gamma} |\psi_{m\boldsymbol{k}}\rangle
	\langle \psi_{m\boldsymbol{k}}|	v_{\beta} |\psi_{n\boldsymbol{k}}\rangle
	)}
{ (E_{n\boldsymbol{k}}- E_{m\boldsymbol{k}})^2},
\end{equation}
where $\mathcal{J}_{\alpha}^{\gamma}=\{s_\gamma,v_\alpha\}/2$ is the spin current operator, $\psi_{n\boldsymbol{k}}$ is an eigenstate of the crystal and $E_{n\boldsymbol{k}}$ its eigenenergy. Whereas the directions $\alpha,\;\beta,\;\gamma$ are mutually orthogonal to each others by definition, the components of the spin Hall conductivity tensor may display significant anisotropy, depending on the crystal symmetries as exemplified in the case of nonmagnetic hcp structures \cite{Freimuth2010,Seeman2015}. By extension, the spin Hall conductivity tensor should also exhibit a modulation as a function of the direction of the magnetic order. Whereas these experiments did not report any influence of the magnetic order, contrary to expectations, a substantial influence of the growth direction was reported in epitaxial MnX crystals \cite{Zhang2015}, as well as in \ce{Mn3Ir} single crystal \cite{Zhang2016m}. The latter result is particularly interesting as \ce{Mn3Ir} displays a noncollinear, coplanar magnetic texture, as mentioned above. \citet{Zhang2016m} reported a "giant" spin Hall angle of 35\% in the (001) growth direction, i.e., comparable to the largest Hall angle obtained in $\beta$-W \cite{Pai2012} and three times larger than when \ce{Mn3Ir} is grown along (111) direction, see Fig. \ref{fig:SHE-af}. Since \ce{Mn3Ir} has a cubic lattice, this observation suggests that the magnetic order, which lies in (111) plane, has dramatic influence on the spin-to-charge conversion efficiency. It is important to emphasize that the spin Hall effect investigated in MnX compounds and in \ce{Mn3Ir} single crystals arise from spin-orbit coupling. In other words, this spin Hall effect is likely to be intrinsic, i.e., associated with the spin-Berry curvature in momentum space induced by the spin-orbit coupling. 

\begin{figure}
	\includegraphics[width=\linewidth]{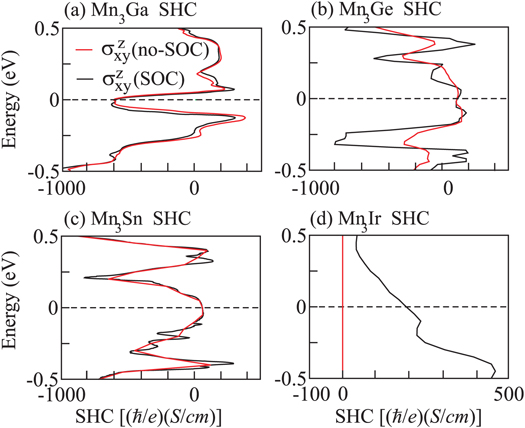}
	\caption{\label{SHCs} 
		Energy dependence of the spin Hall conductivity with and without spin-orbit coupling, for different antiferromagnets with coplanar magnetic orders. From \cite{Zhang2018d}.}
\end{figure}

However, it has been recently realized that spin Hall effect can also survive in noncollinear, coplanar antiferromagnets, even in the {\em absence} of spin-orbit coupling \cite{Zhang2018d}. This spin current is intrinsic and arises from interband transitions: although the spin texture of individual bands remains in the (111) plane (see, e.g. Fig. \ref{fig:FigTexture}), interband transitions induced by the electric field result in a net spin Hall current with a polarization normal to (111). The family of manganese compounds including \ce{Mn3Ir}, \ce{Mn3Ge}, \ce{Mn3Sn} and \ce{Mn3Ga} has been studied for this purpose using \textit{ab initio} calculations. The last three compounds are found to allow for strong spin Hall currents without including spin-orbit coupling. In Fig. \ref{SHCs} the energy dependent spin Hall conductivity is shown for the above mentioned antiferromagnets. The spin currents are calculated with and without spin-orbit coupling. In the case of \ce{Mn3Ir}, the non-trivial magnetic texture does not contribute to the spin Hall effect, due to the symmetry imposed by the cubic lattice structure. Therefore, in the absence of spin-orbit coupling, the spin current signal vanishes. However, the spin conductivities at Fermi level in the three other compounds are $\sigma_{\rm sh}=-613,\;115$ and $90 \frac{\hbar}{e}(\Omega \cdot{\rm cm})^{-1}$ for \ce{Mn3Ge}, \ce{Mn3Sn} and \ce{Mn3Ga} respectively \cite{Zhang2018d}. Interestingly, the strength of the spin Hall effect in these materials allows for designing spin Hall devices without involving heavy elements. Finally, the unusual form of the spin Hall conductivity tensor encountered in noncollinear antiferromagnets has been exploited to generate novel symmetries of spin-orbit torque in \ce{Mn3GaN}/NiFe heterostructures \cite{Nan2019}.

\paragraph{Magnetic spin Hall effect}
\label{subsec:MSHE}

\begin{figure}
	\includegraphics[width=\linewidth]{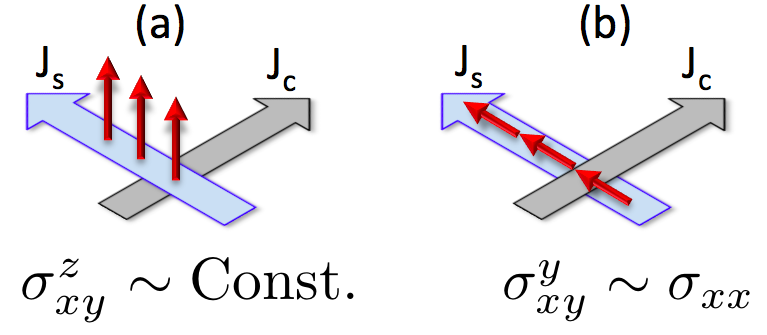}
	\caption{\label{fig:SHE-MSHE} (Color online) Conventional (a) and magnetic (b) spin Hall effects. The polarization of the former is always normal to the scattering plane whereas the polarization of the latter is dictated by the magnetic order and can lie, for instance, in the scattering plane. In addition, the conventional spin Hall effect is intrinsic whereas the magnetic spin Hall effect is extrinsic.}
\end{figure}

The spin Hall effect mentioned in the previous section obeys the conventional symmetry: its spin polarization is perpendicular to both the injected charge current and to the spin current direction [see Fig. \ref{fig:SHE-MSHE}(a)]. In addition, it is "intrinsic", which means that the spin Hall conductivity is independent of the disorder in the weak disorder limit. In a recent work, \citet{Zelezny2017b} discovered that another form of spin Hall effect exists in certain classes of antiferromagnets. This spin Hall current is characterized by a spin polarization that is dictated by the magnetic order, is odd under time-reversal and is of {\em extrinsic} origin [see Fig. \ref{fig:SHE-MSHE}(b)].

Explicitly, the extrinsic spin Hall conductivity reads \cite{Zelezny2017b} 

\begin{equation}
\label{eqn:sigmaodd}
\sigma _{\alpha\beta}^{\gamma}
=-\frac{e \hbar}{\pi} \sum_{m,n} \frac{ \Gamma ^2 \Re (\langle\psi_{n\boldsymbol{k}}|  	\mathcal{J}_{\alpha}^{\gamma} |\psi_{m\boldsymbol{k}}\rangle  
	\langle\psi_{m\boldsymbol{k}}|  	v_{\beta} |\psi_{n\boldsymbol{k}}\rangle
	)}
{ [(E_F - E_{n\boldsymbol{k}})^2 +\Gamma^2]  [(E_F - E_{m\boldsymbol{k}})^2+\Gamma^2 ]  },
\end{equation}
where $\Gamma$ is an effective broadening which accounts for disorder. Under time-reversal symmetry, the conductivity tensor expressed in Eq. \eqref{eqn:sigmaodd} transforms differently from the intrinsic spin Hall conductivity given in Eq. \eqref{eqn:intrinsicSHE}. Since the time-reversal operator is antiunitary, the transformation of the matrix element $\langle \psi_{n\boldsymbol{k}}|\mathcal{J}_{\alpha}^{\gamma}|\psi_{m\boldsymbol{k}}\rangle$ induces an extra minus sign leading to a spin conductivity which is odd under time-reversal symmetry. In this sense the extrinsic contribution to the spin Hall effect has the same symmetry as the anomalous Hall effect discussed previously. According to the discussion in Section \ref{ss:spincond}, the spin conductivity, Eq. \eqref{eqn:sigmaodd}, transforms as in Eq. \eqref{eq:SpinSigmaSymmTR} and, consequently, the transport underlying this effect is of dissipative nature. This particular contribution has been termed {\em magnetic} spin Hall effect because its sign depends on the magnetic order \cite{Zelezny2017b,Kimata2019}.

\begin{figure}
	\includegraphics[width=\linewidth]{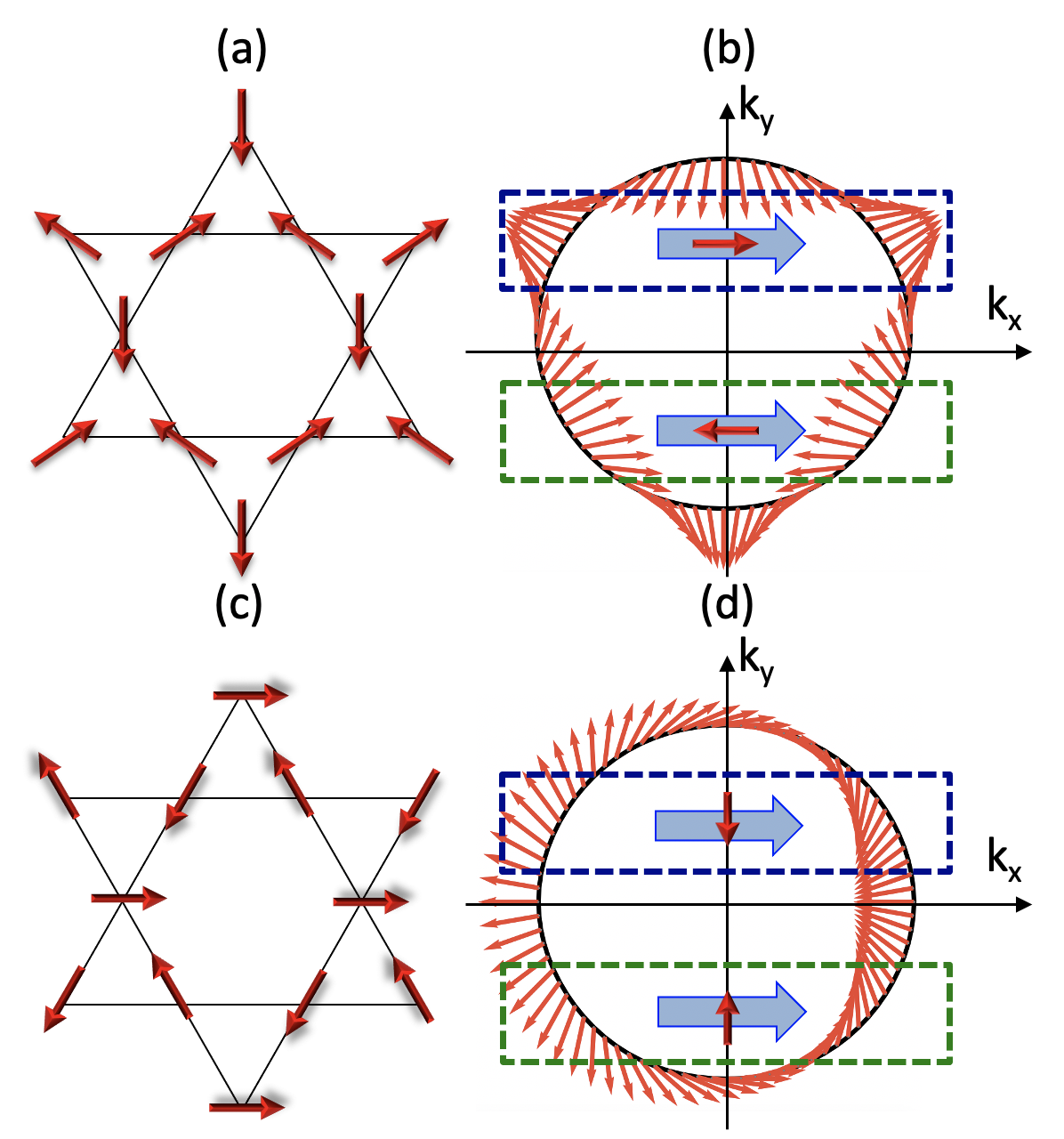}
	\caption{\label{fig:MagneticSpinHall} (Color online) Schematics of the origin of the magnetic spin Hall effect. We consider a noncollinear antiferromagnetic kagom\'e lattice (a), and its corresponding spin texture in momentum space at Fermi level (b). States with $k_y>0$	(blue box) carry a spin current spin-polarized along $+x$ and propagating along $x$, whereas states with $k_y<0$ (green box) carry a spin current spin-polarized along $-x$ and propagating along $x$. At equilibrium, the two spin currents compensate each other, whereas when applying an electric field along $y$, the imbalance between $k_y>0$ and $k_y<0$ states results in a nonequilibrium spin Hall current with in-plane magnetization. Panels (c) and (d) present an alternative case, where the nonequilibrium magnetic spin Hall current is spin-polarized along $y$.}
\end{figure}

Magnetic spin Hall effect can be understood phenomenologically by analyzing the electronic spin texture in {\em momentum} space. Figure \ref{fig:MagneticSpinHall} provides an illustration of this effect in the simplistic case of the noncollinear antiferromagnetic kagom\'e lattice. The real-space lattice is represented on Fig. \ref{fig:MagneticSpinHall}(a), while the corresponding spin texture in momentum space is reported on Fig. \ref{fig:MagneticSpinHall}(b). Let us first consider states with $k_y>0$ [blue box in Fig. \ref{fig:MagneticSpinHall}(b)]. Among these states, those possessing $k_x<0$ are mostly spin-polarized along $-x$ whereas those possessing $k_x>0$ are mostly spin-polarized along $+x$. Therefore, for a given $k_y$, a spin current develops along $x$ that is spin-polarized along $x$. The same scenario occurs for states possessing $k_y<0$ [green box in Fig. \ref{fig:MagneticSpinHall}(b)], but generating a spin current in the opposite direction. When applying an electric field along $y$, the Fermi surface shifts along this direction resulting in an imbalance between states with $k_y<0$ and $k_y>0$. As a result, a nonequilibrium spin current with spin-polarization along $x$ is generated along $x$. We emphasize that the polarization of the magnetic spin Hall current is governed by the magnetic texture in momentum space and therefore depends on the specific magnetic configuration of the antiferromagnet. Hence, taking a different antiferromagnetic configuration, such as the one shown in Figs. \ref{fig:MagneticSpinHall}(c) and (d), results in the generation of a magnetic spin Hall current with spin-polarization along a different direction. An analysis of the magnetic spin Hall effect in terms of spin current vorticity in momentum space has been proposed  \cite{Mook2020}.

\begin{figure}
	\includegraphics[width=\linewidth]{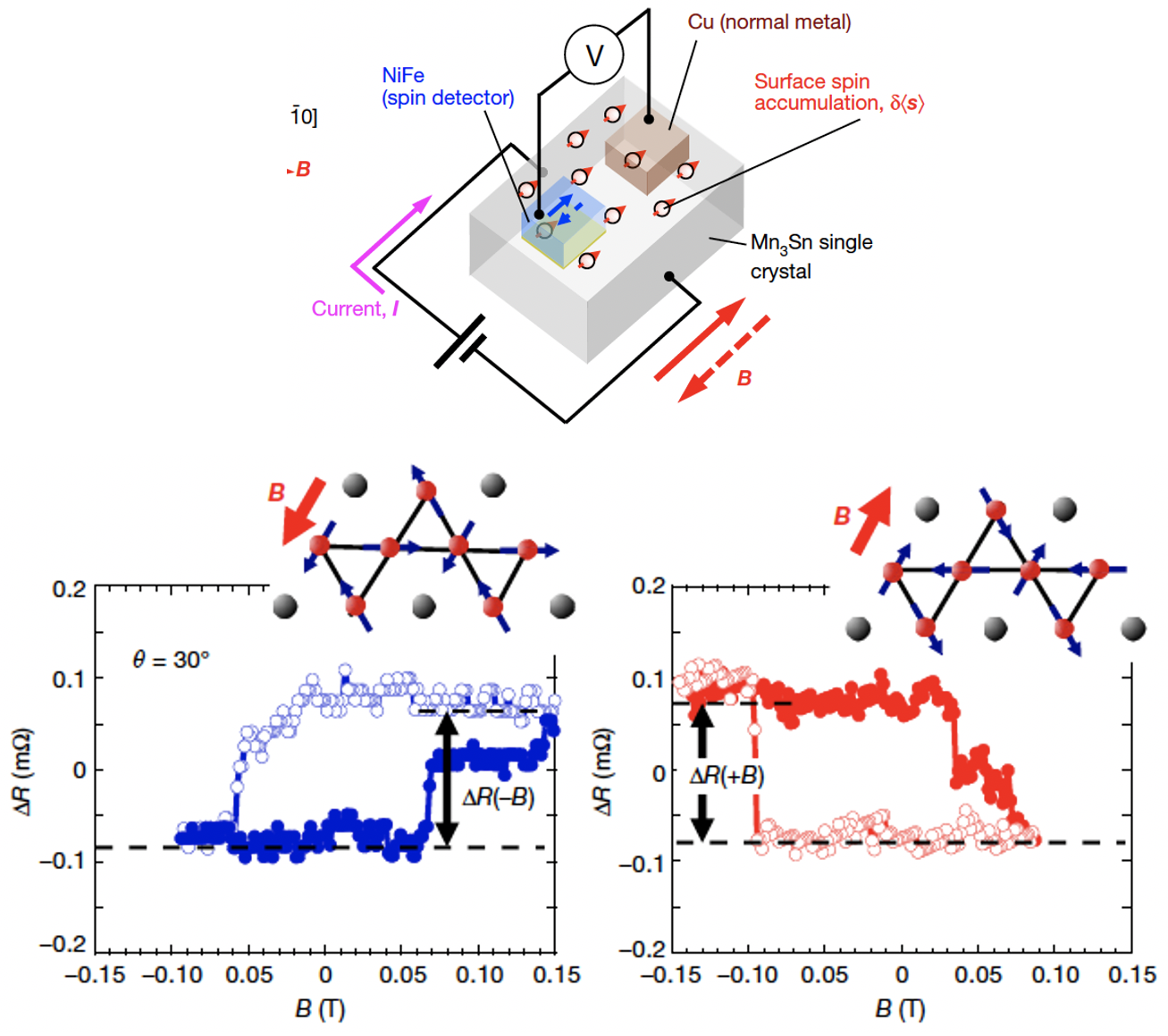}
	\caption{\label{fig:MSE} (Color online) (Top) Schematics of the non-local spin-valve device designed to measure the magnetic spin Hall effect. (Bottom) Non-local magnetoresistive signal revealing the presence of nonequilibrium spin accumulation obeying symmetries that correspond to that of magnetic spin Hall effect. From \cite{Kimata2019}.}
\end{figure}

The magnetic spin Hall effect has been investigated experimentally by \citet{Kimata2019} in \ce{Mn3Sn}. The authors fabricated a non-local magnetic spin-valve and measured the spin accumulation generated by the magnetic spin Hall effect at the interface between \ce{Mn3Sn} and NiFe [see Fig. \ref{fig:MSE}(Top)]. They obtained a magnetoresistive signal whose sign depends on the magnetic configuration of the antiferromagnet [see Fig. \ref{fig:MSE}(Bottom)]. The correlation between the sign of the magnetoresistance and the magnetic order suggests that the spin Hall current that generates the interfacial spin accumulation is odd under time-reversal, which is the hallmark of the magnetic spin Hall effect. The authors also performed the Onsager reciprocal experiment by measuring the inverse magnetic spin Hall effect using spin pumping. In a recent work, \citet{Holanda2020} reported a large current-driven modulation of the magnetic damping of \ce{Ni80Fe20}/\ce{IrMn3} bilayer that strongly depends on the crystal orientation of \ce{IrMn3}. This observation is attributed to magnetic spin Hall effect in \ce{IrMn3}.

\begin{figure}
	\includegraphics[width=\linewidth]{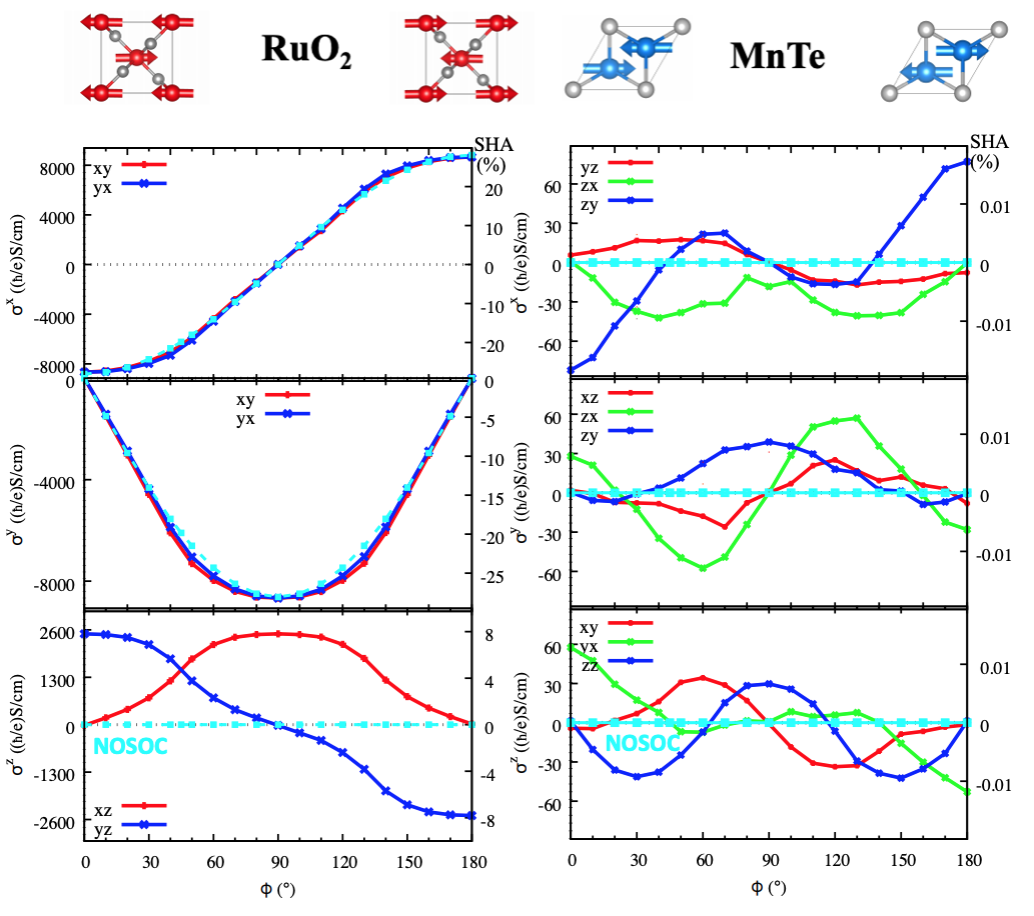}
	\caption{\label{fig:collinearmSHE} 
		Spin Hall angles and dominant components of magnetic spin Hall conductivities  ($\sigma^x$, $\sigma^y$ and $\sigma^z$) as a function of the in-plane N\'eel order orientation with respect to the $x$-axis,  for both \ce{RuO2} (first column) and \ce{MnTe} (second column). Calculations with and without spin-orbit coupling are shown. Adapted from  \onlinecite{Gonzalez-Hernandez2020}.}
\end{figure}

The magnetic spin Hall effect has been also extended to collinear antiferromagnets \cite{Gonzalez-Hernandez2020}. The authors found that the magnetic spin Hall effect is allowed in all systems with either $\mathcal{T}{\bm\tau}_0$ or $\mathcal{PT}$ symmetries and identified two collinear antiferromagnets, \ce{RuO2} and \ce{MnTe}, that exhibit this phenomenon. Figure \ref{fig:collinearmSHE} displays the magnetic unit cell of both crystals (top panel) as well as the three components of the spin Hall conductivity as a function of the in-plane N\'eel order orientation. In the presence of spin-orbit coupling, both \ce{RuO2} and \ce{MnTe} support $\mathcal{T}$-odd spin currents. An analysis of the magnetic point groups of the two systems without spin-orbit interaction shows that the magnetic spin Hall effect signal vanishes in the case of the \ce{MnTe} for any direction of the N\'eel  order. Nevertheless, in \ce{RuO2} some components of the odd spin current do not vanish if the N\'eel order is oriented along specific directions. A Hall angle of about 25\% is predicted for \ce{RuO2}  without spin-orbit interaction.

As a matter of fact, in the presence of either spin-orbit coupling or non collinear order, the electron spin cannot be treated as a good quantum number. This limits the spin diffusion length, and consequently reduces the spin-to-charge conversion efficiency. Nevertheless, in the absence of the relativistic interaction the collinear antiferromagnetic order preserves the spin. This leads to a more efficient spin-to-charge conversion via magnetic spin Hall effect. From a theoretical standpoint we shall emphasize that the predicted magnetic spin Hall conductivities are two orders of magnitude larger than the conventional spin Hall conductivities. This makes the magnetic spin Hall effect very promising in the view of possible applications in spin-orbit torque magnetic random access memories.

\subsection{Anomalous Transport of Magnons \label{ATM}}

\begin{figure}[ht]
	\begin{center}
		\includegraphics[width=0.45\textwidth]{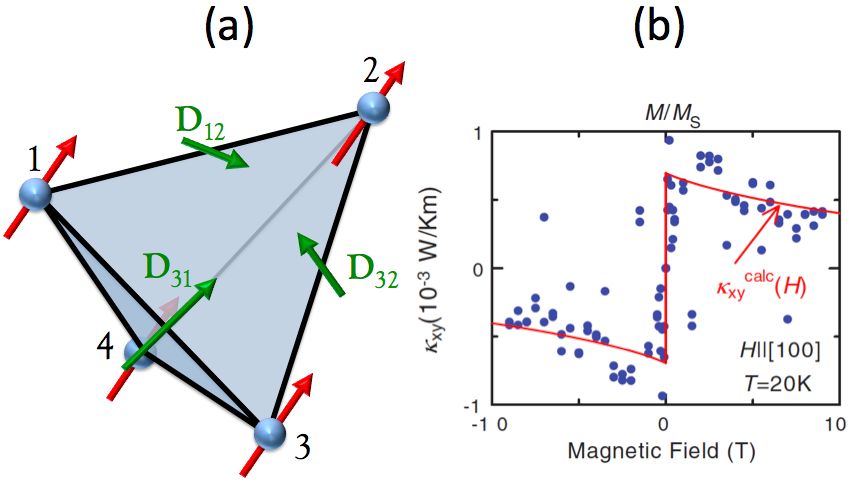}
	\end{center}
	\caption{(Color online) (a) The magnetic unit cell of a ferromagnetic pyrochlore. The red arrows represent the magnetic moments and the green arrows represent the Dzyaloshinskii-Moriya vector. (b) Transverse thermal conductivity measured experimentally as a function of the external magnetic field. From \cite{Onose2010}.}
	\label{Onose}
\end{figure}

\begin{figure}[ht]
	\begin{center}
		\includegraphics[width=0.45\textwidth]{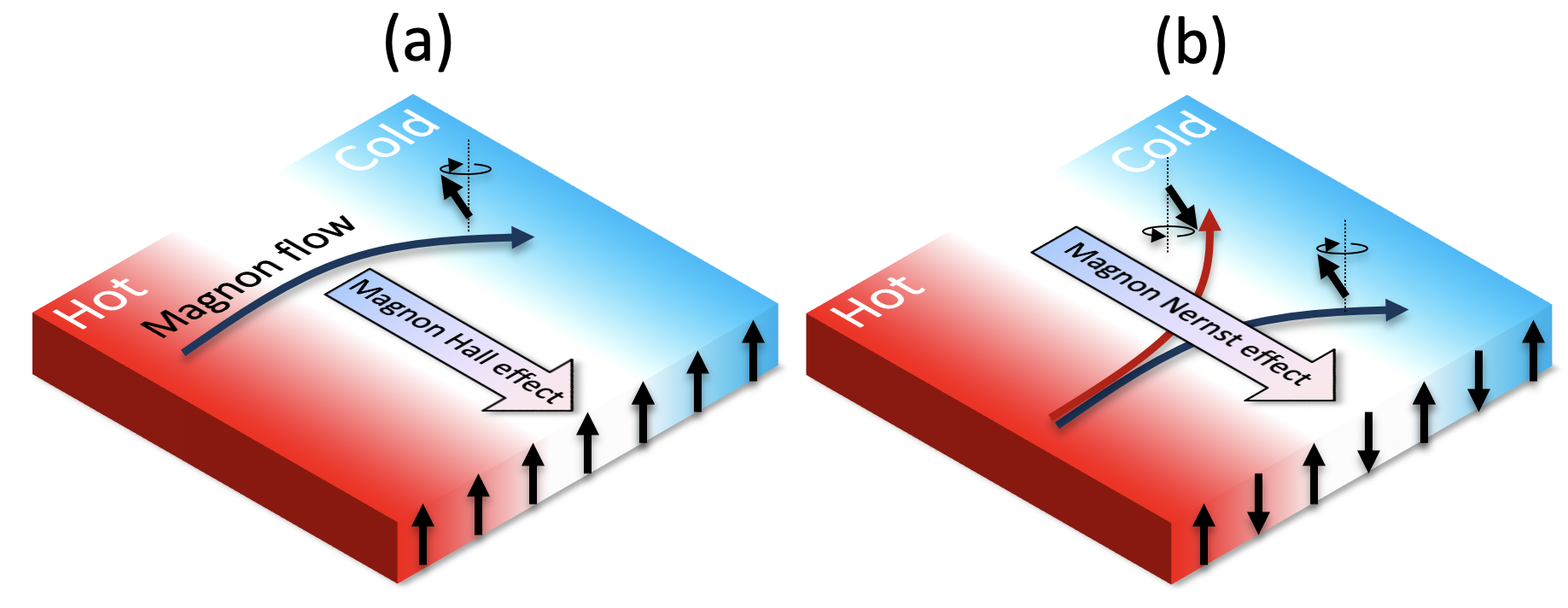}
	\end{center}
	\caption{(Color online) (a) Magnon Hall effect in a ferromagnet. The magnons travel from the hot to the cold region and experience a Hall effect that induces transverse magnon and heat currents, mediated by the magnons. (b) In compensated collinear antiferromagnets, magnon modes of opposite chirality are deviated towards opposite directions, resulting in a pure spin current and a vanishing heat current.}
	\label{fig:mhe}
\end{figure}

The elementary excitations of magnetic crystals, called magnons, are also described by Bloch states and therefore can display anomalous transport properties in the presence of chirality in their band structure. In non-centrosymmetric magnets, for instance, this chirality is associated with the presence of Dzyaloshinskii-Moriya interaction \cite{Dzyaloshinskii1957,Moriya1960}, an antisymmetric exchange interaction that reads
\begin{equation}
H_{DM}=\sum_{\left \langle ij \right \rangle}\mathbf{D}_{ij}\cdot \left (\mathbf{S}_{i}\times \mathbf{S}_{j} \right),
\end{equation}
where $\mathbf{D}_{ij}$ is referred to as the Dzyaloshinskii-Moriya vector. From the expression above, one sees that this interaction tends to cant neighboring spins away from each other and changes sign upon exchanging $\mathbf{S}_{i}$ and $\mathbf{S}_{j}$. This interaction imprints a geometrical phase on the propagating magnon which henceforth experiences an effective Lorentz force. As a result, when applying a thermal gradient on such a non-centrosymmetric magnet, a transverse magnon current develops, called the magnon Hall effect. The magnon Hall effect was first observed experimentally by \citet{Onose2010} in the ferromagnetic insulator Lu$_2$V$_2$O$_7$, whose vanadium sublattice is composed of corner-sharing tetrahedra forming a pyrochlore lattice [see Fig. \ref{Onose}(a)]. Below Curie temperature, $T_C\approx 70$K, the vanadium spins order ferromagnetically in the pyrochlore structure. Because the midpoint between any two apices of a tetrahedron is not an inversion symmetry center, there is a nonzero Dzyaloshinskii-Moriya interaction with $\mathbf{D}_{ij}$ being perpendicular to the vanadium bond and parallel to the surface of the cube containing the tetrahedron. This particular configuration of the Dzyaloshinskii-Moriya vectors imprints a geometrical phase when the magnon Bloch state performs a loop on an elementary triangular plaquette \cite{Katsura2010}, resulting in the Hall effect. Although magnons do not carry charge, they carry both spin and energy and therefore the magnon Hall effect reveals itself as a transverse heat current, as illustrated on Fig. \ref{fig:mhe}(a). The transverse thermal conductivity of Lu$_2$V$_2$O$_7$ reported by \citet{Onose2010} is displayed on Fig. \ref{Onose}(b). The sign reversal of the thermal conductivity upon switching the magnetization confirms that the thermal Hall effect is due to magnons. As a matter of fact, phonons - alike magnons - can also experience ordinary \cite{Strohm2005} and anomalous Hall effect \cite{Sugii2017,Li2020f} and disentangling the two contributions is sometimes challenging. 

The magnon Hall effect discussed above arises from the Berry curvature in momentum space experienced by the magnons in the presence of Dzyaloshinskii-Moriya interaction \cite{Katsura2010}. Alternatively, Matsumoto and Murakami \cite{Matsumoto2011,Matsumoto2011b} demonstrated that such a curvature can also be obtained from the dipolar interaction. Similarly, noncollinear, non-coplanar texture can also induce such a Berry curvature, analogously as for electrons. When the magnon current is driven by the thermal gradient, the transverse thermal conductivity reads
\begin{equation}\label{eq:matsumoto}
\kappa^{xy}=\frac{k_B^2T}{\Omega(2\pi)^2\hbar}\sum_{{\bf k},n}c_2\left(n_B(\varepsilon_{{\bf k},n})\right)\Omega_n^z({\bf k}),
\end{equation}
where $n_B(\varepsilon)$ is the Bose-Einstein distribution, $\varepsilon_{{\bf k},n}$ is the magnonic band dispersion for band $n$ and $\Omega_n^z({\bf k})$ its associated Berry curvature. The function $c_2(x)$ is given by 
\begin{equation}
c_2(z)=(1+x)\left(\ln\frac{1+x}{x}\right)^2-\left(\ln x\right)^2-2 {\rm Li}_2(-x),
\end{equation}
Li$_2(x)$ being the dilogarithm. One readily sees that irrespective of the source of Berry curvature, because magnons are bosons they display non trivial temperature dependence, which can constitute an original signature in experiments. We emphasize that the Chern number associated with band $n$ is of course temperature-independent and reads
\begin{equation}
\mathcal{C}_n=\frac{1}{2\pi\Omega}\sum_{\bf k}\Omega_n^z({\bf k}).
\end{equation}
 In the discussion below, we address the nature of anomalous magnonic transport in various types of antiferromagnets.

\subsubsection{Non-abelian magnonics and pseudospin texture}

Because of the Bloch character of their wave functions, it is tempting to draw a parallel between topological transport of electrons and magnons \cite{Mook2018}. In antiferromagnets though, several aspects need to be carefully considered. First, in the presence of effective time-reversal symmetry, the band degeneracy implies a non-abelian Berry curvature as briefly discussed in Section \ref{s:sym}. For instance, \citet{Daniels2018} have adapted to magnon transport in collinear antiferromagnets the non-abelian wave-packet theory initially developed by \citet{Culcer2004} in the context of multiband electronic transport. In addition, because the magnetic configuration can be quite complex, one needs to define a pseudospin which represents the angular momentum effectively carried by the collective excitation. The magnonic pseudospin possesses properties similar to the spin angular momentum of conduction electrons and can display pseudo-spin momentum locking either due to the noncollinear magnetic arrangement \cite{Okuma2017} or in the presence of Dzyaloshinskii-Moriya interaction \cite{Kawano2019,Kawano2019c}. The former gives a momentum-dependent pseudospin texture in reciprocal space, similar to the one obtained for the conduction electron spin in Fig. \ref{fig:MagneticSpinHall}, and promotes the magnonic counterpart of the magnetic spin Hall effect \cite{Cheng2020}. The latter is the counterpart of the spin texture in momentum space obtained in non-centrosymmetric metals with spin-orbit coupling, leading to the magnonic analog of the Rashba and Dresselhaus spin-momentum locking and its associated Rashba-Edelstein effect \cite{Li2020d}.

\subsubsection{Magnon Hall effect}

Similarly to anomalous Hall effect, the magnon Hall effect only exists in antiferromagnets if the time-reversal symmetry is effectively broken, i.e. if there exists no crystal symmetry that restores it. Concretely, if one considers a bipartite collinear antiferromagnet, such as the one depicted in Fig. \ref{AF-degeneracy}, the two excitation modes possess opposite chirality. Therefore, the Berry curvature of each mode is opposite, which leads to vanishing magnon Hall effect. Nonetheless, as discussed in the next subsection, the analog of spin Hall effect, the magnonic spin Nernst effect, survives [see Fig. \ref{fig:mhe}(b)]. 
\begin{figure}[ht]
	\begin{center}
		\includegraphics[width=0.45\textwidth]{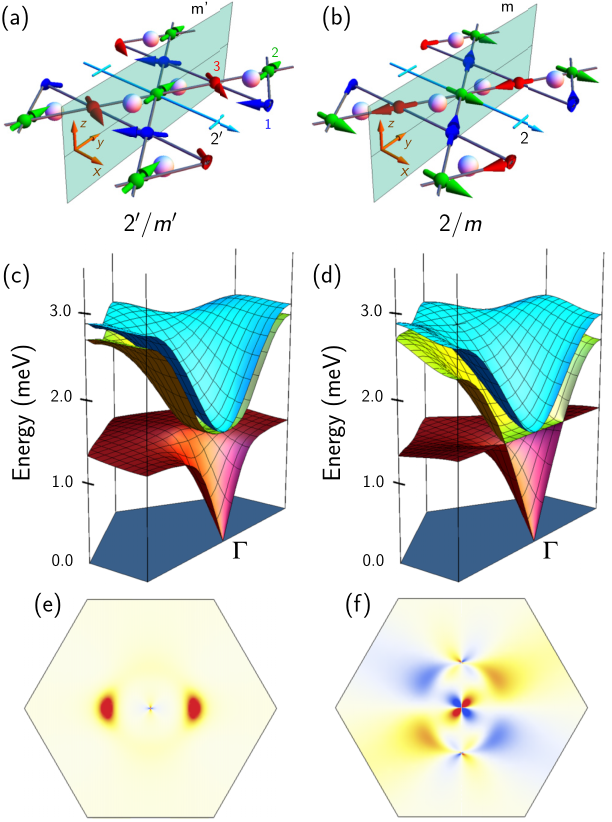}
	\end{center}
	\caption{Top row: Magnetic texture with negative vector chirality on the kagom\'e lattice. Second row: Magnon band structures of the system sketched in (a) and (b). Third row: Berry curvature of the lowest band of (c) and (d). \cite{Mook2019}}
	\label{MOOK}
\end{figure}

Magnon Hall effect has been investigated theoretically in canted collinear antiferromagnets in the presence of Dzyaloshinskii-Moriya interaction (i.e., lacking inversion symmetry)\cite{Owerre2017d,Kawano2019}: the canting explicitly breaks time-reversal symmetry whereas Dzyaloshinskii-Moriya interaction provides the necessary Lorentz force on the magnons. It has also been investigated in noncollinear kagom\'e antiferromagnets with \cite{Laurell2018,Kim2019d} or without canting \cite{Mook2019}, following the same symmetry principles as for electronic transport. All these studies follow the linear spin-wave theory and compute the anomalous transport properties using Eq. \eqref{eq:matsumoto}. In the following, we discuss Ref. \onlinecite{Mook2019} in more details, as an example of magnonic Hall transport in antiferromagnets.

\citet{Mook2019} theoretically predicted thermal magnon Hall effect in noncollinear coplanar insulating antiferromagnets on the kagom\'e lattice. Similar to anomalous Hall effect in noncollinear antiferromagnets \cite{Chen2014,Kubler2014}, they demonstrated two necessary requirements for a nonzero magnon Hall effect within the language of linear spin-wave theory: (1) a broken effective time-reversal symmetry and (2) a magnetic point group compatible with ferromagnetism. The coplanar antiferromagnetic order with negative vector spin chirality is shown in Figs. \ref{MOOK}(a) ($ \alpha=0$) and \ref{MOOK}(b) ($ \alpha=-\pi/2 $), where the global rotation angle $\alpha$ is defined as the angle of the green magnetic moment in Fig. \ref{MOOK}(a) with the $ y $ direction. For $ \alpha=0$, the magnonic band structure displays a band gap between the lowest and the top bands, as shown in Fig. \ref{MOOK}(c), and the Berry curvature of the lowest band [Fig. \ref{MOOK}(e)] remains positive over the Brillouin zone, indicating a nonzero thermal Hall conductivity. The magnon band structure for $ \alpha=-\pi/2 $ exhibits no band gap but two touching points [Fig. \ref{MOOK}(d)]. In Fig. \ref{MOOK}(f), the Berry curvature is antisymmetric over the Brillouin zone, leading to vanishing magnon Hall effect. Any other $ \alpha $ configurations can be understood as an intermediate of these two configurations. Finally, they proposed cadmium kapellasite CdCu$ _{3} $(OH)$ _{6} $(NO$ _{3} $)$ _{2} \cdot$H$ _{2} $O as a feasible material realization. 

\begin{figure}[ht]
	\begin{center}
		\includegraphics[width=0.46\textwidth]{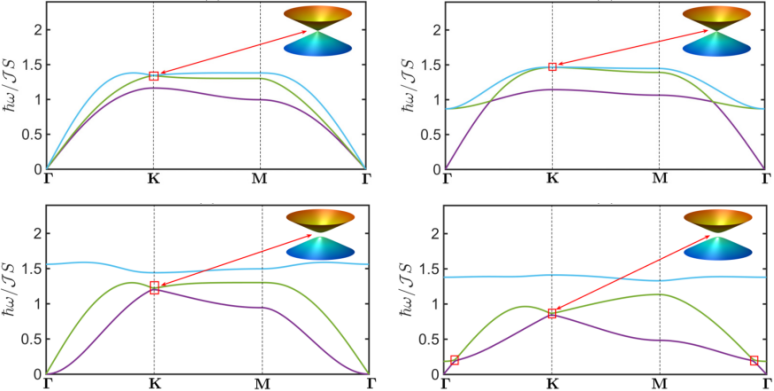}
	\end{center}
	\caption{Top row: Magnon bands at zero magnetic field without (left) and with (right) the out-of-plane Dzyaloshinskii-Moriya interaction. Bottom row: Magnon bands at finite magnetic field without (left) and with (right) the out-of-plane Dzyaloshinskii-Moriya interaction. From \cite{Owerre2017e}.}
	\label{Owe}
\end{figure}

Another canonical example is the thermal Hall effect in frustrated kagom\'e antiferromagnets theoretically studied by \citet{Owerre2017e}. In this system, the magnon Hall effect arises from the scalar spin chirality due to the noncoplanar spin configuration induced by an external magnetic field. The magnon bands of the noncollinear antiferromagnet along the high-symmetry points of the Brillouin zone are shown in Fig. \ref{Owe}. In the absence of magnetic field (top row), two magnon branches cross at the $\textbf{K}$ point, forming a Dirac point. This Dirac point is protected by the combination of mirror reflection and time-reversal symmetries. There is no thermal Hall effect in the absence of magnetic field, irrespective of the presence of Dzyaloshinskii-Moriya interaction. When turning on the magnetic field (bottom row), a gap opens turning on the magnon Hall effect. An experimental realization of this situation could be achieved \cite{Owerre2017e} in KCr$ _{3} $(OH)$ _{6} $(SO$ _{4} $)$ _{2} $ and KFe$ _{3} $(OH)$ _{6} $(SO$ _{4} $)$ _{2}$. Magnon Hall effect has also been investigated in the antiferromagnetic kagome lattice with XXZ anisotropy, in the absence of Dzyaloshinskii-Moriya interaction and in the presence of an external field \cite{GomezAlbarracin2021}. Below a critical temperature, the external field induces a chiral phase transition that favors the magnon Hall effect.

\subsubsection{Magnonic spin Nernst effect}

The spin Nernst effect features the generation of a pure spin current, transverse to a thermal gradient. It is the thermal analog of the electrically-driven spin Hall effect and exists in its electronic \cite{Sheng2017} and magnonic \cite{Shiomi2017} versions. In contrast with the magnon Hall effect, that requires certain symmetries to be broken, the (magnonic) spin Nernst effect, is much more robust and is expected to emerge in pretty much all magnetic materials, as long as a spin Berry curvature is present. Many models have been investigated, considering various types of magnetic configurations and various sources of Berry curvature. 

\citet{Cheng2016b} and \citet{Zyuzin2016b} investigated the magnonic spin Nernst effect in a collinear honeycomb antiferromagnet under thermal gradient. In such a system, the two magnon modes of opposite chirality are degenerate and in the presence of Dzyaloshinskii-Moriya interaction they flow in opposite directions, leading to a pure transverse spin current mediated by magnons, as illustrated in Fig. \ref{fig:mhe}(b). In this honeycomb Heisenberg lattice, the nearest-neighbor antiferromagnetic exchange coupling and the easy-axis anisotropy preserve the combination of time-reversal symmetry with the 180 $ ^{\circ} $ rotation in the spin space while breaking the inversion symmetry. Therefore, it induces a nonzero Berry curvature {\em even in the absence} of the Dzyaloshinskii-Moriya interaction. On the other hand, the second nearest-neighbor Dzyaloshinskii-Moriya interaction term breaks the combined symmetry (time-reversal + 180 $ ^{\circ} $ spin rotation) and makes the energy spectrum of the two magnon branches nondegenerate $\omega\left( \textbf{k}\right) \neq \omega\left( -\textbf{k}\right)$, causing a population imbalance between the $ \textbf{k} $ and $ -\textbf{k} $ states and therefore resulting in a non-zero Berry curvature. Since the Dzyaloshinskii-Moriya interaction correction to $ \omega\left( \textbf{k}\right) $ is opposite for the two branches, the transverse thermal current vanishes identically. Notice that magnonic spin Nernst effect has also been predicted on collinear antiferromagnetic skyrmion crystals \cite{Diaz2019,Daniels2019}. There, the Berry curvature of the magnonic wave function is due to the skyrmion lattice background rather than to Dzyaloshinskii-Moriya interaction directly. 

The experimental observation of magnonic spin Nernst effect has been reported by \citet{Shiomi2017} in MnPS$ _{3}$. In their experiment, the authors investigate the thermoelectric voltage in Pt strips deposited on MnPS$ _{3}$ as a function of the temperature gradient. They attribute the nonmonotonic temperature dependence of the thermoelectric voltage to the magnonic spin Nernst effect in MnPS$ _{3}$. To the best of our knowledge, this experiment remains the only one explicitly addressing the magnonic spin Nernst effect in antiferromagnets and further experimental search along this line is highly desired.

A problem that has received little attention in the past and that is currently being progressively revived is the distinct role of magnetoelastic and dipolar coupling. As a matter of fact, in ferromagnets dipolar coupling is strong and often dominates over magnetoelastic coupling, governing the formation of magnetic domains. In contrast, in antiferromagnets the dipolar coupling is small (but non-vanishing) and is rather dominated by the magnetoelastic coupling. Hence, these two interactions crucially influence the dynamics of the antiferromagnetic order and antiferromagnetic soft modes such as magnons. In fact, dipolar coupling does not conserve the spin angular momentum and can mimic the role of spin-orbit coupling, leading to magnon spin relaxation and magnon spin Nernst effect \cite{Shen2020}. 
On the other hand, the coupling between magnons and phonons can lead to magnon-polaron bands with non-trivial topology \cite{Park2019}, and possibly magnetoelastic spin Nernst effect \cite{Park2020b}. 

\begin{figure}[ht]
	\begin{center}
		\includegraphics[width=0.46\textwidth]{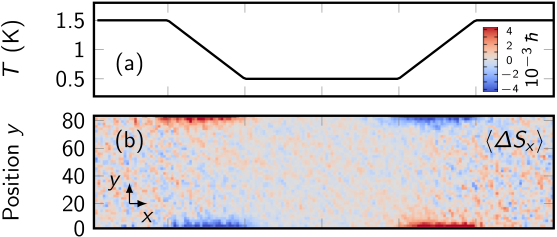}
	\end{center}
	\caption{Nonequilibrium spin accumulation in a single kagom\'e sheet obtained by atomistic spin dynamics simulations. (a) Temperature profile with two opposite gradients. (b) The nonequilibrium spin accumulation at the sample's edges indicates the MSNE. From \cite{Mook2019b}.}
	\label{SNE}
\end{figure}

Besides collinear antiferromagnets, magnonic spin Nernst effect has also been investigated in noncollinear antiferromagnets on kagom\'e lattice \cite{Mook2019b,Li2020e,Cheng2020}. Both references consider the antiferromagnetic insulator KFe$ _{3} $(OH)$ _{6} $(SO$ _{4} $)$ _{2}$ as a platform for their study. This system possesses a 120 $^\circ$ spin configuration with a small out-of-plane canting and exhibits magnonic spin Nernst effect. Figure  \ref{SNE} displays the nonequilibrium magnon spin accumulation obtained at the edges of such an insulator by \citet{Mook2019b} using classical atomistic spin dynamics simulations. The authors also reported the spin Seebeck effect of magnons in this noncollinear antiferromagnet, which features a longitudinal spin current as a response to a temperature gradient.  As mentioned in the previous section, an interesting property of noncollinear antiferromagnetic kagom\'e lattices is the emergence of a spin texture in momentum space. As discussed in Section \ref{subsec:MSHE}, this non-trivial texture gives rise to a "magnetic spin Hall effect", i.e. a transverse spin current whose polarization lies in the scattering plane. Similarly, in the magnonic realm, noncollinear antiferromagnetic kagom\'e lattices also support the magnonic version of the magnetic spin Hall effect, i.e., a magnon-mediated magnetic spin Nernst effect \cite{Mook2019b,Li2020e,Cheng2020}. The symmetry properties of the magnonic spin conductivity tensor are the same as that of the electronic spin conductivity tensor. An interesting idea put forward by \citet{Cheng2020} is that, because of the non-trivial magnon spin texture in momentum space, an unpolarized electron current flowing inside the noncollinear antiferromagnet can generate a magnonic spin current through electron-magnon scattering. We emphasize that these interesting features have been demonstrated for model systems only. The exploration of these effects in noncollinear antiferromagnets of current interest for spintronics, such as IrMn$_3$ \cite{Rezende2020} or SnMn$_3$ \cite{Park2018}, is highly desired.

A last aspect we wish to mention is the possibility to generate pure magnon densities out of equilibrium. This effect emerges in noncentrosymmetric antiferromagnets and is the magnonic analog to the Rashba-Edelstein effect. The latter takes place in noncentrosymmetric metals possessing spin-orbit coupling, i.e., a metallic or a semiconducting interface. The inversion symmetry breaking enables the emergence of a Rashba-like spin-orbit coupling, which promotes the electrical generation of a nonequilibrium (electronic) spin density. Similarly, in antiferromagnets possessing Dzyaloshinskii-Moriya interaction, the initially degenerate magnon states become spin-polarized in the presence a thermal gradient \cite{Li2020d,Zhang2020e}. A parent mechanism, tagged the magnon gravito-magnetoelectric effect, has also been predicted in antiferromagnets \cite{Shitade2019}. This effect follows a similar scenario as the magnonic Edelstein effect: a thermal gradient generates a nonequilibrium magnetization carried by magnons. Nonetheless, whereas the Edelstein effect requires time-reversal symmetry and inversion symmetry breaking, the gravitomagnetoelectric effect requires time-reversal symmetry breaking and therefore involves interband, rather than intraband, transitions.


\subsubsection{Spinon (spin) Nernst effect}

In magnetic materials, not all excitations are bosonic magnons. In fact, in situations where geometric frustration is large, the elementary magnetic excitations are better represented by fermionic or anyonic quasiparticles, as further discussed in Section \ref{s:topexc}. A large thermal Hall conductivity was recently reported in the quantum spin-ice phase of Tb$_2$Ti$_2$O$_7$ and attributed to charge neutral magnetic excitations \cite{Hirschberger2015a}. This thermal Hall effect is induced by applying an external magnetic field to the magnet and is therefore not associated by the intrinsic Berry curvature of the system (in contrast, for instance, with \cite{Onose2010}). Nonetheless, this observation opens interesting perspective because in quantum spin liquids, the elementary excitations are often dressed with a gauge flux, and are therefore likely to exhibit an intrinsic (anomalous or spin) Hall effect of some sort. For instance, \citet{Zhang2018} have investigated the spin Nernst effect of spinons in the paramagnetic regime of a collinear honeycomb antiferromagnetic insulator, similar to \cite{Cheng2016}. They found that even in the absence of magnetic order, a transverse spin current can be mediated by these thermally driven spinons. In a similar spirit, the spinon spin Nernst effect has been investigated in honeycomb Kitaev spin liquid \cite{Gao2019,Carvalho2018}, leading to Majorana edge states observed experimentally \cite{Kasahara2018}. Thermal Hall effect of spinons has also been investigated theoretically in kagom\'e lattice noncollinear magnets \cite{Gao2020c} and honeycomb lattice antiferromagnets \cite{Gao2020d}. Because spinons are bosons, the thermal Hall effect is quantized in the non-trivial phase at zero temperature \cite{Gao2020d}.



\section{Topological antiferromagnets \label{s:topaf}} 

In the previous section, we have seen that the presence of Berry curvature in the momentum space of a crystal induces a geometrical phase on the (electronic or magnonic) Bloch wave functions, which promotes a variety of Hall effects (spin, charge, magnonic, so-called topological etc.). These effects are {\em geometrical} rather than {\em topological} because the driving force is the {\em local} Berry curvature in momentum space that acts like an emergent (possibly spin-dependent) magnetic field. In their foundational work, \citet{Thouless1982} realized that the quantized Hall effect of a metal submitted to an external magnetic field is equal to the sum of the Berry curvatures of each Bloch state present below Fermi level, integrated over the Brillouin zone (the famous TKNN theorem - see Section \ref{ss:Berrycurv}). The integration of the Berry curvature over the momentum space is nothing but the emergent magnetic flux endowed by this Bloch state. Therefore, the quantum Hall effect is associated to the {\em global} Berry phase of the Bloch states and is of {\em topological} origin. A few years later, \citet{Haldane1988} showed that engineering the Berry phase of a material can lead to quantized Hall effect even in the absence of an external magnetic field, thereby initiating the hunt for novel quantum phases of matter.\par

The search for topological materials has blossomed over the past decade \cite{Hasan2010,Qi2010,Moore2010,Ortmann2015} and the continuous expansion of topological phases is regularly fed by the prediction of novel candidate materials \cite{Bansil2016}. A topological material is characterized by a non-local topological order protected by symmetries and indexed by a topological invariant. The same is true for the quantum spin liquids \cite{Wen1991} briefly discussed in Section \ref{s:topexc}. For instance, a time-reversal invariant topological insulator is identified by a $\mathbb{Z}_2$ topological invariant and features insulating bulk states and an odd number of conducting Dirac surface states, protected by time-reversal symmetry ${\cal T}$. The $\mathbb{Z}_2$ invariants can be determined from the knowledge of the parity of the occupied Bloch wave functions at the time-reversal invariant points in the Brillouin zone \cite{Fu2007b}. Another flavor of the topological insulator is the topological crystalline insulator that possesses similar features but protected by a crystalline symmetry \cite{Fu2011,Hsieh2012,Slager2013}. Combining crystal and time-reversal symmetry enriches the potential topological phases \cite{Fang2015} and intense effort has been paid over the past few years to develop a systematic search of such candidate materials \cite{Bradlyn2017,Bradlyn2016,Schindler2018}. In fact, topological materials are not limited to insulators but also encompass a wealth of semimetals with Dirac or Weyl cones located close the Fermi level. Systematic investigations using density functional theory and aiming at establishing an exhaustive catalogue of topological insulators and semimetals have been recently published exploiting either symmetry eigenvalues method \cite{Kruthoff2017}, elementary band representations \cite{Vergniory2019,Zhang2019h} or the theory of symmetry indicators \cite{Po2017,Tang2019a,Tang2019b}. These powerful approaches, which suggests that up to 27\% of all crystalline matter is in fact topological \cite{Vergniory2019}, have been recently extended to magnetic materials \cite{Xu2020b,Elcoro2020}. \par 

In this section, we focus on the materials that support both topological and antiferromagnetic orders. As exposed in Section \ref{s:sym}, antiferromagnets can display a very rich zoo of spin texture, which vastly enlarges the family of topological materials by introducing new symmetries. To date, most of the effort has been paid towards identifying the compatibility conditions between simple, mostly collinear antiferromagnetism and topological materials.


\subsection{Electronic topological materials}
\subsubsection{$\mathbb{Z}_{2}$ topological insulators}

\begin{figure}[ht]
	\begin{center}
		\includegraphics[width=0.48\textwidth]{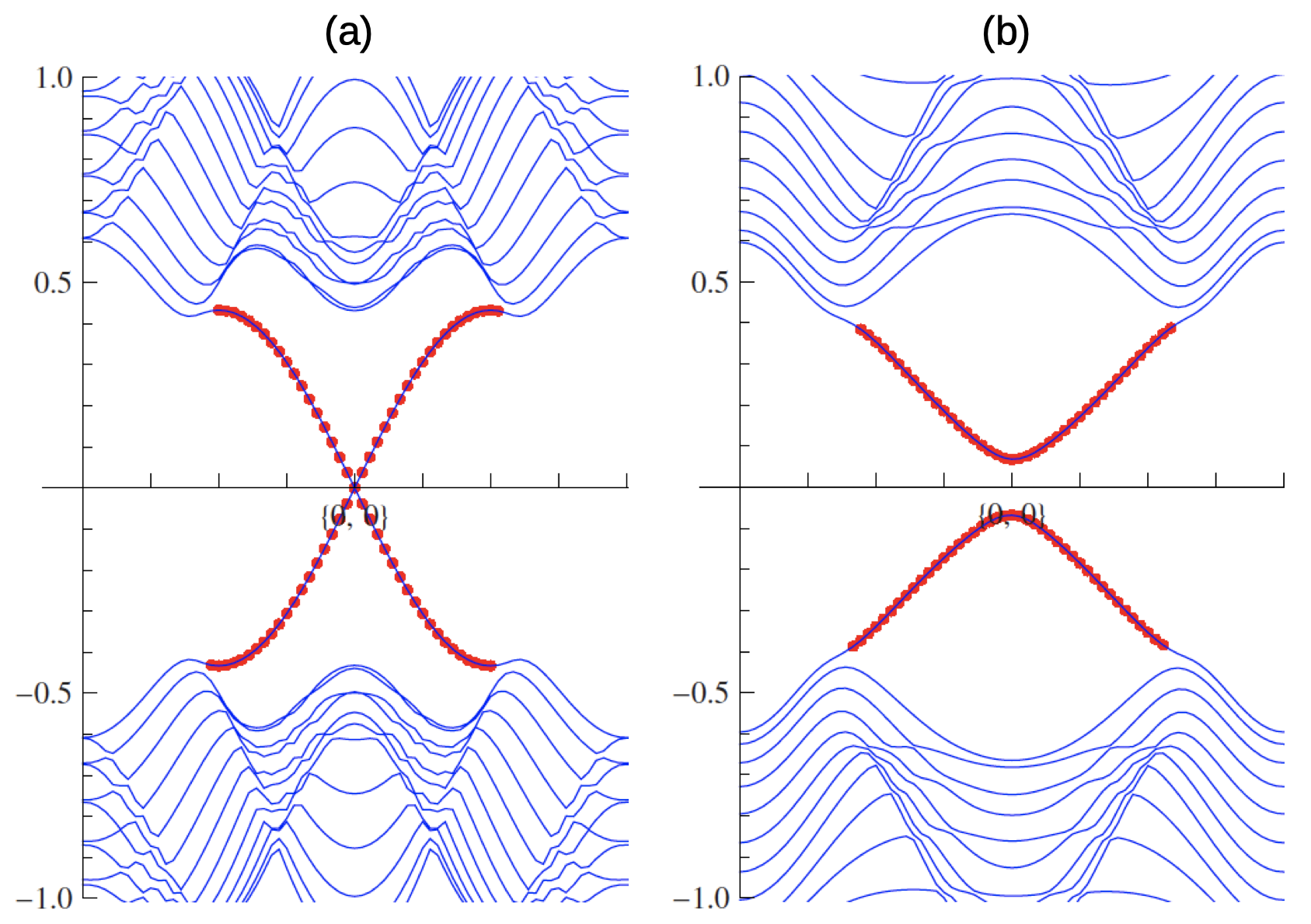}
		\caption{(Color online) Band structure of an antiferromagnetic topological insulator cut along two surfaces. In (a), the surface respects effective time-reversal symmetry whereas in (b), the surface effectively breaks time-reversal. The blue bands refer to the bulk states and the red dots refer to the surface states. From \cite{Mong2010}.}
		\label{Fig:Mong}
	\end{center}
\end{figure}

Because $\mathbb{Z}_{2}$ topological insulators are characterized by time-reversal symmetry, one could naively expect that antiferromagnetism is incompatible with this topology due to their inherent time-reversal symmetry breaking. However, if a crystal symmetry of the antiferromagnet is such that it effectively reverses the magnetic moment back to the original configuration, it is actually possible to obtain topological insulators. In a pioneering theoretical work, \citet{Mong2010} demonstrated that in certain collinear antiferromagnets, time-reversal symmetry combined with sublattice translation ${\cal T}{\bm \tau}_s$ can be preserved so that a topological insulating phase emerges. In this case, the bulk is gapped while the gapless edge state only exists at surfaces obeying the ${\cal T}{\bm \tau}_s$ symmetry, see Fig. \ref{Fig:Mong}. This prediction triggered substantial theoretical effort to extend the classification of topological insulators to antiferromagnets, define the appropriate topological invariant \cite{Fang2013} and propose appropriate model Hamiltonians \cite{Yu2017c}. The robustness of the topological protection of the edge states has been confirmed numerically, as expected \cite{Baireuther2014,Ghosh2017}. Although these predictions are based on the mean-field treatment of the magnetic order, simulations based on dynamic mean-field theory of the electron correlations in a Bernevig-Highes-Zhang model \cite{Bernevig2006b} confirms the emergence of the antiferromagnetic topological insulating phase \cite{Yoshida2013,Miyakoshi2013,Amaricci2018}. The investigation of the correlated system has been extended to Kondo topological insulators \cite{Li2018g}.

In their pioneering work, \citet{Mong2010} suggested that half-Heusler GdPtBi might be an antiferromagnetic topological insulator. As a matter of fact, the Gd magnetic moments form ferromagnetic planes which are stacked antiferromagnetically along the body diagonal \cite{Muller2014}, in a somewhat similar manner as NiO. However, recent experiments indicate that GdPtBi is rather a Weyl semimetal \cite{Hirschberger2016,Suzuki2016} (see below). In fact, much of the experimental effort has been focusing on ${\rm MnBi_2Te_4}$, which is a van der Waals crystal composed of septuple ${\rm MnBi_2Te_4}$ layers stacked along (0001) \cite{Zeugner2019}. The magnetic moments are coupled ferromagnetically along the (0001) planes and weak interlayer antiferromagnetic coupling stabilizes layered antiferromagnetism. Bulk samples exhibit long-range antiferromagnetism below 24 K and the Mn magnetic moments point out-of-plane. A flurry of experiments have recently demonstrated a bulk gap around 200 meV and gapped Dirac surface states \cite{Otrokov2019a,Gong2019b,Vidal2019,Lee2019} around 70-85 meV. We notice that other authors have reported a gapless surface state, unchanged across N\'eel temperature \cite{Hao2019,Swatek2020}, raising the question about possible surface reconstruction and altered magnetism.


The reason layered antiferromagnetic topological insulators are attracting so much attention is because they are the ideal platform to realize the long-sought axion insulator. The axion insulator possesses an insulating bulk, gapped Dirac surface states and is characterized by a half-integer quantum Hall effect and quantized magnetoelectric coupling \cite{Qi2010}. This class of insulator has been recently realized in heterostructures of magnetic topological insulators (typically, ${\rm (Bi,Sb)_2Te_3}$ doped with V or Cr elements), aligned antiferromagnetically with each other \cite{Mogi2017,Grauer2017,Mogi2017b,Xiao2018}. ${\rm MnBi_2Te_4}$ presents an interesting paradigm because of its layered antiferromagnetism, which realizes compensated or uncompensated magnetism depending on the number (odd or even) of septuple layers \cite{Otrokov2019b}. Hence, quantum anomalous Hall effect has been reported for an odd-number of septuple layers \cite{Deng2020} due to the uncompensated ferromagnetism and a transition from quantum anomalous Hall effect to axion state has been reported upon tuning the number of layers \cite{Liu2020,Zhang2019j}. The search for antiferromagnetic topological insulators has been extended to ${\rm MnBi_2Te_4/Bi_2Te_3}$ heterostructures \cite{Rienks2019} and other members of the ${\rm MnBi_2Te_4}$ family \cite{Li2019g,Hu2020a}, including promising results on ${\rm MnBi_4Te_7}$ \cite{Li2019f} and ${\rm MnBi_6Te_{10}}$ \cite{Jo2019b} and the possibility for higher-order topology \cite{Zhang2020b}. Recent first principles calculations also suggest that MnTe-\ce{Bi2(Se, Te)3}-MnTe heterostructure, where MnTe is an antiferromagnet, could display Chern insulating and axion insulating phases \cite{Pournaghavi2021}. \par

Other candidates have been recently uncovered such as ${\rm EuSn_2As_2}$ \cite{Li2019f}, ${\rm MnBi_2Se_4}$ \cite{Chowdhury2019}, but also tetragonal FeS \cite{Hao2017}. ${\rm EuIn_2As_2}$ is also predicted to host high-order topology \cite{Xu2019c}. Noticeably, ${\rm SmB}_6$ is predicted to become a Kondo topological insulator under pressure \cite{Chang2018b,Li2018g}. Finally, it is worth mentioning that the superconductor FeSe, which displays antiferromagnetic order when deposited as a monolayer on SrTiO$_3$ \cite{Huang2017d,Zhou2018}, has been predicted and experimentally proven to be a two-dimensional topological antiferromagnetic insulator \cite{Wang2016i}. The coexistence between topology and superconductivity is particularly appealing for the search of Majorana bound states. In this perspective, higher-order topological states have been predicted in Fe(S,Te) heterostructures \cite{Zhang2019i}. Two-dimensional antiferromagnetic topological insulators have been predicted in ${\rm NiTl_2S_4}$ \cite{Liu2019} and (Sr,Ba)Mn(Sn,Pb) \cite{Niu2019}.

\subsubsection{Chern insulators}

A Chern insulator exhibits quantized anomalous Hall conductance and has been demonstrated in magnetic topological insulators \cite{Mogi2017,Grauer2017,Mogi2017b,Xiao2018}, as discussed above. In order to obtain quantized charge current, time-reversal symmetry must be effectively broken. Several strategies can be pursued to obtain an antiferromagnetic Chern insulator. The rationale is to search for an antiferromagnet that exhibits both anomalous Hall effect in its band structure, as discussed in detail in Section \ref{subsec:AHE}, and an orbital gap. \par

\begin{figure}[ht]
	\begin{center}
		\includegraphics[width=0.48\textwidth]{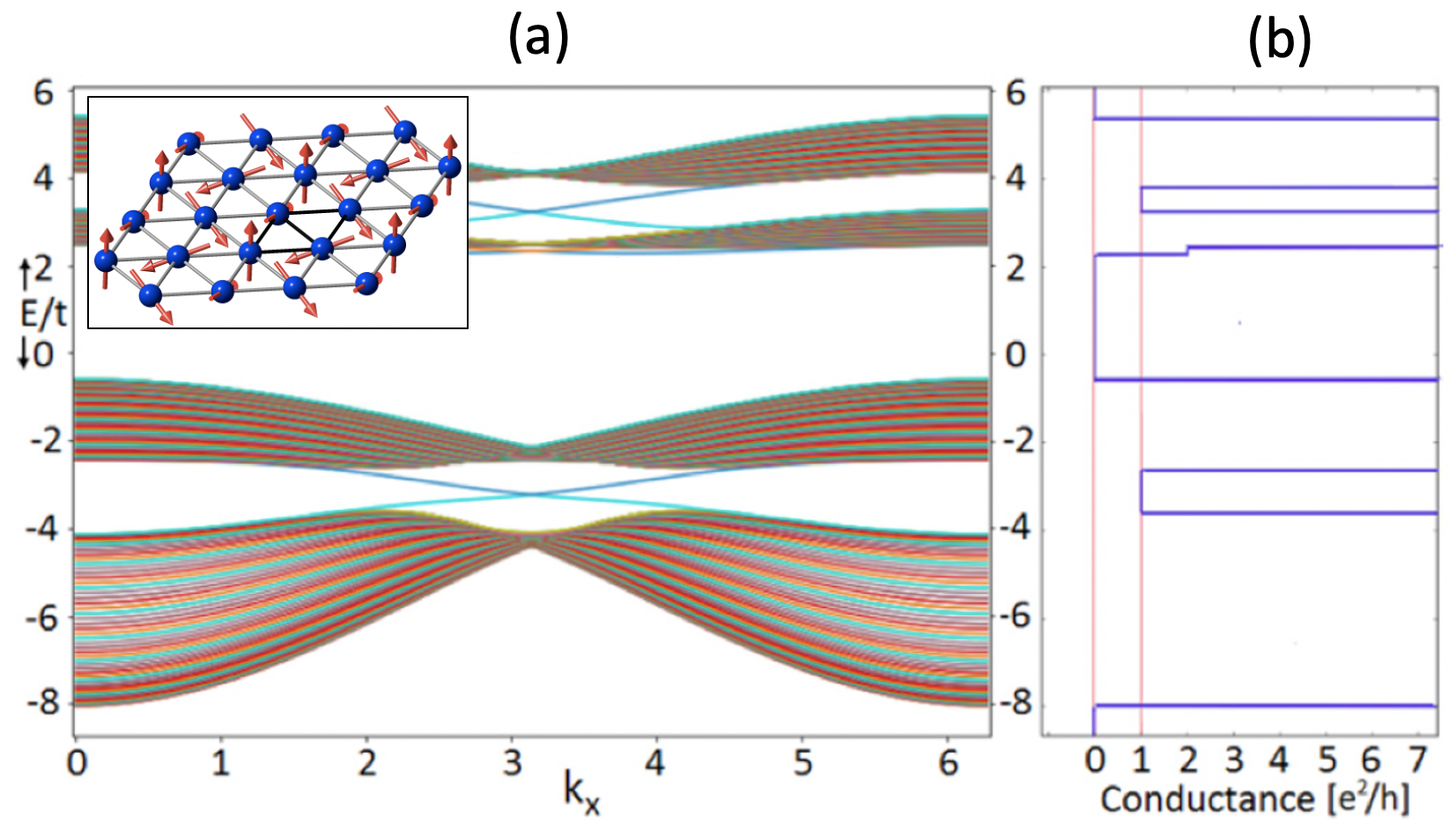}
		\caption{(Color online) (a) Band structure of a triple-Q triangular lattice antiferromagnetic nanoribbon (insert), displaying the topological edge states. (b) Corresponding conductance as a function of the transport energy, showing the quantized plateaus associated with the Chern insulating regime. From \cite{Ndiaye2019}.}
		\label{Fig:Ndiaye2019}
	\end{center}
\end{figure}

The most obvious approach is to exploit the topological Hall effect emerging from the noncollinear, non-coplanar magnetic texture in real space. These non-coplanar magnetic moments promote the emergence of a spin-Berry phase even in the absence of spin-orbit coupling \cite{Taguchi2001} and can experience topological phase transitions \cite{Ohgushi2000,Shindou2001}. Such a situation can be achieved with the Kondo-lattice model on a triangular lattice \cite{Martin2008,Venderbos2012,Ndiaye2019} or with the double-exchange model on a kagom\'e lattice \cite{Ishizuka2013,Chern2014} where noncollinear, non-coplanar antiferromagnetism is stabilized and provides both non-vanishing Berry curvature and orbital gap. An example of quantized edge conductance obtained for such a triple-Q triangular lattice antiferromagnet is displayed on Fig. \ref{Fig:Ndiaye2019}. Recently, it was proposed that ${\rm K_{0.5}RhO_2}$ could be such an antiferromagnetic Chern insulator \cite{Zhou2016}. Another realization of an antiferromagnetic Chern insulator was proposed by \citet{Klinovaja2015} in a model combining stripes with a magnetic texture coupled together with spin-selective coupling. In the presence of strong electron-electron interaction, fractional quantum anomalous Hall effect is obtained.

One could also exploit the anomalous Hall effect obtained in coplanar noncollinear antiferromagnets observed in Mn$_3$Sn \cite{Nakatsuji2015,Nayak2016}, and attributed to the coexistence of the spin-Berry phase of the electronic ground states in the presence of spin-orbit coupling and symmetry breaking due to the noncollinear spin texture \cite{Chen2014,Kubler2014}. Finally, the crystal Hall effect \cite{Smejkal2020} could also be a source of anomalous Hall effect and therefore could lead to an antiferromagnetic Chern insulator, if an orbital gap can be opened. Considering that this field remains in its infancy, these comments are merely speculations but are worth investigating further.

\subsubsection{Dirac semimetals}

In a generic investigation of accidental band degeneracy in crystals, \citet{Herring1937} noticed that twofold degeneracies often occur even in the absence of any symmetry. These degeneracies become particularly meaningful when located close to Fermi level and when the wave functions associated with them are topologically non-trivial \cite{Armitage2018}. Among the various possibilities, Dirac semimetals are characterized by the presence of doubly degenerate Dirac fermions close to Fermi level, protected by ${\cal PT}$ symmetry \cite{Young2012}. As a matter of fact, assuming the existence of a linearly dispersing Weyl point at a given point ${\bf k}$ of the Brillouin zone, time-reversal symmetry ${\cal T}$ requires that another Weyl point with an equal Chern number be present at $-{\bf k}$. Since the total Chern number of the Fermi surface must vanish, there exist two more Weyl points of opposite Chern number at ${\bf k}'$ and $-{\bf k}'$. In the presence of inversion symmetry ${\cal P}$, the Chern number at ${\bf k}$ and $-{\bf k}$ must be opposite and therefore, under ${\cal PT}$ symmetry, there must be two linearly dispersing bands at ${\bf k}$ and two similar bands at $-{\bf k}$. In addition, to avoid that a gap opens at the band crossing, an additional crystal symmetry, such as a threefold or sixfold rotation but also nonsymmorphic symmetries, must be enforced. In other words, the two doubly degenerate bands must have different representations of the crystal symmetry operation \cite{Tang2016}. Depending on the position of the Dirac points, either at high symmetry points of the Brillouin zone or at a generic point on a $C_3$ symmetry axis, the Dirac semimetal is referred to as Type I or Type II. Nonmagnetic Dirac semimetals (Type II) have been reported, for instance, in ${\rm Cd_3As_2}$ \cite{Wang2013h,Neupane2014,Moll2016} and ${\rm Na}_3$Bi \cite{Liu2014,Xiong2015}.\par

\begin{figure}[ht]
	\begin{center}
		\includegraphics[width=0.48\textwidth]{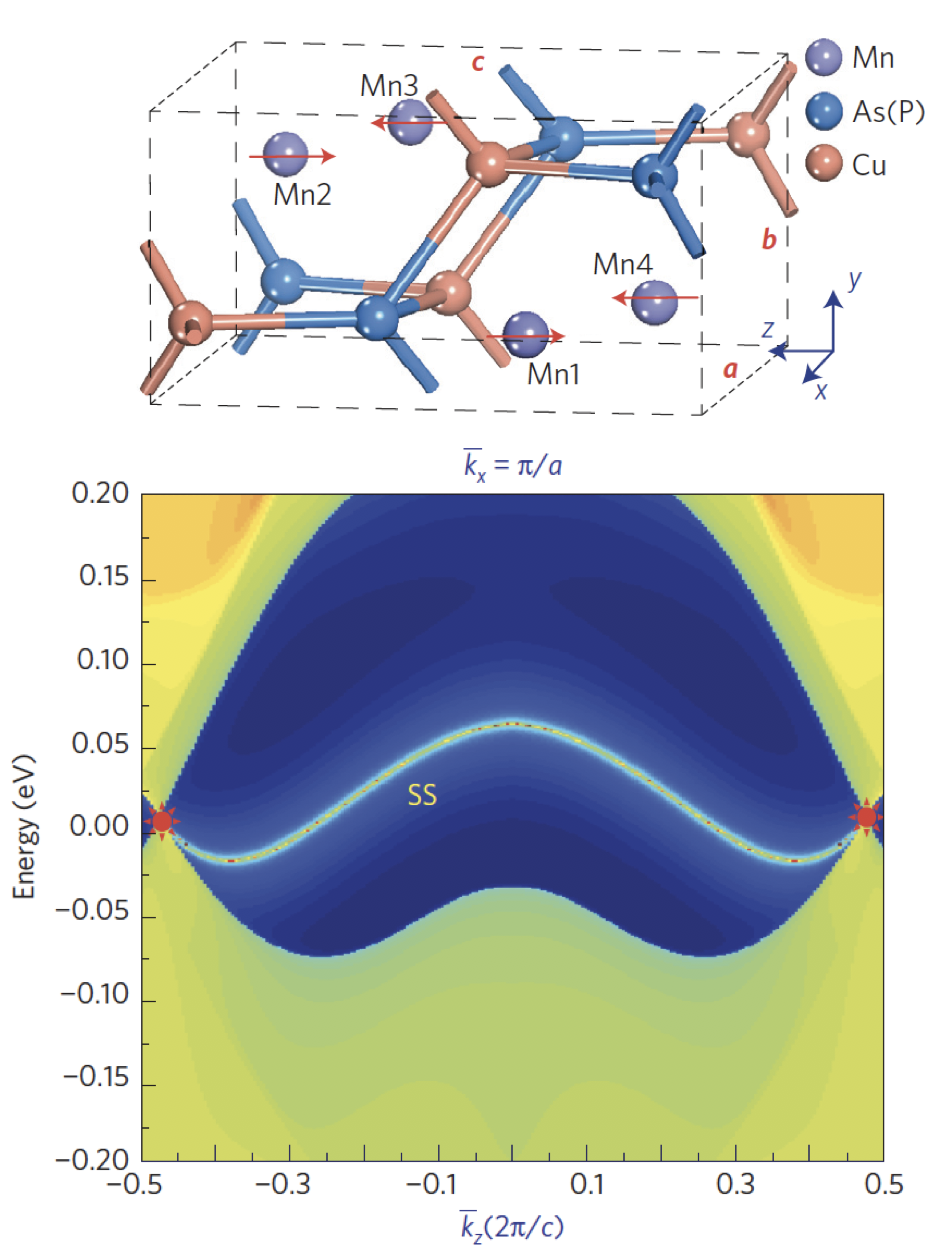}
		\caption{(Color online) (top) Magnetic unit cell of orthorhombic CuMnAs. (bottom) Electronic spectrum on the (010) surface along $\bar{k}_x=\pi/a$. The red stars denote the gapless Dirac points and SS identifies the surface state. From \cite{Tang2016}.}
		\label{Fig:DiracSM}
	\end{center}
\end{figure}

In fact, Dirac semimetals were recently extended to magnetic materials \cite{Brzezicki2017,Young2017,Wang2017j,Hua2018}. Since Kramers degeneracy at point ${\bf k}$ is a requirement to obtain a Dirac point, time- or nonsymmorphic time-reversal symmetry must be preserved. We remind that nonsymmorphic time-reversal symmetry is the combination ${\cal T}{\bm \tau}_s$, where ${\bm \tau}_s$ is a sublattice translation operator. Therefore, Dirac semimetals can be found in Type II magnetic space group \cite{Yang2014c}, in Type III magnetic space group, they can only exist when ${\cal PT}$ is preserved (such as in \cite{Tang2016}) and finally, in Type IV, there is no pure time-reversal symmetry but the nonsymmorphic time-reversal symmetry $\mathcal{T}{\bm\tau}_s$ is preserved. In the latter, if ${\cal PT}{\bm\tau}_s$ is preserved, it is a sufficient condition to obtain Dirac semimetal. However, in general, such band crossing results in a gap due to the level repulsion unless we introduce an additional symmetry to stick the bands together at the band crossing point \cite{Wang2017j}.

In a magnetic material where ${\cal T}$ is broken, the parity must be such that it restores the time reversal, ensuring that ${\cal PT}$ remains preserved. \citet{Tang2016} demonstrated that this happens in the orthorhombic phase of CuMnAs and CuMnP, as shown on Fig. \ref{Fig:DiracSM}. As a matter of fact, CuMnAs possesses the required symmetry: the two antiferromagnetic sublattices are inversion partners, so that, although time-reversal and parity symmetries are individually broken, their combination is preserved. This symmetry allows for the emergence of Dirac points in the bulk, and associated (unprotected) Fermi arcs at the surface. Another interesting candidate is ${\rm EuCd_2As_2}$ \cite{Hua2018,Ma2020}, that breaks ${\cal T}$ but preserves ${\cal P}$ as well as the nonsymmetric time-reversal symmetry ${\cal T}{\bm \tau}_s$. Finally, Dirac points have also been reported in ${\rm EuMg_2Bi_2}$, but above its N\'eel temperature \cite{Kabir2019} (so not related to the magnetic order). 

Dirac points have also been reported in NdSb \cite{Wang2018f} and FeSn \cite{Lin2020}. Progress has also been achieved in two dimensions where two dimensional Dirac points have been observed in iron-pnictide superconductor ${\rm (Ba,Sr)Fe_2As_2}$ \cite{Morinari2010,Chen2017c}, but also predicted in ${\rm TaCoTe_2}$ monolayer \cite{Li2019h}. Besides Dirac points, antiferromagnets can also host topological nodal lines \cite{Wang2017a}, as predicted in ${\rm CrAs}_2$ \cite{Wang2018e} as well as in ${\rm MnPd_2}$ \cite{Shao2019}. Antiferromagnetic Dirac semimetals that break parity symmetry are particularly interesting because their symmetry allows for current-driven staggered magnetic fields mediated by spin-orbit coupling \cite{Zelezny2014}. This spin-orbit torque can be exploited to manipulate the Dirac points or nodal lines, and open gaps using external currents \cite{Smejkal2017,Kim2018d,Shao2019}. The current-driven switching of noncollinear antiferromagnetic order has been demonstrated experimentally on Mn$_3$Sn Weyl semimetal \cite{Tsai2020} (see below), opening appealing prospects for the electrical control of topological features of antiferromagnets.

\subsubsection{Weyl semimetals}

Weyl semimetals are the counterpart of Dirac semimetals in the absence of either ${\cal P}$ or ${\cal T}$ and were initially predicted in pyrochlore iridates exhibiting all-in/all-out antiferromagnetic order, such as ${\rm Y_2Ir_2O_7}$ \cite{Wan2011,Yang2014}. They accommodate Weyl, rather than Dirac, fermions close to Fermi level, which are singular monopoles of Berry curvature. These bulk Weyl points are connected through topologically-protected Fermi arcs at the surface of the material. Once again, they were initially observed in nonmagnetic materials such as Ta(As,P) \cite{Xu2015c,Xu2015d,Inoue2015} or ${\rm (W,Mo)Te_2}$ \cite{Ali2014,Deng2016,Huang2016}, and recently extended to magnetic materials such as \ce{Co3Sn2S2} \cite{Wang2018g,Xu2020c,Belopolski2019,Liu2019d,Morali2019,Yang2020} and \ce{Pr2Ir2O7} with strain \cite{Ohtsuki2020}. The hallmark of Weyl semimetal is the chiral anomaly \cite{Burkov2015b,Nielsen1983}, i.e., the charge pumping between lowest Landau levels of two Weyl nodes of opposite chirality when a magnetic field is applied along the electric field, whose signature is the negative longitudinal magnetoresistance. Henceforth, negative magnetoresistance has been considered as the hallmark of Weyl semimetals.

\begin{figure}[ht]
	\begin{center}
		\includegraphics[width=0.48\textwidth]{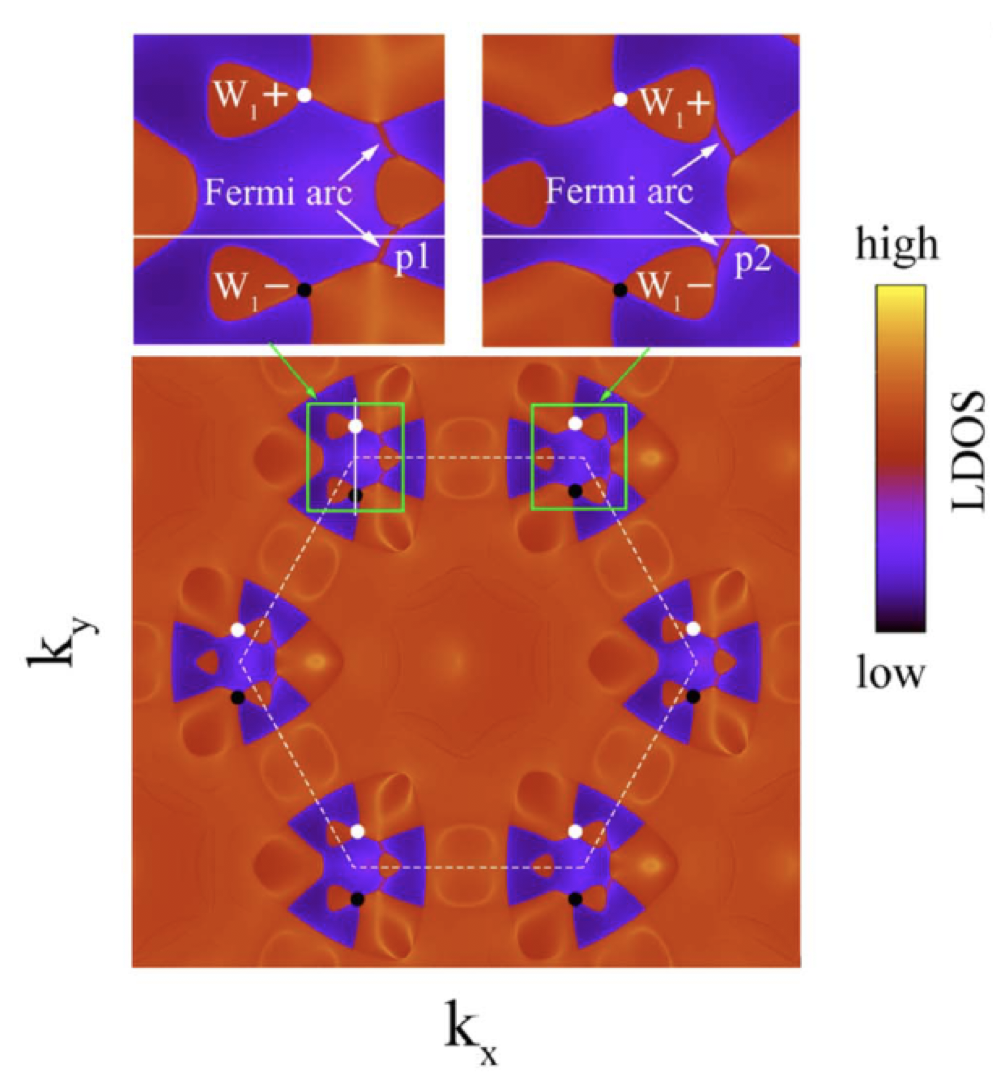}
		\caption{(Color online) Fermi surface of Mn$_3$Sn slab along (001). The top panels show a zoom of the corners of the Brillouin zone, crossing two Weyl points of opposite chirality, W$_1$+ and W$_1$-. Fermi pockets due to bulk states and surface Fermi arcs coexist at Fermi level. From \cite{Yang2017}.}
		\label{Fig:WeylSM}
	\end{center}
\end{figure}

This indirect approach has been used to identify potential Weyl fermions in antiferromagnetic compounds and this signature has been reported in (Gd,Nd)PtBi antiferromagnet \cite{Hirschberger2016, Suzuki2016,Shekhar2018}, ${\rm EuO_3}$ \cite{Takahashi2018} and DyPdBi \cite{Pavlosiuk2019}. We mention that Weyl points have been measured in the antiferromagnet YbPtBi \cite{Guo2018}, but above the N\'eel temperature. Notice that it is also possible to induce Weyl points by applying an external field to antiferromagnets, such as in ${\rm Mn(Bi_{1-x}Sb_x)_2Te_4}$ \cite{Lee2020}. Remarkably, Weyl semimetals have been predicted and reported in the noncollinear antiferromagnets Mn$_3$Sn and Mn$_3$Ge \cite{Yang2017,Kuroda2017}, as shown in Fig. \ref{Fig:WeylSM}. Angle-resolved photoemission spectroscopy (ARPES) measurements and large negative magnetoresistance signal provide strong experimental evidence for the presence of Weyl nodes in Mn$_3$Sn \cite{Kuroda2017}. 

\subsection{Magnonic topological materials}

In less than ten years, a vast literature has addressed the topological properties of magnons using various theoretical models. Whereas the majority of these theories focus on ferromagnets, the interest for antiferromagnetic topological materials is steadily increasing. To date, magnonic analogs to topological insulators and semimetals have been proposed. Since magnons are bosons, the full magnon spectrum contributes to the transport of spin and energy; therefore, magnonic topological insulators and semimetals are never "true" insulators or semimetals. As a consequence, the experimental progress in this fascinating research field has remained difficult due to the outstanding challenge of detecting topologically protected edge states. As a matter of fact, the topological edge states naturally coexist with a large background of trivial bulk states, and a robust experimental methodology able to selectively probe these topological magnonic states remains to be developed. In the present section, we voluntarily keep our discussion of topological ferromagnetic magnonic materials short and rather expand on the antiferromagnetic cases.

\subsubsection{Magnonic topological insulators}

\begin{figure}[ht]
	\begin{center}
		\includegraphics[width=0.48\textwidth]{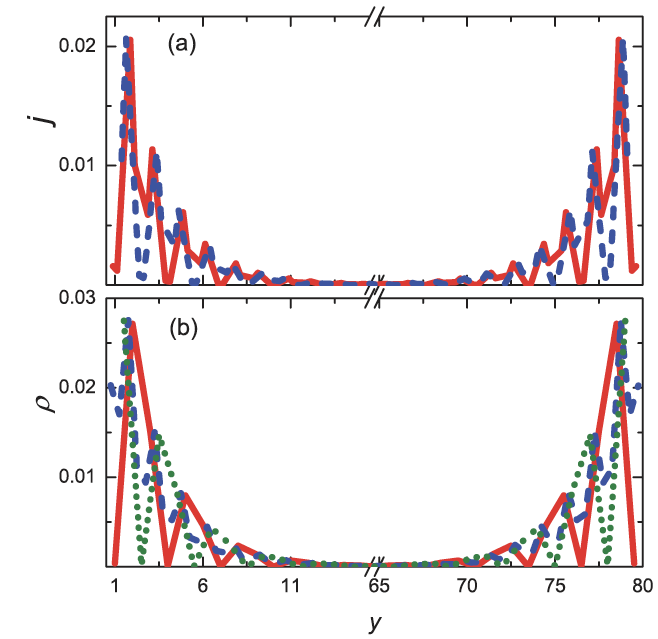}
		\caption{(Color online) Spatial profile of the (a) local energy current and (b) magnon density of states along the transverse direction of magnetic nanoribbon. The solid, dashed, and dotted lines correspond to the local density of states in three different columns in one unit cell. From \cite{Zhang2013f}.}
		\label{Fig:Zhang2013}
	\end{center}
\end{figure}

In their pioneering work, \citet{Matsumoto2011,Matsumoto2011b} remarked that magnons experiencing a Lorentz force (induced by intrinsic Berry curvature as explained in Section \ref{ATM}) are driven towards the boundaries of the magnetic sample and follow its edge, even in the presence of a scattering potential. Although not explicitly stated, this observation suggests that this edge motion is topologically protected. Soon after, \onlinecite{Zhang2013f} and \onlinecite{Shindou2013} proposed the first version of a magnonic topological insulator, characterized by chiral edge states. While both works consider the elementary excitations of a ferromagnet, the former addresses exchange magnons in the presence of Dzyaloshinskii-Moriya interaction in a kagom\'e lattice, whereas the latter investigates magnetostatic spin waves in the presence of dipolar interaction. Both situations give similar results, i.e., magnonic edges states protected by the topology of the bulk states (see Fig. \ref{Fig:Zhang2013}).

The edge modes have been investigated in details in ferromagnetic kagom\'e \cite{Mook2014,Mook2014b,Mook2016a,Malz2019} and honeycomb lattices \cite{Owerre2016c,Owerre2016a,Owerre2016b,Owerre2016d,Wang2017b,Lee2018}. A quantum Hall plateau has also been identified in a spin-ice model governed by dipolar coupling, rather than by Heisenberg exchange \cite{Xu2016b}. Whereas all these theories are developed in the context of the linear spin-wave theory, \onlinecite{Chernyshev2016} pointed out that the Dzyaloshinskii-Moriya interaction breaks magnon conservation and substantially enhances the magnon damping. Therefore the magnonic topological insulators are in fact interacting in nature.

%

The topological insulators described above are the magnonic counterpart of the electronic Chern insulator: time-reversal symmetry is explicitly broken by the net magnetic order and the edge state carries a magnon "charge". This concept has been extended to various antiferromagnetic configurations presenting a net magnetic moment, such as canted collinear antiferromagnet on honeycomb lattices \cite{Owerre2017d} or canted noncollinear antiferromagnet on kagom\'e lattices \cite{Owerre2017a,Owerre2017b,Owerre2017e,Owerre2017c,Laurell2018}. In these examples, the antiferromagnet (originally collinear or in 120$^\circ$ spin configuration), the canting is induced by an external magnetic field, resulting in a net magnetization and spin chirality. Therefore, magnonic edge states can emerge even in the absence of Dzyaloshinskii-Moriya interaction. An elegant example of this situation is given by the all-in all-out state of a pyrochlore iridate. In this case, the net magnetization vanishes but magnonic edge states survive \cite{Laurell2017}.

Most interestingly for the present review, several recent works have demonstrated the realization of $\mathbb{Z} _2$ magnonic topological insulators characterized by helical magnonic edge states protected by time-reversal symmetry. A first example is given by \onlinecite{Nakata2017} who considered a simple (N\'eel) collinear antiferromagnet on a square lattice in the presence of an electric field. The electric field induces a Berry curvature through Aharonov-Casher effect \cite{Aharonov1984}, which in turns induces chirality-dependent Lorentz force on the itinerant antiferromagnetic magnons. The Aharonov-Casher effect enables the control of the helical edge states by an external laser field \cite{Nakata2019}. The edge states resulting from this effect are therefore helical. Alternatively, \onlinecite{Mook2018} proposed a N\'eel collinear antiferromagnet on a square-octagon lattice in the presence of Dzyaloshinskii-Moriya interaction. Finally, magnonic edge states have also been predicted on collinear antiferromagnetic skyrmion crystals \cite{Diaz2019,Daniels2019}. There, the Berry curvature of the magnonic wave function is due to the skyrmion lattice background rather than to Dzyaloshinskii-Moriya interaction directly and can lead to magnon Landau levels \cite{Li2020g}. We remind that in order to obtain a $\mathbb{Z} _2$ topological insulator, a gap must be opened to obtain edge states, uncoupled to bulk states. The $\mathbb{Z} _2$ invariant of magnonic topological insulators has been defined by \citet{Kondo2019,Kondo2020} in a similar way as for electronic topological insulators \cite{Fu2006}.

%

%
%

On the experimental side, \onlinecite{Chisnell2015} have investigated the magnon band structure of a kagom\'e lattice ferromagnet, Cu[1,3-benzenedicarboxylate], using inelastic neutron scattering. The authors reported a magnon flat band that can be gapped from the neighboring dispersive bands by the application of a magnetic field perpendicular to the kagom\'e plane. This feature is consistent with the theory that predicts the emergence of a gap induced by Dzyaloshinskii-Moriya interaction. More recently, \onlinecite{Bao2018} and \onlinecite{Yao2018} reported the band dispersion of a three-dimensional antiferromagnet, Cu$_3$TeO$_6$ (see Table \ref{Table1}), using the same technique (see Fig. \ref{Fig:TopoMagnon} and related discussion). The authors identified magnonic Dirac points protected by U(1) spin-rotation symmetry. More recently, \citet{Bao2020} reported evidence of strong magnon-phonon coupling, suggesting the onset of (potentially topological) magnon-polaron below N\'eel temperature. These observations are experimental evidence of non-trivial band topology in magnetic systems. However, they do not demonstrate the existence of magnonic chiral or helical edge states. To the best of our knowledge, the only experimental evidence of protected edge transport mediated by magnetic excitations remains the observation of quantized transverse heat conductance in $\alpha$-RuCl$_3$ \cite{Kasahara2018} (see Section \ref{ss:kitaev}). Nonetheless, the quasiparticles responsible for the edge transport are Majorana fermions, rather than magnonic bosons. In fact, magnonic edge states are likely to be more difficult to identify experimentally as they systematically coexist with bulk states that are expected to dominate the signal. Therefore, developing new probes to specifically address these elusive edge states is highly desired.
\subsubsection{Magnonic Weyl and Dirac semimetals}

Alike their electronic counterpart, magnons also exhibit Dirac or Weyl dispersion, depending on the system's symmetry. Magnonic Weyl point emerge naturally in ferromagnets where time-reversal symmetry is explicitly broken. Such Weyl points have been obtained in ferromagnetic pyrochlores \cite{Mook2016b,Su2017}

Dirac magnons have been obtained by \onlinecite{Fransson2016} in the case of a magnetic honeycomb lattice with nearest-neighbor exchange only, neglecting magnetic anisotropy and Dzyaloshinskii-Moriya interaction. In the ferromagnetic case, the magnonic Hamiltonian resembles the single electron Hamiltonian of graphene and exhibits magnonic Dirac cones at K and K' points in the reciprocal lattice. By tuning the magnetic configuration from ferromagnetic to ferrimagnetic and eventually antiferromagnetic, the Dirac cones at K and K' points become gapped and a Dirac cone progressively emerges at $\Gamma$ point. The Dirac magnons have also been found in collinear kagom\'e antiferromagnet with anisotropic nearest-neighbor exchange ($J_x=J_y$, $J_z=0$) \cite{Okuma2017}, as well as in 120$^\circ$ noncollinear kagom\'e antiferromagnet \cite{Owerre2017e}.

\begin{figure}[ht]
	\begin{center}
		\includegraphics[width=0.48\textwidth]{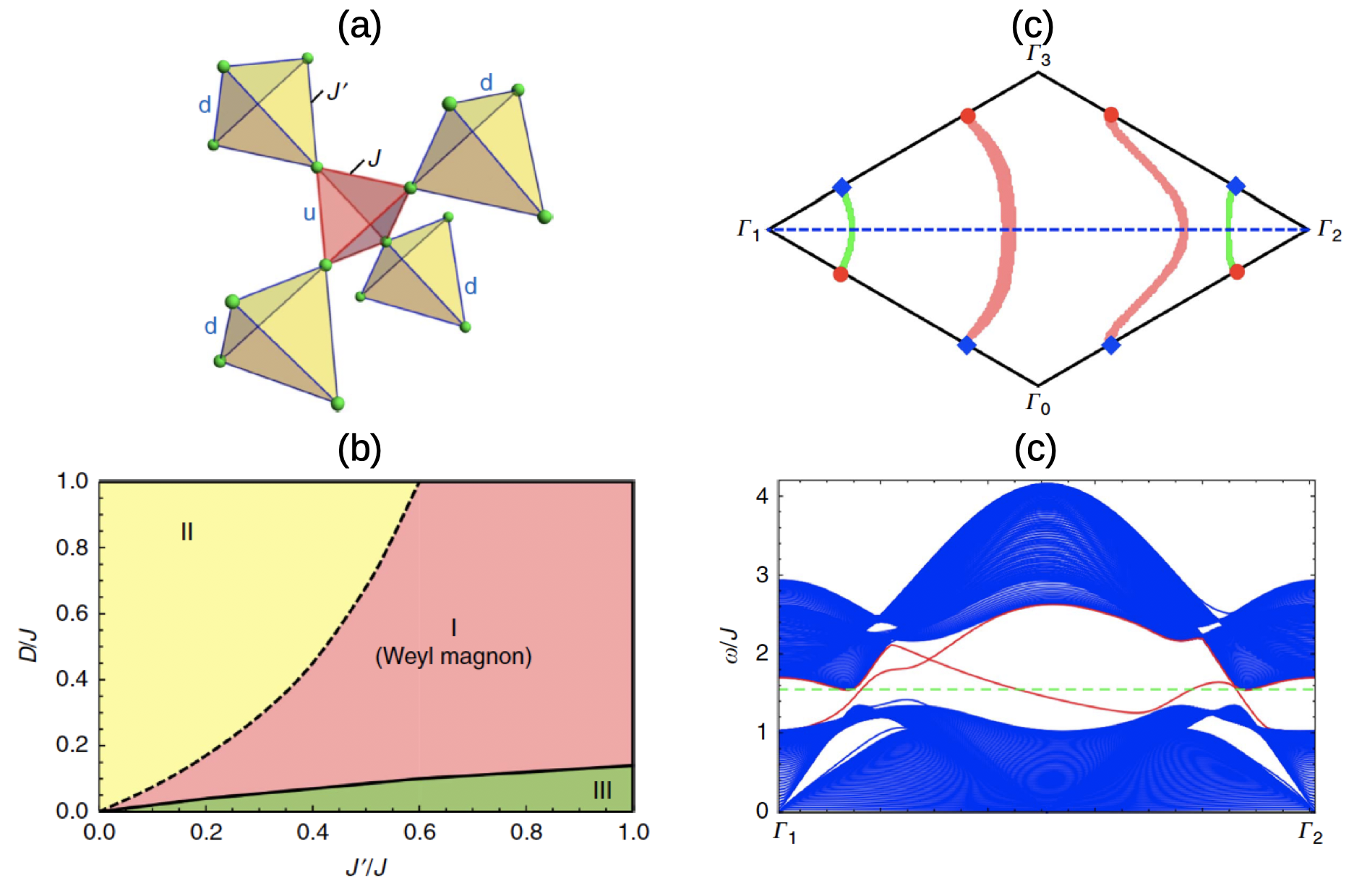}
		\caption{(Color online) (a) The breathing pyrochlore antiferromagnet with all-in all-out spin configuration. The letters u and d refer to the up- and down-pointing tetrahedra. J and J' indicate the nearest-neighbour exchange couplings for u and d tetrahedra. (b) Magnetic phase diagram showing the Weyl magnon region (region II). (c) Surface magnon arcs of a slab normal to the [11${\bar1}$] direction. (d) Corresponding magnon band structure showing bulk (blue) and surface states (red). From \cite{Li2016g}.}
		\label{Fig:MagnonWeyl}
	\end{center}
\end{figure}

Weyl magnons have been first predicted in ferromagnetic pyrochlores \cite{Mook2016b,Su2017} in the presence of nearest-neighbor exchange and Dzyaloshinskii-Moriya interaction. The latter preserves inversion symmetry but breaks the pseudo time-reversal symmetry (time reversal followed by a $\pi$ spin rotation), which enables the existence of Weyl nodes. These Weyl cones are accompanied by magnonic Fermi arcs at the surface of the magnet. These interesting features have also been predicted in breathing pyrochlore antiferromagnets \cite{Li2016g,Jian2018} (see Fig. \ref{Fig:MagnonWeyl}). In this structure, the adjacent tetrahedra possess different sizes, which breaks inversion symmetry and promotes Weyl points and chiral surface states. Realistic calculations have shown that Weyl nodes are expected in Cu$_2$OSeO$_3$ \cite{Zhang2019g}, which also possesses a breathing pyrochlore structure. In the absence of Dzyaloshinskii-Moriya interaction, these points are doubly degenerate and located at R and $\Gamma$ points; when turning on Dzyaloshinskii-Moriya interaction, they split into four Weyl points. Alternatively, \onlinecite{Zyuzin2018} proposed to design Weyl magnonic semimetals by alternatively stacking ferromagnetic and antiferromagnetic honeycomb monolayers, and \onlinecite{Owerre2018} proposed to use a stack of noncollinear, noncoplanar kagom\'e antiferromagnets. Finally, we mention that spin liquids, discussed further in Section \ref{s:topexc}, can also host Weyl excitations. \onlinecite{Hermanns2015} recently predicted that the magnetic excitations of a three-dimensional Kitaev model are Majorana modes that form a Weyl superconductor and proposed $\beta$-Li$_2$IrO$_3$ as a potential candidate.

\begin{figure}[ht]
	\begin{center}
		\includegraphics[width=0.48\textwidth]{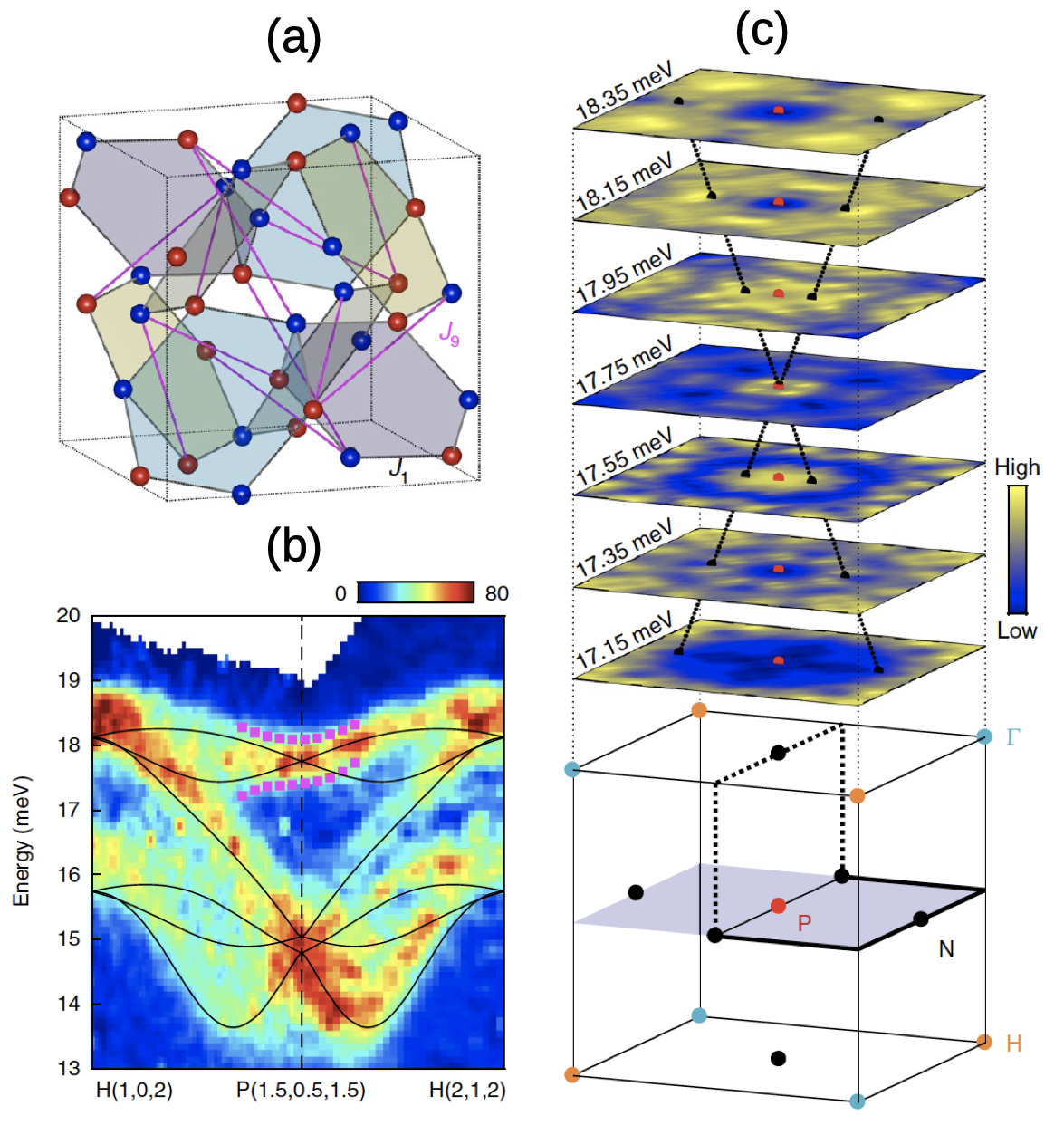}
		\caption{(Color online) (a) Magnetic unit cell of Cu$_3$TeO$_6$. The red and blue spheres refer to Cu$^{2+}$ spin moments of opposite directions. (b) Inelastic neutron scattering intensity spectrum along H-P-H momentum trajectory. The solid lines indicate LSWT-calculated dispersions. (c) Corresponding inelastic neutron scattering intensity distribution in 0.2 meV intervals in Q-space planes that connect P with its four neighbouring N-points and illustrating the Dirac-like magnon. From \cite{Yao2018}.}
		\label{Fig:TopoMagnon}
	\end{center}
\end{figure}

To complete this overview of topological magnonic excitations, nodal-lines and triple-points have been theoretically predicted in several magnetic materials. In collinear antiferromagnets possessing $\mathcal{PT}$ symmetry and U(1) spin rotation symmetry (i.e., preserving $S_z$), the magnon band crossings are Dirac points with integer monopole charges \cite{Li2017c}. On breaking U(1) symmetry while preserving $\mathcal{PT}$ symmetry, e.g. turning on Dzyaloshinskii-Moriya interaction, these points become a nodal line, characterized by a $\pi$-Berry phase and protected by a $\mathbb{Z} _2$ monopole charge. \onlinecite{Li2017c} predicted such magnon Dirac points and nodal lines in Cu$_3$TeO$_6$, the former of which was recently confirmed experimentally \cite{Yao2018}, as reported on Fig. \ref{Fig:TopoMagnon}. Nodal points as well as triply degenerate band crossings, called triple-points, have been predicted in pyrochlore iridates with all-in all-out spin configuration \cite{Hwang2018}.

The realization of Dirac and Weyl magnons offers inspiring perspectives for experimental observation because their properties can be easily (at least theoretically) manipulated using an external magnetic field. These recent theoretical results therefore call for revisiting decades of experimental investigation on frustrated magnets and for developing novel probed to identify magnonic surface states, as further discussed in Section \ref{s:topexc}.


%
%
%
%
%
%


\section{Topological Antiferromagnetic Textures \label{s:textures}}

In the previous sections, we discussed transport effects that are enabled by the presence of a geometrical phase in the momentum space and in real space. In the present section, we concentrate on the static and dynamical properties of the magnetic soft modes of antiferromagnets whose texture in real space is topologically non trivial. We pay particular attention to topological solitons in one and two dimensions as well as the newly predicted antiferromagnetic skyrmions.

\subsection{Antiferromagnetic solitons in one-dimension} 

Solitons are localized, self-reinforcing entities which preserve their shape during propagation and collision, and move with constant velocity in a medium \cite{Korteweg1895, Bona1980}. Solitons are stabilized when the dispersive effects are neutralized by the nonlinearities in the medium. In a mathematical sense, solitons are localized lump or wave-like solutions to nonlinear equations.  In a magnetic system, solitons are obtained as exact solutions of the Landau-Lifshitz equation, continuum Heisenberg model,  nonlinear Schr\"odinger equation and $\sigma$ model equations \cite{Kosevich1990}. Magnetic solitons such as domain walls in one-dimensional systems and skyrmions/vortices in two-dimensional systems are starting to find their applications in the areas of digital information storage and processing. 

The solitons observed in magnetic systems can be classified as topological or as dynamical \cite{Kosevich1990, Affleck1986, Ivanov1995a, Gomonai1990}. Topological solitons have their origin in the overall symmetries of the magnetization field and are robust to perturbations. Examples of topological solitons include domain walls in one-dimensional magnets, skyrmions and vortices in two-dimensional magnetic systems etc. In contrast, dynamical solitons can be reduced to uniform magnetization by continuous deformation \cite{Kosevich1990}. Examples include precessional solitons in magnetic systems. In some cases, topological solitons may possess internal modes such as magnons  or precessing spins. Such two-parameter solitons are termed as dyons \cite{Affleck1986, Ivanov1995a}. 

\begin{figure}[ht]
\begin{center}
\includegraphics[width=8cm]{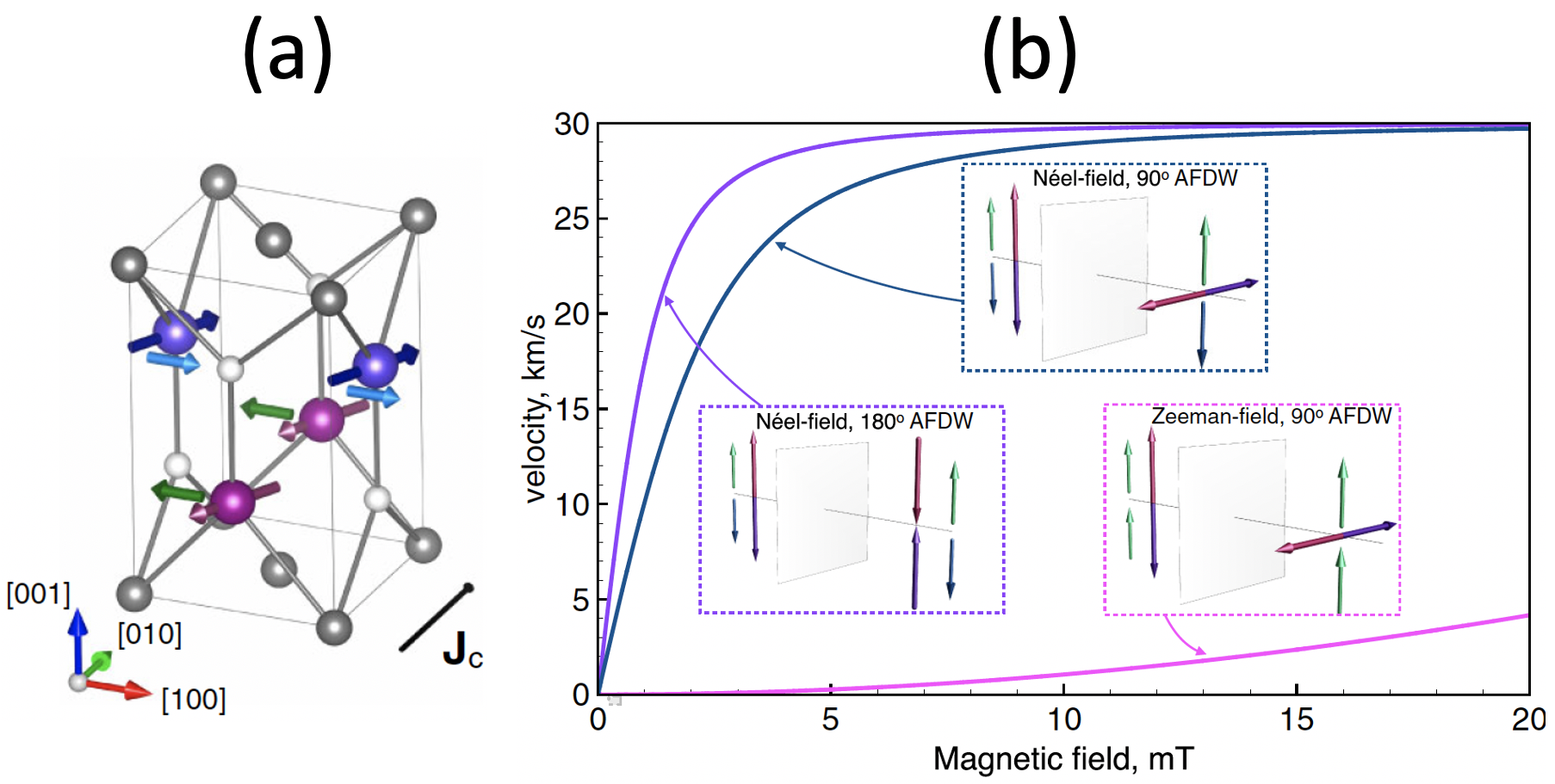}
\caption{(Color online) (a) Crystal structure of CuMnAs. The red and purple spheres represent the two Mn sublattices. The current-induced staggered spin-orbit field (blue and green arrows) is opposite on the two sublattices creating a "N\'eel field". The nonmagnetic As ions (grey spheres) break local inversion symmetry. ${\bf J}_c$ is the applied current. (b) Current-velocity curves for $180^{\circ}$ and $90^{\circ}$ domain walls under a current (N\'eel field) and an external field (Zeeman field). From \cite{Gomonay2016}.}
\label{domain}
\end{center}
\end{figure}

To date, one-dimensional topological solitons in Heisenberg ferromagnets, whose dynamics is governed by Landau-Lifshitz equation for the magnetization vector $\bf{m}$, have been studied in detail. With the recent development of antiferromagnetic spintronics, the static and dynamical properties of magnetic solitons in antiferromagnets have gained renewed interest. The first significant difference between ferromagnetic and antiferromagnetic solitons is their energetics. Because antiferromagnetic domains display small or vanishing demagnetization, in sharp contrast with ferromagnetic ones, their thermodynamical stability is not determined by minimizing the magnetostatic energy but rather by a compromise between exchange energy and entropy \cite{Li1956}. This suggests that antiferromagnetic domain walls are much more sensitive to the magnetic history of a given sample than their ferromagnetic counterparts.

The domain wall stabilized in an antiferromagnet can be a $180^{\circ}$ wall or $90^{\circ}$ wall \cite{Baryakhtar1972,Ivanov1981, Gomonay2016}. In such systems, the strong exchange coupling gives rise to resonance frequencies in the Terahertz range and much larger maximum soliton velocities compared to ferromagnetic solitons. While the maximum velocity for ferromagnetic domain walls is limited by the Walker breakdown \cite{Schryer1974}, it is severely reduced in antiferromagnets due to the vanishing dipolar field. In addition, the exchange enhancement can lead to much higher domain wall velocities in antiferromagnets \cite{Baryakhtar1972, Ivanov1981, Gomonay2016,  Shiino2016}. The dynamics of a soliton in a (possibly uniaxial) N\'eel antiferromagnet is described by the O(3) nonlinear sigma model, a field theory that admits topological solutions that are Lorentz invariant \cite{Haldane1983}. In other words, the antiferromagnetic domain wall experiences a Lorentz contraction when approaching the maximum speed, set by the spin wave velocity, $v_{m}=\sqrt{A/\rho}$, where $A$ is the antiferromagnet exchange constant and $\rho$ quantifies the inertia of the staggered magnetization. In this model, $v_m$ plays the role of the speed of light $c$. Hence, the domain wall profile ${\bm n}_v(t,x)$ at finite velocity $v$ can be deduced from the zero-velocity profile ${\bm n}_0(t,x)$ by \cite{Haldane1983,Kim2014c,Salimath2020} 
\begin{equation}
{\bm n}_v(t,x)={\bm n}_0\left(\frac{t-vx/v_m^2}{\sqrt{1-(v/v_m)^2}},\frac{x-vt}{\sqrt{1-(v/v_m)^2}}\right).
\end{equation}
Since the maximum velocity $v_m$ is set by the magnon velocity in antiferromagnets, it can be very large \cite{Ivanov1981, Gomonay2016}. For instance, when driven by current-induced torques, it can reach up to 30 km/s \cite{Gomonay2016} (see \figref{domain}). 

The $90^{\circ}$ phase appears as an intermediate phase when the antiferromagnet sublattices are inverted in an applied magnetic field \cite{Baryakhtar1972}. In this phase, the antiferromagnet order parameter $\bf{l}$ is perpendicular to the uniaxial anisotropy axis and the magnetic moment $\bf{m}$ is not zero. The $90^{\circ}$ domain wall acts as a transition region between a collinear phase and a flopped phase in a two sublattice antiferromagnet \cite{Ivanov1981}. These domain walls are thermodynamically stable \cite{Ivanov1981}, contrary to $180^{\circ}$ walls that are metastable \cite{Ivanov1981, Li1956}. The dipolar fields play an important role in the dynamics of $90^{\circ}$ domain walls. While the maximum velocity for the $180^{\circ}$ domain wall is limited by the magnon velocity, the $90^{\circ}$ wall velocity is always zero if the magnetic dipolar fields are not taken into consideration \cite{Kosevich1990, Ivanov1981}. Recently, it was shown that the spin-orbit torques in antiferromagnets can drive the $90^{\circ}$ walls at much higher velocities compared to the field driven motion \cite{Gomonay2016} (see \figref{domain}). Also, the velocities are of the same order for both $90^{\circ}$ and $180^{\circ}$ walls under spin-orbit torques. 

Another intriguing concept that is widely studied is the effect of Dzyaloshinskii-Moriya interaction on the domain wall dynamics \cite{Gomonai1990, Shiino2016}. Dzyaloshinskii-Moriya interaction arises from the broken inversion symmetry and is commonly found in antiferromagnets \cite{Dzyaloshinsky1958, Moriya1960}. This interaction introduces gyroscopic terms proportional to $\frac{\partial \bf{l}}{\partial t}$ in the Lagrangian, which significantly modifies the domain wall dynamics \cite{Gomonai1990, Galkina2017}. The Lagrangian describing the antiferromagnet is no longer Lorentz invariant, which may lead to the instability of propagating domain walls and changes in the wall's structure at critical velocities \cite{Gomonai1990}. The inclusion of Dzyaloshinskii-Moriya interaction in the energy functional of the antiferromagnet lowers the symmetry of the moving wall and its critical velocity is significantly reduced \cite{Gomonai1990}. 

Apart from being movable topological solitons, one-dimensional antiferrmagnetic domain walls also possess internal modes such as precessional spins and magnons localized at the soliton center \cite{Kosevich1977, Ivanov1995a}. Such two-parameter solitons referred to as dyons are characterized by their velocity and precession frequency $\omega$. The dynamics of the internal modes are strongly coupled to the translation mode of the soliton. These solitons are dynamical and their stability is determined by the conservation of integral of motion for the antiferromagnet. They are relatively less understood as their existence is limited by the natural dissipation inherent to magnetic systems. However, the phenomenological theory for these internal modes was developed long ago \cite{Kosevich1990}. Many works in the past have been dedicated to understanding the stability of precessional solitons as a function of anisotropy direction \cite{Baryakhtar1983, Ivanov1995a,Kosevich1977}. For a uniaxial antiferromagnet, the precessional solitons are stable for all possible $\omega$. When the uniaxial antiferromagnet is supplemented with symmetry breaking Dzyaloshinskii-Moriya interaction, the precessions were determined to be inhomegeneous in time. In the presence of Dzyaloshinskii-Moriya interaction an oscillating magnetic moment at the domain wall center appears and is akin to the rotational dynamics of $\bf{l}$.  The precessional modes can be excited by energy pumping with spin transfer torque and they find application as nano-oscillators \cite{Mohseni2013}. These modes are associated with quantum signatures and their characteristic frequencies can be determined experimentally \cite{Ivanov1995a, Ivanov1997}.

\subsection{Antiferromagnetic skyrmions} 

\begin{figure}[ht]
	\begin{center}
	\includegraphics[width=0.44\textwidth]{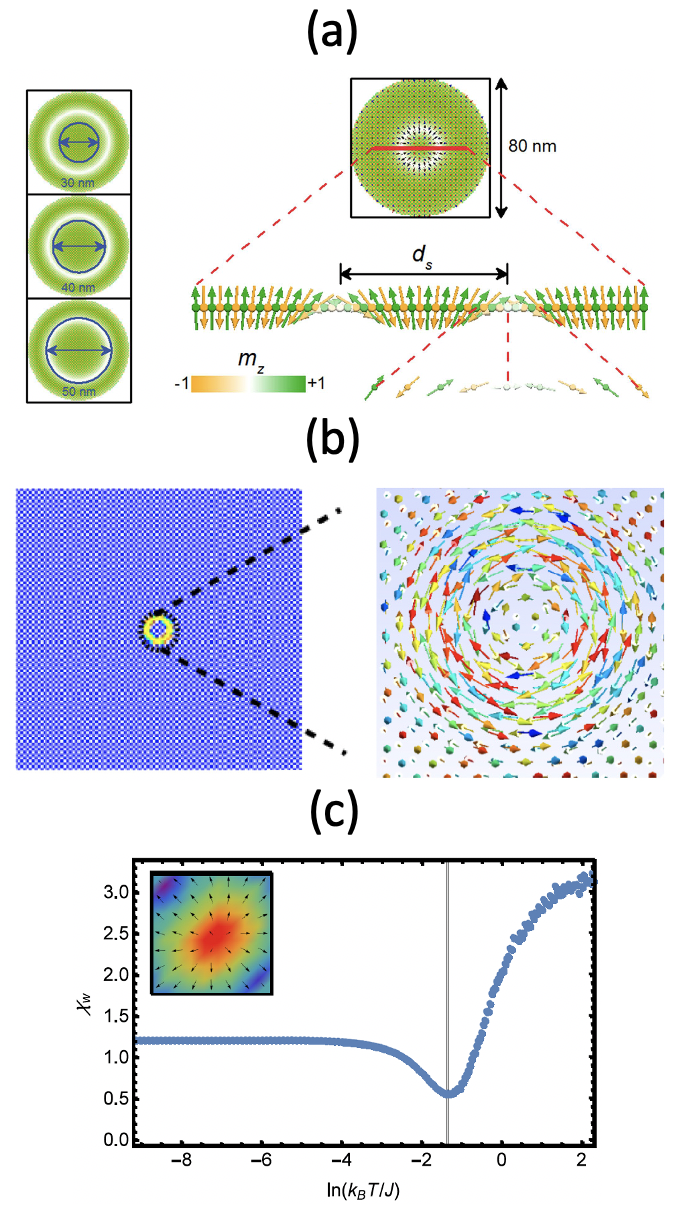}
		\caption{(Color online) (a) Antiferromagnetic skyrmion in the nanodisk nucleated by a spin-polarized current pulse perpendicularly injected into the nanodisk in the marked circle region. From \cite{Zhang2016k}. (b) Antiferromagnetic skyrmion nucleation in a confined  region by a magnetic field pulse. From \cite{Khoshlahni2019}.  (c) Isolated antiferromagnetic skyrmion obtained from Monte-Carlo simulation on a Heisenberg Hamiltonian. From \cite{Keesman2016}.}
		\label{nucleation}
	\end{center}
\end{figure}

Skyrmions were originally proposed as nonlinear excitations in field theories involving baryons \cite{Skyrme1961, Skyrme1962}. Major interest in magnetic skyrmions emerged with the prediction \cite{Bogdanov2001,Roszler2006} and observation of a skyrmion crystal in noncentrosymmetric chiral magnets, MnSi and Fe$_{0.5}$Co$_{0.5}$Si, with bulk Dzyaloshinskii-Moryia interaction \cite{Muhlbauer2009,Pfleiderer2010,Yu2010}. The Dzyaloshinskii-Moriya interaction in MnSi results in Bloch-type skyrmions under an applied magnetic field. Later, it was shown that Dzyaloshinskii-Moriya interaction induced by a broken structural inversion symmetry at the interface of ferromagnet/heavy metal bilayer systems can also stabilize skyrmions of N\'eel type \cite{Dupe2014,Chen2015b,Jiang2015,Moreau-Luchaire2016, Woo2016,Boulle2016}. Due to their topological stability, skyrmions have been proposed to serve as attractive candidates for information storage and processing \cite{Rosch2017, Fert2013}. Since their inception, considerable work has been achieved both experimentally and theoretically to understand the physics and dynamics of an isolated skyrmion under current-driven torques in ferromagnets \cite{Nagaosa2013, Nagaosa2012, Schutte2014}. Their topological signatures make skyrmion detection in ferromagnets rather simple in principle \cite{Akosa2017, Metalidis2006, Ritz2013, Bruno2004}. However, ferromagnetic skyrmions experience current-driven gyroscopic forces, called skyrmion Hall effect, that limit their scalability \cite{ Fert2013, Sampaio2013, Jiang2017b, Litzius2017,Salimath2019} and hence their technological applications. To address skyrmion Hall effect in ferromagnets, which is a limiting factor for information transfer application, several proposals have been put forward in recent years. The use of synthetic antiferromagnets \cite{Zhang2016l, Tomasello2017a} or engineering the skyrmion nanotrack \cite{Purnama2015,Abbout2018} stand out as promising solutions to suppress skyrmion Hall effect. However, they do not serve as technologically viable solutions owing to the fabrication and integration challenges. 

In this context, antiferromagnets serve as ideal active materials for housing skyrmions. In fact, the relativistic Dzyaloshinskii-Moriya interaction necessary for the stabilization of skyrmions is inherently present in antiferromagnets \cite{Dzyaloshinsky1958, Moriya1960} and theories predicting the existence and topological stability of antiferromagnetic  skyrmions existed even before the skyrmions were experimentally observed in chiral ferromagnets \cite{Roszler2006, Bogdanov2002}. Recent numerical investigations consider the simplest case of two sublattice G-type collinear antiferromagnets \cite{Zhang2016k, Barker2016, Shen2018}. In a micromagnetic framework, a G-type antiferromagnet consists of two inter-penetrating ferromagnetic sublattices with strong antiferromagnetic coupling between them. An isolated skyrmion in a G-type antiferromagnet can be nucleated by applying spin polarized current pulse through a circular nanocontact \cite{Zhang2016k, Jin2016b} or Gaussian magnetic field pulse in a confined geometry \cite{Khoshlahni2019} and allowing the system to stabilize (see \figref{nucleation}). Even, artificially flipping the spins in small area on two sublattices and relaxing the system stabilizes metastable skyrmions \cite{Barker2016}. The relaxed structure exhibits two skyrmions, one on each sublattice with an opposite skyrmion number and antiferromagnetically coupled. As a consequence, an antiferromagnetic skyrmion does not carry net topological charge resulting in the suppression of gyroscopic forces. Further, Monte Carlo simulations on a Heisenberg antiferromagnet have determined stable isolated skyrmion configurations even at non zero temperatures \cite{Keesman2016} (see \figref{nucleation}).

The absence of skyrmion Hall effect in antiferromagnetic skyrmions has been considered as a major advantage over their ferromagnetic competitors. The absence of net gyroscopic forces prevents skyrmion annihilation when they are driven along the nanotrack \cite{Zhang2016l, Barker2016}. In addition, since antiferromagnets are insensitive to magnetic field perturbations and do not generate any stray field, device scalability and integration is expected to be greatly improved \cite{Jungwirth2016, Gomonay2014}. Nonetheless, the topological texture of antiferromagnetic skyrmions promotes a topological spin current that can substantially enhance the skyrmion mobility when reducing the skyrmion size \cite{Akosa2018}. It is important to point out that although antiferromagnetic skyrmions are not solutions of the O(3) sigma model, they still experience a reminiscence from Lorentz invariance, which can be detrimental to their transport properties. As a matter of fact, at large velocity, the antiferromagnetic skyrmion experiences a {\em lateral expansion} \cite{Salimath2020,Jin2016b} that can eventually lead to its destruction when reaching the boundaries of the nanotrack (see Section \ref{ss:skymotion}). In spite of their tremendous potential, the progress in antiferromagnetic skyrmions has been mostly theoretical as it is experimentally difficult to detect skyrmions in antiferromagnets. Nonetheless, the recent observation of fractional antiferromagnetic skyrmion lattice stabilized by anisotropic coupling in \ce{MnSc2S4} compound opens inspiring avenues \cite{Gao2020b}.

\subsection{Textures in noncollinear antiferromagnets}

\begin{figure}[ht]
	\begin{center}
		\includegraphics[width=0.42\textwidth]{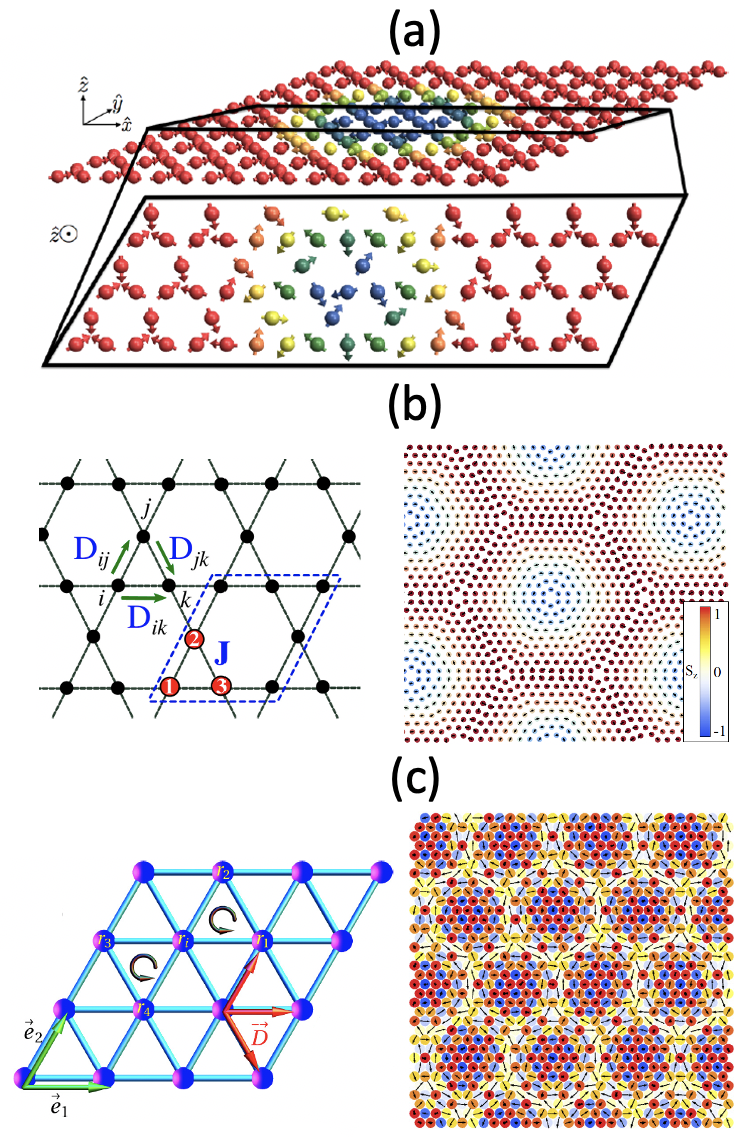}
		\caption{(Color online) Schematics of skyrmion nucleation in noncollinear antiferromagnets: (a) Trivial texture in a two-dimensional kagom\'e antiferromagnet, from \cite{Ulloa2016}; (b) Pseudo skyrmion phase in a kagom\'e antiferromagnet, from \cite{Villalba2019}; (c) Skyrmion phase in a triangular lattice antiferromagnet, from \cite{Rosales2015}}.
		\label{noncollinear}
	\end{center}
\end{figure}

Recently, noncollinear antiferromagnets have garnered significant interest as they exhibit large topological Hall effect, magnetic Hall effect, anomalous Hall effect, anomalous  Nernst effect, spin Hall effect and magneto-optical Kerr effect \cite{Chen2014, Nakatsuji2015, Surgers2014,Guo2017, Feng2015,Higo2018,Higo2018b, Kimata2019}, as described in Section \ref{s:ahe}. noncollinear magnetism arises due to geometric frustration of antiferromagnet interactions. The frustration gives rise to exotic phenomena in these systems and when supplemented with Dzyaloshinskii-Moriya interaction and external magnetic field, topologically non-trivial phases appear \cite{Takagi2018}. \citet{Leonov2015}, theoretically investigated skyrmion phase with arbitrary vorticity and helicity in a triangular lattice magnet with antiferromagnetic next-nearest neighbor exchange interactions. The interaction between isolated skyrmions in these systems was found to be oscillatory. The dynamics of isolated skyrmions showed that the coupling between the helicity and the center of mass dynamics give rise to magnetoelectric effect. The second homotopy group of the order parameter space in a noncollinear antiferromagnet is trivial \cite{Ulloa2016}. Starting from the Heisenberg Hamiltonian for a kagom\'e antiferromagnet with nearest neighbour antiferromagnetic exchange and performing gradient expansion on the energy functional, \onlinecite{Ulloa2016} showed that the skyrmion-like textures in a two-dimensional noncollinear antiferromagnet are not topologically protected and they can be continuously deformed into uniform state (see \figref{noncollinear}(a)). \onlinecite{Villalba2019} studied the effect of Dzyaloshinskii-Moriya interaction on Heisenberg kagom\'e antiferromagnet through Monte Carlo simulation and found pseudo skyrmion phases which are distinguished with the sublattice chirality (see \figref{noncollinear}(b)). The pseudo skyrmion phase is characterized by a canted core and was found to be stable at low enough simulated temperatures and over a broad magnetic field range. However, Monte Carlo simulation study on a Heisenberg triangular lattice antiferromagnet with isotropic antiferromagnetic exchange and Dzyaloshinskii-Moriya interactions have shown stabilization of non-trivial skyrmion phase \cite{Rosales2015}. The skyrmion phase observed under an applied external magnetic field consists of three interpenetrating skyrmions, one on each sublattice, and is found to be stable over a wide range of magnetic field (see \figref{noncollinear}(c)). So, the stability of skyrmion phase in a noncollinear antiferromagnet still remains an open research topic.

\subsection{Current-driven torques and skyrmion motion\label{ss:skymotion}}

\begin{figure}[ht]
	\begin{center}
			\includegraphics[width=0.48\textwidth]{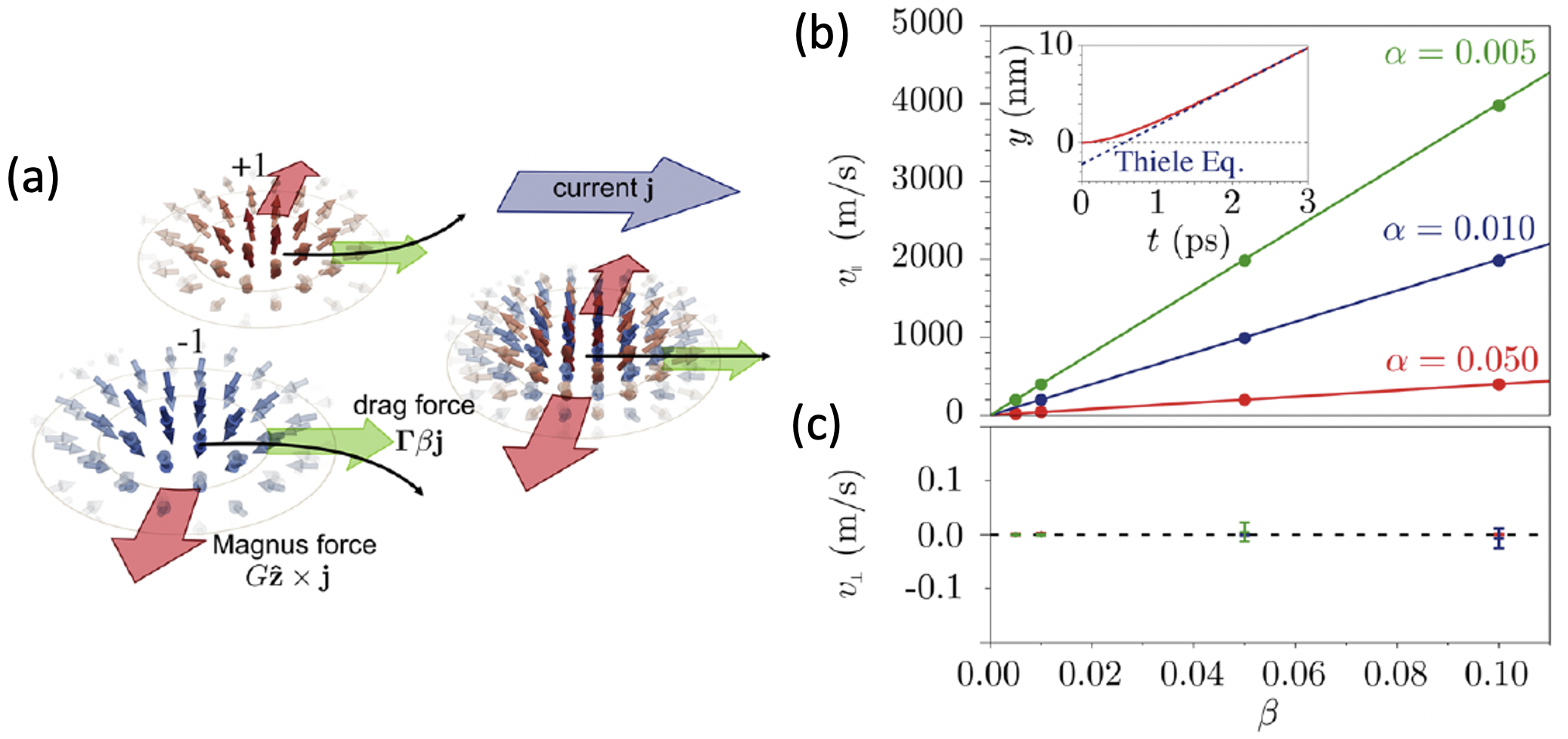}
		\caption{(Color online) (a) Spin transfer torque-driven antiferromagnetic skyrmion motion. The skyrmions with opposite topological charge in two sublattices experience Magnus force in opposite direction resulting in perfect cancellation of the transverse force. (b) Longitudinal and (c) transverse skymion velocity when driven by spin transfer torque as a fonction of the non-adiabaticity $\beta$. From \cite{Barker2016}.}
		\label{Dynamics}
	\end{center}
\end{figure}

The electrical control of magnetic textures' dynamics opens new avenues for technological applications. Recent experiments show that it is possible to electrically control the dynamics of skyrmions in ferromagnets \cite{Jiang2017b, Romming2013, Legrand2017,Litzius2017}. Also, current-in-plane or current-perpendicular-to-plane techniques are employed to electrically detect skyrmions in ferromagnets \cite{Kanazawa2015, Neubauer2009, Crum2015}. However, the challenges involved in detecting antiferromagnetic textures have made electrical control of nucleation and dynamics of antiferromagnetic skyrmions rather elusive. Nevertheless, the recent theories unambiguously predict the possibility of electrically detecting antiferromagnetic skyrmions through topological spin Hall effect \cite{Akosa2018, Goebel2017, Buhl2017} and control their nucleation and dynamics through current induced torques just as in ferromagnets \cite{Zhang2016k, Barker2016}.

For a two-dimensional bipartite antiferromagnet, the current-induced torques on the two sublattices can be introduced in the same spirit as for ferromagnets \cite{Tomasello2017a}. Hence, in the micromagnetic framework the skyrmion motion within sublattice $i$ is governed by the Landau-Lifshitz-Gilbert equation,

\begin{multline}\label{eq1}
	\dot{\mathbf{m}}_i=-\gamma\mu_0{\bf m}_i\times \mathbf{H}^i_{\rm eff}+ \alpha\mathbf{m}_i\times \dot{\mathbf{m}}_i+\tau_{sh}\mathbf{m}_i\times(\mathbf{y}\times \mathbf{m}_i)\\-b_{j}\partial_x\mathbf{m}_i+ \beta b_{j}\mathbf{m}_i \times \partial_x\mathbf{m}_i
\end{multline}%

where $\mu_0\mathbf{H}^i_{\rm eff}=-(1/M_s)\delta W/\delta {\bf m}_i$ is the effective field, $W$ being the magnetic energy density and $M_s$ the saturation magnetization. The effective field includes the contributions from exchange, perpendicular uniaxial anisotropy, and interfacial Dzyaloshinskii-Moriya interaction. The exchange field takes into account both ferromagnetic coupling between neighbors in each sublattice and the antiferromagnetic coupling between the two sublattices. The current-driven torques have contributions from both spin transfer and spin-orbit torques. The third term in Eq. \eqref{eq1} corresponds to the dampinglike spin-orbit torque, modeled by a spin Hall torque with polarization along $y$. The fourth and fifth terms are the contributions from the adiabatic torque and the non-adiabatic torque, and $b_{j}$ is the driving electron velocity proportional to the applied current density. 

When the skyrmion is driven by spin transfer torque alone, the cancellation of gyroscopic forces for the coupled skyrmion motion leads to the skyrmion terminal velocity $v= \frac{\beta b_{j}}{\alpha}$ \cite{Barker2016}, see Fig. \ref{Dynamics}. The velocity is independent of the skyrmion size, increases linearly with the driving current density and depends on the ratio between the damping parameter $\alpha$ and the non-adiabaticity parameter $\beta$, similarly to ferromagnetic domain walls \cite{Thiaville2005}. Similar current-velocity behavior are obtained for antiferromagnetic skyrmions \cite{Velkov2016}. The relation holds true even in the high-current density regime where the skyrmion can undergo lateral deformation due to large gyroscopic forces acting on the skyrmions in two sublattices. In the high-current regime, the skyrmion adopts an elliptic shape with longitudinal and lateral diameters given by \cite{Salimath2020}
\begin{eqnarray}
\Delta_x&=&\frac{\Delta_0}{2}\left[1+\sqrt{1-(v/v_m)^2}\right],\\
\Delta_y&=&\frac{\Delta_0}{2}\left[1+\frac{1}{\sqrt{1-(v/v_m)^2}}\right],
\end{eqnarray}
where $\Delta_0$ is the skyrmion diameter at rest. The lateral deformation is a reminiscence of Lorentz invariance so that when the antiferromagnetic skyrmion's velocity gets close to the magnon velocity, its lateral expansion diverges. The large antiferromagnetic coupling strength can drive the skyrmions at much higher longitudinal velocities than those observed in ferromagnets. Similarly to one-dimensional solitons, the maximum velocity for skyrmions is the magnon velocity.  

\begin{figure}[ht]
	\begin{center}
			\includegraphics[width=0.48\textwidth]{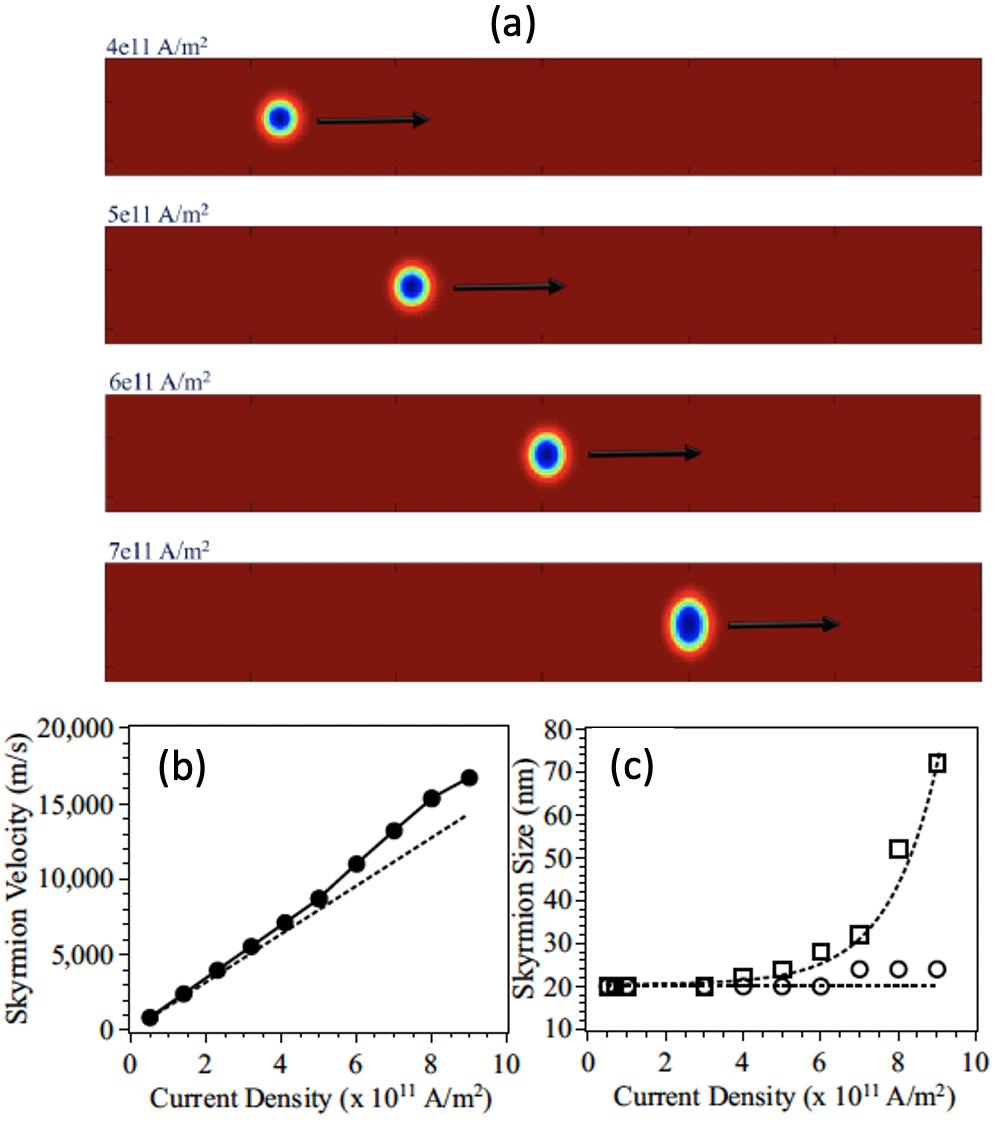}
		\caption{(Color online) (a) Skyrmion motion in an antiferromagnet under spin Hall torque showing the skyrmion deformation at high current densities. (b) Simulated skyrmion velocity-current characteristics (black dots) and (c) skyrmion width variation along x (open circles) and y (filled circles), as a function
of applied current density. From \cite{Salimath2020}.}
		\label{Akshay}
	\end{center}
\end{figure}

Similar investigations with spin-orbit torque driven skyrmion motion in antiferromagnets showed cancellation of gyroscopic forces and a linear current velocity behavior \cite{Jin2016b, Zhang2016k}. The phenomenological theory based on collective coordinate approach shows the velocity, $v_{sk}= [\frac{C_{2}\Delta_0}{2\alpha _{G}}I]j$, to depend on the skyrmion size, $\Delta_0$, in the steady state \cite{Velkov2016}. In the high-current regime where the skyrmion undergoes lateral deformation, its velocity varies nonlinearly with the current \cite{Jin2016b,Salimath2020}, see Fig. \ref{Akshay}. 

In addition to the two types of torques discussed above, recent theories have predicted another contribution to the current driven torques arising from the non trivial topology of the skyrmions \cite{Akosa2018}. This new torque has its origin in the emergent electrodynamics of skyrmion and when supplemented with the spin transfer torque will result in enhanced skyrmion mobility and is discussed in detail in the next section.






\subsection{Topological transport in non-trivial solitons}

\begin{figure}[ht]
	\begin{center}
			\includegraphics[width=0.49\textwidth]{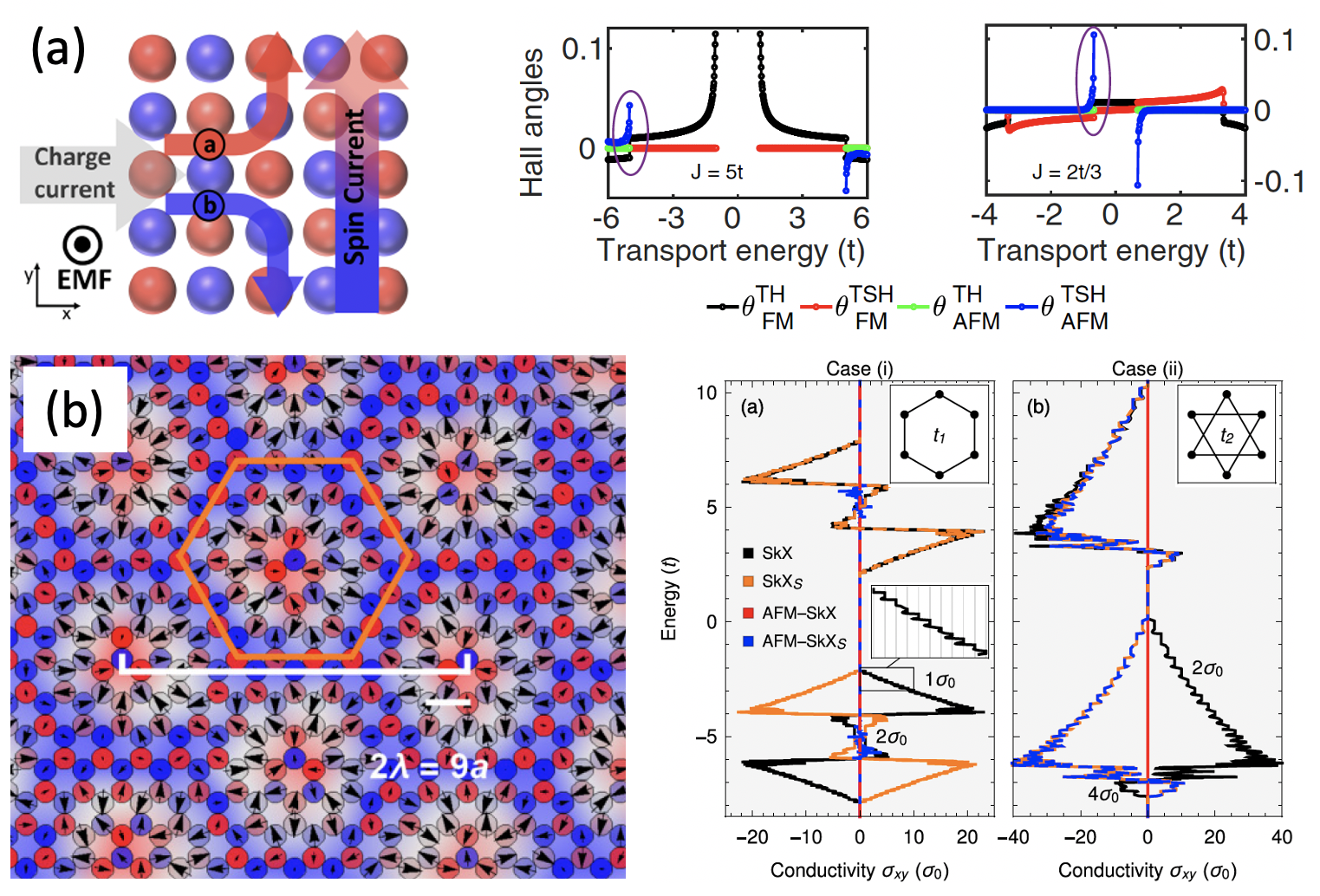}
		\caption{(Color online) (a) Topological spin Hall effect in antiferromagnetic skyrmions arising due to the emergent magnetic field. Computed topological Hall effect and topological spin Hall effect for ferromagnetic and antiferromagnetic skyrmions as a function of the transport energy in the intermediate (left) and strong (right) exchange limits. From \cite{Akosa2018}. (b) Antiferromagnetic skyrmion crystal on a honeycomb lattice. Topological Hall conductivities in ferromagnetic [black: charge conductivity; orange: spin conductivity] and antiferromagnetic skyrmion crystals [red: charge conductivity ; blue: spin conductivity]. From \cite{Goebel2017}.}
		\label{TSHE}
	\end{center}
\end{figure}

The emergent magnetic field associated with the skyrmion texture results in the transverse deflection of electrons when they traverse the texture, inducing topological Hall effect \cite{Neubauer2009b,Nagaosa2013}. Reciprocally, these deviated electrons exert a "topological" torque on the skyrmion, which substantially enhances the non-adiabatic torque \cite{Bisig2016,Akosa2017}. In an antiferromagnet with perfectly matched sublattices, the topological charge currents flowing in the two sublattices are counterpropagating, which results in a vanishing topological Hall signal but a finite topological spin Hall effect in antiferromagnets \cite{Buhl2017}, see \figref{TSHE}. Numerical studies have shown that the magnitude of the topological spin Hall signal is comparable to the topological Hall signal observed in ferromagnets \cite{Buhl2017,Goebel2017,Akosa2018}. This implies that the topological spin currents can be detected experimentally, thus opening new avenues for an experimental observation of skyrmions in antiferromagnets. Of course, applying an external field lifts the spin degeneracy and can lead to topological Hall effect. Such a situation has been demonstrated in an antiferromagnetic skyrmion crystal on triangular lattice with strong Hund coupling, where the topological Hall effect is accompanied by orbital magnetization \cite{Tome2021}.

On a phenomenological level, the emergent electric and magnetic field experienced by carriers with spin $\sigma$ on sublattice $\eta$ is given by,

\begin{eqnarray}
	\mathbf{E}_{em}^{\sigma }&=&\left(\frac{\sigma \hbar}{2e}\right)\mathit{P^\eta_{\sigma }}[(\partial _{t}\mathbf{m} \times \partial _{i}\mathbf{m}).\mathbf{m}]\mathbf{e}_{i}\\
	\label{eq2}
	\mathbf{B}_{em}^{\sigma }&=&-\left(\frac{\sigma \hbar}{2e}\right)\mathit{P^\eta_{\sigma }}[(\partial _{x}\mathbf{m} \times \partial _{y}\mathbf{m}).\mathbf{m}]\mathbf{e}_{z}.
	\label{eq3}
\end{eqnarray}
Here $P^\eta_{\sigma }=(1+\sigma\eta P)/2$ where $P$ is the polarization of the carrier's density of state per sublattice. The emergent magnetic field that is responsible for topological spin Hall effect in antiferromagnets gives rise to additional non-adiabatic torque $T=b_{j}\lambda^{2} {\cal N}_{xy}\partial_y\mathbf{m}$ \cite{Akosa2017, Akosa2018}, where $\lambda$ is the spin dephasing length that does not exceed 1 to 3 nm in ferromagnets. Previous studies demonstrated that the topological torque can explain the very large non-adiabatic torque reported in ferromagnetic vortices \cite{Bisig2016,Akosa2017} and therefore such a torque can also be substantial in antiferromagnetic skyrmions. A key feature of this torque is that it scales inversely to the magnetization gradient and hence increases upon reducing the skyrmion size. It is expected to dominate in the case of small ($<$ 20nm) skyrmions. Reciprocally, the emergent electric field, which is non zero when the skyrmion is propagating, pumps spin current through the action of topological spin motive forces. In the semiclassical limit the pumped current for spin $\sigma$ reads $I^\sigma=\int ds \left ( \mathbf{E}_{em}^{\sigma}\times \mathbf{B}_{em}^{\sigma}+\sigma _{H}^{\sigma}\mathbf{E}_{em}^{\sigma} \right )$ \cite{Abbout2018}, where $\sigma _{H}^{\sigma}$ is the spin Hall conductivity. This pumped current can enhance the mobility of trains of skyrmions significantly.
\begin{figure}
	\begin{center}
		\includegraphics[width=0.49\textwidth]{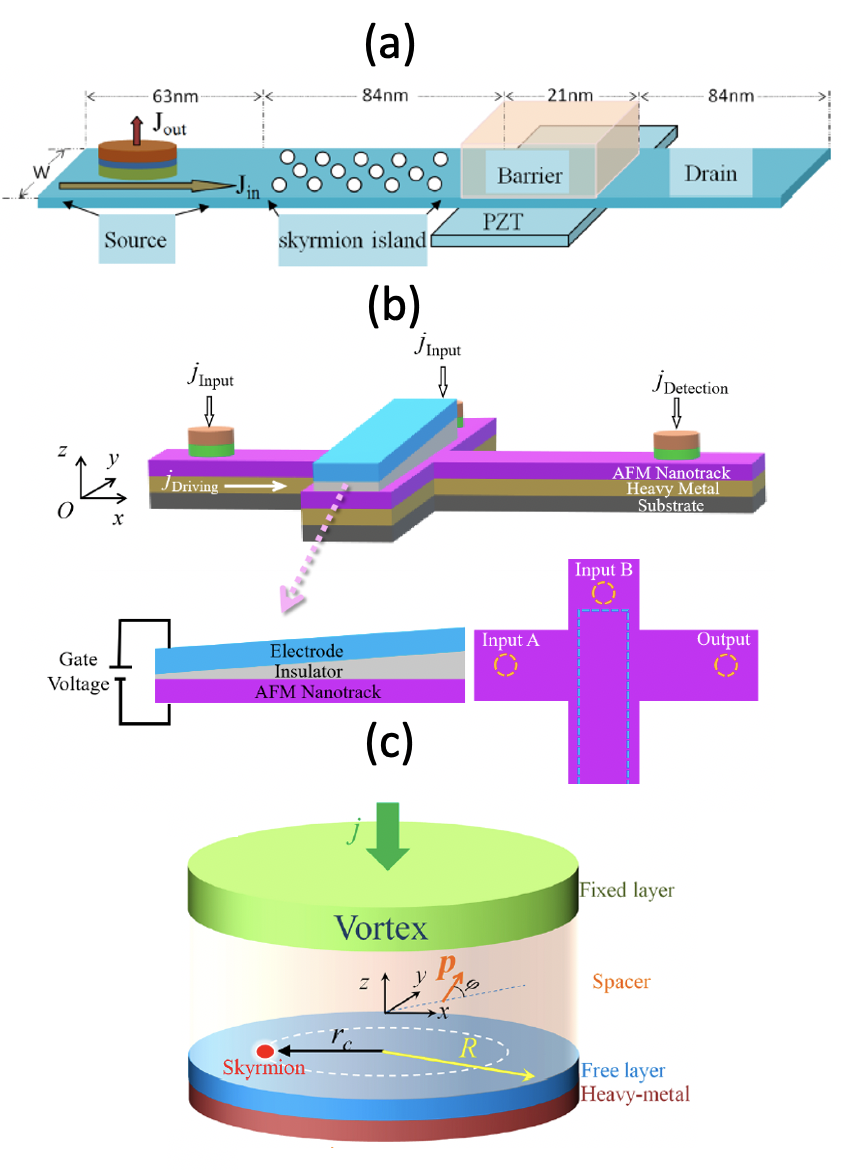}
		\caption{(Color online) Antiferromagnetic skyrmion devices. (a) Single skyrmion transistor. The transistor consists of a source terminal, a skyrmion island, a barrier region, and a drain. The source is a skyrmion generator. The skyrmion island is used to store skyrmions. The barrier acting as the gate terminal controls the skyrmion moving from the skyrmion island to the drain. From \cite{Zhao2018}. (b) Antiferromagnetic skyrmion logic gate. It consists of cross shaped nanotrack and magnetic tunnel junctions as input and output sensors. By controlling the perpendicular anisotropy in transverse nanotrack region through applied voltage we can control the motion of skyrmion bits to the output terminal. From \cite{Liang2019a}. (c) Antiferromagnetic skyrmion spin torque nano-oscillator. The fixed layer with a vortex magnetic configuration is used to generate the spin-polarized current. The heavy-metal layer is necessary to induce the interfacial
			Dzyaloshinskii-Moriya interaction, which stabilizes skyrmion in the nanodisk. From \cite{Shen2019}.}
		\label{Logic}
	\end{center}
\end{figure}

Finally, from the technological standpoint antiferromagnetic skyrmions can be utilized to encode bits of information and their dynamics can be controlled by electric fields and currents. Several device proposals are beginning to emerge. Recently, a three terminal skyrmion field-effect transistor analogous to MOSFET has been been proposed \cite{Zhao2018}. It basically has a source terminal where skyrmions are nucleated by spin currents polarized through a magnetic tunnel junction, a gate terminal where the anisotropy can be modulated to control the motion of the skyrmion and a drain terminal where they can be detected with current-perpendicular-to-plane technique (see \figref{Logic}(a)). Similarly, antiferromagnetic nanotracks can be engineered to realize Boolean logic gates \cite{Liang2019a, Xia2017}. \citet{Liang2019a} have proposed a cross bar type structure to implement universal logic gates (see \figref{Logic}(b)). In this structure, the anisotropy at the junction of the cross bar is modulated by the electric field which controls the motion of skyrmion bits to the output. One can also realize an antiferromagnetic skyrmion-based oscillator as proposed by \citet{Shen2019}. In the proposed structure, the skyrmion circular motion in a nanodisk is controlled by the spin current generated through vortex contact to generate oscillations in several gigahertz frequency range applicable for telecommunication and space technology (see \figref{Logic}(c)).

%




\section{Topological Excitations in Quantum Antiferromagnets \label{s:topexc}} 

\begin{figure}[ht]
	\begin{center}
		\includegraphics[width=0.5\textwidth]{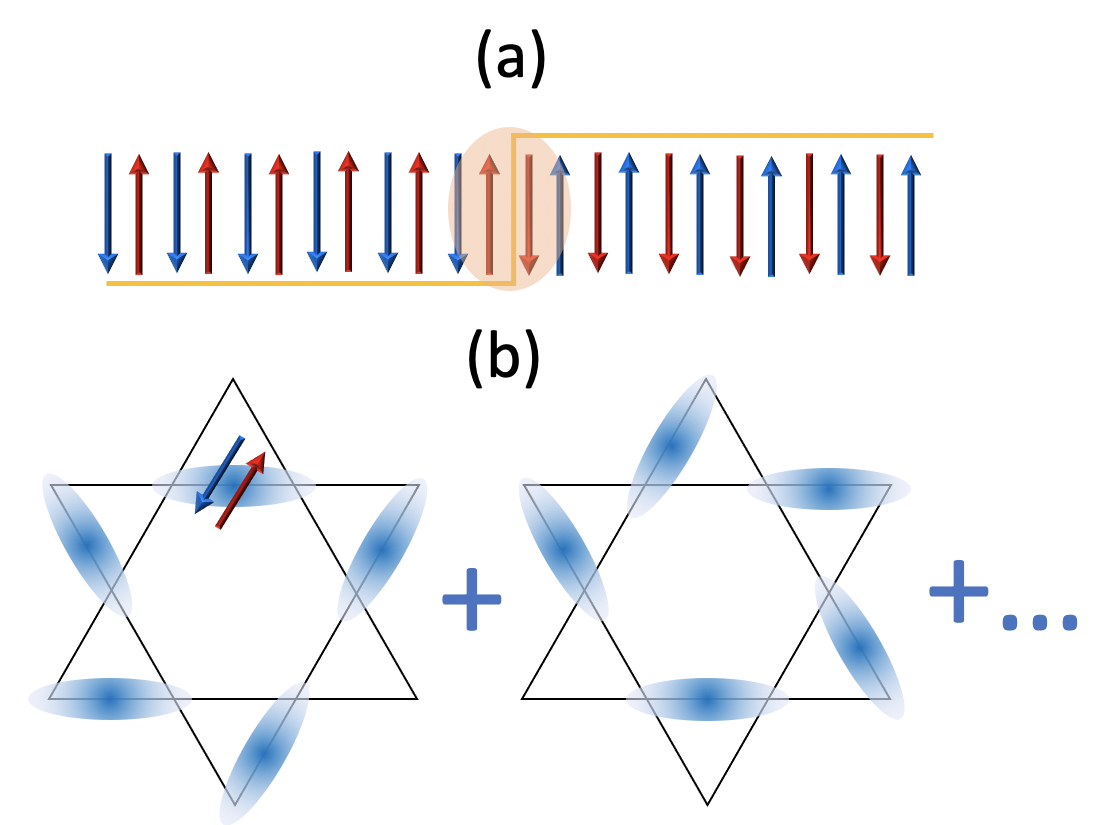}
	\caption{(Color online) Two examples of quantum antiferromagnets. (a) 1-dimensional Ising antiferromagnetic spin chain, displaying spinon excitation (orange shadowed region). (b) Kagom\'e quantum antiferromagnet displaying RVB state, composed of a large superposition of random configurations of spin singlet states distributed over the lattice.}
		\label{QAF-1}
	\end{center}
\end{figure}
In the previous sections, we have considered antiferromagnetic materials whose magnetic moments can be treated classically with a well defined orientation. The cooperation of magnetic configuration, crystal symmetry and possibly spin-orbit coupling is at the origin of a non-trivial Berry curvature that gets imprinted on the carrier wave function. The search for enhanced complexity and richer behavior naturally leads us to consider even more exotic classes of antiferromagnetic materials and reach out the realm of quantum magnetism, where the magnetic moments cannot be considered classical anymore. In this context, quantum antiferromagnets feature an ideal platform.\par

\begin{figure}[ht]
	\begin{center}
		\includegraphics[width=0.5\textwidth]{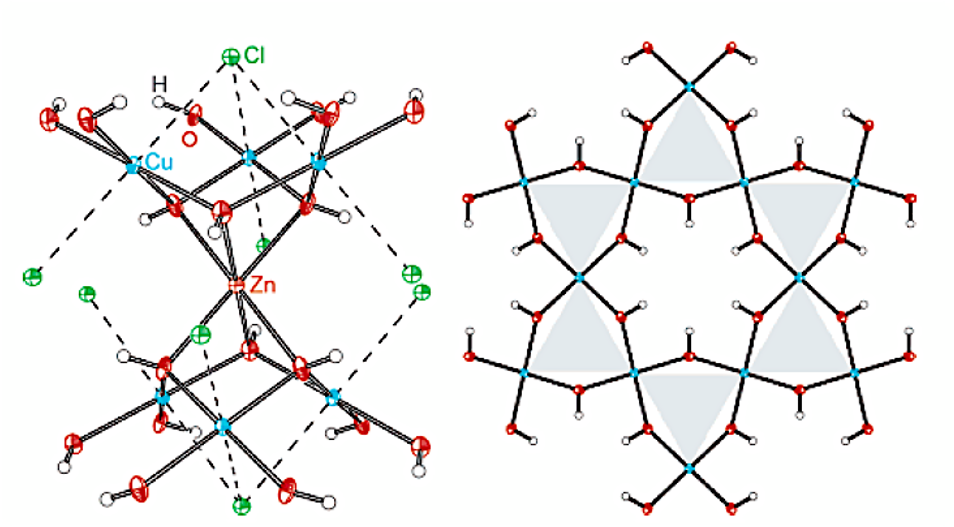}
	\caption{(Color online) (left) Crystal structure of Zn-paratacamite and (right) Cu$_3$(OH)$_6$ kagom\'e lattice, projected perpendicular to
the c axis. From \cite{Shores2005}.}
		\label{QSL-2}
	\end{center}
\end{figure}

As a matter of fact, it has been realized quite early that even the simplest quantum antiferromagnet, the antiferromagnetic Ising chain, does not display any magnetic order at finite temperature due to the large quantum fluctuations occurring in one dimension. In this model system, magnetic excitations take the form of fermionic soliton-like quasiparticles, called spinons, that carry spin-1/2 \cite{Mourigal2013,Lake2005} (see Fig. \ref{QAF-1}(a)). The ability for quantum fluctuations to stabilize novel many-body state of matter in one dimension is exemplified by several paradigms like the spin-1/2 Heisenberg chain, or the Luttinger-Tomonaga liquid \cite{Giamarchi2003}. However, quantum fluctuations are also particularly strong in higher dimensions when geometric frustrations are important. This was first realized by Anderson \cite{Anderson1973} in his search for an alternative state to the long-range classical N\'eel order in triangular antiferromagnets. He proposed that the ground state of such antiferromagnets could rather be described by a linear superposition of all the possible configurations of short-range spin singlets covering the entire lattice (see Fig. \ref{QAF-1}(b)). In contrast to N\'eel antiferromagnetism, the "resonating valence-bond" (RVB) wave function avoids spontaneous symmetry breaking down to zero temperature. This proposal triggered extraordinary creativity in theoretical research attempting to uncover the unconventional properties of such quantum liquids, the prediction of novel quantum phases and their implementation in real materials \cite{Balents2010,Lacroix2011,Knolle2019b}. Quantum spin liquids are defined as materials whose magnetic configuration emerges out of frustration and displays high topological ground-state degeneracy, long-range entanglement, and a fractionalization of the elementary excitations. Confined to theoretical prospects for several decades, this vast field of research has experienced an outstanding burst with the synthesis of the first quantum spin liquid candidate based on a kagom\'e antiferromagnet \cite{Shores2005}, as illustrated on Fig. \ref{QSL-2}.\par

Several excellent reviews are available that address various aspects of this vast field of research \cite{Balents2010,Lacroix2011,Zhou2017,Savary2017,Batista2016,Knolle2019b}. It is not our intention to give a detailed account on the fundamental properties of quantum spin liquids, but rather to inspire novel research directions by bridging this remarkably creative field of research with spintronics. After a short introduction on the relation between frustration, fractionalization and topology, we outline some of the most representative realizations of fractional excitations in quantum antiferromagnets. In the last subsection, we discuss how spintronics can contribute to better understanding spin liquids and how quantum antiferromagnets can expand the scope of spintronics.

\subsection{Frustrations, fractional excitations and topology}

In quantum spin liquids, the ground state is stabilized by frustrated interactions, i.e., pairwise magnetic interactions that cannot be satisfied simultaneously. If frustration is strong enough, then it may be more favorable for the system to break down into spin dimers randomly distributed over the lattice. In the quantum realm, the ground state is then a linear superposition of the different pavings, as suggested by Anderson \cite{Anderson1973}. Whether RVB state (or any other spin liquid state for that matter) competes favorably with an ordered N\'eel state depends on the strength of the quantum fluctuations (i.e., on the dimension) and on the level of geometric frustration. Qualitatively, lattices possessing a high coordination number, such as the edge-sharing triangular lattice that first inspired Anderson, tend to support classical ordered state (and in fact, the triangular Heisenberg antiferromagnet does exhibit 120$^\circ$ ordering \cite{Capriotti1999,White2007}). This is the reason why corner-sharing kagom\'e lattices have been the workhorse of quantum spin liquids \cite{Mendels2016}. In such systems, geometric frustrations imply a large number of competing phases, leading to high degeneracy of the highly-entangled ground state. A direct consequence is that the stability of a given candidate spin liquid phase is very sensitive to perturbations (lattice distortion, anisotropy, defects etc.) and its determination using numerical methods is often challenging, the stable phase depending on minute differences in energy, often measured in tiny fractions of the magnetic exchange. An outstanding difficulty lies in the determination of the ground state and on its stability. Several debates are on-going as exemplified by the kagom\'e spin-1/2 Heisenberg antiferromagnet where different formalisms (Gutzwiller projection, partons theory, field theory) and numerical methods are employed (variational Monte Carlo, multiscale entanglement renormalization ansatz, dynamical matrix renormalization group etc.) and result in different ground states (for a detailed discussion, see Ref. \cite{Mendels2016}).\par

Besides their fascinating ground state properties, the most appealing aspect of quantum spin liquids from a spintronics perspective is the emergence of fractional excitations. Fractionalization is the emergence of quasiparticles with quantum numbers which are fractions of the elementary degrees of freedom (spin, charge etc.). In other words, the elementary excitation decays into fractional quasiparticle excitations that can be separated to infinite distance at finite energy cost, a notion called "deconfinement" \cite{Wen1989b,Wen1991}. For instance, in fractional quantum Hall effect, an electron breaks down into quasiparticles with charge $e q/p$, where $q$ and $p$ are relatively prime integers. For instance in an antiferromagnetic Ising chain, a spin-flip decays into spinon carrying spin-1/2. What makes spin-1/2 spinons remarkable is that they obey Fermi-Dirac statistics, in sharp contrast with the more familiar bosonic magnons discussed in Section \ref{s:ahe}. We emphasize that fermionic spinons are just one type of fractional excitation that can be encountered in quantum spin liquids. Bosonic spinless vortices called visons and anyonic composite particles can also emerge, depending on the specific nature of the interactions \cite{Read1989}.Most importantly, the deconfinement of fractional excitations is accompanied by the liberation of a gauge field, which appears explicitly when treating the spin liquid in the parton theory. This gauge field can be viewed as the equivalent of the electromagnetic field for electrons: it dresses the fractional excitations and mediates long-range interaction between them.\par

Because quantum fluctuations are so strong, they wash out the order parameter \cite{Kitaev2006}. Consequently, quantum spin liquid ground states do not display broken symmetries and they cannot be described by conventional local order parameters. In fact, in a series of foundational papers, Wen \cite{Wen1989,Wen1990,Wen1991} demonstrated that these ground states can be characterized by a non-local, "hidden", topological order, which arises from the high degeneracy of the ground state. A quantum spin liquid can be characterized according to this topological classification \cite{Wen2002}. In fact, topology is a necessary condition to fractionalization \cite{Oshikawa2006}. In the next sections, we briefly describe selected realization of fractionalization in frustrated antiferromagnets.

\subsection{Topological spinons in kagom\'e antiferromagnets}

Kagom\'e lattices, depicted in Fig. \ref{QAF-1}(b), are characterized by their extensive ground state degeneracy, which makes them a promising platform for the search of spin liquids. The first difficulty though is that this highly frustrated system features a large number of competing phases and therefore, several scenarios have been proposed to describe the ground state of kagom\'e lattice spin-1/2 Heisenberg antiferromagnet involving various flavors of short-range or long-range RVB states. The first proposal was that of a valence-bond crystal \cite{Marston1991,Sachdev1992,Nikolic2003,Singh2008}, where the spins are distributed on {\em localized} valence bonds, leading to symmetry breaking and a spin gap between singlet and triplet states. The current debate is now focusing on the competition between a U(1) Dirac spin liquid \cite{Hastings2001}, a $\mathbb{Z}_2$ spin liquid \cite{Yan2011b} or a chiral spin liquid \cite{Messio2012}. The U(1) Dirac spin liquid, also called the algebraic spin liquid, possesses {\em long-range} RVB states whose spinons display a massless Dirac spectrum. This system is gapless and the spin-spin correlation decays as a power law \cite{Ran2007,Hermele2008}. More recently, \citet{Yan2011b,Depenbrock2012,Jiang2012b} proposed that $\mathbb{Z}_2$ spin liquid constitute the ground state. The $\mathbb{Z}_2$ spin liquid, originally introduced by \onlinecite{Wen1991}, is a {\em short-range} RVB spin liquid which possesses a gap. The excitations are spinons, which can be fermions or bosons, as well as singlet excitations which are vortices (or flux) of the $\mathbb{Z}_2$ gauge field, also called "visons". Notice that visons, sometimes called magnetic monopoles, can also emerge in U(1) quantum spin liquids and display topological Hall effect \cite{Chen2016e, Zhang2020g}. A last possibility is the chiral spin liquid first introduced by \onlinecite{Kalmeyer1987,Kalmeyer1989}, which is also a long-range RVB spin liquid stabilized by chiral interactions (e.g., Dzyaloshinskii-Moriya interaction \cite{Elhajal2002}) and with simultaneously and spontaneously broken space reflection and time-reversal symmetries \cite{Yang1993,Messio2012,Zaletel2016}. Because of its non-trivial topology, this spin liquid features topologically protected gapless edge states and bulk excitations with anyonic statistics, called "semions" \cite{Bauer2014}. The question of the gap has been nurturing intense numerical investigations recently \cite{Iqbal2013,Zaletel2016,Liao2017,He2017,Zhu2018} and the debate is not settled yet.\par

Although the kagom\'e lattice spin-1/2 Heisenberg antiferromagnet has been intensively scrutinized theoretically, hardly a handful of candidate materials has been identified and only one of them has seriously been investigated experimentally. As a matter of the fact, the extreme sensitivity of the spin liquid phase to perturbations puts a daring constraint on the nature of the magnetic exchange present in the material. To date, the potential candidates are vesignieite \cite{Boldrin2018}, ${\rm BaCu_3V_2O_8(OH)_2}$, Zn-doped claringbullite ${\rm Cu_3Zn(OH)_6FCl}$ \cite{Feng2019b}, barlowite \cite{Han2014b,Smaha2018} ${\rm Cu_3Zn(OH)_6FBr}$, ${\rm Y_3Cu_9(OH)_{18}OCl_8}$ \cite{Barthelemy2019}, volborthite \cite{Watanabe2016}, ${\rm Cu_3V_2O_7(OH)_2\cdot2H_2O}$, kappellasite \cite{Fak2012}, ${\rm Cu_3Zn(OH)_6Cl_2}$, and herbertsmithite \cite{Shores2005}, ${\rm ZnCu_3(OH)_6Cl_2}$. \par
\begin{figure}[ht]
	\begin{center}
		\includegraphics[width=0.5\textwidth]{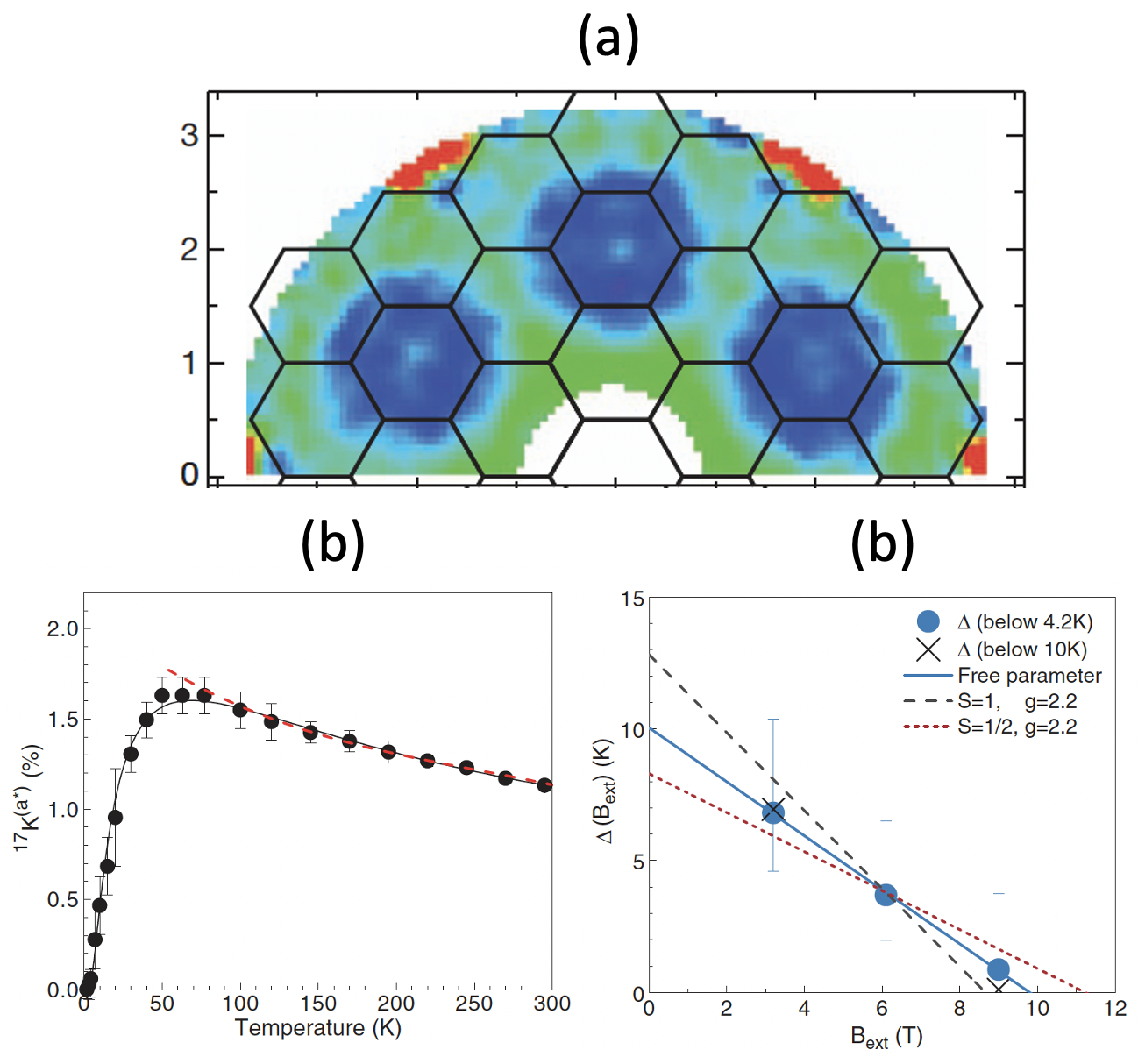}
	\caption{(Color online) (a) Inelastic neutron scattering from the spin excitations of single crystal
sample of ZnCu$_3$(OH)$_6$Cl$_2$ measured at 1.6K. The exceedingly diffusive pattern is consistent with that of short-range RVB state. From \cite{Han2012}. (b) 	Magnetic susceptibility of ZnCu$_3$(OH)$_6$Cl$_2$ showing the absence of magnetic order at low temperature and (c) spin gap deduced from oxygen-17 nuclear magnetic resonance frequency shift. From \cite{Fu2015c}.}
		\label{QSL-3}
	\end{center}
\end{figure}

Herbertsmithite, depicted on Fig. \ref{QSL-2}, is from far the most studied spin liquid candidate \cite{Mendels2016,Norman2016}. It shows no magnetic ordering at low temperature \cite{Mendels2007}, and specific heat measurements suggest gapless states \cite{deVries2008}. In addition, nuclear magnetic resonance on powder samples \cite{Olariu2008} suggests also short-range correlations that comply with the U(1) gapless spin liquid scenario. The advent of single crystal samples has enabled substantial progress in the determination of the spin liquid nature of herbertsmithite. Neutron scattering experiments on such single crystals reported highly diffuse scattering intensity in a range of energy that is two orders of magnitude below the antiferromagnetic exchange, supporting the scenario of short-range RVB states \cite{Han2012}. In addition, the scattering pattern is mostly independent on the energy pointing towards a continuum of excitations, demonstrating the presence of fractionalized excitations. The contribution of impurities in the sample hindered the determination of a spin gap \cite{Han2016b}. Complementary nuclear magnetic resonance experiments \cite{Fu2015c} identified a gap about 0.86 meV, thereby challenging the experiments on powder samples \cite{Olariu2008}. In a recent theoretical work, \citet{Punk2014} attribute the excitation continuum to the presence of topological visons in a $\mathbb{Z}_2$ spin liquid.\par

In the past few years, two additional spin liquid candidates have been successfully synthesized. Volborthite \cite{Yoshida2012b} is another kagom\'e antiferromagnet that displays quantized magnetic plateaus \cite{Ishikawa2015}. Recently, a non-trivial thermal Hall effect has been reported and interpreted as arising from topological spinons \cite{Watanabe2016}. Finally, barlowite \cite{Han2014b} has also been recently claimed to be a $\mathbb{Z}_2$ spin liquid \cite{Wei2019}.



\subsection{Majorana fermions in Kitaev honeycomb lattices\label{ss:kitaev}}

Anyons are quasiparticles that only exist in two-dimensions and whose quantum statistics differs from that of bosons or fermions. In an anyonic gas, exchanging two quasiparticles results in multiplying the total wave function by a phase of $e^{i\varphi}$ where $\varphi=2\pi\nu$, where $\nu$ is a rational number. This fractional phase, distinct from the familiar $\varphi=0,\pi$ case for bosons and fermions, stems from the non-trivial topology of the paths taken by two quasiparticles during the swap. The anyon was introduced by Wilczek \cite{Wilczek1982,Wilczek1982b} who demonstrated that a composite particle made of an integer charge orbiting around a vortex carrying a magnetic flux in two dimensions obeys anyonic statistics. In 2003, Kitaev \cite{Kitaev2003} proposed that anyons could be exploited to develop an entirely new type of "fault-tolerant" topological quantum computing scheme \cite{Nayak2008,Stern2013,DasSarma2015,Field2018}.\par

\begin{figure}[ht]
	\begin{center}
		\includegraphics[width=0.5\textwidth]{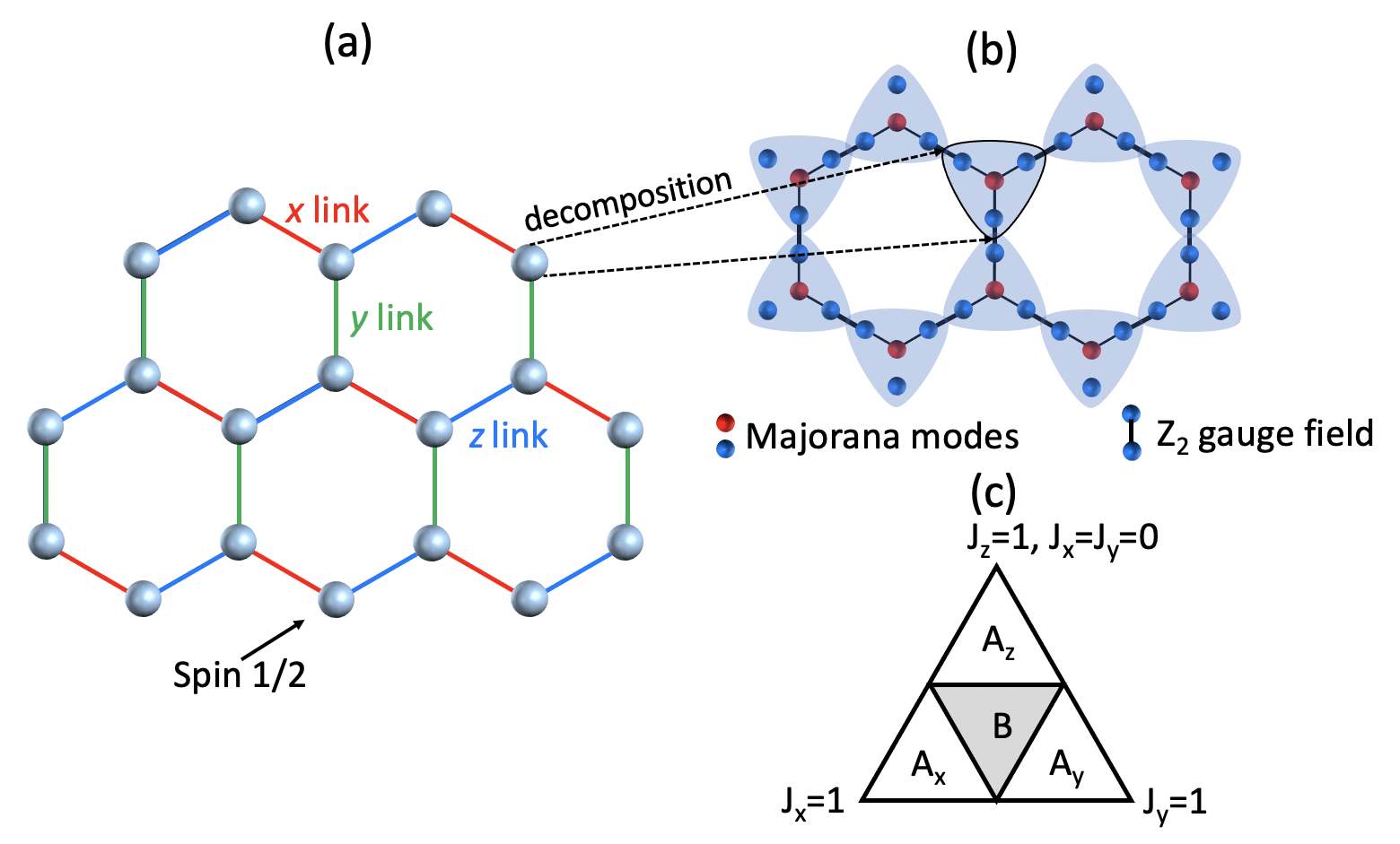}
	\caption{(Color online) (a) Schematics of Kitaev's honeycomb spin liquid. (b) Majorana decomposition of the spin-1/2 operators, resulting in free Majorana fermions (red) and pairs of bound Majorana fermions that can be interpreted as $\mathbb{Z}_2$ gauge fields. (c) Topological phase diagram of Kitaev's model, inspired from \cite{Kitaev2006}. A$_x$, A$_y$, or A$_z$ refer to the gapped A phase whereas B is gapless and can acquire non-trivial topology in the presence of a magnetic field.}
		\label{fig:kitaev1}
	\end{center}
\end{figure}

Inspired by these fascinating prospects, \citet{Kitaev2006} proposed that such anyons could be found as low energy excitations of a special kind of spin liquid, described by a spin-1/2 Hamiltonian on a honeycomb lattice whose exchange interaction depends on the direction of the bonds linking nearest-neighbors. This Hamiltonian belongs to the vast class of {\em compass models}, which typically applies to Mott insulators whose strong spin-orbit coupling results in highly anisotropic exchange interaction \cite{Nussinov2015}. The Hamiltonian explicitly reads
\begin{equation}\label{eq:kitaev1}
{\hat H}=-J_x\sum_{x~{\rm link}}{\hat \sigma}_i^x{\hat \sigma}_j^x-J_y\sum_{y~{\rm link}}{\hat \sigma}_i^y{\hat \sigma}_j^y-J_z\sum_{z~{\rm link}}{\hat \sigma}_i^z{\hat \sigma}_j^z,
\end{equation}
where the geometry is represented on Fig. \ref{fig:kitaev1}(a). Notice that whereas frustration in kagom\'e or triangular quantum Heisenberg antiferromagnets comes from the geometry, in Kitaev model the frustration comes from the anisotropy of the exchange interaction. To solve this problem, Kitaev first points out that Hamiltonian \eqref{eq:kitaev1} commutes with the loop operator
\begin{equation}\label{eq:kitaev2}
{\hat W}_p={\hat \sigma}_1^x{\hat \sigma}_2^y{\hat \sigma}_3^z{\hat \sigma}_4^x{\hat \sigma}_5^y{\hat \sigma}_6^z,
\end{equation}
defined for a given hexagonal plaquette $p$. Therefore, the total Hilbert space can be decomposed as the direct product of the eigenspaces of ${\hat W}_p$, called {\em sectors}. However, this decomposition is not sufficient to solve Hamiltonian \eqref{eq:kitaev1}. In fact, since each plaquette possesses six vertices and each vertex is shared between three plaquettes, the dimension of each sector is only $\sim 2^{n/2}$, $n$ being the number of plaquettes composing the lattice. To solve the problem exactly, Kitaev decomposed each spin-1/2 into two real fermions, called Majorana fermions \cite{Majorana1937}. These particles have the peculiarity to be their own antiparticle. Besides their major interest for nuclear and particle physics \cite{Elliott2015}, Majorana fermions have been intensively hunted in condensed matter systems for the past decade. As a matter of fact, Majorana bound states obey anyon statistics and could be therefore exploited for topological quantum computing. The first glimpses of Majorana bound states have been observed as zero-modes of topological superconducting nanowires (e.g., InSb)\cite{Mourik2012,Nadj-perge2014,Zhang2018e}, as well as one-dimensional topologically-protected state along a line defect in Fe(Se,Te) \cite{Wang2020}.\par

\begin{figure}[ht]
	\begin{center}
		\includegraphics[width=0.3\textwidth]{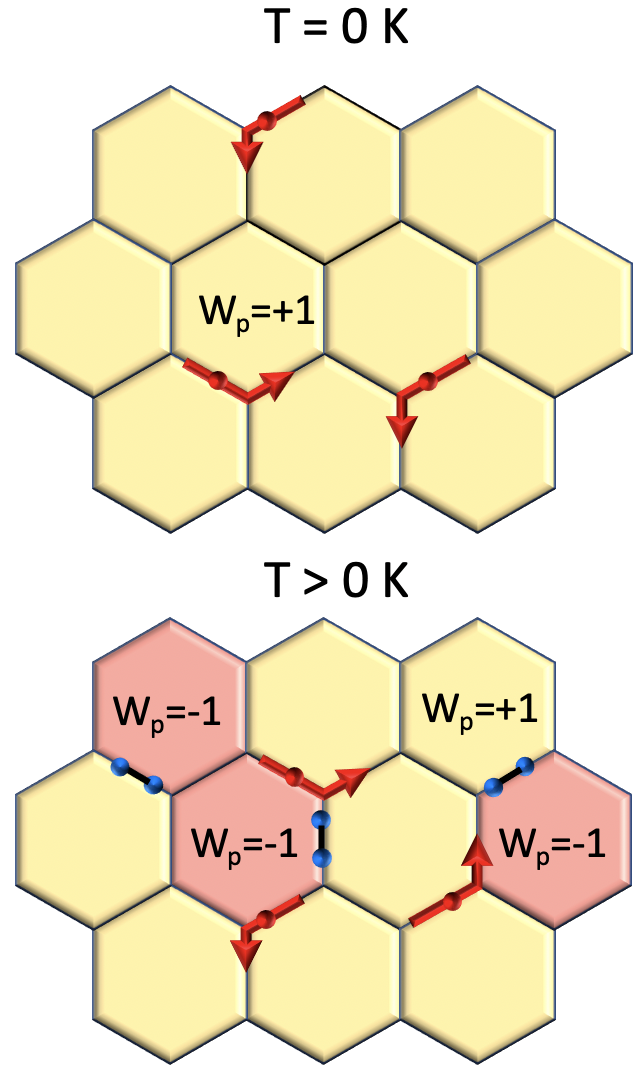}
	\caption{(Color online) (a) Schematics of Kitaev's spin liquid ground state, composed of frozen $\mathbb{Z}_2$ fluxes ($W_p=+1$) and itinerant Majorana fermions (red). (b) At finite temperature, $\pi$ fluxes ($W_p=-1$) get excited and are accompanied by localized Majorana pairs (blue). Inspired from \cite{Do2017}.}
		\label{fig:kitaev2}
	\end{center}
\end{figure}

Coming back to Kitaev's spin liquid, the key idea is to fractionalize the complex wave function of the fermionic spin-1/2 into four real Majorana fermions (two per spin projection $\uparrow$, $\downarrow$). This procedure enlarges the Hilbert space and a gauge flux naturally emerges when projecting the solution on the real physical space. This gauge flux is artificial and introduced by the Majorana representation. The effective Hamiltonian reads
\begin{equation}\label{eq:kitaev3}
{\hat H}=\frac{i}{4}\sum_{\langle i,j\rangle}{\hat A}_{ij}c_ic_j,
\end{equation}
where $c_i$ is the Majorana fermion and ${\hat A}_{ij}=2J_{\mu_{ij}}{\hat u}_{ij}$ with $\mu_{ij}$ being the direction of the $\langle i,j\rangle$ bond and ${\hat u}_{ij}$ is a {\em static} $\mathbb{Z}_2$ gauge field defined by the interaction between the Majorana fermions associated with each site of the bond. This decomposition is illustrated on Fig. \ref{fig:kitaev1}(b). One can associate a gauge flux ${\hat W}_p=\Pi_{\langle i,j\rangle \in p}{\hat u}_{ij}$ to each plaquette. Within the scope of the present review, the rather technical definition of Eq. \eqref{eq:kitaev3} is not as important as its physical meaning. There are therefore two types of excitations, a Majorana fermion "$c$" and a gauge flux ${\hat W}_p=-1$, referred to as a $\pi$ flux \cite{Baskaran2007}. An illustration of this representation is given in Fig. \ref{fig:kitaev2}.\par

The ground state is reached when the system is vortex-free (no  $\pi$ fluxes). In this case, the excitation spectrum of the itinerant Majorana mode is
\begin{equation}\label{eq:kitaev4}
\epsilon_{\bf q}=\pm|J_x e^{i{\bf q}\cdot{\bf a}}+J_y e^{i{\bf q}\cdot{\bf b}}+J_z|,
\end{equation}
where ${\bf a}=(1/2,\sqrt{3}/2)$ and ${\bf b}=(-1/2,\sqrt{3}/2)$. A remarkable aspect of this model is that depending on the relative strength of the exchange parameters, the system can enter two phases, called phase A and phase B, illustrated in Fig. \ref{fig:kitaev1}(c). Phase A is gapped and hosts abelian anyonic excitations, whereas phase B is gapless but acquires a gap in the presence of a magnetic field and hosts non-abelian anyonic excitations, coming from the binding between an itinerant Majorana fermion and a pair of localized $\pi$ fluxes. Phase B is the one that has attracted the most intense theoretical and experimental efforts (see, e.g. \cite{Nayak2008}).

Kitaev's prediction has nurtured impressive enthusiasm in the research community, resulting in vast theoretical developments \cite{Hermanns2018}. Keeping on sight our goal to introduce promising growth directions in the investigation of complex antiferromagnets, we briefly outline some of the remarkable theoretical endeavors achieved recently and rather concentrate on the experimental observations reported in the past few years. The chief characteristic of Kitaev materials is to display bond-directional exchange interactions on a honeycomb lattice. \citet{Jackeli2009} showed that such anisotropic exchange can emerge in 5d and 4d transition metal oxides with the edge-sharing oxygen octahedra, due to the destructive interference between mirror-symmetric MO$_6$ bonds. Iridates, ${\rm A_2IrO_3}$, are ideal candidates as the edge-sharing ${\rm IrO_6}$ octahedra form a honeycomb lattice \cite{Chaloupka2010,Singh2012b}. Among these candidates, ${\rm Na_2IrO_3}$ attracted substantially promises \cite{Chaloupka2013,Rau2014b,Chun2015} but was found to order magnetically \cite{Choi2012}, while spin liquid behavior was reported in ${\rm H_3LiIr_2O_6}$ \cite{Kitagawa2018}. As mentioned in Section \ref{s:topaf}, \onlinecite{Hermanns2015} recently predicted that the magnetic excitations of a three-dimensional Kitaev model, $\beta$-Li$_2$IrO$_3$, are Majorana modes that form a Weyl superconductor.

\begin{figure}[ht]
	\begin{center}
		\includegraphics[width=0.5\textwidth]{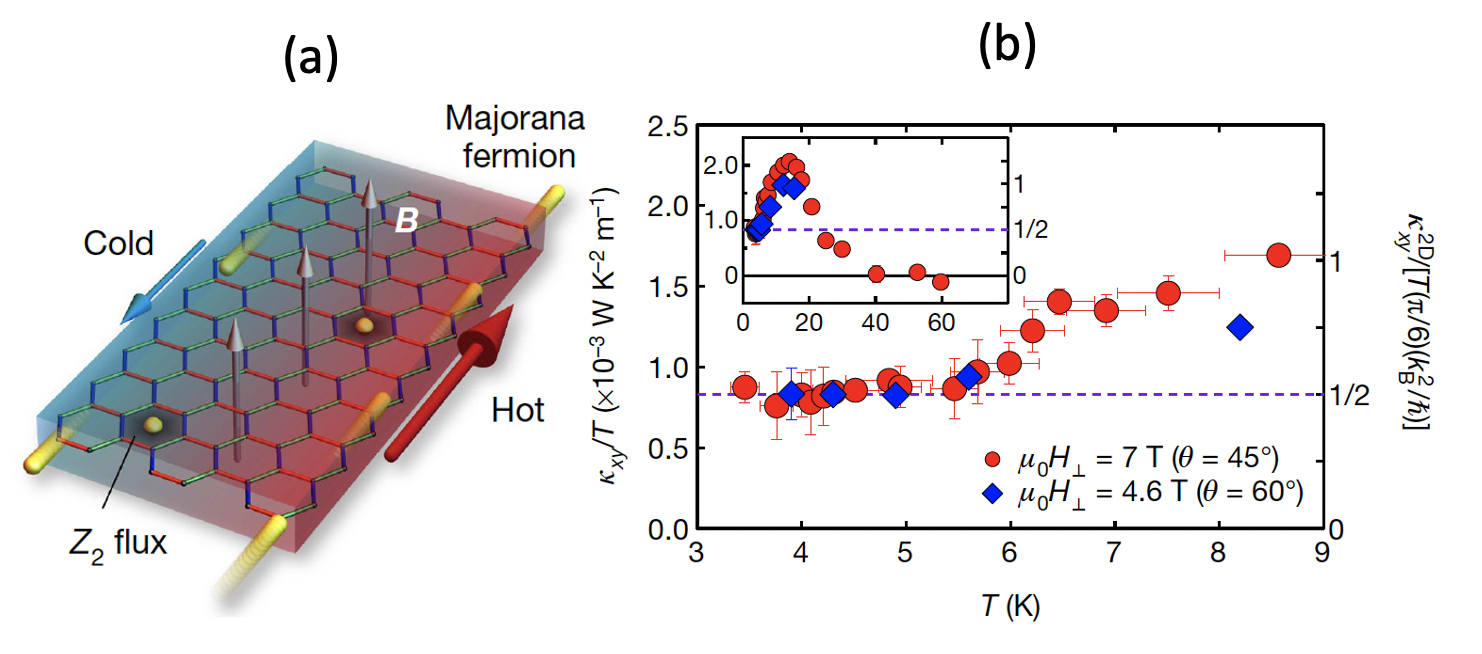}
	\caption{(Color online) (a) Illustration of the Majorana edge transport in the B phase of Kitaev's phase diagram in the presence of an external field. (b) Temperature dependence of the thermal Hall effect. On the right axis, the thermal conductance is normalized to the quantum of heat conductance. From \cite{Kasahara2018}.}
		\label{fig:kitaev3}
	\end{center}
\end{figure}

To date, it seems that the most successful experiments have been performed on $\alpha-{\rm RuCl_3}$ which fulfills Kitaev's requirements \cite{Sears2015,Ran2017}. Raman scattering study reveals a broad continuum \cite{Sandilands2015}, attributed to fractionalized fermionic excitations \cite{Nasu2016}. Neutron scattering \cite{Banerjee2016,Banerjee2017} reveals a continuum response at $\Gamma$-point that is incompatible with spin-wave theory and whose main features are consistent with that of Kitaev model \cite{Knolle2014}. Nuclear magnetic resonance has also identified a spin excitation gap that is cubic in applied magnetic field, consistent with the behavior of Majorana fermions in phase B of Kitaev's model, as well as a zero-field contribution consistent with gauge flux excitations \cite{Jansa2018}. Remarkably, exploiting a combination of heat capacity and neutron scattering experiments, \citet{Do2017} reported the progressive localization of fractional excitations upon reducing the temperature, accompanied by the progressing quenching of the gauge flux, in agreement with a scenario proposed by \citet{Nasu2015}. Upon increasing the temperature, the system evolves from a spin liquid phase without itinerant carriers and associated with the Kitaev ground state (quenched gauge flux excitations) towards a Kitaev paramagnet characterized by the coexistence of localized and itinerant Majorana fermions, as well as fluctuating gauge flux excitations. This scenario is illustrated on Fig. \ref{fig:kitaev2}. Notice that this interpretation is challenged by an alternative theory that involves more conventional breaking down of magnons \cite{Winter2017b} (see also the discussion given in \cite{Perkins2017}). Finally, \citet{Kasahara2018} recently reported quantized thermal Hall effect in $\alpha-{\rm RuCl_3}$, upon a tilted external magnetic field, see Fig. \ref{fig:kitaev3}. This quantized thermal Hall effect is the hallmark of fractional topologically protected edge states and evidences the existence of topological Majorana modes in this spin liquid. In a recent extension of this pioneering work, \citet{Yokoi2020} reported the observation of quantized thermal Hall effect even in the absence of perpendicular magnetic field. We close this discussion by pointing out that beyond the Kitaev spin liquid phase, $\alpha-{\rm RuCl_3}$ displays high-field-induced quantum spin liquid phase that host gapless excitations \cite{Zheng2017}.
\subsection{Spinons in triangular antiferromagnets}

The triangular antiferromagnet has been the first paradigm for quantum spin liquid. As mentioned above, in spite of Anderson's inspirational proposal the quantum fluctuations are not strong enough to completely destabilize the N\'eel order in triangular antiferromagnets. Therefore, substantial effort has been paid on the theory side to discover under which conditions a spin liquid, and if so which type, could be stabilized. The competition between nearest neighbor, next-nearest neighbor, anisotropy, Dzyaloshinskii-Moriya interaction, four-spin interaction etc. leads to very different types of quantum spin liquids and therefore, the triangular antiferromagnet is probably one of the most studied frustrated systems, at least theoretically.\par

For instance, when both nearest-neighbor and next-nearest neighbor interactions are present, a quantum spin liquid phase develops at the boundary between the 120$^\circ$ and row-wise N\'eel states \cite{Hu2015b,Wu2015}. Introducing anisotropic exchange can discriminate between the competing phases and stabilize $\mathbb{Z}_2$ spin liquid \cite{Hu2015b,Wu2015}, while \citet{Iqbal2016} obtained a U(1) gapless spin liquid. As mentioned at the beginning of this section, because the energy difference between competing phases is a tiny fraction of the magnetic exchange, different numerical approaches tend to provide different results and therefore extreme scrutiny must be exercised to settle the debate for S=1/2 triangular antiferromagnets. For S=1 triangular antiferromagnets, a chiral spin liquid with spinon and holon excitations with spins larger than 1/2 obeying non-abelian statistics has also been predicted \cite{Greiter2009}.

On the experimental side, two quasi-two dimensional organic salts have emerged as excellent candidates: $\kappa-{\rm(ET)_2Cu_2(CN_3)}$ and ${\rm ETMe_3Sb[Pd(dmi)_2]_2}$, which are both on the verge of metal-insulator transition. The antiferromagnetic exchange scales around $J\sim 20$ meV but does not stabilize long-range antiferromagnetic order. Linear-in-temperature heat capacity and heat conductivity have been reported \cite{Yamashita2009,Yamashita2010}, suggesting free fermionic heat carriers. Unfortunately, no neutron scattering can be achieved due to the strong contribution of the hydrogen atoms to scattering. In $\kappa-{\rm(ET)_2Cu_2(CN_3)}$, \citet{Yamashita2009} showed the possibility of a gap.

%
%
%
%
%
%
%
%
%

\subsection{Monopoles and Dirac strings in magnetic pyrochlores}

\begin{figure}[ht]
	\begin{center}
		\includegraphics[width=0.5\textwidth]{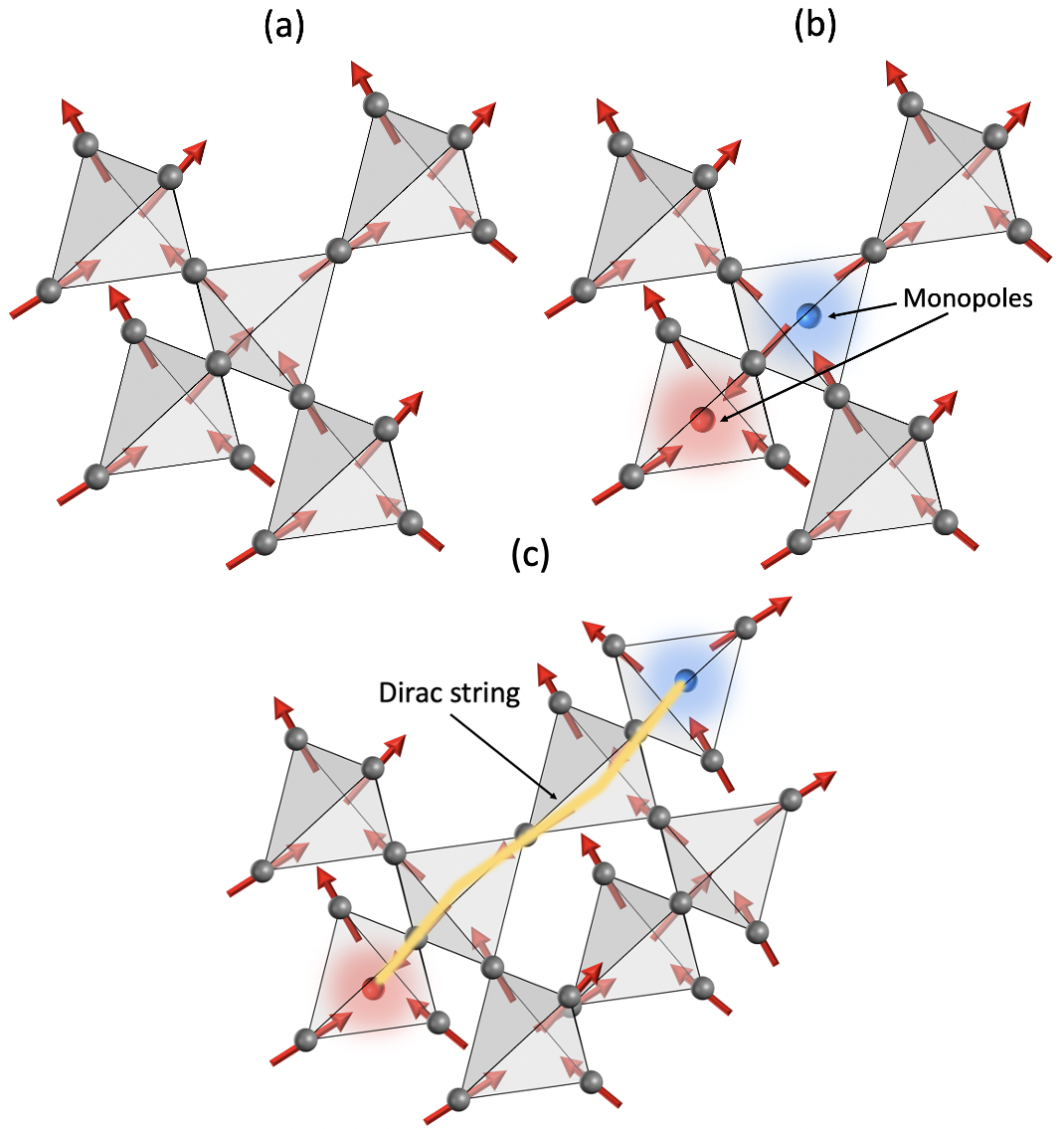}
	\caption{(Color online) (a) Schematics of a pyrochlore lattice with two-in two-out magnetic configuration. (b) Flipping one spin induces two monopoles of opposite polarity. (c) The successive splitting of magnetic moments along a Dirac string occurs at no additional energy cost no matter the separation between the monopoles.}
		\label{fig:monopole}
	\end{center}
\end{figure}

As a last example of fractional excitation in quantum magnets, we briefly discuss the emergence of magnetic monopoles in spin ice. This is slightly off-topic since the spin ice is not an antiferromagnet, but rather a highly frustrated ferromagnet characterized by a zero-point entropy \cite{Ramirez1999,Bramwell2001} $s \approx k_{\rm B}{\rm ln}(3/2)$, similar to the one obtained by Pauling on water ice \cite{Pauling1935}. In the case of water ice, this entropy is reached under the condition that two protons are near to and two protons are away from each oxide ion in order to minimize Coulomb interactions. In magnetic pyrochlores, composed of a lattice of corner-sharing tetrahedra with a magnetic moment at each vertex, this rule is enforced by dipolar rather than Coulomb interaction and translates into two magnetic moments pointing in and two magnetic moments pointing out of the tetrahedron, as illustrated in Fig. \ref{fig:monopole}(a). Therefore, although the nearest-neighbor interaction is ferromagnetic, the magnetic unit cell of a spin ice resembles that of a (highly frustrated and noncollinear) antiferromagnet.\par

The fundamental excitation in spin ice corresponds to a spin flip in the tetrahedral magnetic unit cell, hereby violating Pauling's ice rule locally. This spin flip introduces a magnetic monopole-antimonopole pair \cite{Castelnovo2008}, see Fig. \ref{fig:monopole}(b). As a matter of fact, flipping one spin creates an excess of magnetic charge in one tetrahedron and a depletion of magnetic charge in the adjacent tetrahedron. These two magnetic charges interact via long-range dipolar interaction $\sim 1/r$ and therefore can be taken away from each other at finite energy cost by successively flipping the magnetic moments along a given path, called a Dirac string, depicted on Fig. \ref{fig:monopole}(c). The spin ice ground state is a "soup" of Dirac strings of various sizes and extension, forming either closed loops or extended strings opened at the boundaries. Because of the finite energy cost for creating a Dirac string, the monopole-antimonopole pair is truly a deconfined fractional magnetic excitation. A remarkable property of magnetic monopoles is that they obey Gauss law, ${\bm\nabla}\cdot{\bf M}=\rho_m$ (where $\rho_m$ is the monopole density), and can be manipulated by external magnetic field exactly like electric charges are controlled by electric fields. The spin ice belongs to the larger class of Coulomb phase materials, which are systems whose excitations feature effective charges interacting via effective Coulomb interaction \cite{Henley2010}.\par
\begin{figure}[ht]
	\begin{center}
		\includegraphics[width=0.4\textwidth]{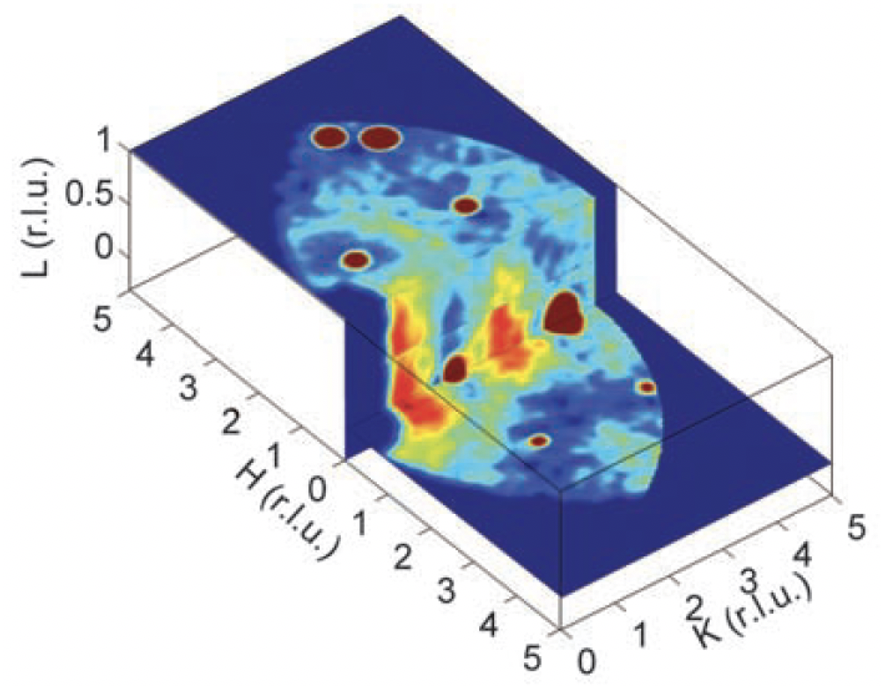}
	\caption{(Color online) Neutron diffraction data showing cone scattering appearing at the (020) Bragg peak and characteristic of two-dimensional random walk Dirac string. From \cite{Morris2009}.}
		\label{fig:monopole2}
	\end{center}
\end{figure}

Spin ices have been found in the rich family of pyrochlore oxides, which also displays various long-range orders together with spin-glass and spin-liquid behavior \cite{Gardner2010}. Magnetic monopoles have been observed a decade ago in both ${\rm Dy_2Ti_2O_7}$ \cite{Morris2009,Bramwell2009,Kadowaki2009} and ${\rm Ho_2Ti_2O_7}$ \cite{Fennell2009}. These two materials are characterized by large ($\sim 10\mu_B$) classical Ising spins pointing along the local $\langle111\rangle$ directions and displaying a macroscopically degenerate ground state obeying the ice rule. As a matter of fact, a first clue was obtained by \citet{Snyder2004} by investigating the spin freezing dynamics of ${\rm Dy_2Ti_2O_7}$, and later interpreted by \citet{Jaubert2009} as revealing the diffusive motion of monopoles. The equivalence of magnetic monopoles to effective electric charges was revealed in an elegant experiment by \citet{Bramwell2009}. Using muon spin relaxation experiments, they demonstrated that the monopole dynamics is not described by Ohm's law, but rather by Onsager-Wien mechanism of carrier dissociation and recombination, observed for instance in electrolytes \cite{Onsager1934}. In the presence of thermal fluctuations, a bound monopole-antimonopole pair is created when applying an external magnetic field and this pair is broken into two free monopoles when increasing the field. The production rate is governed by a non-Ohmic law
\begin{equation}
\frac{\kappa(B)}{\kappa(0)}=1+\frac{1}{2}\left(\frac{\mu_0Q^3B}{8\pi k_B^2T^2}\right)+\frac{1}{24}\left(\frac{\mu_0Q^3B}{8\pi k_B^2T^2}\right)^2+...
\end{equation}
where $Q$ is the effective magnetic charge and $B$ is the applied field. The same year, it was also demonstrated that the heat capacity of ${\rm Dy_2Ti_2O_7}$ \cite{Morris2009,Kadowaki2009} corresponds to the Debye-H\"uckel theory of a gas of monopoles with Coulomb interactions \cite{Castelnovo2011}. The existence of Dirac strings, which carry local magnetism and can be controlled by an external magnetic field, has been uncovered by neutron scattering \cite{Morris2009,Fennell2009}, see Fig. \ref{fig:monopole2}.\par

Recent experiments have shown the possibility to long-lived magnetic currents with a lifetime up to a minute \cite{Giblin2011}, as well as the central role of quantum spin tunneling and monopole interactions during the Brownian motion \cite{Bovo2013}. Current direction covers the growth of pyrochlore thin films \cite{Leusink2014, Bovo2014}, which displays rich phase transitions \cite{Bovo2019,Barry2019} and even possibly surface ordering leading to the stabilization of a square spin ice in the monolayer limit \cite{Jaubert2017}. Artificial spin ice \cite{Nisoli2013,Hugli2012,Skjaervo2020} consisting of two-dimensional (square or honeycomb) arrangement of single domain nanomagnets mutually interacting via dipolar field have also led to the observation of classical Dirac strings and monopole dynamics \cite{Mengotti2011,Perrin2016}.

We conclude this brief presentation by mentioning recent progress in the field. It was indeed realized that in spite of the large magnetic moment of the rare-earth element, some members of the pyrochlore family experience large quantum fluctuations that favor novel quantum-entangled phases at the boundary between classical N\'eel and spin ice state, tagged quantum spin ice \cite{Molavian2007,Gingras2014}. This quantum spin ice state is characterized by quantum strings which are a superposition of flipped spin paths of different lengths, with magnetic monopoles at their end. This phase of matter has been searched in ${\rm Tb_2Ti_2O_7}$ and ${\rm Yb_2Ti_2O_7}$. In the latter, it was proposed that string-like excitations emerge at low fields using time-domain terahertz spectroscopy \cite{Pan2014b}. These excitations are identified as fluctuating quantum strings connecting monopole pairs \cite{Wan2012}. In a follow-up work, the same authors reported the observation of massive monopole gas \cite{Pan2016}. 

\subsection{A new playground for spintronics}

The discussion developed in the previous paragraphs outlines the vast opportunities offered by frustrations. Although intimately related, the fields of frustrated magnetism and spintronics have evolved almost independently of each other, relying on quite different techniques. Whereas the experimental methodology used to probe the ground state and exotic excitations of bulk frustrated magnets are mostly volume sensitive (neutron scattering, Raman scattering, nuclear magnetic resonance etc.), spintronics has taken off with the advent of thin films fabrication and submicron scale lithography, enabling the electrical injection and detection of spin currents. The itinerant nature of the fractional excitations of quantum magnets, be it a Majorana fermion, a spinon or a monopole, makes them particularly appealing for the development of quantum spintronics. These fractionalized excitations could be injected, controlled and detected via local electric probes and could serve as elementary information carrier, like electrons and magnons, but obeying much more exotic laws, as discussed above. In turn, using standard spin-charge conversion techniques (spin pumping and spin Hall effect) could bring new insights on the physics of these exotic excitations.\par

It is now well established that injecting a spin current into a magnetic material induces a spin-transfer torque on the local magnetic moments \cite{Ralph2008,Brataas2012}. This torque can lead to current-driven switching \cite{Slonczewski1996} and, most importantly for us, excite thermal magnons \cite{Berger1996} and even Bose-Einstein condensates \cite{Bender2012}. Reciprocally, magnetic excitations can pump a spin current into an adjacent metallic layer \cite{Brataas2002,Uchida2010,Bauer2012}. The latter effect is particularly interesting as the injected spin current can be analyzed to probe the magnetic phase transition of the insulator \cite{Qiu2016b}. Along the same line of thoughts, one expects that the spin current pumped by the magnetic excitation in the insulator can provide information about the fractional excitations. An example of such an approach has been recently reported by \citet{Hirobe2016}, who measured the one-dimensional spinon spin transport in ${\rm Sr_2CuO_3}$. A few theoretical endeavors have proposed to use spin pumping to probe spinon \cite{Chen2013g,Chatterjee2015} or Majorana fermions \cite{Carvalho2018} for instance. The presence of interfacial transport such as the one observed in $\alpha-{\rm RuCl_3}$ could lead to unique characteristic in the spin current that remain to be determined. We emphasize that the spin current pumped by these excitations can display interesting features in their dc and ac components, but also in their noise \cite{Matsuo2018}. This field of research requires further investigation to establish clear signature of fractional excitations and could open fascinating perspective for experiments.

\section{Perspectives \label{s:persp}}

This review has highlighted several important topological aspects of antiferromagnets: (i) the emergence of unconventional anomalous Hall transport of electrons, magnons and other types of magnetic excitations; (ii) the realization of topological states of matter such as topological insulators and semimetals; (iii) the existence of magnetic entities with non-trivial magnetic texture in real space and their unusual dynamical properties; (iv) the advent of fractional quasiparticles arising from the quantum topological order. These four lively research directions illustrate the potential of antiferromagnets for expanding the realm of topological materials science. Here, we intend to briefly comment on future cross-disciplinary developments. 

Whereas anomalous and spin Hall effects have been intensively studied in all sorts of materials over the past twenty years, most of the theoretical developments achieved considered the conventional form of the (spin) conductivity tensor expressed in Eqs. \eqref{eq:jch}-\eqref{eq:jsh}. The extension of anomalous and spin Hall effect to antiferromagnets, exposed in Section \ref{s:ahe}, has led researchers to realize that the (spin) conductivity tensor is in fact much richer in these materials. In particular, the spin polarization of the "magnetic" spin Hall effect is not necessarily perpendicular to the scattering plane but points in a direction dictated by the magnetic texture in the unit cell. In addition, it was found that anomalous transport is not necessarily "intrinsic", i.e., governed by the Berry curvature of the ground state, but can be in fact "extrinsic" as in the case of the magnetic spin Hall effect. Contemplating the tremendous acceleration of the research in this field, we expect more surprises in the coming years in the ever-expanding family of spin Hall effects. We notice that these discoveries are particularly interesting in the context of applied spintronics as they unlock novel forms of spin currents with spin polarization that can be easily tuned using external knobs by manipulating the magnetic texture of the antiferromagnet. The interplay between these anomalous currents and the magnetic dynamics remains an open question.

The recent realization of antiferromagnetic topological materials (insulators and semimetals) discussed in Section \ref{s:topaf} opens an exciting area of research as they feature topological states that are governed by the magnetic order itself. Therefore, we expect that these states - be it a Dirac cone at the surface of the topological insulator or a Fermi arc at the surface of a Weyl semimetal - can be also manipulated via the magnetic order. For instance, one could think of exciting the magnetic order to dynamically pump spin currents at the surface without altering the topological protection of the states. Reciprocally, injecting charges through the surface states could result in magnetic excitations in the bulk. On the other hand, we acknowledge that the experimental progress in this direction has been limited due to the difficulty to synthetize the proposed candidate materials. 

The same is true for magnonic topological insulators and semimetals. Despite the vast number of model theoretical studies, no experimental realization has been achieved so far, in part due to the absence of realistic predictions. An additional difficulty is to devise a robust procedure to probe the presence of magnonic surface states. Close collaboration between theorists and experimentalists is needed to overcome these obstacles and conventional methods inherited from spintronics, such as spin-charge conversion, need to be revisited and adapted to these new materials in order to properly address the existence of magnonic surface states. 

Theoretical and numerical search of topological solitons and magnetic skyrmions in antiferromagnets have been very active in the past few years, as presented in Section \ref{s:textures}. However, the detection of these particle-like entities in real materials remains difficult because of the absence of a net magnetization. From this standpoint, antiferromagnets that exhibit anomalous Hall effect and magnetooptical Kerr effect could prove a powerful platform for the observation of topological magnetic textures. Another interesting challenge from the materials standpoint is the design of a system (bulk antiferromagnet or heterostructure) that displays the appropriate form and strength of Dzyaloshinskii-Moriya interaction. To do so, first principles calculations are needed to explore materials' combinations and magnetic ground states' competition.

A last research direction that could substantially benefit from progress in spintronics is the field of frustrated magnetism and quantum spin liquids. As exposed in Section \ref{s:topexc}, these complex antiferromagnets host exotic excitations with very unusual characteristics such as fermionic or anyonic statistics and the coexistence of a gauge field. Conventional spintronics methods could be adapted to study the physics of these excitations and assess their ability to transmit and store information. This low-temperature mechanisms could be of interest in the broader context of quantum computing from two viewpoints: first, these excitations can serve for topological quantum computing and identifying efficient ways to control them individually is central to achieve braiding. Second, topological computers working at very low temperature require low temperature conventional circuitry to support and complement the quantum operations. Quantum spin liquids and other quantum magnets could be used as spintronics components working at such low temperatures. This research direction is rather far-fetched but could lead to major interdisciplinary advances in the near future.

Finally, we close this perspective by mentioning an adjacent field of research that was completely overlooked in the present review: superconductivity. As a matter of fact, superconductors such as copper oxides or iron pnictide emerge from deviation from an antiferromagnetic Mott insulating state \cite{Lee2006,Monthoux2007}. Although magnetism usually disappears in the superconducting phase, the coexistence of superconductivity and antiferromagnetism has been observed in some systems such as (Li$_{0.8}$Fe$_{0.2}$)OHFeSe \cite{Lu2015c} and FeSe \cite{Wang2016h} for instance. It is therefore natural to expect that the vast topological physics described in the present review also finds its ways in selected high-T$_c$ superconductors. Thermal Hall effect has progressively become an interesting way to probe the normal phase of high-$T_c$ superconductors and \citet{Grissonnanche2019} revealed that the sign of the thermal Hall effect might be related to the existence of the pseudogap phase. Another promising research direction is the exploitation of antiferromagnets, being topologically trivial or non-trivial, for the realization of topological superconductivity. In a recent proposal, \citet{Lado2018,Lado2020} suggest that a trival antiferromagnetic insulator in proximity with a conventional superconductor with spin-orbit coupling can stabilize topological solitons, whereas \citet{Peng2019} propose to use an antiferromagnetic topological insulator to induce Majorana hinge modes. These examples illustrate how the unique spin transport properties of antiferromagnets can be used to either induce topological superconducting states or serve as a probe to better understand the emergence of strong correlations.


%
%
%
%

\section*{Acknowledgements}
A. M. acknowledges support from the Excellence Initiative of Aix-Marseille Universit\'e - A*Midex, a French "Investissements d'Avenir" program. B. V., F. Z.,  and O. L. were supported by the King Abdullah University of Science and Technology (KAUST). A. A. acknowledges the support provided by the Deanship of Scientific Research at King Fahd University of Petroleum and Minerals (KFUPM) through Project No.  SR191021. A. M. acknowledges fruitful discussions with J. Zelezny.

\bibliography{Biblio_resub}


\end{document}